\documentclass[letterpaper,12pt,titlepage,twoside,openright,final]{report}

\pdfoutput=1

\usepackage{fancyhdr}
\usepackage{array}
\usepackage{subfigure}

\usepackage{amsthm}
\usepackage{amssymb}
\usepackage{amsmath}
\usepackage{amsfonts}
\usepackage{amstext}

\usepackage{enumerate}

\usepackage[sort,numbers]{natbib}

\usepackage{epic}

\usepackage{color}

\usepackage{ifthen}
\newboolean{ElectronicVersion}
\setboolean{ElectronicVersion}{true}

\newboolean{DRAFT}
\setboolean{DRAFT}{false}

\newboolean{MARK}
\setboolean{MARK}{false}

\usepackage{geometry}

\usepackage{nomencl}
\usepackage{makeidx} 
\makeindex

\usepackage[nohints]{minitoc}
\dominitoc

\usepackage{pxfonts}
\usepackage{eulervm}

\newcommand{\tinyspace}{\mspace{1mu}}

\newcommand*{\bra}[1]{\langle #1|}
\newcommand*{\ket}[1]{|#1\rangle}
\newcommand{\braket}[2]{\langle #1|#2\rangle}

\DeclareRobustCommand\openone{\leavevmode\hbox{\upshape\small1\normalsize\kern-.33em1}}%
\newcommand{\tensor}{\otimes}
\newcommand{\tprod}{\otimes}
\newcommand{\smallnorm}[1]{\bigl\lVert\tinyspace#1\tinyspace\bigr\rVert}
\newcommand{\norm}[1]{\left\lVert\tinyspace#1\tinyspace\right\rVert}
\newcommand{\tnorm}[1]{\norm{#1}_{\mathrm{tr}}}
\newcommand{\dnorm}[1]{\norm{#1}_{\diamond}}
\newcommand{\threenorm}[1]{\left\lvert\left\lvert\left\lvert\tinyspace
  #1\tinyspace\right\rvert\right\rvert\right\rvert}
\newcommand{\smthreenorm}[1]{\bigl\lvert\bigl\lvert\bigl\lvert\tinyspace #1\tinyspace\bigr\rvert\bigr\rvert\bigr\rvert}
\newcommand{\abs}[1]{\left\lvert\tinyspace #1 \tinyspace\right\rvert}

\newcommand{\dm}[1]{\dim \mathcal{#1}}
\newcommand{\mc}[1]{\mathcal{#1}}
\newcommand{\linear}[1]{\mathbf{L}(\mathcal{#1})}
\newcommand{\density}[1]{\mathbf{D}(\mathcal{#1})}
\newcommand{\unitary}[1]{\mathbf{U}(\mathcal{#1})}
\newcommand{\unitarynomc}[1]{\mathbf{U}(#1)}

\newcommand{\transform}[1]{\mathbf{T}(\mathcal{#1})}
\newcommand{\id}{\openone}
\newcommand{\identity}[1]{\id_{\mathcal{#1}}}
\newcommand{\nidentity}[1]{\tilde{\id}_{\mathcal{#1}}}
\newcommand{\tidentity}[1]{I_{\mathcal{#1}}}

\newcommand{\rank}{\operatorname{rank}}
\newcommand{\tr}{\operatorname{tr}}
\newcommand{\F}{\operatorname{F}}
\newcommand{\opv}{\nu}

\newcommand{\ptr}[1]{\tr_\mathcal{#1}}
\newcommand{\ceil}[1]{\left\lceil #1 \right\rceil}
\newcommand{\floor}[1]{\left\lfloor #1 \right\rfloor}

\newcommand{\prob}[1]{\textup{\textsc{#1}}}
\newcommand{\class}[1]{\textup{\textbf{#1}}}

\DeclareMathOperator{\lspan}{span}

\newtheorem{theorem}{Theorem}[chapter]
\newtheorem{lemma}[theorem]{Lemma}
\newtheorem{proposition}[theorem]{Proposition}
\newtheorem{corollary}[theorem]{Corollary}
\theoremstyle{definition}
\newtheorem{defn}[theorem]{Definition}
\newtheorem{proto}[theorem]{Protocol}
\newtheorem{problem}[theorem]{Problem}

\ifthenelse{\boolean{DRAFT}}
{ \usepackage[pagewise,mathlines,switch]{lineno}
 \newcommand*\patchAmsMathEnvironmentForLineno[1]{%
   \expandafter\let\csname old#1\expandafter\endcsname\csname
   #1\endcsname
   \expandafter\let\csname oldend#1\expandafter\endcsname\csname
   end#1\endcsname
   \renewenvironment{#1}%
     {\linenomath\csname old#1\endcsname}%
     {\csname oldend#1\endcsname\endlinenomath}}% 
   \newcommand*\patchBothAmsMathEnvironmentsForLineno[1]{%
     \patchAmsMathEnvironmentForLineno{#1}%
     \patchAmsMathEnvironmentForLineno{#1*}}%
   \AtBeginDocument{%
     \patchBothAmsMathEnvironmentsForLineno{equation}%
     \patchBothAmsMathEnvironmentsForLineno{align}%
     \patchBothAmsMathEnvironmentsForLineno{flalign}%
     \patchBothAmsMathEnvironmentsForLineno{alignat}%
     \patchBothAmsMathEnvironmentsForLineno{gather}%
     \patchBothAmsMathEnvironmentsForLineno{multline}%
   }
   \linenumbers}
{}

\ifthenelse{\boolean{ElectronicVersion}}
{\geometry{letterpaper, margin=1in, inner=1.125in, outer = 1.125in}}
{\geometry{letterpaper, margin=1in,inner=1.25in, outer = 1in}}

\let\origdoublepage\cleardoublepage
\newcommand{\clearemptydoublepage}{%
  \clearpage{\pagestyle{empty}\origdoublepage}}
\let\cleardoublepage\clearemptydoublepage

\ifthenelse{\boolean{DRAFT}} {
\ifthenelse{\boolean{MARK}}
  {\pagestyle{fancyplain}%
   \lhead{}%
   \rhead{\scriptsize{\color{red}{\underline{\textsc{DRAFT: \today}}}}}}%
{}
}{}

\RequirePackage{ifthen} 
\renewcommand{\nomgroup}[1]{% 
  \ifthenelse{\equal{#1}{G}}{\item[\textbf{Quantum Circuits}]}{% 
  \ifthenelse{\equal{#1}{F}}{\item[\textbf{Operators}]}{% 
  \ifthenelse{\equal{#1}{B}}{\item[\textbf{Classes of Operators}]}{% 
  \ifthenelse{\equal{#1}{P}}{\item[\textbf{Computational Problems}]}{% 
  \ifthenelse{\equal{#1}{C}}{\item[\textbf{Complexity Classes}]}{}}}}}}

\begin{document}

%=============================================================================%

% T I T L E   P A G E
% -------------------
% The title page is counted as page `i' but we need to suppress the
% page number.  We also don't want any headers or footers.
\pagestyle{empty}
\pagenumbering{roman}

% The contents of the title page are specified in the ``titlepage''
% environment.
\begin{titlepage}
        \ifthenelse{\boolean{MARK}}{\thispagestyle{fancyplain}}{}

        \begin{center}
        \vspace*{1.0cm}

        \Huge

        {\bf Computational Distinguishability of Quantum Channels}

        \vspace*{1.0cm}

        \normalsize
        by \\

        \vspace*{1.0cm}

        \Large
        William Rosgen \\

        \vspace*{3.0cm}

        \normalsize
        A thesis \\
        presented to the University of Waterloo \\ 
        in fulfillment of the \\
        thesis requirement for the degree of \\
        Doctor of Philosophy \\
        in \\
        Computer Science \\

        \vspace*{2.0cm}

        Waterloo, Ontario, Canada, 2009 \\

        \vspace*{1.0cm}

        \copyright\ William Rosgen 2009
        \end{center}
\end{titlepage}

% The rest of the front pages should contain no headers and be
% numbered using roman numerals starting with `ii'
\pagestyle{plain}
\setcounter{page}{2}

% D E C L A R A T I O N   P A G E
% -------------------------------
\ifthenelse{\boolean{DRAFT}}{}{
\ifthenelse{\boolean{ElectronicVersion}}{
  % The following is the sample Delaration Page as provided by the GSO
  % December 13th, 2006.  It is designed for an electronic thesis.
  \noindent
  I hereby declare that I am the sole author of this thesis.  This is
  a true copy of the thesis, including any required final revisions,
  as accepted by my examiners.

  \bigskip
  
  \noindent
  I understand that my thesis may be made electronically available to
  the public.}{
 % The following text was what was required back when the GSO acceped
 % printed versions and you may want to continue to use it for your
 % printed version.
 \noindent
 I hereby declare that I am the sole author of this thesis.

 \smallskip

 \noindent
 I authorize the University of Waterloo to lend this thesis to other
 institutions or individuals for the purpose of scholarly research.

 \bigskip

 \noindent
 I further authorize the University of Waterloo to reproduce this
 thesis by photocopying or by other means, in total or in part,
 at the request of other institutions or individuals for the purpose
 of scholarly research.
}
\newpage
}

% A B S T R A C T
% ---------------

\begin{center}\textbf{Abstract}\end{center}

The computational problem of distinguishing two quantum channels is
central to quantum computing.  It is a generalization of the
well-known satisfiability problem from classical to quantum
computation.  This problem is shown to be surprisingly hard: it is
complete for the class \class{QIP} of problems that have quantum
interactive proof systems, which implies that it is hard for the class
\class{PSPACE} of problems solvable by a classical computation in
polynomial space.

Several restrictions of distinguishability are also shown to be hard.
It is no easier when restricted to quantum
computations of logarithmic depth, to mixed-unitary channels, to
degradable channels, or to antidegradable channels.  These hardness
results are demonstrated by finding reductions between these classes
of quantum channels.  These techniques have applications outside the
distinguishability problem, as the construction for mixed-unitary
channels is used to prove that the additivity problem for the
classical capacity of quantum channels can be equivalently restricted
to the mixed unitary channels.
\cleardoublepage

% A C K N O W L E D G E M E N T S
% -------------------------------

\begin{center}\textbf{Acknowledgements}\end{center}

I would like to thank my supervisor John Watrous for years of
guidance, support, and insight.  Without his help this would not have
been possible.  I would also like to thank the rest of my committee,
Richard Cleve, Stephen Fenner, Achim Kempf, and Ben Reichardt, for
providing helpful comments on an earlier draft of this thesis.  I
would also like to thank Lana for putting up with me during the
writing of this thesis and supporting me throughout the process.

\cleardoublepage

% % D E D I C A T I O N
% % -------------------

% \begin{center}\textbf{Dedication}\end{center}

% This is dedicated to the one I love.
% \newpage

% T A B L E   O F   C O N T E N T S
% ---------------------------------
\setcounter{tocdepth}{1}
\tableofcontents
\cleardoublepage

% % L I S T   O F   T A B L E S
% % ---------------------------
% \listoftables
% \addcontentsline{toc}{chapter}{List of Tables}
% \mtcaddchapter
% \newpage

% L I S T   O F   F I G U R E S
% -----------------------------
\ifthenelse{\boolean{ElectronicVersion}}{\phantomsection}{}
\listoffigures
\addcontentsline{toc}{chapter}{List of Figures}
\mtcaddchapter
\cleardoublepage

% L I S T   O F   S Y M B O L S
% -----------------------------
\ifthenelse{\boolean{ElectronicVersion}}{\phantomsection}{}
\renewcommand{\nomname}{List of Symbols}
\addcontentsline{toc}{chapter}{List of Symbols}
\mtcaddchapter
\printnomenclature[2in]
\cleardoublepage

% Change page numbering back to Arabic numerals
\pagenumbering{arabic}

%%% Local Variables: 
%%% mode: latex
%%% TeX-master: "thesis"
%%% End: 

\chapter{Introduction}\label{chap-intro}

Distinguishing two quantum channels is one of the most important tasks in
quantum information.  This is the problem of determining if there is
an input state on which the two channels to produce output states
that are distinguishable.  When this is phrased as a computational
problem it is complete for the complexity class
\class{QIP} of problems that have quantum interactive proof systems.
This problem seems to be computationally much more difficult than
other variants of the problem, such as distinguishing classical
circuits or distinguishing unitary quantum circuits.

In light of this hardness, it is natural to consider restricted
versions of the problem.  Many of these special cases are also hard:
reductions can be found to some of the more interesting 
classes of quantum channels.  These results suggest that this problem
is not likely to be tractable even on many of the restricted channels
that can be realized by experiment.  This is, however, not a surprise:
distinguishing two channels is a restricted version of quantum process
tomography, which is computationally intractable for large systems.

These reductions provide simulations of general quantum channels by
channels in restricted classes.  While these simulations do not
accurately model all aspects of the original channel, the constructed
simulations do share many properties with the original channel.  Many of
these results can be applied outside the narrow focus of
distinguishing quantum channels: it is hoped that these techniques
will prove useful for a number of problems in quantum information theory.

\minitoc

\section{Overview}\label{scn-intro-overview}

This thesis studies the computational problem of distinguishing
quantum channels.  This problem asks, given
two implementations of quantum channels, is there an input on which
the implemented channels behave distinctly?  One of the main results
of the thesis is that this problem is in general extremely difficult:
it is complete for the complexity class \class{PSPACE} of problems
that can be solved with a polynomially bounded amount of memory.
Since this problem is intractable in general it remains to understand those classes of
channels for which the problem has an efficient solution and those classes
on which it remains hard.  This problem is becoming more significant for
quantum computing: as larger practical systems are being studied it
becomes more difficult to verify that the implemented transformation
is close to an ideal transformation.

One of the other problems considered in the thesis is the question of the
additivity of the Holevo capacity for quantum channels.  If this
quantity were additive, it would significantly simplify the tasks of
encoding and decoding for the transmission of classical information
through a quantum channel.  Specifically, this question asks
whether two uses of a channel can send more than twice
the classical information that can be sent with only one use of the
channel.  That this might be possible is because entanglement 
may be present between the two inputs to the channel.  
This question stood open for several years until it was recently
shown that there exist channels for which the Holevo capacity is
super-additive~\cite{Hastings09}.  The classes of channels that
are known to have the additivity property are generally quite
restricted.  It is important to better understand which classes of
channels are additive as the use of quantum channels for sending
classical information is an important application of quantum
information.

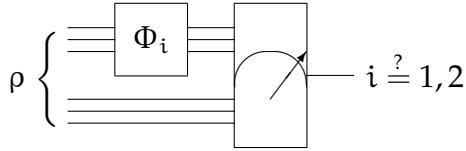
\begin{figure}
  \begin{center}
    \setlength{\unitlength}{3947sp}%
\begingroup\makeatletter\ifx\SetFigFont\undefined%
\gdef\SetFigFont#1#2#3#4#5{%
  \reset@font\fontsize{#1}{#2pt}%
  \fontfamily{#3}\fontseries{#4}\fontshape{#5}%
  \selectfont}%
\fi\endgroup%
\begin{picture}(2280,924)(2911,-448)
\put(5176,-61){\makebox(0,0)[lb]{\smash{{\SetFigFont{12}{14.4}{\rmdefault}{\mddefault}{\updefault}{\color[rgb]{0,0,0}$i \stackrel{?}{=} 1,2$}%
}}}}
\thinlines
{\color[rgb]{0,0,0}\put(4051,314){\line( 1, 0){300}}
}%
{\color[rgb]{0,0,0}\put(4351,239){\line(-1, 0){300}}
}%
{\color[rgb]{0,0,0}\put(4051,164){\line( 1, 0){300}}
}%
{\color[rgb]{0,0,0}\put(3301,-286){\line( 1, 0){1050}}
}%
{\color[rgb]{0,0,0}\put(3601, 14){\framebox(450,450){$\Phi_i$}}
}%
{\color[rgb]{0,0,0}\put(3301,-211){\line( 1, 0){1050}}
}%
{\color[rgb]{0,0,0}\put(3301,-136){\line( 1, 0){1050}}
}%
{\color[rgb]{0,0,0}\put(4351,-436){\framebox(450,900){}}
}%
{\color[rgb]{0,0,0}\put(4576,-136){\vector( 3, 4){225}}
}%
{\color[rgb]{0,0,0}\put(4801, 14){\line( 1, 0){300}}
}%
{\color[rgb]{0,0,0}\put(3301,314){\line( 1, 0){300}}
}%
{\color[rgb]{0,0,0}\put(3301,239){\line( 1, 0){300}}
}%
{\color[rgb]{0,0,0}\put(3301,164){\line( 1, 0){300}}
}%
\put(2926,-61){\makebox(0,0)[lb]{\smash{{\SetFigFont{12}{14.4}{\rmdefault}{\mddefault}{\updefault}{\color[rgb]{0,0,0}$\rho \; \Bigg\{$}%
}}}}
{\color[rgb]{0,0,0}\put(4576,-61){\oval(450,450)[tr]}
\put(4576,-61){\oval(450,450)[tl]}
}%
\end{picture}%
  \end{center}
  \caption[Optimal strategy for channel distinguishability]{The
    optimal strategy for determining which channel $\Phi_1, \Phi_2$ is the
    unknown channel $\Phi_i$.  The strategy is to prepare some state
    $\rho$ on which the two channels output maximally distinguishable
    states, send this state through the channel $\Phi_i$, and then make an
    optimal measurement of the result.}\label{intro-fig-dist}
\end{figure}

The problem of distinguishing channels can be equivalently rephrased
as: given a single use of an unknown channel that is one of two known
channels $\Phi_1$ and $\Phi_2$, what is the probability that the optimal
strategy can detect which of the two channels it is?  In general the
best strategy in this case is to prepare an input state $\rho$ on which the
output states of $\Phi_1$ and $\Phi_2$ are maximally distinguishable, send
$\rho$ through the unknown channel, and then attempt to solve the
distinguishability problem for the output states of the channels.  In
general this strategy requires the preparation of a state on some
larger system, only part of which is sent through the unknown
channel.  This strategy is illustrated in
Figure~\ref{intro-fig-dist}.
One of the main results of the thesis is that this problem, properly
formalized, is complete for the complexity class \class{QIP} of
problems that have quantum interactive proof systems.  This is a
surprising result: the same problem restricted to deterministic
classical circuits is a restatement of the canonical
\class{NP}-complete problem satisfiability.  These complexity classes
are discussed in more detail in Chapter~\ref{chap-complexity}.  For the
reader unfamiliar with computational complexity theory, it is
important only to know that the class \class{QIP} contains the class
\class{NP} and it is thought that \class{QIP} is much larger than
\class{NP}.

This hardness result is found in
Chapter~\ref{chap-distinguishability}, where it is shown using a Karp
reduction from the problem of determining if two quantum channels can
be made to output states that are close together, where a Karp
reduction is simply an efficient procedure that transforms instances
of one problem into equivalent instances of another problem.  A
problem that is the target of a Karp reduction is thus shown to be at
least as hard as the starting problem.

The Close Images problem is
easily derived from the definition of the class~\class{QIP}, which
implies that it is also~\class{QIP}-complete.  This
derivation can be found in Chapter~\ref{chap-close-images}, and is
originally due to Kitaev and Watrous~\cite{KitaevW00}.

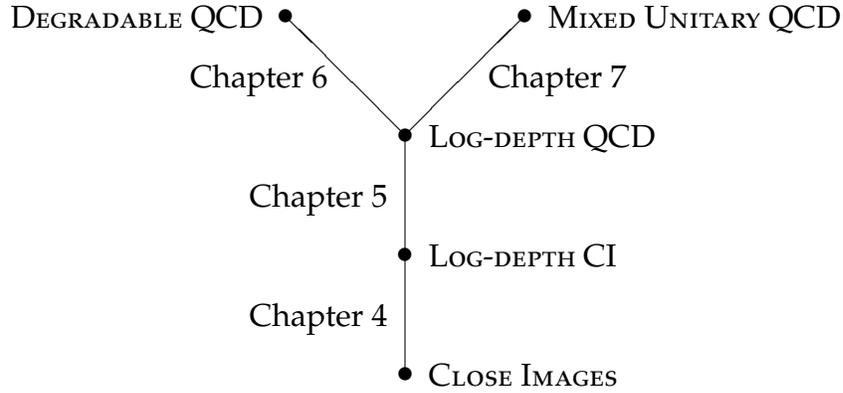
\begin{figure}
  \begin{center}
    \setlength{\unitlength}{3947sp}%
\begingroup\makeatletter\ifx\SetFigFont\undefined%
\gdef\SetFigFont#1#2#3#4#5{%
  \reset@font\fontsize{#1}{#2pt}%
  \fontfamily{#3}\fontseries{#4}\fontshape{#5}%
  \selectfont}%
\fi\endgroup%
\begin{picture}(6174,2574)(2014,-5473)
\put(2626,-3136){\makebox(0,0)[lb]{\smash{{\SetFigFont{12}{14.4}{\rmdefault}{\mddefault}{\updefault}{\color[rgb]{0,0,0}\prob{Degradable QCD}}%
}}}}
{\color[rgb]{0,0,0}\thinlines
\put(5101,-3811){\circle*{76}}
}%
{\color[rgb]{0,0,0}\put(5101,-5311){\circle*{76}}
}%
{\color[rgb]{0,0,0}\put(4351,-3061){\circle*{76}}
}%
{\color[rgb]{0,0,0}\put(5851,-3061){\circle*{76}}
}%
{\color[rgb]{0,0,0}\put(5101,-4561){\line( 0, 1){750}}
}%
{\color[rgb]{0,0,0}\put(5101,-3811){\line( 1, 1){750}}
}%
{\color[rgb]{0,0,0}\put(5101,-5311){\line( 0, 1){750}}
}%
{\color[rgb]{0,0,0}\put(5101,-3811){\line(-1, 1){750}}
}%
\put(5251,-4636){\makebox(0,0)[lb]{\smash{{\SetFigFont{12}{14.4}{\rmdefault}{\mddefault}{\updefault}{\color[rgb]{0,0,0}\prob{Log-depth CI}}%
}}}}
\put(5251,-5386){\makebox(0,0)[lb]{\smash{{\SetFigFont{12}{14.4}{\rmdefault}{\mddefault}{\updefault}{\color[rgb]{0,0,0}\prob{Close Images}}%
}}}}
\put(5251,-3886){\makebox(0,0)[lb]{\smash{{\SetFigFont{12}{14.4}{\rmdefault}{\mddefault}{\updefault}{\color[rgb]{0,0,0}\prob{Log-depth QCD}}%
}}}}
\put(4126,-4261){\makebox(0,0)[lb]{\smash{{\SetFigFont{12}{14.4}{\rmdefault}{\mddefault}{\updefault}{\color[rgb]{0,0,0}Chapter \ref{chap-distinguishability}}%
}}}}
\put(4126,-5011){\makebox(0,0)[lb]{\smash{{\SetFigFont{12}{14.4}{\rmdefault}{\mddefault}{\updefault}{\color[rgb]{0,0,0}Chapter \ref{chap-close-images}}%
}}}}
\put(6001,-3136){\makebox(0,0)[lb]{\smash{{\SetFigFont{12}{14.4}{\rmdefault}{\mddefault}{\updefault}{\color[rgb]{0,0,0}\prob{Mixed Unitary QCD}}%
}}}}
\put(5626,-3511){\makebox(0,0)[lb]{\smash{{\SetFigFont{12}{14.4}{\rmdefault}{\mddefault}{\updefault}{\color[rgb]{0,0,0}Chapter \ref{chap-mixed-unitary}}%
}}}}
\put(3751,-3511){\makebox(0,0)[lb]{\smash{{\SetFigFont{12}{14.4}{\rmdefault}{\mddefault}{\updefault}{\color[rgb]{0,0,0}Chapter \ref{chap-degr}}%
}}}}
{\color[rgb]{0,0,0}\put(5101,-4561){\circle*{76}}
}%
\end{picture}%
  \end{center}
  \caption[Reductions presented in the thesis]{Reductions presented in
    the thesis. Problems are reduced to those problems above them.
    \prob{CI} and \prob{QCD} are shorthand for \prob{Close Images} and
    \prob{Quantum Circuit Distinguishability}.  Edges are marked with
    the chapter the reduction appears in.}\label{intro-fig-reductions}
\end{figure}

Given that the distinguishability problem is intractable, much of the remainder of the
thesis is a study of several restricted classes of quantum channels,
with a focus on the hardness of this distinguishability problem on
them.  For many of these classes Karp reductions are found from the
general problem to the problem on the restricted class.  These
reductions prove that these restricted versions of the
distinguishability problem are also \class{QIP}-complete.  
The reductions found in the thesis are
illustrated in Figure~\ref{intro-fig-reductions}.

The problems shown to be \class{QIP}-complete using these reductions
are also hard for the more familiar class \class{PSPACE}.  In fact, it
has recently been shown that $\class{QIP} = \class{PSPACE}$, which
implies that these two classes are the same~\cite{JainJ+09}.  These
problems on quantum channels then provide an interesting
characterization of a fundamental classical complexity class.  Despite
this equivalence, the class is referred to as \class{QIP} throughout
the thesis, as the hardness results presented here all follow from the
definition of the class in terms of quantum interactive proof systems.

The first of these reductions, in Chapter~\ref{chap-close-images}
concerns not the distinguishability problem, but the close images
problem.  It is shown that this problem can be equivalently restricted
to the channels implemented by circuits of logarithmic depth.  These
channels are an important class: they can be implemented in parallel
in a logarithmic amount of time.  This makes these channels interesting for a
practical perspective, as a quantum system implementing one of these
channels needs to be protected from decoherence for only a short time.

The second reduction presented in the thesis is the focus of
Chapter~\ref{chap-distinguishability}.  This is the reduction from the
close images problem to the problem of distinguishing two channels,
where the channels are given as input to the problem in the form of
circuits.  This reduction proves the hardness of the
distinguishability problem for general quantum channels.  One other
important property of this reduction is that it adds only logarithmic
overhead to the depth of the circuits.  This implies that even the
log-depth circuit distinguishability problem is \class{QIP}-complete,
which provides powerful evidence that this problem, a restricted case
of quantum process tomography, is likely to be difficult in practice even for
computations that can be performed in a very limited amount of time.

Chapter~\ref{chap-degr} extends the hardness of the distinguishability
problem in a different direction: to the channels known as the
degradable channels.  These channels can be thought of as
the channels that preserve most of the information about the input,
since there exists a second channel that maps the output of a
degradable channel to the state of the environment.
These channels are described in more detail
Section~\ref{intro-scn-classes}.  This result implies that
distinguishing these channels that do not lose very much information t
the environment remains hard.

In the other direction, Chapter~\ref{chap-degr} also contains a
reduction of the distinguishability problem to the antidegradable
channels.  These are the channels for which there exists a
second map that takes the environment state to the output state.  In
particular, this means that an eavesdropper with sufficient resources
can reconstruct the output of the channel.  At an intuitive level,
this implies that an antidegradable channel loses more information to
the environment than it preserves in the output state.  The fact that
distinguishing these channels is \class{QIP}-hard is evidence that
even channels that do not preserve very much information are hard to distinguish.

The final reduction presented in the thesis, in
Chapter~\ref{chap-mixed-unitary}, is a transformation that
approximates a general channel by one that is mixed-unitary.  The
mixed-unitary channels are those channels that can be expressed as the
convex mixture of unitary (i.e.\ noise-free and reversible) channels.
The mixed unitary channels have several nice properties that make them
interesting in quantum information theory.  The approximate
simulation of a channel by a mixed-unitary channel 
performs well only for measures of quality based on the
maximized purity of the output of the channel.  This, however, suffices to
reduce the distinguishability problem to the mixed-unitary channels.
This technique is also used to show that the additivity of the Holevo
capacity of a general channel can be approximately restated in terms of a
mixed-unitary channel.  
The Holevo capacity, which can be used to measure the amount of classical
information that can be sent through a quantum channel, 
is introduced in detail in Chapter~\ref{chap-meas}.
This technique allows, for instance, the observation
that this quantity is additive for all channels if and only if it is
additive for mixed-unitary channels (which has recently been shown by
Hastings~\cite{Hastings09} through the construction of a mixed-unitary
channel that is not additive).

Taken together, these reductions demonstrate the hardness of the
distinguishability problem on several distinct classes of channels.
As this problem is one of the most interesting problems in quantum
computation, these hardness results point to cases of the problem that
are not able to be efficiently solved, under the usual complexity
theoretic assumptions.  It is also hoped that
these complete problems for \class{QIP} will provide a way to further
understand this class.  The problems
shown in Figure~\ref{intro-fig-reductions} are among only a few
problems in quantum information that are known to be complete for \class{QIP}.

The technique of reducing a problem to a restricted class by
simulating general channels by those of a restricted
class can also have applications outside of quantum computational
complexity.  For instance, the reduction to mixed-unitary channels in
Chapter~\ref{chap-mixed-unitary} was
initially constructed for the distinguishability problem, but the same
construction has implications for the additivity of certain
capacities.  These techniques are powerful and general: any problem
defined on quantum channels is a candidate for reduction to these
restricted classes.  This does not work in general, as these
reductions produce channels that do not simulate the general channel
in every sense, but for any problem defined using similar notions of
distance on quantum channels, these reductions apply.  These
techniques provide not only a method for the study of the
distinguishability problem, the primary application studied
in this thesis, but a tool for the more general study of quantum
channels and their properties.

\section{Quantum information}\label{intro-scn-qinfo}

In this section the necessary mathematical framework for the problems
outlined in the previous section is introduced.  The concepts and
notation used here are relatively standard.  This is not a complete
introduction to quantum information.

More background on quantum information can be found in the
books~\cite{BengtssonZ06, NielsenC00}.  Background on much of the
linear algebra introduced here, including a thorough discussion of the tensor
product, can be found in~\cite{Roman05}.   A good general reference for
results from functional analysis that are occasionally useful in quantum
information is~\cite{Conway90}, while the books~\cite{Bhatia97,
  HornJ85, HornJ91} provide a more focused treatment of the types of
operators often found in quantum information

\nomenclature[adeltaij]{$\delta_{ij}$}{Kronecker delta function: $\delta_{ij} = 1$ if $i=j$ and $0$ otherwise}%

\subsection{Hilbert spaces}

\index{Hilbert space}%
\nomenclature[aH,K]{$\mathcal{H,K,\ldots}$}{finite dimensional Hilbert spaces}%

The fundamental backdrop for quantum information is the complex
Hilbert space.  These spaces are the complete vector spaces over
$\mathbb{C}$ with inner product, where the completeness of the
space is with respect to the topology induced by the inner product.
All such spaces considered in this thesis are finite dimensional and denoted
by calligraphic letters $\mathcal{H, K, \ldots}$.
Elements of a Hilbert space $\mathcal{H}$ of dimension $d$ space can
be represented as vectors
in $\mathbb{C}^d$.  These vectors are denoted $\ket \phi$.  Elements
of the dual space $\mathcal{H}^*$, which are (complex) linear
functionals on the space $\mathcal{H}$, are denoted $\bra{\phi} =
(\ket{\phi})^*$.  The inner product on these Hilbert spaces is defined,
for two vectors $\ket\phi$ with elements $u_i$ and $\ket\psi$ with
elements $v_i$,  by
\index{inner product}%
\begin{equation*}
  \langle \ket{\phi}, \ket{\psi} \rangle
  = \braket{\phi}{\psi}
  = \sum_i \bar{u}_i v_i,
\end{equation*}
where $\bar{u}$ denotes the complex conjugate of $u$.  This inner
product is linear in the second argument and conjugate linear in the
first: this is the usual convention in physics, but it is opposite to
the way things are typically defined in mathematics.  The
\emph{dimension} of a space has been mentioned several times: this is
simply the maximum number of elements in a pairwise orthogonal set.  When
the elements of a $d$-dimensional space are viewed as vectors with
complex entries, the dimension of the space coincides with the length
of these vectors.

Norms are a fundamental tool in quantum information.  They provide a
means to define a notion of size on quantum states and quantum
channels.  Throughout the thesis, we will be most interested in using
norms to bound the distance between two objects.
\index{norm}%
\begin{defn}\label{intro-defn-norm}
  A norm $\threenorm{\cdot}$ on some linear space
  $V$ (over the field $\mathbb{C}$) is a function from $V$ to
  $\mathbb{R}$ satisfying three basic properties, for all $x,y \in V$:
  \begin{align}
    \text{Nonnegativity: } 
    & \threenorm{x} \geq 0 \text{ with equality if and only if } x = 0 
    \label{intro-eqn-norm-nonnegativity} \\
    \text{Homogeneity: }
    & \threenorm{ c x } = \abs{c} \threenorm{x} \text{ for all } c \in
    \mathbb{C} 
    \label{intro-eqn-norm-homogeneity}\\
    \text{Triangle Inequality: }
    & \threenorm{x + y} \leq \threenorm{x} + \threenorm{y}
    \label{intro-eqn-norm-triangle-inequality}
  \end{align}
\end{defn}

The standard Euclidean norm on a Hilbert space can be defined in terms
of the inner product given above.
This is the norm of a vector $\ket\phi$ with elements $v_i$ given by
\begin{equation*}
  \norm{ \ket\phi } = \sqrt{ \braket{\phi}{\phi} } = \sqrt{ \sum_i \abs{v_i}^2 }.
\end{equation*}
A vector $\ket\phi$ is called normalized if $\norm{\ket\phi} = 1$.

The standard basis of the Hilbert space $\mathcal{H}$ of dimension $d$
is given by the set of orthonormal (i.e.\ normalized and pairwise
orthogonal) vectors $\{ \ket 0, \ket 1, \ldots, \ket{d-1} \}$.  The
vector $\ket i$ viewed as a vector in $\mathbb{C}^d$ is simply the
vector with a one in position $i+1$ and zeroes in all other
positions.  This basis is also known as the \emph{computational
  basis}.
\index{computational basis}%
When no confusion will arise, this basis will also be labelled $\{\ket
1, \ket 2, \ldots \ket d\}$.

Two finite dimensional Hilbert spaces  $\mathcal{H,K}$ are isomorphic if they are both
of the same dimension.  This is written $\mathcal{H} \cong
\mathcal{K}$.  In such a case, the canonical isomorphism between the two spaces
simply maps the computational basis of $\mathcal{H}$ to the
computational basis of $\mathcal{K}$.  When two spaces are isomorphic,
the isomorphism between them will often be used implicitly to consider
vectors in one Hilbert space as being vectors in the other space.

Quantum systems of large dimension are often built up of many smaller
dimensional system.  If $\mathcal{H}$ and $\mathcal{K}$ are Hilbert
spaces of dimension $\dm{H} = d_{\mathcal{H}}$ and $\dm{K} =
d_{\mathcal{K}}$, then the Hilbert space of dimension $d_{\mathcal{H}}
d_{\mathcal{K}}$ formed by combining them is denoted $\mathcal{H
  \tprod K}$.  Similarly, the element $\ket \phi \tprod \ket \psi \in
\mathcal{H \tprod K}$ is formed by combining the two elements
$\ket\phi \in \mathcal{H}$ and $\ket \psi \in \mathcal{K}$.  When viewed as
complex vectors, these elements are given by the Kronecker product
\begin{align*}
  \ket\phi &= 
  \begin{pmatrix}
    u_1 \\ u_2 \\ \vdots \\ u_{d_{\mathcal{H}}}
  \end{pmatrix},&
  \ket\psi &= 
  \begin{pmatrix}
    v_1 \\ v_2 \\ \vdots \\ v_{d_{\mathcal{K}}}
  \end{pmatrix},&
  \ket\phi \tprod \ket\psi &= 
  \begin{pmatrix}
    u_1 \ket\psi \\ u_2 \ket\psi \\ \vdots \\ u_{d_{\mathcal{H}}} \ket\psi
  \end{pmatrix}.
\end{align*}
The notation $\ket\phi \tprod \ket\psi$ will often be abbreviated
$\ket\phi \ket\psi$ or even $\ket{\phi \psi}$ where no confusion is
likely to arise.

The space $\mathcal{H \tprod K}$ does not consist solely of elements
of the form $\ket\phi \ket\psi$: it also contains linear combinations
of these elements.  As an example, an element of the
tensor product of two systems of dimension two is
$\ket{00} + \ket{11}$.  This element cannot
be written in tensor product form.
A basis for $\mathcal{H \tprod K}$ can be formed by taking the
pairwise tensor product of basis elements for the two subsystems,
i.e.\ the set $\{ \ket{i} \ket{j} : 0 \leq i < d_{\mathcal{H}},
0 \leq j < d_{\mathcal{K}} \}$ is a basis for $\mathcal{H \tprod K}$.
When convenient we will also use the standard basis $\{ \ket{i} : 0 \leq i <
d_{\mathcal{H}} d_{\mathcal{K}} \}$ for this space.

\subsection{Pure states}

The state of a quantum system is described, up to a phase $e^{i \theta}$, by a normalized vector
$\ket\phi \in \mathcal{H}$, known as a \emph{pure state}.  
\index{pure state}%
Provided that there is no uncertainty about the system, these pure states
suffice to completely describe the state of a quantum system.  For
this reason they are of fundamental importance in quantum
information.  Any such state on a $d$-dimensional Hilbert space can be expressed as
\begin{equation*}
  \ket\phi = \sum_{i=0}^{d-1} a_i \ket{i},
\end{equation*}
where the \emph{amplitudes} $a_i$ are complex numbers satisfying
$\sum_i \abs{a_i}^2 = 1$, which is simply a restatement of the
normalization requirement.

\index{qubit}%
The smallest system of interest in quantum computation is the two
dimensional Hilbert space.  Such a system is often referred to as a
\emph{qubit}.  On such a system, the two standard basis states are
$\ket 0$ and $\ket 1$.  A second basis that is often extremely useful
is given by the two orthogonal states
\begin{align*}
  \ket{+} &= \frac{1}{\sqrt{2}} (\ket 0 + \ket 1), &
  \ket{-} &= \frac{1}{\sqrt{2}} (\ket 0 - \ket 1).
\end{align*}
 
\index{entanglement}%
As was previously mentioned, there are elements in a composite Hilbert
space $\mathcal{H \tprod K}$ that cannot be decomposed into a tensor
product of an element of $\mathcal{H}$ and an element of
$\mathcal{K}$.  When quantum states have this property they are called
\emph{entangled}.  Up to normalization, we have already met the
maximally entangled state.  This is the state $\ket{\phi_+} \in \mathcal{H} \tprod
\mathcal{H}$, where $d = \dm{H}$, given by
\begin{equation*}
  \ket{\phi_+} = \frac{1}{\sqrt{d}} \sum_{i=0}^{d-1} \ket i \ket i.
\end{equation*}
\index{separable state}%
Any state that is not entangled is called \emph{separable}.

\index{Schmidt decomposition}%
An important representation of pure states of a composite system
$\mathcal{H \tprod K}$ is the \emph{Schmidt decomposition}.  Any
state $\ket \phi \in \mathcal{H \tprod K}$ may be expressed as
\begin{equation}\label{intro-eqn-schmidt}
  \ket \phi = \sum_{i=1}^{r} \lambda_i \ket{a_i} \ket{b_i}.
\end{equation}
In this decomposition the sets $\{\ket{a_i}\}$ and $\{\ket{b_i}\}$
form orthonormal sets in $\mathcal{H}$ and $\mathcal{K}$,
respectively, and the coefficients $\lambda_i$ are all positive and real.
The number $r$ in Equation~\eqref{intro-eqn-schmidt} satisfies $r \leq
\min\{\dm{H}, \dm{K}\}$ and is known as the \emph{Schmidt rank} of
$\ket \phi$.  The numbers $\lambda_i$ are 
known as the \emph{Schmidt coefficients}.  They satisfy $\sum_i
\lambda_i^2 = 1$.  Notice that a pure state has Schmidt rank one if
and only if it is separable: this is only one example of the utility
of this decomposition in quantum information theory.

\subsection{Linear operators}

In order to introduce how states evolve during a quantum computation,
we must take a detour through some of the different spaces of linear
operators that act on a Hilbert space $\mathcal{H}$.  The most general
\index{$\linear{H,K}$}%
of these is $\linear{H,K}$, which is the set of all linear operators
that map elements of $\mathcal{H}$ to elements of $\mathcal{K}$.  As
we assume that all Hilbert spaces appearing in the thesis are finite dimensional, linearity
implies boundedness which in turn implies continuity: this space is
often referred to as $\mathbf{B}(\mathcal{H,K})$ for this reason.
When the Hilbert space that a linear operator acts on is viewed as the
space of vectors $\mathbb{C}^{\dm{H}}$, the set $\linear{H,K}$ is
exactly the set of $\dm{K}$ by $\dm{H}$ complex matrices.  
The notation $\linear{H}$ is shorthand for $\linear{H,H}$.

For $A \in \linear{H,K}$, the operator $A^* \in \linear{K,H}$ is the
adjoint of $A$, in the sense that $A^*$ is the unique operator that,
for any $\ket\phi \in \mathcal{H}$ and any $\ket\psi \in \mathcal{K}$,
satisfies
\begin{equation*}
  \langle \ket \psi , A \ket \phi \rangle
  = \bra{\psi} A \ket{\phi}
  = \langle A^*\ket\psi, \ket\phi \rangle.
\end{equation*}
When $A$ is represented by a matrix, $A^*$ is the conjugate
transpose of $A$.  Given such a representation, the complex conjugate
of $A$ is denoted $\bar{A}$, and the transpose of $A$ is denoted $A^\top$.

There are a few more classes of operators that are extremely important
in quantum information.
\index{Hermitian}%
One these classes of operators is the
class of \emph{Hermitian} operators.  These are those operators
$A \in \linear{H}$ such that $A = A^*$.
\index{positive}%
An important subclass of the Hermitian operators is the set of
\emph{positive}, or \emph{positive semidefinite}, operators.  These
are the Hermitian operators $A \in \linear{H}$ such that for any
$\ket\phi \in \mathcal{H}$
\begin{equation*}
  \bra{\phi} A \ket{\phi} \geq 0.
\end{equation*}
The positive operators can be equivalently characterized as those
operators $A \in \linear{H}$ that can be expressed as $A = B^* B$ for
some $B \in \linear{H}$.  The notation $A \geq B$ is used to denote
that the operator $A - B$ is positive, with the special case $A \geq
0$ used to state that $A$ is positive.
\index{unitary}%
The other important operators are the \emph{unitary} operators.  These
are the invertible operators $U \in \linear{H}$ with $U^* = U^{-1}$.
It follows from this property that applying a unitary matrix to a
pair of element of $\mathcal{H}$ does not change their inner product, which
further implies that unitaries do not not change the norm.  This property
implies that the unitary operators are exactly the invertible operators that
preserve the pure states, a property which makes them extremely
important for quantum computing.  One other important characterization
is the that unitary operators are exactly those operators that map
orthonormal bases to orthonormal bases.  The set of all unitary
operators in $\linear{H}$ is denoted $\unitary{H}$.
The notion of unitarity can be extended to $V \in \linear{H,K}$ with $\dm{K}
\geq \dm{H}$ by considering those $V$ with the property that $V^*V =
\identity{H}$.  Such an operator is called an \emph{isometry}, and the
set of all such operators is denoted $\unitary{H,K}$.  These operators
embed the elements of the space $\mathcal{H}$ into the
larger space $\mathcal{K}$.

One of the most important operators in this space is $\identity{H}$,
which is the identity operator on $\mathcal{H}$.
\index{$\identity{H}$}%
As a matrix, this operator has ones on the main diagonal and zeroes in
all other positions.
\index{Pauli operators}\index{$X$}\index{$Y$}\index{$Z$}%
\nomenclature[FX]{$X$}{Pauli $X$ matrix}%
\nomenclature[FY]{$Y$}{Pauli $Y$ matrix}%
\nomenclature[FZ]{$Z$}{Pauli $Z$ matrix}%
When restricted to qubits, the identity is one of the four \emph{Pauli
  matrices}.  These four matrices belonging to $\linear{H}$, where
$\dm{H} = 2$, are defined by
\begin{align*}
  \identity{H} & =
  \begin{pmatrix}
    1 & 0 \\
    0 & 1
  \end{pmatrix},&
  X & =
  \begin{pmatrix}
    0 & 1 \\
    1 & 0
  \end{pmatrix},&
  Y & =
  \begin{pmatrix}
    0 & -i \\
    i & 0
 \end{pmatrix},&
  Z & =
  \begin{pmatrix}
    1 & 0 \\
    0 & -1
  \end{pmatrix}.
\end{align*}
One other matrix that will be consistently useful is the
\emph{Hadamard} matrix. \nomenclature[FH]{$H$}{Hadamard matrix}%
 \index{$H$}%
This is the unitary operator that converts the basis $\{\ket 0, \ket
1\}$ to the basis $\{ \ket +, \ket - \}$ and vice versa.  This
operator can be expressed in matrix form as
\begin{equation*}
  H = \frac{1}{\sqrt{2}}
  \begin{pmatrix}
    1 & 1 \\
    1 & -1
  \end{pmatrix}.
\end{equation*}
  
\index{trace}%
An extremely important function on linear operators is the \emph{trace}.  This
is the operation $\tr \colon \linear{H} \to \mathbb{C}$ that, on a
matrix representation of an operator $A$, is simply the sum of the
main diagonal.  One of the most important properties of the trace is
that it is \emph{cyclic}, i.e.\ for operators $A,B,C$ we have
\begin{equation*}
  \tr( ABC ) = \tr( BCA ) = \tr( CAB ),
\end{equation*}
whenever the products in the above equation are defined.  Note that
the trace is not stable under more general commutation of the arguments,
i.e.\ there are operators $A,B,C$ such that $\tr(ABC) \neq \tr(CBA)$.

The space $\linear{H,K}$ equipped with the inner product given by
\begin{equation*}
  \langle A, B \rangle = \tr(A^*B)
\end{equation*}
is also a Hilbert space.  This implies that $\linear{H} \tprod
\linear{K}$ is well defined.  In fact, it is the case that
\begin{equation*}
  \linear{H,K} \tprod \linear{A,B} = \linear{H \tprod A, K \tprod B}.
\end{equation*}
When the tensor product is extended to operators it behaves in the
same way as it does on vectors.  For $A \in \linear{H}$ with elements
$a_{ij}$ for $1 \leq i,j \leq d$ and $B \in \linear{K}$, the operator
$A \tprod B$ has matrix representation given by the block matrix
\begin{equation*}
  A \tprod B =
  \begin{pmatrix}
    a_{11} B & a_{12} B & \cdots & a_{1d} B \\
    a_{21} B & a_{22} B & \cdots & a_{2d} B \\
    \vdots & \vdots &  \ddots & \vdots \\
    a_{d1} B & a_{d2} B & \cdots & a_{dd} B
  \end{pmatrix}.
\end{equation*}

\nomenclature[F1H]{$\identity{H}$}{identity operator on $\linear{H,H}$}%
\nomenclature[BL(H,K)]{$\linear{H,K}$}{set of all linear operators from $\mathcal{H}$ to $\mathcal{K}$}
\nomenclature[BU(H)]{$\unitary{H}$}{unitary operators on $\mathcal{H}$}
\nomenclature[BU(H,K)]{$\unitary{H,K}$}{isometries mapping $\mathcal{H}$ to $\mathcal{K}$}
\nomenclature[aiso]{$\cong$}{isomorphism between Hilbert spaces}

As $\linear{H}$ is itself a Hilbert space, we can find bases of
operators for it.
An important orthogonal basis for
this space is given by the discrete Weyl operators, also
known as the generalized Pauli operators.
\index{discrete Weyl operators}%
\index{Weyl operators|see{discrete Weyl operators}}%
\index{Pauli operators!generalized}%
These operators extend the Pauli operators to the $d$ dimensional
space $\mathcal{H}$, keeping the properties of orthogonality and unitarity, but
losing Hermiticity.  As these operators will be essential to several
of the arguments in the thesis, they are introduced in detail.  The
discrete Weyl operators are based on
generalizations of the $X$ and $Z$ operations, given by
\begin{align*}
  X &= \sum_{j=1}^d \ket {j + 1} \bra {j} \\
  Z &= \sum_{j=1}^d \omega_d^j \ket {j} \bra {j},
\end{align*} 
where $\omega_d$ is a $d$-th primitive root of unity (such as
$e^{2i\pi/d}$), and in the definition of $X$ the operator
$\ket{d+1}\bra{d}$ is taken to be $\ket 1 \bra d$.  
The operator $X$ simply advances each state of the computational basis
to the next, and the operator $Z$ applies a different phase to each
basis state.  It is clear from the definition that $XX^* = \identity{H} =
ZZ^*$, which implies that these operators are unitary.  It is also
clear that $X$ and $Z$ fail to commute:
\begin{equation}\label{intro-eqn-XZcom}
  ZX = \omega_d XZ.
\end{equation}
Using these operators, the discrete Weyl operator
with index $(a,b) \in \mathbb{Z}_d \times \mathbb{Z}_d$ is given by
\begin{align*}
  W_{a,b} = X^a Z^b.
\end{align*}
For two dimensional systems, these operators are, up to phases,
exactly the usual Pauli matrices.  These operators are unitary,
since they are products of the unitary operators $X$ and $Z$.
Equation~\ref{intro-eqn-XZcom} can be directly extended to these
operators to obtain
\begin{equation}\label{intro-eqn-weyl-commute}
  W_{a,b} W_{e,f} 
  = X^a Z^b X^e Z^f 
  =  \omega_d^{be - af}X^e Z^f X^a Z^b
  =  \omega_d^{be - af} W_{e,f} W_{a,b}.
\end{equation}
To see that these operators form an orthogonal basis for $\linear{H}$,
notice that, by the cyclic property of the trace
\begin{align}\label{intro-eqn-weyl-basis}
  \tr{ W_{a,b}^* W_{e,f} }
  &= \tr Z^{-b} X^{-a} X^e Z^f
    = \tr X^{e-a} Z^{f-b}
    = \begin{cases}
      d & \text{if $a = e$ and $b = f$}, \\
      0 & \text{otherwise}.
    \end{cases}
\end{align}
These operators can be normalized to obtain an orthonormal basis for $\linear{H}$, but
this comes at the cost of unitarity, so we will not do this here.
The discrete Weyl operators will be used in
Chapter~\ref{chap-mixed-unitary} to show that several transformations
on quantum states can be realized as convex mixtures of unitary transformations.

\index{normal operator}%
A linear operator $A \in \linear{H}$ is \emph{normal} if $A^*A =
AA^*$.  By definition the Hermitian and unitary operators are normal.
\index{spectral decomposition}%
Any normal operator $A \in \linear{H}$ has a \emph{spectral decomposition}, which is a
representation as
\begin{equation}\label{intro-eqn-spectral}
  A = \sum_i^{\dm{H}} \lambda_i \ket{\phi_i} \bra{\phi_i},
\end{equation}
where the vectors $\{\ket{\phi_i}\}$, called the \emph{eigenvectors}
of $A$, are an orthonormal basis for $\mathcal{H}$.  The associated
complex numbers $\lambda_i$ are called the \emph{eigenvalues} of $A$.
The space spanned by the eigenvectors of $\mathcal{A}$ corresponding
to nonzero eigenvalues is called the \emph{support} of $A$. \index{support}%
This space has dimension equal to the rank of $A$.

The classes of operators that we have previously encountered can be
characterized in terms of the spectral decomposition.  A normal matrix $U$ is
unitary if and only if all of its eigenvalues are have norm one,
i.e.\ if $\abs{\lambda_i} = 1$ for all $i$.  This also implies that $U$
is full rank.  A normal operator is Hermitian if and only if all of its
eigenvalues are real.  This can be seen by considering the adjoint of
the representation in Equation~\eqref{intro-eqn-spectral}.  As a
further restriction, an operator is positive if and only if all of
it is normal and has only nonnegative real eigenvalues.
In addition to this, if $A$ is an operator with eigenvalues
$\lambda_i$, then the trace of $A$ is given by $\tr A = \sum_i
\lambda_i$.  This characterization of the trace is extremely useful.

The spectral decomposition also allows functions on the complex numbers to
be extended to operators in $\linear{H}$.  An example of this is
square root of a positive matrix.  This is defined, for any positive
operator $A$, by taking the square roots of the eigenvalues, i.e.\
\begin{equation*}
  \sqrt{A} = \sum_i \sqrt{\lambda_i} \ket{\phi_i} \bra{\phi_i},
\end{equation*}
where $A$ has spectral decomposition $A = \sum_i \lambda_i
\ket{\phi_i} \bra{\phi_i}$.  It is easy to see from this definition
that the operator square root satisfies $\sqrt{A} \sqrt{A} = A$.  It
is less obvious that $\sqrt{A}$ defined in this way is the unique
square root of the operator $A$, but this is indeed the case. 
This technique can be extended to
define $\log(A)$ for a positive matrix $A$ as well as $e^A$ and
$\abs{A}$ for general normal matrices.  

The absolute value of an operator has a different definition when $A
\in \linear{H,K}$ is not normal (and potentially not square).  In this
case, $\abs{A} = \sqrt{A^*A}$, relying on the fact that for any
operator $A$, the operator $A^*A$ is positive.  The eigenvalues of
$\abs{A}$ play a central role in another important decomposition of an
operator.
\index{singular value decomposition}%
The \emph{singular value decomposition} of $A \in \linear{H,K}$ gives
a representation of $A$ that mimics the spectral decomposition, but
exists even when $A$ is not normal.  This representation is
\begin{equation*}
  A = \sum_{i=1}^d s_i \ket{\phi_i} \bra{\psi_i}
\end{equation*}
where $d = \min\{\dm{H}, \dm{K}\}$.  The values $s_i$ are nonnegative
and real, these are called the \emph{singular values} of $A$.  They
are equal to the eigenvalues of $\abs{A}$.  The vectors
$\{\ket{\phi_i} \}$ and $\{ \ket{\psi_i} \}$ form orthogonal sets in
$\mathcal{K}$ and $\mathcal{H}$, respectively.  The singular values
will be very important in Chapter~\ref{chap-meas} where we consider a
collection of operator norms that depend solely on them.

\subsection{Mixed states}

Pure states suffice to  model the behaviour of a quantum system in a
known state, but they do not completely capture the situation when
there is uncertainty about exactly which state a system is in.  As an
example, if the state a two-dimensional system is $\ket 0$ or $\ket 1$
each with probability one-half, then the behaviour of the system is
identical to one in which the state is a uniform mixture of $\ket +$
or $\ket -$, yet these two descriptions differ.

This problem is resolved by resorting to density operators.  Given a system
that is in the state $\ket{\phi_i}$ with probability $p_i$ (this is
called the \emph{ensemble} $\{ (p_i, \ket{\phi_i}) \}$), the density
operator associated with the system is given by
\begin{equation*}
  \rho = \sum_i p_i \ket{\phi_i} \bra{\phi_i}.
\end{equation*}
A given density matrix may, in general, have an infinite set of
ensembles that generate it.
This notation resolves the earlier example, since we have,
 for any two orthonormal bases $\ket{\phi_i}$ and $\ket{\psi_i}$ of
 $\mathcal{H}$
\begin{equation*}
  \frac{1}{\dm{H}} \sum_i \ket{\phi_i} \bra{\phi_i}
  = \frac{1}{\dm{H}} \sum_i \ket{\psi_i} \bra{\psi_i}
  = \frac{\identity{H}}{\dm{H}}
  = \nidentity{H}
\end{equation*}
\index{$\nidentity{H}$}%
where the symbol $\nidentity{H}$ denotes $\identity{H}/\dm{H}$, the
normalized identity operator on $\mathcal{H}$.
\nomenclature[a1nH]{$\nidentity{H}$}{completely mixed state on $\mathcal{H}$, $\nidentity{H} = \identity{H} / \dm{H}$}%
An element $\rho \in \linear{H}$ is called a \emph{density operator}
(or equivalently a density matrix) if and only if it satisfies the two properties
\index{density operator}%
\begin{enumerate}
  \item $\rho$ is positive,
  \item $\tr \rho = 1$.
\end{enumerate}
\index{$\density{H}$}%
The set of all such density operators on $\mathcal{H}$ is denoted $\density{H}$.
\nomenclature[BD(H)]{$\density{H}$}{density operators on $\mathcal{H}$}

\index{mixed state}%
The density operators are also referred to as the \emph{mixed states},
as they provide a complete description of a quantum system.  Pure
states also fit into this framework: the state $\ket{\phi}$
corresponds to the density operator $\ket{\phi} \bra{\phi}$, and these
two notions of state will be used interchangeably in this case.
Notice also that the
set of density matrices $\density{H}$ is both compact and convex.  The extreme
points of this set are simply the rank one projectors $\ket \phi \bra
\phi$ corresponding to pure states in $\mathcal{H}$.

A mixed state in $\density{H \tprod K}$ is called separable if it is
the convex combination of a set of separable states in $\mathcal{H
  \tprod K}$.  If a mixed state cannot be decomposed in this way,
i.e.\ any ensemble contains an entangled pure state, then it is called
entangled.

\subsection{State evolution and measurement}

The evolution of a quantum system in the state $\rho \in \density{H}$
is determined by the action of a unitary operator $U \in \unitary{H}$.
The state of the system after this evolution is $U \rho U^*$.  As we
shall see in the next section, this does not capture every quantum
process, but it is an important special case.  As $U^{-1} = U^*$ is
also unitary, this implies that any unitary evolution is, in
principle, reversible.

Measurements provide a method for retrieving information from a
quantum system.  The simplest case of measurement is given by a
\emph{projective measurement}, which is a set $\{\Pi_i\}$ of
orthogonal projectors in $\linear{H}$ with the property that $\sum_i \Pi_i =
\identity{H}$.  When this measurement is performed on a state $\rho
\in \density{H}$ the outcome is $i$ with probability $p_i = \tr(\Pi_i \rho)$
and the state after measurement is $(\Pi_i \rho \Pi_i)/p_i$.
In the case that the outcome of the measurement is unknown, i.e.\ it is
discarded or forgotten, the resulting state is given by $\sum_i \Pi_i
\rho \Pi_i$, i.e.\ the weighted mixture of all of the measurement outcomes.

There is a special case of projective measurement that is of
particular importance in quantum computing.  This is known as
measurement in the computational basis, which is given by the complete
set of projectors $\{ \ket i \bra i : 0 \leq i < \dm{H} \}$.  Any
measurement using orthogonal rank one projectors can be derived from
this measurement by rotating the state $\rho$ to be measured using
some unitary operation $U$.

\index{POVM}%
Projective measurements are not the only case allowed by quantum
mechanics.  More generally, a $POVM$ measurement is given by a set
$\{E_i\}$ of positive operators that sum to $\identity{H}$.  The
outcome of such a measurement is $i$ with probability $p_i = \tr( E_i
\rho )$.  While the state after measurement can be defined as in the
case of projective measurements, a simpler model will suffice for the results in
this thesis.  In this model the outcome after measurement is the state $\ket i$
when the result is $i$.  This form of measurement can be quite
convenient to work with.  POVM measurements can always be realized by
projective measurements on the state $\rho \tprod \ket 0 \bra 0$ in a
larger Hilbert space.  This result is known as Naimark's theorem~\cite{Naimark43}.

\subsection{Channels}

We have already seen two types of evolution for quantum states:
unitary evolution and measurement.  Both of these types of evolution
are special cases of the most general type of transformation on
quantum states.  These are the linear transformations that map density matrices
to density matrices, known as quantum \emph{channels}.
\index{quantum channel|see{channel}}\index{channel}%
Such a map can capture any process allowed by quantum mechanics.
These maps can also be characterized as the linear operators $\Phi$ from
$\linear{H}$ to $\linear{K}$ that satisfy two properties
\begin{align*}
  \text{Trace preserving: } 
  & \tr \Phi(X) = \tr X \text{ for all $X \in \linear{H}$} \\
  \text{Complete positivity: }
  & \text{If $X \geq 0$ then }
  (\Phi \tprod \tidentity{K})(X) \geq 0 \text{ for all
    $\mathcal{K}$, $X \in \linear{H \tprod K}$.}
\end{align*}
\index{$\tidentity{H}$}%
\nomenclature[FIH]{$\tidentity{H}$}{identity transformation on $\linear{H}$}%
In the above definition, $\tidentity{K}$ is the identity
transformation on $\linear{K}$ and the map $\Phi \tprod \Psi$ is
simply the map that applies $\Phi$ and $\Psi$ on their respective
Hilbert spaces.  The set of all channels from $\linear{H}$ to
$\linear{K}$ is denoted $\transform{H,K}$.
\nomenclature[BT(H,K)]{$\transform{H,K}$}{set of all channels from $\linear{H}$ to $\linear{K}$}%
\index{completely positive}%
A linear map taking $\linear{H}$ to $\linear{K}$ that is not necessarily a channel
will occasionally be referred to as a super-operator.

\index{partial trace}%
\nomenclature[Ftrk]{$\ptr{K}$}{Partial trace over the system in $\mathcal{K}$}%
One of the most important quantum channels is the operation known as the \emph{partial trace}.  This
is the channel in $\transform{H \tprod K, H}$ that traces out the
system in the space $\mathcal{K}$.  This map is defined on
$X \tprod Y$ with $X \in \linear{H}$ and $Y \in \linear{K}$ by
\begin{equation*}
  \ptr{K} X \tprod Y = (\tr Y) X,
\end{equation*}
and extended to the whole space $\linear{H \tprod K}$ by linearity.
This is the operation that discards the system in $\mathcal{K}$, and
as such, is very useful in quantum information.  The partial trace can
also be expressed by explicitly writing out the trace over
$\mathcal{K}$.  Let $\{ \ket{\phi_i} \}$ be any orthonormal basis for
$\mathcal{K}$, then for any $X \in \linear{H \tprod K}$, the partial
trace over $\mathcal{K}$ is
\begin{equation*}
  \ptr{K} X = \sum_i ( \identity{H} \tprod \bra{ \phi_i }) X
  (\identity{H} \tprod \ket{\phi_i}).
\end{equation*}

One important feature of mixed states is that they can always be
viewed as part of a pure state on a larger Hilbert space.  Any $\rho
\in \density{H}$ can be expressed as a pure state  $\ket \phi \in \mathcal{H}
\tprod \mathcal{K}$, where $\dm{K} \geq \rank \rho$, as
\begin{equation*}
  \rho = \ptr{K} \ket\phi\bra\phi.
\end{equation*}
 \index{purification}%
The state $\ket\phi$ is referred to as a \emph{purification} of $\rho$.
These purifications will form an integral part of many of the proof
techniques used in this thesis.  It is also important that any two
purifications $\ket\phi, \ket\psi \in \mathcal{H \tprod K}$ of a state
$\rho \in \density{H}$ are related by a unitary operation on the space
$\mathcal{K}$ alone, i.e.\ there exists a $U \in \unitary{K}$ such that
\begin{equation*}
  (\identity{H} \tprod U) \ket\phi = \ket\psi.
\end{equation*}
This fact will be used in the definition of the fidelity in Chapter~\ref{chap-meas}.

There are two convenient representations of quantum channels that will
be needed.  The first of these is the representation of a completely
positive map $\Phi$ by a set of Kraus operators, which are matrices
$A_i$ such that
\[ \Phi(X) = \sum_i A_i X A_i^*. \]\index{Kraus operators}%
This representation is due to Choi~\cite{Choi75}.
If, in addition, $\Phi$ is trace preserving, then the operators $A_i$
satisfy the property
\[ \sum_i A_i^* A_i = \identity{}. \]
If $\Phi \in \transform{H,K}$ then the number of Kraus operators in a
minimal 
Kraus decomposition is at most $(\dm{H}) (\dm{K})$.

The second representation of importance is known as the Stinespring
Dilation Theorem, after the 1955 work of
Stinespring~\cite{Stinespring55}, though the precise statement of the
result we use here is given by Hellwig and Kraus~\cite{HellwigK70}.  This
theorem states that any quantum channel can be
represented as a unitary operation on a larger space, some of which is
traced out.  
More formally, for a channel $\Phi \in \transform{H,K}$ there 
are spaces $\mathcal{A,B}$ and a $U \in \unitary{H \tensor A, K \tensor
  B} \cong \unitary{H \tprod A}$ such that
\[ \Phi(X) = \ptr{B} U ( X \tensor \ket 0 \bra 0)
U^*. \]\index{Stinespring representation}
Where $\mathcal{A}$ can be chosen so that $\dim \mathcal{A} \leq \dim
\mathcal{H} \dim \mathcal{K}$.  Such a representation is unique up to
an isometry on the space that is traced out.
This representation can be used to recover a Kraus representation:
see~\cite{Schumacher96} for an overview of this result.  

The Stinespring representation implies that in order to model a
quantum channel, we need worry about only three parts: introducing
ancillary qubits in a known pure state, implementing unitary
operations, and implementing the partial trace.  This will be
extremely helpful for the reductions in this thesis that seek to
simulate general quantum channels with channels from restricted
classes.  More details can be found in
Section~\ref{compl-scn-circuit-model} where the model of circuits used
in the thesis is formally defined.

\section{Classes of quantum channels}\label{intro-scn-classes}

This section provides an overview of the different classes of quantum
channels that will be encountered in this thesis.  This overview will
be kept somewhat brief, as the classes that will receive detailed
treatment are reintroduced more thoroughly in the chapters
where they appear.

The classes of channels presented here place different restrictions on
the set of channels.  Some of these restrictions come from
practical notions, like the channels that can be implemented in a
small amount of time, and some of the restrictions come from more
theoretical concerns, such as the antidegradable channels.  The
restricted classes studied here are largely incomparable: this is
because one of the aims of this thesis is to present simplified versions of
the distinguishability problem that are nevertheless just as hard as the
general case.  In order that these results cover more of the
quantum channels that are likely to arise in practice, it is helpful
if the distinguishability problem is shown to be hard on several
unrelated classes of channels.

The material that appears in this section makes several
references to the material that follows in Chapter~\ref{chap-meas}.
Because this material is only used in a superficial way in this
section, it is not necessary to have read Chapter~\ref{chap-meas}
first, though a familiarity with quantum information will help.

\subsection{Circuit restrictions}

Quantum circuits are a convenient way to provide a quantum
channel as input to a computational problem, such as the problem of
distinguishing quantum channels.  One of the advantages of this
representation
is that, given a quantum computer, it allows the channel to be
evaluated, but it remains computationally infeasible to
find a matrix representation for all but the simplest channels.  The
circuit model used here is presented in detail in
Section~\ref{compl-scn-circuit-model}.  This level of detail will not
be necessary to introduce the classes of channels defined by placing
restrictions on this model.

The circuit representation for quantum channels allows restricted classes of
channels to be defined by placing restrictions on the types of
circuits that are allowed.  
These channels can be much simpler than general channels.
An example of this is the class of channels with \emph{stabilizer
  circuits}, which are the circuits defined on a
restricted set of quantum gates.  Given such a circuit, a
channel can be efficiently simulated using a deterministic classical
computer~\cite{AaronsonG04}, but it is expected that this is not
possible for quantum channels given as general quantum circuits, as
this would imply the equivalence of classical and quantum computation.

Restricting the input circuits to the distinguishability problem
mentioned in Section~\ref{scn-intro-overview}
can lead to simpler variants of the
problem.  One such restriction
is to the class of channels that implement unitary operations.  These
channels can be obtained as a circuit restriction by eliminating the
non-unitary gates from the circuit model.
Distinguishing these
circuits appears to be easier than general mixed-state circuits~\cite{JanzingW+05}.

One of the more interesting circuit restrictions is the requirement that the
input circuits to the distinguishability problem have depth logarithmic
in the number of input qubits.  These are  the circuits that can be
performed in logarithmic time with a parallel model of quantum
computing.  Such as model not unreasonable in many implementation
schemes for quantum computing.
These circuits are interesting as they limit the length of time that quantum information
needs to be protected from decoherence.  For this reason, much of
experimental quantum computing is concerned with very short
computations, and log depth circuits are an interesting generalization
of such computations.  Many important quantum algorithms are known to
have log depth circuits, such as the approximate quantum Fourier
transform~\cite{CleveW00} and the encoding and decoding operations 
for many quantum error correcting codes~\cite{MooreN02}.

One of the results on the thesis is that distinguishing log depth
quantum mixed-state circuits is complete for \class{QIP}, i.e.\ just as
hard as the general case.  This is
shown by reducing the close images problem, studied in Chapter~\ref{chap-close-images},
to a log depth version of itself.  The essential idea behind this
reduction is to simulate a general quantum circuit by a log-depth one
by slicing the circuit into log depth pieces that are performed in
parallel.  This circuit will perform the same computation as the
original circuit only if the input to one piece matches the output of
the previous piece.  To ensure that this is the case for the circuits
constructed in the reduction, tests are applied to force this to be
the case for any outputs of the two circuits that can potentially have close
images.
This reduction shows that the close images problem
remains \class{QIP}-complete when restricted to log depth circuits.
Extending this to the distinguishability problem on log-depth circuits
follows from the fact that the reduction in
Chapter~\ref{chap-distinguishability} preserves the log-depth
restriction of the circuits.

\subsection{Degradable and antidegradable channels}

The degradable channels are those channels $\Phi$ for which there exists a
second channel that maps the output of $\Phi$ to the environment of
$\Phi$, i.e.\ the space that is traced out in a Stinespring
representation.  These channels were introduced by Shor and
Devetak~\cite{DevetakS05} and can be thought of as the channels where
the environment contains no information that is not also present in
the output of the channel.  A more formal definition is: a channel
$\Phi \in \transform{H,K}$ expressed as 
$\Phi(\rho) = \ptr{B} U (\rho \tensor \ket 0 \bra 0) U^*$ is
\emph{degradable} if there exists a channel $D$ such that
\begin{equation*}
  \ptr{K} U (\rho \tensor \ket 0 \bra 0) U^* = D(\Phi(\rho)).
\end{equation*}
Stinespring representations are not unique, but any two differ by an
isometry on the environment space, 
and this isometry can be absorbed into the channel
$D$, which implies that the notion of degradability does not depend on the choice of
representation.

The channel $\Phi^C$ given by tracing out the output space $\mathcal{K}$ and
not the environment space $\mathcal{B}$ is often referred to as the complementary (or
conjugate) channel to $\Phi$, with the caveat that it is only defined
up to the Stinespring representation chosen for $\Phi$.  A channel is
\emph{antidegradable} if the complementary channel is degradable.  More
plainly, a channel $\Phi$ is antidegradable if there exists a second
channel that maps the environment of $\Phi$ to the output of $\Phi$.
Antidegradable channels are also well-defined, since as in the case of
degradability, the choice of Stinespring representation can be
absorbed into the degrading map.
A thorough discussion of the degradable and antidegradable
channels can be found in~\cite{CubittR+08}.

These channels are discussed in Chapter~\ref{chap-degr}, where it is
shown that the problem of computationally distinguishing two channels
is made no easier when the channels are  promised to be degradable or
antidegradable.  This is done using a construction similar to one
found in~\cite{CubittR+08} that is used to reduce the additivity of
the classical capacity to the degradable case.

\subsection{Entanglement-breaking channels}

An entanglement-breaking channel $\Phi$ is a channel for which the
output $(\Phi \tensor \tidentity{H})(\rho)$ is separable for any input
state $\rho$.
This class of channels contains many of the commonly used channels, such as
the completely depolarizing channel and the complete dephasing channel.
It is helpful to state a
few alternate characterizations of the entanglement-breaking channels.
\begin{proposition}[Horodecki, Shor, and Ruskai~\cite{Horodecki+03}]
\label{intro-prop-eb-char}
  Let  $\Phi \in \transform{H,K}$.
  The following are equivalent
  \begin{enumerate}
    \item $\Phi$ is entanglement-breaking,
    \item $(\Phi \tensor \tidentity{H})(\ket \psi \bra \psi)$ is
      separable, for $\ket \psi$ a maximally entangled state on
      $\mathcal{H \tensor H}$,
    \item $\Phi$ has a Kraus decomposition using only rank one
      operators,
    \item $\Phi$ can be written as
      \[ \Phi(\rho) = \sum_k \sigma_k \tr ( E_k \rho ), \]
      where the $\sigma_k$ are density matrices and the set $\{E_k\}$
      forms a POVM.
  \end{enumerate}
\end{proposition}
Another property of these channels is that all
entanglement-breaking channels are antidegradable~\cite{CubittR+08}.

The distinguishability problem
on quantum circuits, considered in Chapter~\ref{chap-distinguishability}, 
is defined in terms of distinguishability with access to a reference
system.  This method, distinguishing channels by observing their
action on part of a larger space, is the most general method for
distinguishing channels.  There are channels known for which this
reference system is required to obtain an optimal
distinguishing strategy~\cite{Watrous08distinguishing}.
This reference system allows for entangled inputs to aid in
distinguishing the two channels and it appears to be essential to
problem.  It might be expected that this entanglement cannot help
distinguish entanglement-breaking channels as the output is always separable, but this is not the case.
An example on qubit channels has been provided by
Sacchi~\cite{Sacchi05entanglement, Sacchi05optimal}.  When this
example is generalized to channels on a $d$-dimensional space,
however, the amount that this entanglement assists the
distinguishability goes to zero quadratically with $d$.  This is in
contrast to the large difference that this reference system can make
in the distinguishability of general channels.
Examples of entanglement-breaking channels
with this property are not known.  Whether or not this is a roadblock
to extending the hardness of the distinguishability problem to these
channels is an interesting open problem.

It is simple to show that the problem of distinguishing two channels
is \class{QIP}-hard for entanglement-breaking channels that are
exponentially close together using a straightforward reduction from
the general problem.  This can be achieved by simply mixing the
channels in a given instance of distinguishability with enough of the
completely depolarizing channel that the resulting channels are
entanglement-breaking.  That this occurs is a consequence of the fact
that there exists a ball of separable states around the completely
mixed state~\cite{GurvitsB02}.  Unfortunately this ball has
radius that is exponentially small in the log of the dimension
(i.e.\ the number of qubits in the original circuits), and so the
resulting entanglement-breaking channels are exponentially close
together.  The polarization technique that can be applied in the general
case, which is discussed in Section~\ref{meas-scn-polarization},
cannot be applied to these circuits as they are
too close together, and so this reduction can only be used to show the
hardness of distinguishing circuits that are exponentially close
together, which is perhaps not a terribly surprising result.

\subsection{Unital channels}

\index{unital}\index{channel!unital}%
A superoperator $\Phi \colon \linear{H} \to \linear{K}$ is \emph{unital} if $\Phi(\identity{H}) = \identity{K}$.  The unital
channels are often called \emph{doubly stochastic} as in
addition to being unital they are also trace preserving.  The trace
preserving property of quantum channels requires that any unital
channel have input and output spaces of the same dimension.
The unital channels have the interesting property that the entropy of
the output of the channel is always at least at large as the entropy
of the input, as noted by King and Ruskai~\cite{KingR01}.  

This property makes the unital channels interesting from the
perspective of the additivity of the Holevo capacity, 
as channels that do not reduce entropy can
be used as a natural noise model.  Fukuda has shown how to construct a
unital channel from a general channel, without changing the minimum
output entropy or the maximum output $p$-norm~\cite{Fukuda07}.  This
implies that for a specific class of channels the question of
additivity can be restricted to a subclass of the unital channels.

Mendl and Wolf have recently characterized the unital channels as the
quantum channels that can be decomposed into the affine combination of a
set of unitary channels~\cite{MendlW08}.  More explicitly, they have shown
that a channel $\Phi$ is unital if and only if there exist unitaries $U_i$ and
$\lambda_i \in \mathbb{R}$ with $\sum_i \lambda_i = 1$ such that
\[ \Phi(X) = \sum_{i=1} \lambda_i U_i X U_i^*. \]
The form of this decomposition is very similar to the next class of
channels that we consider.

\subsection{Mixed-unitary channels}

A quantum channel is \emph{mixed-unitary} if it can be decomposed
into the probabilistic application of a set of unitary
operations.  These channels are often referred to as the \emph{random
  unitary} channels, but this is avoided here because this name often
causes confusion with the channels defined by drawing unitary
operators from the Haar measure.
More formally, $\Phi$ is mixed-unitary if there exist
unitary operators $U_1, \ldots, U_n$ and a probability distribution
$p_1, \ldots, p_n$ such that
\begin{equation}\label{intro-eqn-rand-unitary}
  \Phi(X) = \sum_{i=1}^n p_i U_i X U_i^*.
\end{equation}
It has been shown by Gregoratti
and Werner~\cite{GregorattiW03} that the mixed-unitary channels
describe exactly the noise processes that can be corrected using
classical information obtained by measuring the environment.
Audenaert and Scheel have recently
provided necessary and sufficient conditions for a channel to be 
mixed-unitary~\cite{AudenaertS08}.  
Buscemi has also provided an upper bound on the number of unitaries
needed for a mixed-unitary decomposition~\cite{Buscemi06}.

The set of mixed-unitary channels is contained in the set of all unital
channels; this is a simple consequence of
Equation~\eqref{intro-eqn-rand-unitary}.
For channels on qubits these two sets of channels coincide,
but for larger dimensions this is not the
case~\cite{Tregub86,KuemmererM87,LandauS93}.
It is known, however, that there exists in the set of unital channels a ball
of mixed-unitary channels around the completely depolarizing
channel~\cite{Watrous09mixing}, which is the channel that maps all
input states to the completely mixed state.
In the case of qubit mixed-unitary
channels, both additivity of the Holevo capacity and multiplicativity
or the maximum output $p$-norm are known to
hold~\cite{King02}.  For general mixed-unitary channels,
both additivity~\cite{Hastings09} and 
multiplicativity~\cite{HaydenW08} are known to fail.  These properties
are considered in more detail in Chapter~\ref{chap-meas}.

The fact that these properties do not hold in general for
mixed-unitary channels does not completely eliminate interest in the
additivity properties of specific mixed-unitary channels.  One of the
contributions of this thesis is a method to approximate a general
quantum channel with a mixed-unitary one.  This approximation can be
made arbitrarily good in the minimum output entropy or the maximum
output $p$-norm by increasing the dimension of the ancillary space
used by the approximation.  This can be used to show that the
additivity and multiplicativity problems for a general channel can be
reduced to the same problem on a mixed-unitary approximation, where
the approximation error can be made arbitrarily small.  Results on
this approximation may be easier to prove; mixed-unitary channels
have been essential to finding counterexamples to both the additivity
and multiplicativity conjectures.  These results can then be applied
to the original channel by sending the approximation error to zero.

The method for approximating a general channel by a mixed-unitary one
is discussed in Chapter~\ref{chap-mixed-unitary}.  Starting with a channel in Stinespring
form $\Phi(X) = \ptr{B} U (X \tensor \ket 0 \bra 0) U^*$ there are
only two operations that are not mixed-unitary: the partial trace on
the system $\mathcal{B}$ and the introduction of the auxiliary system in the
$\ket 0$ state.  The partial trace is the easy operation to simulate
with a mixed-unitary channel as it may be directly replaced by the completely
depolarizing channel on the space $\mathcal{B}$.  The auxiliary system
in the $\ket 0$ state is more difficult to replace.  
The strategy employed is to add this extra system
to the input space of the channel and test that the input in this
space is close to $\ket 0$.  If this auxiliary input is close to $\ket 0$ the
channel proceeds exactly as does the original channel.  If the
auxiliary input
is far from $\ket 0$ then the testing procedure sends the input state
very close to the maximally mixed state, which results in the output
of the channel having very high entropy.  As we are concerned with
approximating the minimum output entropy
this ensures that any input state achieving the minimum is very close
to $\ket 0$ in the auxiliary space.  This construction produces a
channel with similar minimum output entropy and maximum output
$p$-norm to the original channel, and so it can be used to reduce
problems of additivity and multiplicativity to the mixed-unitary
case.

This construction can also be performed on circuits in time polynomial
in the size of the circuit and so it also has implications for the
problem of distinguishing quantum circuits.  This can be used to show
that the problem of distinguishing two mixed-unitary circuits is as
hard as distinguishing two general circuits, which is a
\class{QIP}-complete problem.  This result is found in
Chapter~\ref{chap-mixed-unitary}.

%%% Local Variables: 
%%% mode: latex
%%% TeX-master: "thesis"
%%% End: 

\chapter{Quantum Computational Complexity}\label{chap-complexity}

This chapter lays the complexity theoretic groundwork for the
remainder of the thesis.  This includes a definition of the circuit
model that is used throughout the thesis as well as a brief overview
of some of the complexity classes that will be encountered later.

The circuit model used here is the mixed-state circuit model of
Aharonov et al.~\cite{AharonovK+98} that allows measurements and other
non-unitary operatations to take place during a computation.  This
model will be essential to the problems considered in the thesis: the
distinguishability problem appears to be strictly
more difficult on this circuit model than it is on the model of
unitary circuits.  This is despite the fact that both of these models
are computationally equivalent, in the sense that any problem solvable
by a circuit in one model can also be solved in the other.  This
equivalence does not extend to problems that take these circuits as input.

The wide array of complexity classes often encountered in theoretical
computer science is not particularly useful or relevant to the results
of the thesis.  For this reason, the introduction of complexity
classes is kept quite brief, with only a few of the most important
classes introduced.

\minitoc

\section{Quantum circuits}\label{compl-scn-circuit-model}

Many questions on quantum channels can be extended to computational
problems. This extension leaves one difficulty: what is the correct
way to encode a quantum channel as input to a computational problem?
One obvious choice is to provide the Kraus operators or the unitary
matrix from a Stinespring dilation. Such a representation allows for
any quantum channel to be represented approximately, as these
matrices can only be specified up to some precision.  Viewed
computationally, however, this representation is
unsatisfying. The reason for this is that the description of the
channel is polynomial in the input and output dimensions, which are
exponential in the number of qubits needed to represent the input and
output. This representation is similar to modelling any classical
process as a table of inputs and outputs -- this form is convenient,
but often exponentially larger than necessary.

Taking a hint form classical complexity theory, we will represent
quantum channels using circuits. These circuits will allow for the
simulation of a quantum channel from the circuit description, but they
will not, in general, allow the efficient solution to most of the computational
problems on these channels. This is equivalent to the classical case,
where the circuit satisfiability problem is used to represent the
problem of determining if a computation can be made to accept.
Providing a complete table of outputs as the input to this problem
trivializes it, as in the case of a polynomial size circuit, the table
encodes the information in the circuit in an exponentially larger
description. The problems on quantum computations that we consider in
this thesis are similarly trivialized by a representation of quantum
channels as matrices of size exponential in the number of input and
output qubits.

The most widely used model of quantum computation is the unitary
circuit model. In this model a computation is represented by a
directed acyclic graph, where the edges represent qubits and the nodes
represent gates. In order for a circuit to implement a valid quantum
operation, each gate is labelled with a quantum channel that maps the
state of the input qubits to the state of the output qubits. The
operations that can appear as gates in a circuit depend on exactly
which model of quantum computation is being used. As one final
restriction, no isolated vertices in the graph are allowed, because these
would correspond to gates in the circuit that neither take input nor
produce output, and so they cannot affect the computation being
performed.

There are two important quantities related to a circuit: \emph{size}
and \emph{depth}. If a circuit is represented as a graph, the size of
the circuit is the number of vertices, i.e.\ the number of gates in the
circuit. \index{circuit!size}%
This definition leaves the possibility of very small circuits acting
on a large number of input qubits -- this undesirable feature is
avoided by taking the size of a circuit to be the maximum of the
number of gates and the number of qubits that the circuit acts on.
Using this definition, the size of a circuit is essentially equivalent to
the number of bits needed to represent the circuit, so long as the
number of different types of gates available is constant.

The depth of a circuit is the length of the longest directed path in the
graph. \index{circuit!depth}%
As circuits are acyclic, the depth of a circuit can be efficiently
computed from a description. Since the transformations implemented by
gates acting on different qubits commute, they can be performed in
parallel. This implies that the depth of a circuit represents
essentially the minimum amount of time used by an implementation of
the circuit, provided that gates acting on disjoint sets of qubits can
be performed in parallel.

\begin{figure}
  \begin{center}
    \setlength{\unitlength}{3947sp}%
\begingroup\makeatletter\ifx\SetFigFont\undefined%
\gdef\SetFigFont#1#2#3#4#5{%
  \reset@font\fontsize{#1}{#2pt}%
  \fontfamily{#3}\fontseries{#4}\fontshape{#5}%
  \selectfont}%
\fi\endgroup%
\begin{picture}(1824,1374)(2989,-2323)
\thinlines
{\color[rgb]{0,0,0}\put(4051,-1411){\framebox(450,450){$b$}}
}%
{\color[rgb]{0,0,0}\put(3001,-2086){\line( 1, 0){300}}
}%
{\color[rgb]{0,0,0}\put(3751,-1936){\line( 1, 0){300}}
}%
{\color[rgb]{0,0,0}\put(4501,-1936){\line( 1, 0){300}}
}%
{\color[rgb]{0,0,0}\put(3001,-1486){\line( 1, 0){1800}}
}%
{\color[rgb]{0,0,0}\put(4051,-2161){\framebox(450,450){$d$}}
}%
{\color[rgb]{0,0,0}\put(3301,-2311){\framebox(450,750){$c$}}
}%
{\color[rgb]{0,0,0}\put(3001,-1186){\line( 1, 0){300}}
}%
{\color[rgb]{0,0,0}\put(3301,-1411){\framebox(450,450){$a$}}
}%
{\color[rgb]{0,0,0}\put(3751,-1186){\line( 1, 0){300}}
}%
{\color[rgb]{0,0,0}\put(3001,-1786){\line( 1, 0){300}}
}%
\end{picture}%
  \end{center}
  \caption[An example quantum circuit]{An example quantum circuit.}
 \label{compl-fig-circuit}
\end{figure}
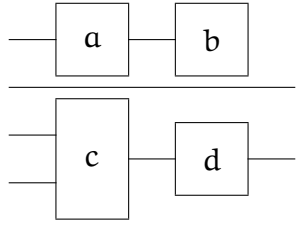

As an example of size and depth, the
circuit in Figure~\ref{compl-fig-circuit} takes four qubits as input,
produces two qubits as output, and has size four and depth two.
In this figure, and in all the circuit diagrams that will appear in
the thesis,
the gates in the circuit are represented by boxes and the edges by the
lines connecting them.  The edges in circuits are directed, but by
convention, the edges in the diagrams appearing here are always
directed from left to right, so that time flows left to right during
the evaluation of the circuit.  The circuit in the example maps four
qubits to two qubits.  This can be thought of as a channel
$\Phi \in \transform{H,K}$, where $\dm{H} =
2^4 = 16$ and $\dm{K} = 2^2 = 4$.  An alternate view of this circuit
is, where $\mathcal{A}$ is a Hilbert space of dimension two, as a
transformation in $\transform{A^{\tprod \text{$4$}}, A^{\tprod \text{$2$}}}$.  We will
take whichever view is most convenient, as these two sets of
transformations are isomorphic.

As a further notational convenience, throughout the thesis, a circuit
$C$ will be identified with the transformation $C \in \transform{H,K}$
that it implements, so that for a state $\rho \in \density{H}$, the
state $C(\rho)$ is the output of the circuit when executed on the
input state $\rho$.  Each circuit specifies exactly one
transformation, but the converse is not true: any quantum channel has
(infinitely) many circuit implementations, and so given only a
transformation, a circuit implementing it will need to be carefully
constructed.  In most of the cases we will encounter this is not difficult to do.

Circuits as a model for quantum computation are significantly easier
to work with than earlier models of computation, such as the model of quantum
Turing machines introduced in~\cite{BernsteinV97}.  When the transformations
that can be used as gates are restricted to the right set, it is known
that the circuit model of computation is equivalent to the quantum
Turing machine model~\cite{Yao93}.  For this reason, we will use the
circuit model of quantum computation, though for most of the results
in this thesis the exact model of computation will not be important.

\subsection{Unitary circuits}\label{compl-scn-circuit-unitary}

The most commonly used model of quantum circuits is the unitary
circuit model.  In this model every gate implements a unitary
transformation on one or two qubits.  Unitarity implies that all the
gates of this model have the same number of input and output qubits.

\begin{figure}
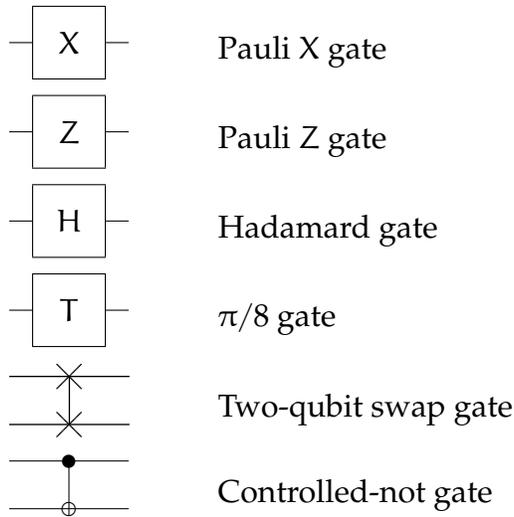

  \begin{center}
    \begin{tabular}{m{0.15\linewidth}m{0.6\linewidth}}
        \input{figures/gate-X.latex} & Pauli $X$ gate \\
        \input{figures/gate-Z.latex} & Pauli $Z$ gate \\
        \input{figures/gate-H.latex} & Hadamard gate \\      
       \input{figures/gate-T.latex} & $\pi/8$ gate \\
        \input{figures/gate-swap.epic} & Two-qubit swap gate \\
        \input{figures/gate-cnot.latex} & Controlled-not gate
   \end{tabular}
    \caption[Gates in the unitary circuit model]{Gates in the unitary
      circuit model.  The Pauli $X$ and $Z$ gates, and the swap
      gate are not required for universality, but they are included for
      convenience.  The $\pi/8$ gate is needed for
      universality, but will not be used in any of the circuits
      constructed outside of this section.}
    \label{compl-fig-unitary-gates}
  \end{center}
\end{figure}

It is known that any unitary operation can be approximately
represented using only a finite set of one- and two-qubit unitary
gates.  The set of gates we will use is given in
Figure~\ref{compl-fig-unitary-gates}, and the proof that it is
(approximately) universal is due to Boykin et al.~\cite{BoykinM+00}.
Different universality proofs for slightly different sets of gates
can be found in~\cite{Shor96, KnillL+98, AharonovB-O08}.  An excellent
overview of this and other universal sets of gates, as well a proof
that the gate set used here is universal can be found
in~\cite{NielsenC00}.

\nomenclature[GX]{\setlength{\unitlength}{3947sp}%
\begingroup\makeatletter\ifx\SetFigFont\undefined%
\gdef\SetFigFont#1#2#3#4#5{%
  \reset@font\fontsize{#1}{#2pt}%
  \fontfamily{#3}\fontseries{#4}\fontshape{#5}%
  \selectfont}%
\fi\endgroup%
\begin{picture}(774,474)(4189,2)
\thinlines
{\color[rgb]{0,0,0}\put(4351, 14){\framebox(450,450){$X$}}
}%
{\color[rgb]{0,0,0}\put(4351,239){\line(-1, 0){150}}
}%
{\color[rgb]{0,0,0}\put(4801,239){\line( 1, 0){150}}
}%
\end{picture}%
}{\raisebox{10pt}{Pauli $X$ gate}}%
\nomenclature[GZ]{\setlength{\unitlength}{3947sp}%
\begingroup\makeatletter\ifx\SetFigFont\undefined%
\gdef\SetFigFont#1#2#3#4#5{%
  \reset@font\fontsize{#1}{#2pt}%
  \fontfamily{#3}\fontseries{#4}\fontshape{#5}%
  \selectfont}%
\fi\endgroup%
\begin{picture}(774,474)(4189,2)
\thinlines
{\color[rgb]{0,0,0}\put(4351, 14){\framebox(450,450){$Z$}}
}%
{\color[rgb]{0,0,0}\put(4351,239){\line(-1, 0){150}}
}%
{\color[rgb]{0,0,0}\put(4801,239){\line( 1, 0){150}}
}%
\end{picture}%
}{\raisebox{10pt}{Pauli $Z$ gate}}%
\nomenclature[GH]{\setlength{\unitlength}{3947sp}%
\begingroup\makeatletter\ifx\SetFigFont\undefined%
\gdef\SetFigFont#1#2#3#4#5{%
  \reset@font\fontsize{#1}{#2pt}%
  \fontfamily{#3}\fontseries{#4}\fontshape{#5}%
  \selectfont}%
\fi\endgroup%
\begin{picture}(774,474)(4189,2)
\thinlines
{\color[rgb]{0,0,0}\put(4351, 14){\framebox(450,450){$H$}}
}%
{\color[rgb]{0,0,0}\put(4351,239){\line(-1, 0){150}}
}%
{\color[rgb]{0,0,0}\put(4801,239){\line( 1, 0){150}}
}%
\end{picture}%
}{\raisebox{10pt}{Hadamard gate}}%
\nomenclature[GW]{\setlength{\unitlength}{3947sp}%
\begingroup\makeatletter\ifx\SetFigFont\undefined%
\gdef\SetFigFont#1#2#3#4#5{%
  \reset@font\fontsize{#1}{#2pt}%
  \fontfamily{#3}\fontseries{#4}\fontshape{#5}%
  \selectfont}%
\fi\endgroup%
\begin{picture}(774,624)(4189,-148)
\thinlines
{\color[rgb]{0,0,0}\put(4351,-136){\framebox(450,600){$W$}}
}%
{\color[rgb]{0,0,0}\put(4801,239){\line( 1, 0){150}}
}%
{\color[rgb]{0,0,0}\put(4351,389){\line(-1, 0){150}}
}%
{\color[rgb]{0,0,0}\put(4951,389){\line(-1, 0){150}}
}%
{\color[rgb]{0,0,0}\put(4951,314){\line(-1, 0){150}}
}%
{\color[rgb]{0,0,0}\put(4351,314){\line(-1, 0){150}}
}%
{\color[rgb]{0,0,0}\put(4351, 89){\line(-1, 0){150}}
}%
{\color[rgb]{0,0,0}\put(4351, 14){\line(-1, 0){150}}
}%
{\color[rgb]{0,0,0}\put(4351,-61){\line(-1, 0){150}}
}%
{\color[rgb]{0,0,0}\put(4951,-61){\line(-1, 0){150}}
}%
{\color[rgb]{0,0,0}\put(4951, 14){\line(-1, 0){150}}
}%
{\color[rgb]{0,0,0}\put(4951, 89){\line(-1, 0){150}}
}%
{\color[rgb]{0,0,0}\put(4351,239){\line(-1, 0){150}}
}%
\end{picture}%
}{\raisebox{14pt}{Multi-qubit swap gate $W(\ket a \ket b) = \ket b \ket a$}}%
\nomenclature[GSwap]{\hspace{-7pt}\input{figures/gate-swap.epic}}{\raisebox{10pt}{Two-qubit swap gate}}%
\nomenclature[GCont]{\setlength{\unitlength}{3947sp}%
\begingroup\makeatletter\ifx\SetFigFont\undefined%
\gdef\SetFigFont#1#2#3#4#5{%
  \reset@font\fontsize{#1}{#2pt}%
  \fontfamily{#3}\fontseries{#4}\fontshape{#5}%
  \selectfont}%
\fi\endgroup%
\begin{picture}(774,807)(4189,-223)
{\color[rgb]{0,0,0}\thinlines
\put(4576,539){\circle*{76}}
}%
{\color[rgb]{0,0,0}\put(4351,-211){\framebox(450,450){$U$}}
}%
{\color[rgb]{0,0,0}\put(4351, 14){\line(-1, 0){150}}
}%
{\color[rgb]{0,0,0}\put(4951, 14){\line(-1, 0){150}}
}%
{\color[rgb]{0,0,0}\put(4201,539){\line( 1, 0){750}}
}%
{\color[rgb]{0,0,0}\put(4576,539){\line( 0,-1){300}}
}%
\end{picture}%
}{\raisebox{24pt}{Controlled $U$ gate}}%
\nomenclature[GCnot]{\setlength{\unitlength}{3947sp}%
\begingroup\makeatletter\ifx\SetFigFont\undefined%
\gdef\SetFigFont#1#2#3#4#5{%
  \reset@font\fontsize{#1}{#2pt}%
  \fontfamily{#3}\fontseries{#4}\fontshape{#5}%
  \selectfont}%
\fi\endgroup%
\begin{picture}(774,399)(1114,-865)
{\color[rgb]{0,0,0}\thinlines
\put(1501,-511){\circle*{76}}
}%
{\color[rgb]{0,0,0}\put(1501,-811){\circle{76}}
}%
{\color[rgb]{0,0,0}\put(1501,-511){\line( 0,-1){342}}
}%
{\color[rgb]{0,0,0}\put(1126,-511){\line( 1, 0){750}}
}%
{\color[rgb]{0,0,0}\put(1126,-811){\line( 1, 0){750}}
}%
\end{picture}%
}{\raisebox{10pt}{Controlled-not gate}}%

\begin{figure}
  \begin{center}
   \input{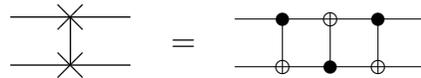} 
    \hspace{1ex}\raisebox{2ex}{$=$}\hspace{2ex}
   \raisebox{2pt}{\setlength{\unitlength}{3947sp}%
\begingroup\makeatletter\ifx\SetFigFont\undefined%
\gdef\SetFigFont#1#2#3#4#5{%
  \reset@font\fontsize{#1}{#2pt}%
  \fontfamily{#3}\fontseries{#4}\fontshape{#5}%
  \selectfont}%
\fi\endgroup%
\begin{picture}(1224,408)(889,-865)
\thinlines
{\color[rgb]{0,0,0}\put(901,-811){\line( 1, 0){1200}}
}%
{\color[rgb]{0,0,0}\put(1501,-511){\circle{76}}
}%
{\color[rgb]{0,0,0}\put(1501,-811){\line( 0, 1){342}}
}%
{\color[rgb]{0,0,0}\put(1201,-511){\circle*{76}}
}%
{\color[rgb]{0,0,0}\put(1201,-811){\circle{76}}
}%
{\color[rgb]{0,0,0}\put(1201,-511){\line( 0,-1){342}}
}%
{\color[rgb]{0,0,0}\put(1801,-511){\circle*{76}}
}%
{\color[rgb]{0,0,0}\put(1801,-811){\circle{76}}
}%
{\color[rgb]{0,0,0}\put(1801,-511){\line( 0,-1){342}}
}%
{\color[rgb]{0,0,0}\put(901,-511){\line( 1, 0){1200}}
}%
{\color[rgb]{0,0,0}\put(1501,-811){\circle*{76}}
}%
\end{picture}%
}
  \end{center}
  \caption[Simulation of the swap gate with three controlled-not
  gates]{Simulation of the swap gate with three controlled-not
    gates.}\label{compl-fig-swap-cnot}
\end{figure}

We have seen a few of these gates before: the Pauli $X$ and $Z$ and
Hadamard gates simply apply the corresponding unitary operators to
the qubits they act on, where the operators $X,Z,$ and $H$ are as
defined in Section~\ref{intro-scn-qinfo}.  A few of these operators
are new.  In matrix form, the swap, controlled-not ($CNOT$), and
$\pi/8$ ($T$) gates
are given by
\begin{align*}
  W &= \begin{pmatrix} 1 & 0 & 0 & 0 \\ 0 & 0 & 1 & 0 \\ 0 & 1 & 0 & 0 \\ 0 & 0 & 0 & 1 \end{pmatrix},&
 CNOT  &= \begin{pmatrix} 1 & 0 & 0 & 0 \\ 0 & 1 & 0 & 0 \\ 0 & 0 & 0 & 1 \\ 0 & 0 & 1 & 0 \end{pmatrix},&
 T &= \begin{pmatrix} 1 & 0 \\ 0 & e^{i \pi / 4} \end{pmatrix}.
\end{align*}
For the sake of convenience we have added a few gates to the circuit
model.  The Pauli $X$ and $Z$ and swap gates are not needed for a
universal set of gates.  They can, however, be constructed exactly
using gates from the standard set.  Two of these gates are
simple to build from the standard set
\begin{align*}
  Z &= T^4, &
  X &= HZH = H T^4 H.
\end{align*}
The third unnecessary gate, the swap gate, can be implemented in the standard
model using no gates at all!  This is because the unitary operation
that swaps two qubits can be introduced into a circuit by simply
redirecting the edges in the underlying graph.  
In many practical models of computation it is nontrivial to connect
gates together in arbitrary directed graphs.  One such model is the
nearest-neighbour model, where a qubit can only interact with the
qubits immediately adjacent to it.  This model (with polynomial
depth and size overhead) can simulate the more permissive model if the swap gate
included in the circuit model, since the required qubits for any
operation can always be swapped together.  The swap gate can be
implemented as a series of three controlled-not gates in this model,
as shown in Figure~\ref{compl-fig-swap-cnot}.
We will use $W$ to represent the gate that swaps the input
systems, so that $W \ket a \ket b = \ket b \ket a$ even when the
dimension of the systems to be swapped is larger than two.  This gate
can be implemented using several two-qubit swap gates.
The introduction of these gates does not change the circuit model, as they
can be exactly implemented in the model of Boykin et
al.~\cite{BoykinM+00} using a constant number of gates.

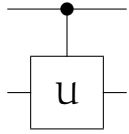
\begin{figure}
  \begin{center}
    \setlength{\unitlength}{3947sp}%
\begingroup\makeatletter\ifx\SetFigFont\undefined%
\gdef\SetFigFont#1#2#3#4#5{%
  \reset@font\fontsize{#1}{#2pt}%
  \fontfamily{#3}\fontseries{#4}\fontshape{#5}%
  \selectfont}%
\fi\endgroup%
\begin{picture}(774,807)(4189,-223)
{\color[rgb]{0,0,0}\thinlines
\put(4576,539){\circle*{76}}
}%
{\color[rgb]{0,0,0}\put(4351,-211){\framebox(450,450){$U$}}
}%
{\color[rgb]{0,0,0}\put(4351, 14){\line(-1, 0){150}}
}%
{\color[rgb]{0,0,0}\put(4951, 14){\line(-1, 0){150}}
}%
{\color[rgb]{0,0,0}\put(4201,539){\line( 1, 0){750}}
}%
{\color[rgb]{0,0,0}\put(4576,539){\line( 0,-1){300}}
}%
\end{picture}%
  \end{center}
  \caption[Controlled-$U$ gate]{Controlled-$U$ gate.}\label{compl-fig-contU}
\end{figure}

It is often very useful in a quantum circuit to control the
application of some unitary operation based on the value of an
additional qubit.  For a unitary $U$, this is the operation commonly
known as a controlled-$U$ gate.  We have already encountered one such
gate: the controlled-not gate in the standard model is exactly the
controlled application of the Pauli $X$ gate.  Given a unitary
operation $U$ (as a circuit), the controlled-$U$ operation
is the unitary operation 
that applies $U$ if the control qubit is $\ket 1$ and does nothing if
the control qubit is in the $\ket 0$ state.  The representation of
this gate in the circuit model is shown in
Figure~\ref{compl-fig-contU}.
For a unitary $U \in \unitary{H}$, this gate is represented in block matrix form as
\begin{equation*}
  \Lambda(U) =
  \begin{pmatrix}
    \identity{H} & 0 \\
    0 & U
  \end{pmatrix}.
\end{equation*}
Given a circuit for $U$, it is simple to construct one for the
controlled-$U$ operation.  Each gate in the circuit can be replaced by
a controlled version, so that all of the gates are applied (i.e.\ $U$
is applied) or none of the gates are applied.  The controlled versions
of each gate in the basis need to be constructed, but these are
guaranteed to (approximately) exist by the fact that we are using a
complete basis of gates.  Notice, however, that this construction may
add significantly to the depth of the circuit: the single control
qubit is used many times.  A more depth-conscious construction is
presented in Section~\ref{compl-scn-circuit-log}.

\subsection{Mixed-state circuits}

Circuits in the unitary model can (approximately) represent any unitary computation.
If these circuits are allowed access to ancillary qubits in a known
pure state, then they can perform any efficient quantum computation.  There
is, however,  a drawback to this model: unitary circuits cannot simulate
an arbitrary quantum channel.  This is because a general completely
positive and trace preserving operation may discard information, it
may make measurements in the middle of a computation, or it may
introduce ancillary qubits in a mixed state.  Many of these operations
are impossible to implement in the unitary model.

\begin{figure}
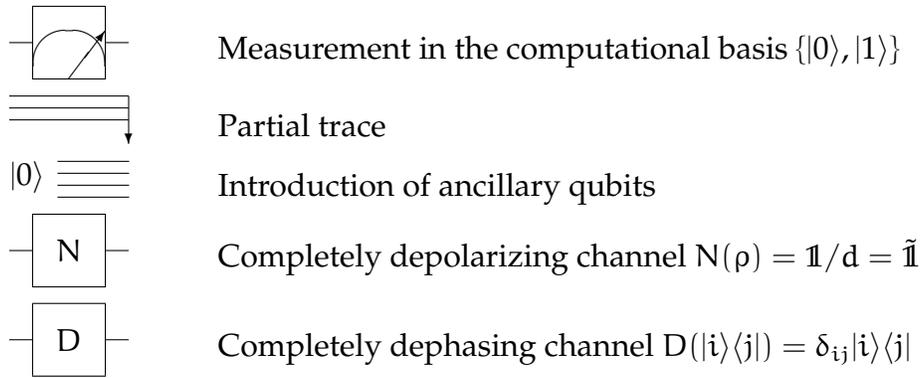

  \begin{center}
    \begin{tabular}{m{0.15\linewidth}m{0.6\linewidth}}
        \setlength{\unitlength}{3947sp}%
\begingroup\makeatletter\ifx\SetFigFont\undefined%
\gdef\SetFigFont#1#2#3#4#5{%
  \reset@font\fontsize{#1}{#2pt}%
  \fontfamily{#3}\fontseries{#4}\fontshape{#5}%
  \selectfont}%
\fi\endgroup%
\begin{picture}(774,474)(4189,2)
\thinlines
{\color[rgb]{0,0,0}\put(4576, 89){\oval(450,450)[tr]}
\put(4576, 89){\oval(450,450)[tl]}
}%
{\color[rgb]{0,0,0}\put(4576, 14){\vector( 3, 4){225}}
}%
{\color[rgb]{0,0,0}\put(4351, 14){\framebox(450,450){}}
}%
{\color[rgb]{0,0,0}\put(4351,239){\line(-1, 0){150}}
}%
{\color[rgb]{0,0,0}\put(4801,239){\line( 1, 0){150}}
}%
\end{picture}%
        & Measurement in the computational basis $\{\ket 0, \ket 1 \}$  \\

        \input{figures/gate-traceout.latex}
        & Partial trace \\
        
        \setlength{\unitlength}{3947sp}%
\begingroup\makeatletter\ifx\SetFigFont\undefined%
\gdef\SetFigFont#1#2#3#4#5{%
  \reset@font\fontsize{#1}{#2pt}%
  \fontfamily{#3}\fontseries{#4}\fontshape{#5}%
  \selectfont}%
\fi\endgroup%
\begin{picture}(777,249)(286,-973)
\thinlines
{\color[rgb]{0,0,0}\put(601,-961){\line( 1, 0){450}}
}%
{\color[rgb]{0,0,0}\put(601,-886){\line( 1, 0){450}}
}%
{\color[rgb]{0,0,0}\put(601,-811){\line( 1, 0){450}}
}%
{\color[rgb]{0,0,0}\put(601,-736){\line( 1, 0){450}}
}%
\put(301,-886){\makebox(0,0)[lb]{\smash{{\SetFigFont{12}{14.4}{\familydefault}{\mddefault}{\updefault}{\color[rgb]{0,0,0}$\ket 0$}%
}}}}
\end{picture}%
        & Introduction of ancillary qubits \\

        \input{figures/gate-N.latex}
        & Completely depolarizing channel $N(\rho) = \id/d = \nidentity{}$ \\

        \input{figures/gate-D.latex}
        & Completely dephasing channel 
        $D(\ket i \bra j) = \delta_{ij} \ket i \bra j$ \\
    \end{tabular}
    \caption[Non-unitary gates in the mixed state circuit model]{Non-unitary gates in the mixed state circuit model.}
    \label{compl-fig-nonunitary-gates}
  \end{center}
\end{figure}

For this reason we will use the mixed-state circuit model of Aharonov,
Kitaev, and Nisan~\cite{AharonovK+98}.  Circuits in this model can
(approximately) represent any quantum channel.  This can be thought of
as a probabilistic model of quantum computation, as the state of the
computation can be a mixed state, whereas the unitary model can be
thought of as deterministic computation, since the state during the
computation is always pure.  The gates available in this model are the
standard gates from the unitary model, as well as the additional gates
shown in Figure~\ref{compl-fig-nonunitary-gates}.
\nomenclature[Gmeas]{\setlength{\unitlength}{3947sp}%
\begingroup\makeatletter\ifx\SetFigFont\undefined%
\gdef\SetFigFont#1#2#3#4#5{%
  \reset@font\fontsize{#1}{#2pt}%
  \fontfamily{#3}\fontseries{#4}\fontshape{#5}%
  \selectfont}%
\fi\endgroup%
\begin{picture}(774,474)(4189,2)
\thinlines
{\color[rgb]{0,0,0}\put(4576, 89){\oval(450,450)[tr]}
\put(4576, 89){\oval(450,450)[tl]}
}%
{\color[rgb]{0,0,0}\put(4576, 14){\vector( 3, 4){225}}
}%
{\color[rgb]{0,0,0}\put(4351, 14){\framebox(450,450){}}
}%
{\color[rgb]{0,0,0}\put(4351,239){\line(-1, 0){150}}
}%
{\color[rgb]{0,0,0}\put(4801,239){\line( 1, 0){150}}
}%
\end{picture}%
}{\raisebox{10pt}{Measurement in the computational basis $\{\ket 0, \ket 1 \}$}}%
\nomenclature[GN]{\setlength{\unitlength}{3947sp}%
\begingroup\makeatletter\ifx\SetFigFont\undefined%
\gdef\SetFigFont#1#2#3#4#5{%
  \reset@font\fontsize{#1}{#2pt}%
  \fontfamily{#3}\fontseries{#4}\fontshape{#5}%
  \selectfont}%
\fi\endgroup%
\begin{picture}(774,474)(4189,2)
\thinlines
{\color[rgb]{0,0,0}\put(4351, 14){\framebox(450,450){$N$}}
}%
{\color[rgb]{0,0,0}\put(4351,239){\line(-1, 0){150}}
}%
{\color[rgb]{0,0,0}\put(4801,239){\line( 1, 0){150}}
}%
\end{picture}%
}{\raisebox{10pt}{Completely depolarizing channel $N(\rho) = \id/d$}}%
\nomenclature[GD]{\setlength{\unitlength}{3947sp}%
\begingroup\makeatletter\ifx\SetFigFont\undefined%
\gdef\SetFigFont#1#2#3#4#5{%
  \reset@font\fontsize{#1}{#2pt}%
  \fontfamily{#3}\fontseries{#4}\fontshape{#5}%
  \selectfont}%
\fi\endgroup%
\begin{picture}(774,474)(4189,2)
\thinlines
{\color[rgb]{0,0,0}\put(4351, 14){\framebox(450,450){$D$}}
}%
{\color[rgb]{0,0,0}\put(4351,239){\line(-1, 0){150}}
}%
{\color[rgb]{0,0,0}\put(4801,239){\line( 1, 0){150}}
}%
\end{picture}%
}{\raisebox{10pt}{Completely
    dephasing channel $D(\ket i \bra j) = \delta_{ij} \ket i \bra j$}}%
\nomenclature[Gptr]{\setlength{\unitlength}{3947sp}%
\begingroup\makeatletter\ifx\SetFigFont\undefined%
\gdef\SetFigFont#1#2#3#4#5{%
  \reset@font\fontsize{#1}{#2pt}%
  \fontfamily{#3}\fontseries{#4}\fontshape{#5}%
  \selectfont}%
\fi\endgroup%
\begin{picture}(804,324)(4189,377)
\thinlines
{\color[rgb]{0,0,0}\put(4201,689){\line( 1, 0){750}}
}%
{\color[rgb]{0,0,0}\put(4201,614){\line( 1, 0){750}}
}%
{\color[rgb]{0,0,0}\put(4201,539){\line( 1, 0){750}}
}%
{\color[rgb]{0,0,0}\put(4951,689){\vector( 0,-1){300}}
}%
\end{picture}%
}{\raisebox{10pt}{Partial trace}}%
\nomenclature[Ganc]{\setlength{\unitlength}{3947sp}%
\begingroup\makeatletter\ifx\SetFigFont\undefined%
\gdef\SetFigFont#1#2#3#4#5{%
  \reset@font\fontsize{#1}{#2pt}%
  \fontfamily{#3}\fontseries{#4}\fontshape{#5}%
  \selectfont}%
\fi\endgroup%
\begin{picture}(777,249)(286,-973)
\thinlines
{\color[rgb]{0,0,0}\put(601,-961){\line( 1, 0){450}}
}%
{\color[rgb]{0,0,0}\put(601,-886){\line( 1, 0){450}}
}%
{\color[rgb]{0,0,0}\put(601,-811){\line( 1, 0){450}}
}%
{\color[rgb]{0,0,0}\put(601,-736){\line( 1, 0){450}}
}%
\put(301,-886){\makebox(0,0)[lb]{\smash{{\SetFigFont{12}{14.4}{\familydefault}{\mddefault}{\updefault}{\color[rgb]{0,0,0}$\ket 0$}%
}}}}
\end{picture}%
}{\raisebox{4pt}{Introduction of ancillary qubits in $\ket 0$ state}}%

\begin{figure}
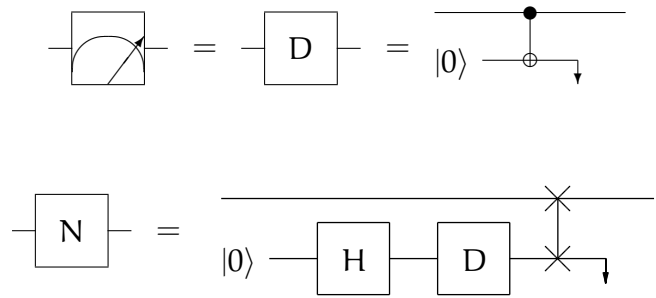

  \begin{center}
   \setlength{\unitlength}{3947sp}%
\begingroup\makeatletter\ifx\SetFigFont\undefined%
\gdef\SetFigFont#1#2#3#4#5{%
  \reset@font\fontsize{#1}{#2pt}%
  \fontfamily{#3}\fontseries{#4}\fontshape{#5}%
  \selectfont}%
\fi\endgroup%
\begin{picture}(774,474)(4189,2)
\thinlines
{\color[rgb]{0,0,0}\put(4576, 89){\oval(450,450)[tr]}
\put(4576, 89){\oval(450,450)[tl]}
}%
{\color[rgb]{0,0,0}\put(4576, 14){\vector( 3, 4){225}}
}%
{\color[rgb]{0,0,0}\put(4351, 14){\framebox(450,450){}}
}%
{\color[rgb]{0,0,0}\put(4351,239){\line(-1, 0){150}}
}%
{\color[rgb]{0,0,0}\put(4801,239){\line( 1, 0){150}}
}%
\end{picture}% 
    \hspace{1ex}\raisebox{2ex}{$=$}\hspace{1ex}
    \input{figures/gate-D.latex}
    \hspace{1ex}\raisebox{2ex}{$=$}\hspace{1ex}
    \setlength{\unitlength}{3947sp}%
\begingroup\makeatletter\ifx\SetFigFont\undefined%
\gdef\SetFigFont#1#2#3#4#5{%
  \reset@font\fontsize{#1}{#2pt}%
  \fontfamily{#3}\fontseries{#4}\fontshape{#5}%
  \selectfont}%
\fi\endgroup%
\begin{picture}(1227,507)(4936,-973)
\thinlines
{\color[rgb]{0,0,0}\put(4951,-511){\line( 1, 0){1200}}
}%
{\color[rgb]{0,0,0}\put(5551,-811){\circle{76}}
}%
{\color[rgb]{0,0,0}\put(5551,-511){\line( 0,-1){342}}
}%
{\color[rgb]{0,0,0}\put(5551,-511){\circle*{76}}
}%
{\color[rgb]{0,0,0}\put(5251,-811){\line( 1, 0){600}}
}%
{\color[rgb]{0,0,0}\put(5851,-811){\vector( 0,-1){150}}
}%
\put(4951,-886){\makebox(0,0)[lb]{\smash{{\SetFigFont{12}{14.4}{\rmdefault}{\mddefault}{\updefault}{\color[rgb]{0,0,0}$\ket 0$}%
}}}}
\end{picture}%

    \vspace{3em}

    \raisebox{2ex}{\input{figures/gate-N.latex}}
    \hspace{1ex}\raisebox{4ex}{$=$}\hspace{1ex}
   \input{figures/compl-depol-sim.epic}
 \end{center}
 \caption[Simulations of three of the gates in the mixed-state
  circuit model]{Simulations of the measurement gate, the completely
    dephasing gate, and the completely depolarizing gate with the
    other gates in the circuit model.}
  \label{compl-fig-nonunitary}
\end{figure}

As in the unitary model, we have included a few unnecessary gates for
the sake of convenience.  The only two gates that are actually
required are the gate that introduces ancillary qubits in the $\ket 0$
state and the gate that traces out a qubit.  The gate that makes a
measurement in the computational basis can be implemented using an
ancillary qubit and a controlled not gate, as shown in
Figure~\ref{compl-fig-nonunitary}.  
The output of this measurement
gate is viewed as a mixed quantum state in the following way.  If the
measurement outcome is 0 with probability $p$ and 1 with probability
$1-p$, the outcome of the measurement gate is $p \ket 0 \bra 0 + (1-p)
\ket 1 \bra 1$.  This density matrix is diagonal, and so it may be
thought of as a classical probability distribution, but there is no
loss in generality in encoding this distribution as a mixed quantum
state.  Controlled operations will function in exactly the same way
regardless of whether they are controlled classically or by the
density matrix corresponding to the same probability distribution.
This measurement gate, as it turns out, performs exactly the same
transformation as the decoherence gate $D$ included in the gate set.
The decoherence gate in Figure~\ref{compl-fig-nonunitary-gates} applies
to systems of an arbitrary number of qubits, but this operation has
the property that when applied individually to a number of qubits, the
result is exactly the same as if it had been applied to all of them at
once.  The completely depolarizing channel also has this property, and
so it is sufficient to give an implementation for the channels acting only on a
single qubit, as is done in Figure~\ref{compl-fig-nonunitary}.

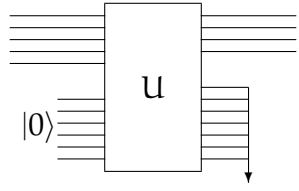
\begin{figure}
  \begin{center}
    \setlength{\unitlength}{3947sp}%
\begingroup\makeatletter\ifx\SetFigFont\undefined%
\gdef\SetFigFont#1#2#3#4#5{%
  \reset@font\fontsize{#1}{#2pt}%
  \fontfamily{#3}\fontseries{#4}\fontshape{#5}%
  \selectfont}%
\fi\endgroup%
\begin{picture}(1824,1149)(589,-1198)
\put(676,-886){\makebox(0,0)[lb]{\smash{{\SetFigFont{12}{14.4}{\rmdefault}{\mddefault}{\updefault}{\color[rgb]{0,0,0}$\ket 0$}%
}}}}
\thinlines
{\color[rgb]{0,0,0}\put(2401,-349){\line( 0, 1){  0}}
}%
{\color[rgb]{0,0,0}\put(2401,-417){\line( 0, 1){  0}}
}%
{\color[rgb]{0,0,0}\put(2401,-486){\line( 0, 1){  0}}
}%
{\color[rgb]{0,0,0}\put(2401,-830){\line( 0, 1){  0}}
}%
{\color[rgb]{0,0,0}\put(2401,-899){\line( 0, 1){  0}}
}%
{\color[rgb]{0,0,0}\put(2401,-967){\line( 0, 1){  0}}
}%
{\color[rgb]{0,0,0}\put(2401,-1036){\line( 0, 1){  0}}
}%
{\color[rgb]{0,0,0}\put(2401,-211){\line( 0, 1){  0}}
}%
{\color[rgb]{0,0,0}\put(2401,-761){\line( 0, 1){  0}}
}%
{\color[rgb]{0,0,0}\put(2401,-692){\line( 0, 1){  0}}
}%
{\color[rgb]{0,0,0}\put(901,-811){\line( 1, 0){300}}
}%
{\color[rgb]{0,0,0}\put(901,-886){\line( 1, 0){300}}
}%
{\color[rgb]{0,0,0}\put(901,-961){\line( 1, 0){300}}
}%
{\color[rgb]{0,0,0}\put(901,-1036){\line( 1, 0){300}}
}%
{\color[rgb]{0,0,0}\put(901,-736){\line( 1, 0){300}}
}%
{\color[rgb]{0,0,0}\put(901,-661){\line( 1, 0){300}}
}%
{\color[rgb]{0,0,0}\put(601,-211){\line( 1, 0){600}}
}%
{\color[rgb]{0,0,0}\put(601,-286){\line( 1, 0){600}}
}%
{\color[rgb]{0,0,0}\put(601,-361){\line( 1, 0){600}}
}%
{\color[rgb]{0,0,0}\put(601,-436){\line( 1, 0){600}}
}%
{\color[rgb]{0,0,0}\put(601,-136){\line( 1, 0){600}}
}%
{\color[rgb]{0,0,0}\put(1801,-811){\line( 1, 0){300}}
}%
{\color[rgb]{0,0,0}\put(1801,-886){\line( 1, 0){300}}
}%
{\color[rgb]{0,0,0}\put(1801,-961){\line( 1, 0){300}}
}%
{\color[rgb]{0,0,0}\put(1801,-1036){\line( 1, 0){300}}
}%
{\color[rgb]{0,0,0}\put(1801,-736){\line( 1, 0){300}}
}%
{\color[rgb]{0,0,0}\put(1801,-661){\line( 1, 0){300}}
}%
{\color[rgb]{0,0,0}\put(1801,-586){\line( 1, 0){300}}
}%
{\color[rgb]{0,0,0}\put(1201,-1111){\framebox(600,1050){$U$}}
}%
{\color[rgb]{0,0,0}\put(1801,-211){\line( 1, 0){600}}
}%
{\color[rgb]{0,0,0}\put(1801,-286){\line( 1, 0){600}}
}%
{\color[rgb]{0,0,0}\put(1801,-361){\line( 1, 0){600}}
}%
{\color[rgb]{0,0,0}\put(1801,-136){\line( 1, 0){600}}
}%
{\color[rgb]{0,0,0}\put(2101,-586){\vector( 0,-1){600}}
}%
{\color[rgb]{0,0,0}\put(2401,-280){\line( 0, 1){  0}}
}%
\end{picture}%
  \end{center}
  \caption[Simulating a channel with a unitary circuit]{The unitary operation $U$ simulates the admissible operation $\Phi$.}
  \label{compl-fig-unitary-simulation}
\end{figure}

Since the standard model of quantum
computation is inherently probabilistic, as we will see in
Section~\ref{compl-scn-complexity-classes}, it is not hard to show that the
mixed-state model is equivalent in computational power to the unitary
circuit model~\cite{AharonovK+98}.
The central idea behind the equivalence of the unitary and mixed-state
models is the fact that any quantum channel can be implemented in
\emph{Stinespring form},
 \index{circuit!Stinespring form}%
which is the introduction of ancillary qubits first, followed by a
unitary operation on the now larger space, and finally tracing out any
qubits that are not part of the output.  For a channel $\Phi \in
\transform{H,K}$, this is exactly the Stinespring representation
\[ \Phi(\rho) = \ptr{B} U (\rho \tprod \ket 0 \bra 0) U^* \]
where the unitary $U$ is implemented by a unitary circuit.
An example of this is illustrated in Figure~\ref{compl-fig-unitary-simulation}.
This equivalence is noted in~\cite{AharonovK+98}, making use of what
is known about quantum channels in the physics
literature~\cite{Stinespring55, HellwigK70}.

Despite this computational equivalence these two models are not
identical.  The distinguishability problem discussed in
Chapter~\ref{chap-distinguishability}
seems to be significantly harder than the distinguishability
problem for unitary computations.  If this is not the case, there are
unexpected consequences in complexity theory~\cite{Vyalyi03}.
Even stronger evidence is provided by the close images problem studied
in Chapter~\ref{chap-close-images}.  This problem involves
determining the distance between the images of two transformations.
If these transformations are unitary, their images always intersect,
rendering the problem trivial.
For these reasons, the standard model of quantum computations used
throughout the thesis is the mixed-state circuit model.  Any quantum
channel that is given as the input of a computational problem will be
in the form of a classical description of a circuit in this model.

\subsection{Short quantum circuits}\label{compl-scn-circuit-log}

A significant challenge in the experimental realization of a quantum
computation is the need to keep a quantum system from interacting with
the environment.  The decoherence caused by these interactions
in practice provides a time limit for the computation.  One way to
ameliorate this difficulty is to find low-depth circuits that solve
the problems we are interested in.

Short quantum circuits have been found for several important problems,
such as the approximate quantum Fourier transform~\cite{CleveW00} and
encoding and decoding operations for many error correcting
codes~\cite{MooreN02}.  These examples show the significant power of
circuits that have depth logarithmic in the size of the circuits.
More evidence for the power of short circuits is provided by Terhal
and DiVincenzo~\cite{TerhalD04} and improved by Fenner et
al.~\cite{FennerG+05} who show that exactly computing the acceptance
probabilities for constant-depth quantum circuits is as hard as
simulating general quantum computation.  Fenner et al. also show that,
under certain restrictions,
the acceptance probabilities for these circuits can be efficiently
approximated.

The purpose of this section is to give a construction for the
controlled version of a log-depth circuit on $n$ qubits
that results in a depth $O(\log n)$ circuit.  It is not immediately
clear how a controlled operation on $n$ qubits, such as a
controlled-swap operation can be performed in depth logarithmic in
$n$.  The straightforward implementation outlined in
Section~\ref{compl-scn-circuit-unitary} requires using one control
qubit to control each of the gates in the operation, resulting in a
linear depth circuit.
Moore and Nilsson~\cite{MooreN02} use a construction from
reversible computing to reduce the depth of this technique.
\begin{proposition}[Moore and Nilsson~\cite{MooreN02}]\label{compl-prop-log-control}
  Any log-depth operation on $n$ qubits controlled by one qubit
  can be implemented in $O(\log n)$ depth with $O(n)$ ancillary qubits.
\end{proposition}
\index{log-depth controlled operations}%
Moore and Nilsson prove this only for the case of constant-depth
operations, but the
proof technique used also applies to the log-depth
case.  They prove this proposition using a tree of $\log n$
controlled-not operations to `duplicate' the control qubit onto $n$
ancillary qubits.  These copies only capture the information in the
computational basis, but this is the exactly same information that is
used by the controlled gates.  These extra control qubits
can are then used to control the remaining operations, with each
control qubit used a logarithmic number of times.  
Finally, the tree of controlled-not
operations is reversed to clean up the ancillary qubits so that they
can be traced out without decohering the system.
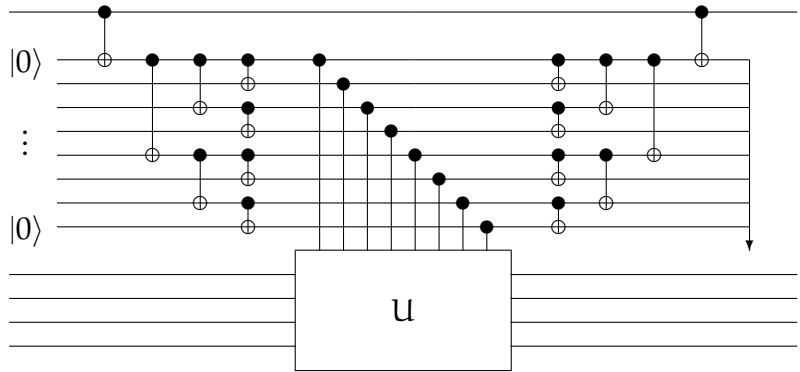
\begin{figure}
  \begin{center}
    \setlength{\unitlength}{3947sp}%
\begingroup\makeatletter\ifx\SetFigFont\undefined%
\gdef\SetFigFont#1#2#3#4#5{%
  \reset@font\fontsize{#1}{#2pt}%
  \fontfamily{#3}\fontseries{#4}\fontshape{#5}%
  \selectfont}%
\fi\endgroup%
\begin{picture}(4977,2307)(886,-2773)
\put(976,-1411){\makebox(0,0)[lb]{\smash{{\SetFigFont{12}{14.4}{\familydefault}{\mddefault}{\updefault}{\color[rgb]{0,0,0}$\vdots$}%
}}}}
{\color[rgb]{0,0,0}\thinlines
\put(1801,-811){\circle*{76}}
}%
{\color[rgb]{0,0,0}\put(1801,-811){\line( 0,-1){642}}
}%
{\color[rgb]{0,0,0}\put(2101,-811){\circle*{76}}
}%
{\color[rgb]{0,0,0}\put(2101,-1411){\circle*{76}}
}%
{\color[rgb]{0,0,0}\put(2101,-1111){\circle{76}}
}%
{\color[rgb]{0,0,0}\put(2101,-1711){\circle{76}}
}%
{\color[rgb]{0,0,0}\put(2101,-811){\line( 0,-1){342}}
}%
{\color[rgb]{0,0,0}\put(2101,-1411){\line( 0,-1){342}}
}%
{\color[rgb]{0,0,0}\put(2401,-811){\circle*{76}}
}%
{\color[rgb]{0,0,0}\put(2401,-1111){\circle*{76}}
}%
{\color[rgb]{0,0,0}\put(2401,-1411){\circle*{76}}
}%
{\color[rgb]{0,0,0}\put(2401,-1711){\circle*{76}}
}%
{\color[rgb]{0,0,0}\put(2401,-1861){\circle{76}}
}%
{\color[rgb]{0,0,0}\put(2401,-1561){\circle{76}}
}%
{\color[rgb]{0,0,0}\put(2401,-1261){\circle{76}}
}%
{\color[rgb]{0,0,0}\put(2401,-961){\circle{76}}
}%
{\color[rgb]{0,0,0}\put(2401,-811){\line( 0,-1){192}}
}%
{\color[rgb]{0,0,0}\put(2401,-1111){\line( 0,-1){192}}
}%
{\color[rgb]{0,0,0}\put(2401,-1411){\line( 0,-1){192}}
}%
{\color[rgb]{0,0,0}\put(2401,-1711){\line( 0,-1){192}}
}%
{\color[rgb]{0,0,0}\put(1501,-511){\circle*{76}}
}%
{\color[rgb]{0,0,0}\put(1501,-811){\circle{76}}
}%
{\color[rgb]{0,0,0}\put(1501,-511){\line( 0,-1){342}}
}%
{\color[rgb]{0,0,0}\put(5251,-511){\circle*{76}}
}%
{\color[rgb]{0,0,0}\put(5251,-811){\circle{76}}
}%
{\color[rgb]{0,0,0}\put(5251,-511){\line( 0,-1){342}}
}%
{\color[rgb]{0,0,0}\put(4351,-811){\circle*{76}}
}%
{\color[rgb]{0,0,0}\put(4351,-1111){\circle*{76}}
}%
{\color[rgb]{0,0,0}\put(4351,-1411){\circle*{76}}
}%
{\color[rgb]{0,0,0}\put(4351,-1711){\circle*{76}}
}%
{\color[rgb]{0,0,0}\put(4351,-1861){\circle{76}}
}%
{\color[rgb]{0,0,0}\put(4351,-1561){\circle{76}}
}%
{\color[rgb]{0,0,0}\put(4351,-1261){\circle{76}}
}%
{\color[rgb]{0,0,0}\put(4351,-961){\circle{76}}
}%
{\color[rgb]{0,0,0}\put(4351,-811){\line( 0,-1){192}}
}%
{\color[rgb]{0,0,0}\put(4351,-1111){\line( 0,-1){192}}
}%
{\color[rgb]{0,0,0}\put(4351,-1411){\line( 0,-1){192}}
}%
{\color[rgb]{0,0,0}\put(4351,-1711){\line( 0,-1){192}}
}%
{\color[rgb]{0,0,0}\put(4651,-811){\circle*{76}}
}%
{\color[rgb]{0,0,0}\put(4651,-1411){\circle*{76}}
}%
{\color[rgb]{0,0,0}\put(4651,-1111){\circle{76}}
}%
{\color[rgb]{0,0,0}\put(4651,-1711){\circle{76}}
}%
{\color[rgb]{0,0,0}\put(4651,-811){\line( 0,-1){342}}
}%
{\color[rgb]{0,0,0}\put(4651,-1411){\line( 0,-1){342}}
}%
{\color[rgb]{0,0,0}\put(4951,-1411){\circle{76}}
}%
{\color[rgb]{0,0,0}\put(4951,-811){\circle*{76}}
}%
{\color[rgb]{0,0,0}\put(4951,-811){\line( 0,-1){642}}
}%
{\color[rgb]{0,0,0}\put(2851,-811){\circle*{76}}
}%
{\color[rgb]{0,0,0}\put(3001,-961){\circle*{76}}
}%
{\color[rgb]{0,0,0}\put(3151,-1111){\circle*{76}}
}%
{\color[rgb]{0,0,0}\put(3301,-1261){\circle*{76}}
}%
{\color[rgb]{0,0,0}\put(3451,-1411){\circle*{76}}
}%
{\color[rgb]{0,0,0}\put(3601,-1561){\circle*{76}}
}%
{\color[rgb]{0,0,0}\put(3751,-1711){\circle*{76}}
}%
{\color[rgb]{0,0,0}\put(3901,-1861){\circle*{76}}
}%
{\color[rgb]{0,0,0}\put(1201,-1261){\line( 1, 0){2250}}
}%
{\color[rgb]{0,0,0}\put(1201,-1111){\line( 1, 0){2250}}
}%
{\color[rgb]{0,0,0}\put(1201,-961){\line( 1, 0){2250}}
}%
{\color[rgb]{0,0,0}\put(1201,-811){\line( 1, 0){2250}}
}%
{\color[rgb]{0,0,0}\put(1201,-1561){\line( 1, 0){2250}}
}%
{\color[rgb]{0,0,0}\put(1201,-1711){\line( 1, 0){2250}}
}%
{\color[rgb]{0,0,0}\put(1201,-511){\line( 1, 0){2250}}
}%
{\color[rgb]{0,0,0}\put(901,-511){\line( 1, 0){300}}
}%
{\color[rgb]{0,0,0}\put(1201,-1411){\line( 1, 0){2250}}
}%
{\color[rgb]{0,0,0}\put(1201,-1861){\line( 1, 0){2250}}
}%
{\color[rgb]{0,0,0}\put(901,-2161){\line( 1, 0){1800}}
}%
{\color[rgb]{0,0,0}\put(901,-2311){\line( 1, 0){1800}}
}%
{\color[rgb]{0,0,0}\put(901,-2461){\line( 1, 0){1800}}
}%
{\color[rgb]{0,0,0}\put(901,-2611){\line( 1, 0){1800}}
}%
{\color[rgb]{0,0,0}\put(2701,-2761){\framebox(1350,750){$U$}}
}%
{\color[rgb]{0,0,0}\put(3451,-961){\line( 1, 0){2100}}
}%
{\color[rgb]{0,0,0}\put(3451,-1111){\line( 1, 0){2100}}
}%
{\color[rgb]{0,0,0}\put(3451,-1261){\line( 1, 0){2100}}
}%
{\color[rgb]{0,0,0}\put(3451,-1411){\line( 1, 0){2100}}
}%
{\color[rgb]{0,0,0}\put(3451,-1561){\line( 1, 0){2100}}
}%
{\color[rgb]{0,0,0}\put(3451,-1711){\line( 1, 0){2100}}
}%
{\color[rgb]{0,0,0}\put(3451,-1861){\line( 1, 0){2100}}
}%
{\color[rgb]{0,0,0}\put(3451,-511){\line( 1, 0){2400}}
}%
{\color[rgb]{0,0,0}\put(3451,-811){\line( 1, 0){2100}}
}%
{\color[rgb]{0,0,0}\put(4051,-2161){\line( 1, 0){1800}}
}%
{\color[rgb]{0,0,0}\put(4051,-2311){\line( 1, 0){1800}}
}%
{\color[rgb]{0,0,0}\put(4051,-2461){\line( 1, 0){1800}}
}%
{\color[rgb]{0,0,0}\put(4051,-2611){\line( 1, 0){1800}}
}%
{\color[rgb]{0,0,0}\put(2851,-811){\line( 0,-1){1200}}
}%
{\color[rgb]{0,0,0}\put(3001,-961){\line( 0,-1){1050}}
}%
{\color[rgb]{0,0,0}\put(3151,-1111){\line( 0,-1){900}}
}%
{\color[rgb]{0,0,0}\put(3301,-1261){\line( 0,-1){750}}
}%
{\color[rgb]{0,0,0}\put(3451,-1411){\line( 0,-1){600}}
}%
{\color[rgb]{0,0,0}\put(3601,-1561){\line( 0,-1){450}}
}%
{\color[rgb]{0,0,0}\put(3751,-1711){\line( 0,-1){300}}
}%
{\color[rgb]{0,0,0}\put(3901,-1861){\line( 0,-1){150}}
}%
{\color[rgb]{0,0,0}\put(5551,-811){\vector( 0,-1){1200}}
}%
\put(901,-886){\makebox(0,0)[lb]{\smash{{\SetFigFont{12}{14.4}{\familydefault}{\mddefault}{\updefault}{\color[rgb]{0,0,0}$\ket 0$}%
}}}}
\put(901,-1936){\makebox(0,0)[lb]{\smash{{\SetFigFont{12}{14.4}{\familydefault}{\mddefault}{\updefault}{\color[rgb]{0,0,0}$\ket 0$}%
}}}}
{\color[rgb]{0,0,0}\put(1801,-1411){\circle{76}}
}%
\end{picture}%
  \end{center}
  \caption{Log-depth implementation of controlled operation on $n$
    qubits}
  \label{compl-fig-log-control}
\end{figure}
This procedure is demonstrated in Figure~\ref{compl-fig-log-control}.
This implies, as an example, that the
$2n$-qubit controlled swap gate can be implemented in depth $O(\log
n)$.  This will be critical to the construction used in Chapter~\ref{chap-close-images}.

If the unbounded fan-out gate is allowed into the standard basis of gates, then
the depth overhead added in this construction can be reduced to a constant.
This gate performs a controlled-not operation from one control qubit to
any number of target qubits in one computational step.  This gate is
not in the standard basis for mixed-state quantum computing: it
requires a linear number of gates and a logarithmic depth circuit to
implement in the standard gate model.
Fan-out in classical circuits is simply the operation that copies the
value from one bit to several other bits, and is often included in the
standard circuit model.
When such a gate is included in the usual quantum circuit models, many
tasks become much simpler.  As an example, this gate allows operations
such as sorting, phase estimation, and the quantum Fourier transform
to be approximated with constant depth circuits~\cite{HoyerS05}.  This
gate will not generally be included in the model, but some of the
results in the thesis can be strengthened when it is.

To see how the scheme for implementing controlled
operations can be implemented in constant depth using this gate,
notice that tree structure of
Figure~\ref{compl-fig-log-control} can be replaced with a single fan-out
gate.
This allows $n$ `copies' of the control qubit to be created, which can
then be used to control each of the $n$ operations.
A final application of the fan-out gate to restore the ancillary
qubits to the $\ket 0$ state.
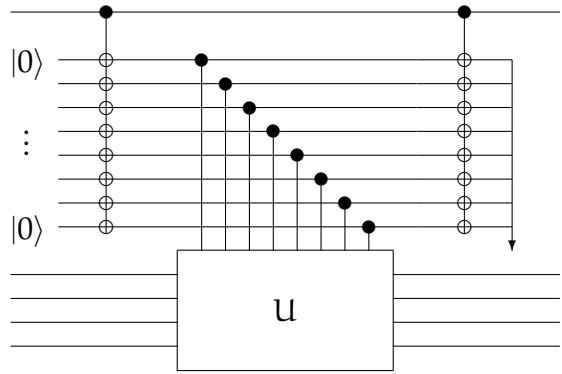
\begin{figure}
  \begin{center}
    \setlength{\unitlength}{3947sp}%
\begingroup\makeatletter\ifx\SetFigFont\undefined%
\gdef\SetFigFont#1#2#3#4#5{%
  \reset@font\fontsize{#1}{#2pt}%
  \fontfamily{#3}\fontseries{#4}\fontshape{#5}%
  \selectfont}%
\fi\endgroup%
\begin{picture}(3477,2307)(886,-2773)
\put(976,-1411){\makebox(0,0)[lb]{\smash{{\SetFigFont{12}{14.4}{\familydefault}{\mddefault}{\updefault}{\color[rgb]{0,0,0}$\vdots$}%
}}}}
{\color[rgb]{0,0,0}\thinlines
\put(1501,-1111){\circle{76}}
}%
{\color[rgb]{0,0,0}\put(1501,-1711){\circle{76}}
}%
{\color[rgb]{0,0,0}\put(1501,-1861){\circle{76}}
}%
{\color[rgb]{0,0,0}\put(1501,-1561){\circle{76}}
}%
{\color[rgb]{0,0,0}\put(1501,-1261){\circle{76}}
}%
{\color[rgb]{0,0,0}\put(1501,-961){\circle{76}}
}%
{\color[rgb]{0,0,0}\put(1501,-511){\circle*{76}}
}%
{\color[rgb]{0,0,0}\put(1501,-811){\circle{76}}
}%
{\color[rgb]{0,0,0}\put(1501,-1711){\line( 0,-1){192}}
}%
{\color[rgb]{0,0,0}\put(1501,-511){\line( 0,-1){1200}}
}%
{\color[rgb]{0,0,0}\put(901,-2161){\line( 1, 0){1050}}
}%
{\color[rgb]{0,0,0}\put(901,-2311){\line( 1, 0){1050}}
}%
{\color[rgb]{0,0,0}\put(901,-2461){\line( 1, 0){1050}}
}%
{\color[rgb]{0,0,0}\put(901,-2611){\line( 1, 0){1050}}
}%
{\color[rgb]{0,0,0}\put(2101,-811){\circle*{76}}
}%
{\color[rgb]{0,0,0}\put(2251,-961){\circle*{76}}
}%
{\color[rgb]{0,0,0}\put(2401,-1111){\circle*{76}}
}%
{\color[rgb]{0,0,0}\put(2551,-1261){\circle*{76}}
}%
{\color[rgb]{0,0,0}\put(2701,-1411){\circle*{76}}
}%
{\color[rgb]{0,0,0}\put(2851,-1561){\circle*{76}}
}%
{\color[rgb]{0,0,0}\put(3001,-1711){\circle*{76}}
}%
{\color[rgb]{0,0,0}\put(3151,-1861){\circle*{76}}
}%
{\color[rgb]{0,0,0}\put(1951,-2761){\framebox(1350,750){$U$}}
}%
{\color[rgb]{0,0,0}\put(2101,-811){\line( 0,-1){1200}}
}%
{\color[rgb]{0,0,0}\put(2251,-961){\line( 0,-1){1050}}
}%
{\color[rgb]{0,0,0}\put(2401,-1111){\line( 0,-1){900}}
}%
{\color[rgb]{0,0,0}\put(2551,-1261){\line( 0,-1){750}}
}%
{\color[rgb]{0,0,0}\put(2701,-1411){\line( 0,-1){600}}
}%
{\color[rgb]{0,0,0}\put(2851,-1561){\line( 0,-1){450}}
}%
{\color[rgb]{0,0,0}\put(3001,-1711){\line( 0,-1){300}}
}%
{\color[rgb]{0,0,0}\put(3151,-1861){\line( 0,-1){150}}
}%
{\color[rgb]{0,0,0}\put(3301,-2161){\line( 1, 0){1050}}
}%
{\color[rgb]{0,0,0}\put(3301,-2311){\line( 1, 0){1050}}
}%
{\color[rgb]{0,0,0}\put(3301,-2461){\line( 1, 0){1050}}
}%
{\color[rgb]{0,0,0}\put(3301,-2611){\line( 1, 0){1050}}
}%
{\color[rgb]{0,0,0}\put(3751,-1411){\circle{76}}
}%
{\color[rgb]{0,0,0}\put(3751,-1111){\circle{76}}
}%
{\color[rgb]{0,0,0}\put(3751,-1711){\circle{76}}
}%
{\color[rgb]{0,0,0}\put(3751,-1861){\circle{76}}
}%
{\color[rgb]{0,0,0}\put(3751,-1561){\circle{76}}
}%
{\color[rgb]{0,0,0}\put(3751,-1261){\circle{76}}
}%
{\color[rgb]{0,0,0}\put(3751,-961){\circle{76}}
}%
{\color[rgb]{0,0,0}\put(3751,-511){\circle*{76}}
}%
{\color[rgb]{0,0,0}\put(3751,-811){\circle{76}}
}%
{\color[rgb]{0,0,0}\put(3751,-1711){\line( 0,-1){192}}
}%
{\color[rgb]{0,0,0}\put(3751,-511){\line( 0,-1){1200}}
}%
{\color[rgb]{0,0,0}\put(3451,-961){\line( 1, 0){600}}
}%
{\color[rgb]{0,0,0}\put(3451,-1111){\line( 1, 0){600}}
}%
{\color[rgb]{0,0,0}\put(3451,-1261){\line( 1, 0){600}}
}%
{\color[rgb]{0,0,0}\put(3451,-1411){\line( 1, 0){600}}
}%
{\color[rgb]{0,0,0}\put(3451,-1561){\line( 1, 0){600}}
}%
{\color[rgb]{0,0,0}\put(3451,-1711){\line( 1, 0){600}}
}%
{\color[rgb]{0,0,0}\put(3451,-1861){\line( 1, 0){600}}
}%
{\color[rgb]{0,0,0}\put(3451,-811){\line( 1, 0){600}}
}%
{\color[rgb]{0,0,0}\put(1201,-1261){\line( 1, 0){2250}}
}%
{\color[rgb]{0,0,0}\put(1201,-1111){\line( 1, 0){2250}}
}%
{\color[rgb]{0,0,0}\put(1201,-961){\line( 1, 0){2250}}
}%
{\color[rgb]{0,0,0}\put(1201,-811){\line( 1, 0){2250}}
}%
{\color[rgb]{0,0,0}\put(1201,-1561){\line( 1, 0){2250}}
}%
{\color[rgb]{0,0,0}\put(1201,-1711){\line( 1, 0){2250}}
}%
{\color[rgb]{0,0,0}\put(1201,-511){\line( 1, 0){2250}}
}%
{\color[rgb]{0,0,0}\put(901,-511){\line( 1, 0){300}}
}%
{\color[rgb]{0,0,0}\put(1201,-1411){\line( 1, 0){2250}}
}%
{\color[rgb]{0,0,0}\put(1201,-1861){\line( 1, 0){2250}}
}%
{\color[rgb]{0,0,0}\put(3451,-511){\line( 1, 0){900}}
}%
{\color[rgb]{0,0,0}\put(4051,-811){\vector( 0,-1){1200}}
}%
\put(901,-886){\makebox(0,0)[lb]{\smash{{\SetFigFont{12}{14.4}{\familydefault}{\mddefault}{\updefault}{\color[rgb]{0,0,0}$\ket 0$}%
}}}}
\put(901,-1936){\makebox(0,0)[lb]{\smash{{\SetFigFont{12}{14.4}{\familydefault}{\mddefault}{\updefault}{\color[rgb]{0,0,0}$\ket 0$}%
}}}}
{\color[rgb]{0,0,0}\put(1501,-1411){\circle{76}}
}%
\end{picture}%
  \end{center}
  \caption[Constant depth implementation of controlled operation on
    $n$ qubits]{Constant depth implementation of controlled operation on $n$
    qubits using the unbounded fan-out gate.}
  \label{compl-fig-const-control}
\end{figure}
\index{constant-depth controlled operations}%
This is demonstrated in Figure~\ref{compl-fig-const-control}, which
implies the following proposition.
\begin{proposition}\label{compl-prop-const-control}
  If the unbounded fan-out gate is in the basis of gates, any constant
  depth operation on $n$ qubits controlled by one qubit can be
  implemented in $O(1)$ depth with $O(n)$ ancillary qubits.
\end{proposition}

\section{Quantum complexity classes}\label{compl-scn-complexity-classes}

\begin{figure}
  \begin{center}
    \setlength{\unitlength}{3947sp}%
\begingroup\makeatletter\ifx\SetFigFont\undefined%
\gdef\SetFigFont#1#2#3#4#5{%
  \reset@font\fontsize{#1}{#2pt}%
  \fontfamily{#3}\fontseries{#4}\fontshape{#5}%
  \selectfont}%
\fi\endgroup%
\begin{picture}(1380,2044)(811,-1409)
\put(1726,464){\makebox(0,0)[lb]{\smash{{\SetFigFont{12}{14.4}{\familydefault}{\mddefault}{\updefault}{\color[rgb]{0,0,0}\class{EXP}}%
}}}}
{\color[rgb]{0,0,0}\thinlines
\put(2101,-811){\circle*{76}}
}%
{\color[rgb]{0,0,0}\put(1201,-811){\circle*{76}}
}%
{\color[rgb]{0,0,0}\put(1651, 89){\circle*{76}}
}%
{\color[rgb]{0,0,0}\put(1651,539){\circle*{76}}
}%
{\color[rgb]{0,0,0}\put(1651,-361){\circle*{76}}
}%
{\color[rgb]{0,0,0}\put(1651,-1261){\line( 1, 1){450}}
}%
{\color[rgb]{0,0,0}\put(1651,-1261){\line(-1, 1){450}}
}%
{\color[rgb]{0,0,0}\put(1651, 89){\line( 0, 1){450}}
}%
{\color[rgb]{0,0,0}\put(1201,-811){\line( 1, 1){450}}
}%
{\color[rgb]{0,0,0}\put(2101,-811){\line(-1, 1){450}}
}%
{\color[rgb]{0,0,0}\put(1651,-361){\line( 0, 1){450}}
}%
\put(1726,-1336){\makebox(0,0)[lb]{\smash{{\SetFigFont{12}{14.4}{\familydefault}{\mddefault}{\updefault}{\color[rgb]{0,0,0}\class{P}}%
}}}}
\put(2176,-886){\makebox(0,0)[lb]{\smash{{\SetFigFont{12}{14.4}{\familydefault}{\mddefault}{\updefault}{\color[rgb]{0,0,0}\class{BQP}}%
}}}}
\put(1726,-436){\makebox(0,0)[lb]{\smash{{\SetFigFont{12}{14.4}{\familydefault}{\mddefault}{\updefault}{\color[rgb]{0,0,0}\class{QMA}}%
}}}}
\put(826,-886){\makebox(0,0)[lb]{\smash{{\SetFigFont{12}{14.4}{\familydefault}{\mddefault}{\updefault}{\color[rgb]{0,0,0}\class{NP}}%
}}}}
\put(1726, 14){\makebox(0,0)[lb]{\smash{{\SetFigFont{12}{14.4}{\familydefault}{\mddefault}{\updefault}{\color[rgb]{0,0,0}\class{QIP}}%
}}}}
{\color[rgb]{0,0,0}\put(1651,-1261){\circle*{76}}
}%
\end{picture}%
  \end{center}
  \caption[Known relationships between complexity classes]{Known relationships
    between the major quantum and classical complexity classes.
    Classes are contained in the classes written above them.  Only the
    containment $\class{P} \subsetneq \class{EXP}$ is known to be
    proper.}\label{compl-fig-classes}
\end{figure}
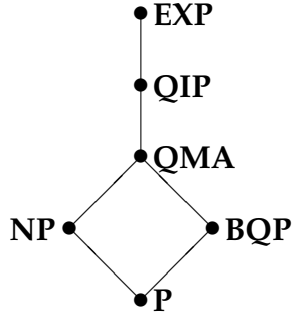

This section provides a brief overview of the quantum complexity
classes related to the topic of this thesis.
Many of the technical details related to the definitions of these classes
are omitted, as a detailed understanding of complexity theory is not
essential to the results that follow.  
For a more complete reference, see the recent survey of
Watrous~\cite{Watrous09complexity}.  The known relationships between the classes
discussed here and some of the more well known classical complexity
classes are illustrated in Figure~\ref{compl-fig-classes}.

\nomenclature[CP]{\class{P}}{classicaly efficiently solvable problems}%
\nomenclature[CPSPACE]{\class{PSPACE}}{classicaly solvable problems in
  polynomial space}%
\nomenclature[CEXP]{\class{EXP}}{classicaly solvable problems in exponential time}%
\nomenclature[CNP]{\class{NP}}{classicaly efficiently verifiable problems}%

\class{BQP}, defined in~\cite{BernsteinV97}, is the quantum complexity
class of primary importance. This class is informally the set of all
decision problems that are efficiently solvable with a quantum
computer. As quantum computation can involve measurements that have
inherently probabilistic outcomes, the quantum computation that solves
a problem in \class{BQP} is permitted to fail with some bounded
probability. This probability can be made arbitrarily small by using
the standard trick of repeating the computation several times in
parallel and taking the majority. Error reduction for this class is
exactly as for probabilistic classical computations: this is because
the inputs and outputs to the decision problems are classical strings
that may be copied any number of times.

More formally, \class{BQP} is the set of all
languages $L$ for which there exists a uniform family $Q$ of polynomial
size quantum circuits, one for each input length, such that
\index{BQP@\class{BQP}}%
\nomenclature[CBQP]{\class{BQP}}{quantum efficiently solvable problems}%
\begin{enumerate}
  \item if $x \in L$, then $\tr(\Pi Q(x)) \geq \frac{2}{3}$,
  \item if $x \not\in L$, then $\tr(\Pi Q(x)) \leq \frac{1}{3}$,
\end{enumerate}
where $\Pi$ is the projector onto the subspace where the first
output qubit of $Q$ is $\ket 1$, i.e.\ the projector onto the accepting
subspace for the circuit.  The error bounds $2/3$ and $1/3$ here are
not significant: they can be replaced with any $a > b$ that have at
least an inverse polynomial gap between them (in the size of the
input string $x$) as noted above.

Several \class{BQP}-complete promise problems are known.  The most
famous of these is probably the approximation of the Jones polynomial,
which is in \class{BQP} by an algorithm of Aharonov et
al.~\cite{AharonovJ+06} (or by earlier works of Freedman et
al.~\cite{FreedmanK+02}) and is complete for \class{BQP} by a result of
Freedman et al.~\cite{FreedmanL+02}.  These problems give an important
method for the study of quantum computation that is not necessarily
connected to quantum information.

Extending the definition of \class{BQP} to include a single message
from a computationally unbounded prover results in the class
\class{QMA}, which is the quantum analogue of \class{NP}. This concept
was first considered in~\cite{Knill96}, first defined
in~\cite{Kitaev99}, and first studied in~\cite{Watrous00}. \class{QMA}
is the class of all problems that can be verified by a polynomial-time
quantum verifier with access to a quantum proof. This proof is a
quantum state on a polynomial number of qubits and may depend on the
input. More formally, a language $L$ is in \class{QMA} if there is a
family of circuits $Q$ such that
\index{QMA@\class{QMA}}%
\nomenclature[CQMA]{\class{QMA}}{quantum efficiently verifiable problems}%
\begin{enumerate}
  \item if $x \in L$, then there exists $\rho$ such that $\tr(\Pi
    Q(x,\rho)) \geq \frac{2}{3}$,
  \item if $x \not\in L$, then for any $\rho$, $\tr(\Pi Q(x,\rho))
    \leq \frac{1}{3}$,
\end{enumerate}
where once again $\Pi$ is the projector onto the accepting subspace
of the output of $Q$.  As in the case of \class{BQP} the error
parameters $2/3$ and $1/3$ are not significant.  Replacing these with
$a,b$ such that $\abs{a - b}$ is at least inverse polynomial in $n$ is also possible in this
case, though the argument is not simple~\cite{KitaevS+02, MarriottW05}.

Similar to \class{BQP}, the class \class{QMA} has complete promise
problems.  The simplest of these is the 2-local Hamiltonian problem,
which is informally the quantum version of the satisfiability problem
for unitary circuits with gates of constant size.  A formal description
of this problem, as well as a proof that the 5-local Hamiltonian
problem is \class{QMA}-complete can be found in~\cite{KitaevS+02}.
The improvement of this result to the 2-local case is due to Kempe,
Kitaev, and Regev~\cite{KempeK+06}.

Extending \class{BQP} further by allowing multiple rounds of
interaction with a prover results in the complexity class \class{QIP},
first defined in~\cite{Watrous03}.  This class is
the quantum analogue of the classical class \class{IP}, which is equal
to the more familiar class \class{PSPACE}~\cite{LundF+92, Shamir92} of
problems solvable with a polynomial amount of space.  A recent result
has also shown that $\class{QIP} = \class{PSPACE}$~\cite{JainJ+09}, 
resolving a major open problem in quantum computational complexity.

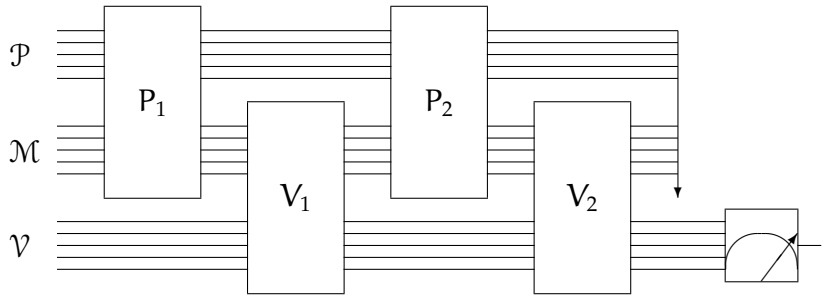
\begin{figure}
  \begin{center}
    \setlength{\unitlength}{3947sp}%
\begingroup\makeatletter\ifx\SetFigFont\undefined%
\gdef\SetFigFont#1#2#3#4#5{%
  \reset@font\fontsize{#1}{#2pt}%
  \fontfamily{#3}\fontseries{#4}\fontshape{#5}%
  \selectfont}%
\fi\endgroup%
\begin{picture}(5127,1824)(586,-2173)
\put(601,-736){\makebox(0,0)[lb]{\smash{{\SetFigFont{12}{14.4}{\rmdefault}{\mddefault}{\updefault}{\color[rgb]{0,0,0}$\mathcal{P}$}%
}}}}
\thinlines
{\color[rgb]{0,0,0}\put(1801,-661){\line( 1, 0){1200}}
}%
{\color[rgb]{0,0,0}\put(1801,-511){\line( 1, 0){1200}}
}%
{\color[rgb]{0,0,0}\put(1801,-586){\line( 1, 0){1200}}
}%
{\color[rgb]{0,0,0}\put(1801,-736){\line( 1, 0){1200}}
}%
{\color[rgb]{0,0,0}\put(1801,-1411){\line( 1, 0){300}}
}%
{\color[rgb]{0,0,0}\put(2101,-1261){\line(-1, 0){300}}
}%
{\color[rgb]{0,0,0}\put(1801,-1111){\line( 1, 0){300}}
}%
{\color[rgb]{0,0,0}\put(1801,-1186){\line( 1, 0){300}}
}%
{\color[rgb]{0,0,0}\put(1801,-1336){\line( 1, 0){300}}
}%
{\color[rgb]{0,0,0}\put(901,-811){\line( 1, 0){300}}
}%
{\color[rgb]{0,0,0}\put(901,-661){\line( 1, 0){300}}
}%
{\color[rgb]{0,0,0}\put(901,-511){\line( 1, 0){300}}
}%
{\color[rgb]{0,0,0}\put(901,-586){\line( 1, 0){300}}
}%
{\color[rgb]{0,0,0}\put(901,-736){\line( 1, 0){300}}
}%
{\color[rgb]{0,0,0}\put(901,-1411){\line( 1, 0){300}}
}%
{\color[rgb]{0,0,0}\put(901,-1261){\line( 1, 0){300}}
}%
{\color[rgb]{0,0,0}\put(901,-1111){\line( 1, 0){300}}
}%
{\color[rgb]{0,0,0}\put(901,-1186){\line( 1, 0){300}}
}%
{\color[rgb]{0,0,0}\put(901,-1336){\line( 1, 0){300}}
}%
{\color[rgb]{0,0,0}\put(901,-2011){\line( 1, 0){1200}}
}%
{\color[rgb]{0,0,0}\put(901,-1861){\line( 1, 0){1200}}
}%
{\color[rgb]{0,0,0}\put(901,-1711){\line( 1, 0){1200}}
}%
{\color[rgb]{0,0,0}\put(901,-1786){\line( 1, 0){1200}}
}%
{\color[rgb]{0,0,0}\put(901,-1936){\line( 1, 0){1200}}
}%
{\color[rgb]{0,0,0}\put(2701,-1411){\line( 1, 0){300}}
}%
{\color[rgb]{0,0,0}\put(3001,-1261){\line(-1, 0){300}}
}%
{\color[rgb]{0,0,0}\put(2701,-1111){\line( 1, 0){300}}
}%
{\color[rgb]{0,0,0}\put(2701,-1186){\line( 1, 0){300}}
}%
{\color[rgb]{0,0,0}\put(2701,-1336){\line( 1, 0){300}}
}%
{\color[rgb]{0,0,0}\put(3601,-1411){\line( 1, 0){300}}
}%
{\color[rgb]{0,0,0}\put(3901,-1261){\line(-1, 0){300}}
}%
{\color[rgb]{0,0,0}\put(3601,-1111){\line( 1, 0){300}}
}%
{\color[rgb]{0,0,0}\put(3601,-1186){\line( 1, 0){300}}
}%
{\color[rgb]{0,0,0}\put(3601,-1336){\line( 1, 0){300}}
}%
{\color[rgb]{0,0,0}\put(4501,-1411){\line( 1, 0){300}}
}%
{\color[rgb]{0,0,0}\put(4801,-1261){\line(-1, 0){300}}
}%
{\color[rgb]{0,0,0}\put(4501,-1111){\line( 1, 0){300}}
}%
{\color[rgb]{0,0,0}\put(4501,-1186){\line( 1, 0){300}}
}%
{\color[rgb]{0,0,0}\put(4501,-1336){\line( 1, 0){300}}
}%
{\color[rgb]{0,0,0}\put(4501,-2011){\line( 1, 0){600}}
}%
{\color[rgb]{0,0,0}\put(5101,-1861){\line(-1, 0){600}}
}%
{\color[rgb]{0,0,0}\put(4501,-1711){\line( 1, 0){600}}
}%
{\color[rgb]{0,0,0}\put(4501,-1786){\line( 1, 0){600}}
}%
{\color[rgb]{0,0,0}\put(4501,-1936){\line( 1, 0){600}}
}%
{\color[rgb]{0,0,0}\put(2701,-2011){\line( 1, 0){1200}}
}%
{\color[rgb]{0,0,0}\put(2701,-1861){\line( 1, 0){1200}}
}%
{\color[rgb]{0,0,0}\put(2701,-1711){\line( 1, 0){1200}}
}%
{\color[rgb]{0,0,0}\put(2701,-1786){\line( 1, 0){1200}}
}%
{\color[rgb]{0,0,0}\put(2701,-1936){\line( 1, 0){1200}}
}%
{\color[rgb]{0,0,0}\put(3601,-811){\line( 1, 0){1200}}
}%
{\color[rgb]{0,0,0}\put(3601,-661){\line( 1, 0){1200}}
}%
{\color[rgb]{0,0,0}\put(3601,-511){\line( 1, 0){1200}}
}%
{\color[rgb]{0,0,0}\put(3601,-586){\line( 1, 0){1200}}
}%
{\color[rgb]{0,0,0}\put(3601,-736){\line( 1, 0){1200}}
}%
{\color[rgb]{0,0,0}\put(5326,-2011){\oval(450,450)[tr]}
\put(5326,-2011){\oval(450,450)[tl]}
}%
{\color[rgb]{0,0,0}\put(5326,-2086){\vector( 3, 4){225}}
}%
{\color[rgb]{0,0,0}\put(5101,-2086){\framebox(450,450){}}
}%
{\color[rgb]{0,0,0}\put(5551,-1861){\line( 1, 0){150}}
}%
{\color[rgb]{0,0,0}\put(1201,-1561){\framebox(600,1200){$P_1$}}
}%
{\color[rgb]{0,0,0}\put(2101,-2161){\framebox(600,1200){$V_1$}}
}%
{\color[rgb]{0,0,0}\put(3001,-1561){\framebox(600,1200){$P_2$}}
}%
{\color[rgb]{0,0,0}\put(3901,-2161){\framebox(600,1200){$V_2$}}
}%
{\color[rgb]{0,0,0}\put(4801,-511){\vector( 0,-1){1050}}
}%
\put(601,-1936){\makebox(0,0)[lb]{\smash{{\SetFigFont{12}{14.4}{\rmdefault}{\mddefault}{\updefault}{\color[rgb]{0,0,0}$\mathcal{V}$}%
}}}}
\put(601,-1336){\makebox(0,0)[lb]{\smash{{\SetFigFont{12}{14.4}{\rmdefault}{\mddefault}{\updefault}{\color[rgb]{0,0,0}$\mathcal{M}$}%
}}}}
{\color[rgb]{0,0,0}\put(1801,-811){\line( 1, 0){1200}}
}%
\end{picture}%
  \end{center}
  \caption[A three message quantum interactive proof system]{A three
    message quantum interactive proof system.  The verifier's
    polynomial time transformations are $V_1$ and $V_2$ and the
    prover's transformations are given by $P_1$ and $P_2$. All
    messages are sent though the message space $\mathcal{M}$ and the
    verifier does not have access to the prover's private space
    $\mathcal{P}$.  At the end of the interaction, a measurement of
    the verifier's private space $\mathcal{V}$ determines the
    acceptance of the computation.  All three of these spaces start in
    the $\ket 0$ state.  No restrictions are made on the size of the
    space $\mathcal{P}$, but the spaces $\mathcal{M}$ and
    $\mathcal{V}$ do not contain more than a polynomial number of
    qubits.  The circuits $V_1, V_2, P_1, P_2$ may depend on the input
    $x$.  The circuits $V_1, V_2$ must be generated from the input $x$
    in polynomial time, but the circuits $P_1, and P_2$ are not so restricted.}
  \label{compl-fig-qip}
\end{figure}

As a more formal definition, for any language $L \in \class{QIP}$, there is
a polynomial time quantum algorithm $V$, known as the verifier, that
exchanges quantum messages with a prover $P$.  Both the prover and the
verifier receive the input string $x$ before the start of the
computation.  The verifier's algorithm $V$ must be generated from the
input string $x$ in polynomial time, but the prover's algorithm is not
constrained in this way.
Given a pair $(V,P)$
the verifier $V$ will accept that $x \in L$ with some probability after interacting
with the prover $P$.  An example of this interaction is shown in
Figure~\ref{compl-fig-qip}, with the Hilbert spaces available to each
party illustrated.
For any input $x$, the verifier $V$ in a \class{QIP} protocol
satisfies
\index{QIP@\class{QIP}}%
\nomenclature[CQIP]{\class{QIP}}{quantum efficiently interactively verifiable problems}%
\begin{enumerate}
  \item if $x \in L$, then there exists a prover $P$ such that, 
    $(V,P)$ accepts with probability at least $\frac{2}{3}$.
  \item if $x \not\in L$, then for any prover $P$, $(V,P)$ accepts with
    probability at most $\frac{1}{3}$.\label{compl-enum-fail}
\end{enumerate}

Once again, the exact parameters used in this definition are not
significant.  It is known how to use parallel repetition in this model of computation
for error reduction, so that as long as the probabilities are an inverse
polynomial apart and the probability of acceptance in
condition~\ref{compl-enum-fail} is nonzero, the resulting class of
problems does not change~\cite{KitaevW00}.

An interesting property of quantum interactive proof systems is that
any quantum interactive proof system can be simulated by one using
only three messages~\cite{KitaevW00}.  This is in contrast to the
classical case, where constant round proof systems seem to be
much weaker than polynomial round proof systems.  For this reason, we
may assume that any problem in \class{QIP} has a proof system as shown
in Figure~\ref{compl-fig-qip}, in which each of the prover and
verifier each perform exactly two transformations, with the verifier
acting last.

It is easy to see from the definitions that $\class{QMA} \subseteq
\class{QIP}$, as interactive proofs with three messages can only be
stronger than those using only one message.
It is expected that this containment is strict: if not, unexpected
things happen to classical complexity classes (it would imply that
$\class{PH} \subseteq \class{PP}$)~\cite{Vyalyi03}.
There is, however, no proof that this cannot happen, as it would
resolve a long-standing open problem complexity (showing that
\class{NP} is properly contained in \class{EXP}).
The case of two message quantum interactive proofs is even more
interesting, as quite little is known about this class.  It it
known that the class of problems two message proofs is contained in
\class{PSPACE}~\cite{JainU+09}, but this result has been subsumed by
the result that $\class{QIP} = \class{PSPACE}$ using similar
techniques~\cite{JainJ+09}.

The class \class{QIP} also has complete (promise) problems.
The \prob{Close Images} problem is the first such problem known.  This
problem is implicitly defined and shown to be complete for
\class{QIP} in~\cite{KitaevW00} where it is used to show that
$\class{QIP} \subseteq \class{EXP}$.  This problem is the focus of
Chapter~\ref{chap-close-images}, where a formal definition can be
found.

This thesis adds several new problems to the list of problems
that are complete for \class{QIP}.  \prob{Close Images} is effectively a
restatement of the acceptance conditions for a quantum interactive
proof system, as we will see in Section~\ref{clim-scn-intro}, and so
these new complete problems provide a method for studying \class{QIP}
that involve quantum information that are not strongly tied to the model of quantum interactive proof systems.

The three quantum complexity classes \class{BQP}, \class{QMA}, and
\class{QIP} are the only classes that will be encountered in this
thesis.  With the exception of Section~\ref{clim-scn-intro} where
\prob{Close Images} is shown to be complete for \class{QIP} from the
definition, it is not essential to have a deep understanding of these
definitions.  More important is to maintain the intuitive picture that
problems in \class{BQP} are easy, problems in \class{QMA} are difficult,
and problems in \class{QIP} are even more difficult.

%%% Local Variables: 
%%% mode: latex
%%% TeX-master: "thesis"
%%% End: 

\chapter{Measures for Quantum Information}\label{chap-meas}

Distances and other measures give a quantitative method to evaluate
how close two states are together, how mixed a state is, or how well a
quantum channel can be used to transmit information.  These are all
tasks that are central to the problems discussed in the thesis, and so
we introduce several different techniques for measuring these
quantities.  It is important that there are several such measures, as
most of these measures have an operational meaning that can help to
ground the otherwise abstract problems that we consider.

The primary quantities discussed in this chapter include the entropy,
the Schatten $p$-norms, and the trace norm on quantum states.  Also
included is an overview of some of the extensions of these quantities
to the case of channels.  A brief overview of the problems related to
the additivity of the classical capacity is also provided.

This chapter is largely a collection of these measures, with proofs of
the important properties that they have.  The results in
Sections~\ref{meas-scn-dnorm-diff} and~\ref{meas-scn-polarization},
are the product of joint work with John Watrous~\cite{RosgenW05}, and
the remainder of the results discussed here are not new and can
be found in several of the standard sources.

\minitoc

\section{Entropy}\label{meas-scn-entropy}

The entropy of a quantum state can be viewed as a measure of the amount of uncertainty
about the value of the state.  In support of this intuitive
picture, the entropy of a pure state is zero as this represents the
case where (in principle) complete knowledge of the state is present.
The other extreme is the completely mixed state $\nidentity{}$, where
nothing at all is known about the system, which corresponds to the maximum entropy.

In the case of a classical probability distribution $p$, the entropy is
defined to be
\[ S( p ) = - \sum_x p(x) \log p(x), \]
where the logarithm is taken base two.
\nomenclature[alog]{$\log$}{base-two logarithm}%
The entropy was first used in an information theoretic context by
Shannon in 1948~\cite{Shannon48}, who derived it from axioms that
he felt that any such measure of uncertainty should satisfy.  By
convention $0 \log 0$ is defined to be $0$ in this equation.

This quantity has a generalization to quantum systems that was first
developed by von Neumann in 1927~\cite{vonNeumann27}.  This
version of the entropy, applied to a density operator $\rho$ with
eigenvalues $\lambda_i$, is given by
\begin{equation}\label{meas-eqn-entropy}
  S(\rho) = - \tr \rho \log \rho = - \sum_i \lambda_i \log \lambda_i.
\end{equation}
\index{entropy}%
\nomenclature[aS]{$S(\rho)$}{von Neumann entropy of the state $\rho$}%
This quantity is often called the von Neumann entropy.
In the case of a probability distribution $p$ encoded as a diagonal
matrix with diagonal entries $p(x)$, these two quantities agree, which
is why the Shannon entropy is usually thought of as a special case of Equation~\eqref{meas-eqn-entropy}.

Returning to the previous examples, notice that since a pure state
expressed as a density operator has exactly one nonzero eigenvalue, the
entropy is given by
\[ S(\ket \psi \bra \psi) = - 1 \log 1 = 0. \]
In the case of the completely mixed state $\nidentity{H}$ on a space
$\mathcal{H}$ of dimension $d$, there are $d$ eigenvalues, each with
value $1/d$, computing the entropy, we obtain
\[ S(\nidentity{H}) 
    = -\sum_{i=1}^d \frac{1}{d} \log \frac{1}{d}
    = \log d. \]
A good reference on the properties of the entropy and many
of the quantities derived from it can be found in~\cite{BengtssonZ06}
and~\cite{NielsenC00}.  The exposition of the 
entropy presented here follows these sources.

One property of the entropy that is useful is that it is additive
with respect to the tensor product.  To see this, let
$\rho = \sum_i \lambda_i \ket{\psi_i}\bra{\psi_i}$ and $\sigma = \sum_i \gamma_i
\ket{\phi_i}\bra{\phi_i}$ be two density operators.  Expanding the definition
of the entropy, we have
\index{entropy!additivity}%
\begin{equation}\label{meas-eqn-entropy-additive}
  S(\rho \tprod \sigma)
  = - \sum_{ij} \lambda_i \gamma_j \log \lambda_i \gamma_j
  = - \sum_{ij} \lambda_i \gamma_j \log \lambda_i
      - \sum_{ij} \lambda_i \gamma_j \log \gamma_j
  = S(\rho) + S(\sigma),
\end{equation}
where we have made use of the fact that since $\rho, \sigma$ are
density operators $\sum_i \lambda_i = \sum_j \gamma_j = 1$.  This
implies that for a multiparty quantum system that is not entangled,
the entropy of the complete system can be determined locally.  This is
not true for entangled systems: if two parties each share half of a
maximally entangled state, the local entropies are maximized, but the
global state is pure, so it has zero entropy.

It is easy to see that the entropy is always nonnegative: the
eigenvalues of a density matrix are always in the range $[0,1]$ as all
density matrices are by definition positive operators with unit trace.
It is more difficult to see that on a Hilbert space of
dimension $d$, no state can have entropy greater than $\log d$.  One
way to see this intuitively is to notice that since the logarithm is concave,
Equation~\eqref{meas-eqn-entropy} is maximized when there are as many
eigenvalues as possible, each of them being as small as possible.
Formalizing this argument will require the use of
Klein's inequality.
The inequality stated here is a special case of Klein's 1931 result, but it
is all that will be needed.
\index{Klein's inequality|see{entropy}}%
\index{entropy!Klein's inequality}%
\begin{theorem}[Klein's Inequality~\cite{Klein31}]\label{meas-thm-Kleins}
  Let $\rho,\sigma \in \density{H}$, then
  \[ \tr ( \rho \log \rho ) - \tr ( \rho \log \sigma ) \geq 0, \]
  with equality only when $\rho = \sigma$.
\end{theorem}

This inequality immediately implies that the completely mixed state is
the unique state with maximum entropy.
\begin{proposition}\label{meas-prop-entropy-bounds}
  Let$\mathcal{H}$ be a Hilbert space of dimension $d$, then for any
  $\rho \in \density{H}$
  \[ 0 \leq S(\rho) \leq \log d. \]
  Furthermore, $S(\rho) = 0$ implies that $\rho$ is a pure state, and
  $S(\rho) = \log d$ implies that $\rho = \nidentity{H}$.

  \begin{proof}
    The first inequality is simple: if $\rank(\rho) > 1$ then, by the
    strict concavity of the logarithm on positive values, $S(\rho) >
    0$.  On the other hand, if $\rho$ is pure a
    simple calculation reveals that $S(\rho) = 0$.

    The second inequality is a direct consequence of Klein's
    inequality (Theorem~\ref{meas-thm-Kleins}) applied to
    $\rho$ and $\nidentity{H}$:
    \[ 0 \leq \tr ( \rho \log \rho - \rho \log \nidentity{H})
    = \log d - S(\rho). \]
    This implies that $S(\rho) \leq \log d$, with equality only when
    $\rho = \nidentity{H}$, by the equality condition of Klein's inequality.
  \end{proof}
\end{proposition}

Klein's inequality can be used to prove another important property of
the entropy: \emph{concavity}.  This property is similar to the
triangle inequality, except that the inequality goes in the opposite
direction.
\index{entropy!concavity}%
\begin{proposition}\label{meas-prop-entropy-concavity}
  Let $\rho, \sigma, \xi \in \density{H}$, with $\rho = q \sigma +
  (1-q) \xi$, where $0 \leq q \leq 1$.  Then
  \[ S(\rho) \geq q S(\sigma) + (1-q) S(\xi). \]

  \begin{proof}
    Expanding the definition of $S$, we have
    \begin{align}
      S(\rho) 
      &= - \tr \rho \log \rho
      = - q \tr \sigma \log \rho - (1-q) \tr \xi \log \rho. \label{meas-eqn-econc1}
    \end{align}
    Klein's inequality (Theorem~\ref{meas-thm-Kleins}) implies that
    $\tr \sigma \log \sigma \geq \tr \sigma \log \rho$, and
    similarly for $\xi$ in place of $\sigma$.  Together with
    Equation~\eqref{meas-eqn-econc1}, this implies that
    \begin{align*}
      S(\rho) 
      &\geq - q \tr \sigma \log \sigma - (1-q) \tr \xi \log \xi
      = q S(\sigma) + (1-q) S(\xi),
    \end{align*}
    which is the desired inequality.
  \end{proof}
\end{proposition}

\subsection{Minimum output entropy}

The entropy can be extended to quantum channels in a straightforward
way: by minimizing the entropy over the output states of the channel.
The resulting quantity is known
as the \emph{minimum output entropy}, which is defined for a transformation $\Phi \in \transform{H,K}$ by
\begin{equation}\label{meas-eqn-smin-defn}
  S_{\min}(\Phi) = \min_{\rho \in \density{H}} S(\Phi(\rho)).
\end{equation}%
\index{entropy!minimum output entropy}%
\index{minimum output entropy|see{entropy}}%
\nomenclature[aS_min]{$S_{\min}(\Phi)$}{minimum output entropy of
  the channel $\Phi$}%
This extension of the entropy to quantum channels, as well as the
properties of the entropy that have been demonstrated here 
will be essential to the results of Section~\ref{mu-scn-add},
concerning the additivity of the minimum output entropy on the tensor
product of two channels.  This is closely related to the capacity of a
quantum channel for the transmission of classical information.  This is discussed in Section~\ref{meas-scn-additivity}

\section{Schatten \texorpdfstring{$p$}{p}-norms}\label{meas-scn-pnorms}

One of the more useful distance measures that can be defined on
quantum states comes from the Schatten $p$-norms~\cite{Schatten60}.
This is the extension to operators in $\linear{H,K}$ 
of the usual $\ell_p$ norm of a sequence
$\{x_i\}$, which for $1 \leq p < \infty$ is defined by
\begin{equation}\label{meas-eqn-lp-norm}
\norm{x}_p = \left( \sum_i \abs{x_i}^p \right)^{\frac{1}{p}}.
\end{equation}
For $p = \infty$, this norm is given by $\norm{x}_\infty = \sup_i
\abs{x_i}$, which can be obtained by taking the limit of
Equation~\eqref{meas-eqn-lp-norm} as $p \to \infty$.
The extension of this norm to an
operator $A \in \linear{H,K}$ is done by taking this norm on the
singular values of $A$, so that for $1 \leq p < \infty$
\begin{equation}\label{meas-eqn-pnorm}
  \norm{A}_p = \left( \tr \abs{A}^p \right)^{\frac{1}{p}} = \norm{s(A)}_p,
\end{equation}%
\index{Schatten $p$-norm|see{$p$-norm}}%
\index{p-norm@$p$-norm}%
\nomenclature[anormp]{$\norm{\cdot}_p$}{Schatten $p$-norm}%
where $s(A)$ is the (finite) sequence of singular values of $A$.  The
extension to the case $p = \infty$ is, as in the vector case, 
given by $\norm{A}_\infty = \max_i s_i(A)$.
Two of these norms are widely used in quantum information: the case $p
= 1$ corresponds to the trace norm, considered in more detail in
Section~\ref{meas-scn-tnorm}, and the case of $p = \infty$ corresponds
to the usual operator norm on $\linear{H,K}$.  This section discusses
some of the most important properties of these norms.  A more complete
overview of the properties of these norms, as well as the properties
of more general classes of norms, can be found in~\cite{Bhatia97}.

Recall from Definition~\ref{intro-defn-norm} that any norm satisfies the
three properties of nonnegativity, homogeneity, and the triangle
inequality.  The first two of these properties are easily to verified for
these norms.  Equation~\eqref{meas-eqn-pnorm} is only zero when all
the singular values are zero, which establishes nonnegativity
(Equation~\eqref{intro-eqn-norm-nonnegativity}).  Homogeneity
(Equation~\eqref{intro-eqn-norm-homogeneity}) follows directly from
the definition of the absolute value of a matrix and the
linearity of the trace.

Verifying the triangle inequality
(Equation~\eqref{intro-eqn-norm-triangle-inequality}) for this norm is nontrivial, and so
only a brief overview of this result is presented here.  The most
important part of this proof is a 1951 result of Ky Fan~\cite{Fan51},
which is a majorization relation on the singular values of $A+B$ in
terms of the singular values of $A$ and $B$.  As is common in this
thesis, the finite-dimensional result that is presented is a
considerable simplification of the known result, which holds in
the infinite dimensional case.
\begin{theorem}[Ky Fan~\cite{Fan51}]\label{meas-thm-kyfan}
  Let $A, B$ be $n \times n$ matrices, and let $s(A)$ denote the
  sequence of singular values of $A$ in decreasing order, then for all
  $k \in \{1, \ldots, n\}$
  \begin{equation}\label{meas-eqn-kyfan}
    \sum_{i=1}^k s_i(A + B) 
    \leq \sum_{i=1}^k s_i(A) + \sum_{i=1}^k s_i(B),
  \end{equation}
  and more generally, if $\tau$ is any symmetric gauge function,
  \[ \tau(s(A + B)) \leq \tau(s(A)) + \tau(s(B)). \]
\end{theorem}
The triangle inequality for $\norm{\cdot}_p$ in the cases that $p = 1,
\infty$ is a direct consequence of Equation~\eqref{meas-eqn-kyfan} for
$k = n, 1$.  The triangle inequality for the remaining values of $p$
follow from Fan's theorem and the fact that for a vector $x \in
\mathbb{R}^n$, the function $\tau(x) = (\sum_{i=1}^n
\abs{x_i}^p)^{1/p}$ is a \emph{symmetric gauge function}.  This
is a strengthened version of the property that the function
$\tau(\cdot)$ is a norm (in the case, the $\ell_p$ norm).  More
details, as well as detailed arguments that $\norm{\cdot}_p$ is a norm
using the theory of symmetric gauge functions can be found
in~\cite{Bhatia97,HornJ91}.

The $p$-norm satisfies two more properties that will be essential to
the results in Chapter~\ref{chap-mixed-unitary}.  The first of these
is~\emph{unitary invariance}
\index{p-norm@$p$-norm!unitary invariance}%
This is the property that for any unitary operators $U, V$
\begin{equation}\label{meas-eqn-pnorm-unitary-invar}
  \norm{U A V}_p = \norm{A}_p.
\end{equation}
It is easy to prove that this property holds.  Consider a singular
value decomposition of $A$ given by
$\sum_i s_i \ket{\phi_i} \bra{\psi_i},$ then
a singular value decomposition for $UAV$ is given by
\[ \sum_i s_i (U \ket{\phi_i}) (\bra{\psi_i} V), \]
from which it can be seen that both $A$ and $UAV$ have the same
singular values.
Then, as the $p$-norm is defined in Equation~\eqref{meas-eqn-pnorm}
solely in terms of the singular values of $A$, the $p$-norms of $A$
and $UAV$ must be identical.

The second important property of the $p$-norm is that it is
\emph{multiplicative} with respect to the tensor product of two
operators.  That is, for operators $A,B$
\begin{equation}\label{emas-eqn-pnorm-mult}
  \norm{A \tprod B}_p = \norm{A}_p \norm{B}_p.
\end{equation}
\index{p-norm@$p$-norm!multiplicativity on states}%
This property follows directly from the properties of the tensor
product.  Let singular value decompositions of $A$ and $B$ be given by $A = \sum_i
s_i \ket{\phi_i} \bra{\psi_i}$ and $B = \sum_i t_i \ket{\gamma_i}
\bra{\nu_i}$, then the a singular value decomposition of $A \tprod B$
is
\begin{equation}\label{meas-eqn-pnorm-mult}
  \left( \sum_i s_i \ket{\phi_i} \bra{\psi_i} \right) \tprod
  \left( \sum_j t_j \ket{\gamma_j} \bra{\nu_j} \right) 
  = \sum_{ij} s_i t_j \ket{\phi_i} \bra{\psi_i} \tprod
  \ket{\gamma_j} \bra{\nu_j},
\end{equation}
so that the singular values of the tensor product $A \tensor B$ are
simply the products of the singular values of $A$ and $B$.  From this
relationship Equations~\eqref{meas-eqn-lp-norm}
and~\eqref{meas-eqn-pnorm} imply the desired property.

\subsection{Maximum output \texorpdfstring{$p$}{p}-norm}

These norms have an important extension to channels.  This is the
\emph{maximum output $p$-norm}, which, for a channel $\Phi \in
\transform{H,K}$ is denoted
$\opv_p(\Phi)$, or sometimes simply $\norm{\Phi}_p$.  For 
$1 \leq p \leq \infty$, this norm is given by
\begin{equation}\label{meas-eqn-maxout-pnorm-defn}
  \opv_p (\Phi) = \max_{\rho \in \density{H}} \norm{\Phi(\rho)}_p.
\end{equation}%
\index{p-norm@$p$-norm!maximum output $p$-norm}%
\nomenclature[anu]{$\opv(\cdot)$}{Maximum output $p$-norm}%
This quantity is normally defined by taking the maximum over all
inputs $X \in \linear{H}$ with $\norm{X}_1 =1$, but a result of
Amosov and Holevo~\cite{AmosovH03} implies that in the case of $\Phi$ completely
positive, this maximization can be restricted to the
density operators.  Note that this simplification cannot be made in
the case of the \emph{difference} of two channels, as the resulting
operation is not completely positive, by a counterexample found
in~\cite{Watrous05}.  In this thesis the maximum output $p$-norm will
only be applied to channels, never the difference of two channels, so
the simplification of the definition in
Equation~\eqref{meas-eqn-maxout-pnorm-defn} is justified.

In the next section, this quantity is related to the capacity of a
quantum channel for the transmission of classical information, the
exact specification of which is currently an important problem in
quantum information.

\section{The classical capacity of a quantum channel}\label{meas-scn-additivity}

The additivity of the capacity for a quantum channel to communicate
classical information is one of the most important unresolved problems in
quantum information theory.  Informally, the additivity problem is:
given two uses of a quantum channel, is it possible to send more than
twice the classical information that could be sent with a single use?
This is a common oversimplification; the classical capacity is defined
in terms of the average amount of information sent per channel use,
asymptotically with the number of uses of the channel~\cite{Holevo98,
  SchumacherW97}.  A more correct statement of the problem is: when
encoding for the transmission of classical information, is
entanglement across multiple uses of the channel necessary to achieve
the best communication rate?  This refined problem stood open for over 10
years before a counterexample was recently found by Hastings~\cite{Hastings09}.

In this section this problem is given a formal definition and the
closely related problems of the additivity of the minimum output
entropy and the multiplicativity of the maximum output $p$-norm are
discussed.  The minimum output entropy and maximum output $p$-norm
both involve maximizing the purity of the output of a channel, a
problem that is intuitively related to the classical capacity by the
notion that a channel that is less noisy should be able to send more
information.  A recent survey of these problems can be found
in~\cite{Holevo06}, though it does not include the recent
counterexamples to both additivity~\cite{Hastings09} and the related
problem of the multiplicativity of the maximum output $p$-norm~\cite{HaydenW08}.

The classical capacity of channel $\Phi$, when the input to multiple
uses of the channel is restricted to product states,
is given by the $\chi$-capacity~\cite{Holevo98, SchumacherW97}
\begin{equation}\label{meas-eqn-chi-capacity}
  C_\chi(\Phi) = \max \big[ S( \sum_i p_i \Phi(\rho_i) ) - \sum_i p_i
  S(\Phi(\rho_i)) \big],
\end{equation}
where the maximum is taken over all convex mixtures $\sum_i p_i
\rho_i$ of quantum states.  This quantity is also referred to as the
Holevo capacity or the ``one-shot'' or ``one-step'' capacity of $\Phi$.  The question
of the additivity of this quantity, i.e.\ can entangling inputs across multiple uses of
the channel be required to increase the capacity,  was first raised
in~\cite{BennettF+97}, and the until recently standing conjecture
was that
\begin{equation}\label{meas-eqn-chi-additivity}
  C_\chi(\Phi \tensor \Psi) \stackrel{?}{=} C_\chi(\Phi) + C_\chi(\Psi).
\end{equation}
This is the statement that entangled inputs do not increase the
classical information carrying capacity of quantum channels.  This
conjecture has recently been shown false in general by a result on the
additivity of the minimum output entropy~\cite{Hastings09}.
This result implies that the maximum
rate that classical information can be transmitted using a channel
$\Phi$ is given by
\[ C(\Phi) = \lim_{n \rightarrow \infty} \frac{1}{n} C_\chi(\Phi^{\tprod n}). \]
If it were that case that $C_\chi$ were additive, this formula would
simplify to $C(\Phi) = C_\chi(\Phi)$, but it is now known that this is
not the case.  This leaves open the question on many restricted
classes of channels: a
survey of some of these special cases can be found in~\cite{Holevo06}.

The $\chi$-capacity captures exactly the amount of classical information
that can be transmitted per use of the channel when encoding with product states,
but it is somewhat awkward to work
with.  In the effort to resolve the additivity question it has been
related to both the minimum output entropy and the maximum output $p$-norm.

\subsection{Relation to the minimum output entropy}\label{meas-scn-smin-hcap}

The minimum output entropy, defined by Equation~\eqref{meas-eqn-smin-defn},
is simpler and often easier to work with
than the $\chi$-capacity.
The additivity of this quantity, given by
\begin{equation}\label{intro-eqn-smin-additivity}
  S_{\min}(\Phi \tensor \Psi) \stackrel{?}{=} S_{\min}(\Phi) + S_{\min}(\Psi),
\end{equation}
was first conjectured by King and Ruskai~\cite{KingR01}, who attribute
the conjecture to Shor.
The additivity of this quantity is connected to the additivity of the
$\chi$-capacity by a result of Shor~\cite{Shor04} that shows that both of these
conjectures are globally equivalent to a third conjecture:
the strong superadditivity of the entanglement of formation.
Hastings has recently given a probabilistic construction that shows that this
conjecture is false in general~\cite{Hastings09}, which also implies the
non-additivity of the $\chi$-capacity.

One direction of Shor's construction in~\cite{Shor04} to show that the
additivity of $C_\chi$ is equivalent to the additivity of $S_{\min}$
is very complicated, but the other direction is quite simple.
\begin{theorem}[Shor~\cite{Shor04}]
  The additivity of $C_\chi$ implies the additivity of $S_{\min}$.

  \begin{proof}
    Let $\Phi_1, \Phi_2$ be arbitrary channels in $\Phi \in
    \transform{H,K}$.  We will construct channels $\Phi_1',
    \Phi_2'$ in the larger space $\transform{C \tprod H, K}$.  The
    channel $\Phi_i'$ uses the input space $\mathcal{C} \cong
    \mathcal{K} \times \mathcal{K}$ to determine which of the discrete
    Weyl operators to apply to the output of $\Phi_i$, i.e.\
    \[ \Phi_i'(\ket{a,b} \bra{a,b} \tprod \rho) = W_{a,b} \Phi_i(\rho) W_{a,b}^*, \]
    where the unitaries $W_{a,b}$ are the discrete Weyl operators
    introduced in Section~\ref{intro-scn-qinfo}.
    This process is applied decoherently, which is to say that
    $\Phi_i'$ measures the space $\mathcal{C}$ in the computational
    basis to decide which operator to apply.  As we will later show
    that the uniform mixture of the discrete Weyl operators forms the
    completely depolarizing channel
    (Proposition~\ref{mu-prop-depol-mu}), the result of placing a
    completely mixed state in the input space $\mathcal{C}$ results in
    the completely mixed state as output, which is
    \[ \Phi_i'( \nidentity{C} \tprod \rho ) = \nidentity{K}. \]

    Let $\rho_1$ and $\rho_2$ be states achieving the minimum output
    entropy for $\Phi_1$ and $\Phi_2$, respectively.  We will assume
    that $C_\chi$ is additive, and show that if
    $S_{\min}$ is not additive on $\Phi_1 \tprod \Phi_2$ then we can
    find a contradiction.
    To do this
    we compute $C_\chi(\Phi_i')$ with the maximization in
    Equation~\eqref{meas-eqn-chi-capacity} restricted to 
    inputs of the form $\nidentity{C} \tprod \rho_i$, which is given by
    \begin{align}
       S( \nidentity{K} ) - \frac{1}{\dm C} \sum_{a,b} S(W_{a,b} \Phi_i(\rho_i) W_{a,b}^*)
       & = \log \dm{K} - \frac{1}{\dm C} \sum_{a,b} S(\Phi_i(\rho_i))\nonumber\\
       & = \log \dm{K} - S(\Phi_i(\rho_i)) \nonumber\\
       & = \log \dm{K} - S_{\min}(\Phi_i). \label{meas-eqn-shor-1}
    \end{align}
    Notice that this is the optimal value of $C_\chi(\Phi_i')$, since
    the first term is maximized and the second term is minimized.
    This implies that restricting to inputs of the form $\nidentity{C}
    \tprod \rho_i$ does not reduce the value of $C_\chi(\Phi_i')$.
    As we have assumed that $C_\chi$ is additive, this expression also gives the
    optimal value of $C_\chi(\Phi_1' \tprod \Phi_2')$.  However, if
    $S_{\min}(\Phi_1 \tprod \Phi_2)$ is not additive, then any state
    $\sigma$ on which
    \[ S((\Phi_1 \tprod \Phi_2)(\sigma)) < S_{\min}(\Phi_1) + S_{\min}(\Phi_2) \]
    can be used to increase $C_\chi(\Phi_1' \tprod \Phi_2')$, exactly
    as in the derivation of Equation~\eqref{meas-eqn-shor-1},
    contradicting the (assumed) additivity of $C_\chi$.
  \end{proof}
\end{theorem}

This result of Shor~\cite{Shor04}, coupled with Hastings'
results~\cite{Hastings09} shows that $C_\chi$ is non-additive in
general.  As the proof is constructive, it also shows that if $S_{\min}$
is not additive on a class of channels, then $C_\chi$ is also not
additive for the related class of channels given by applying the
construction in the proof to all the channels in the class.
This result is included here as it is of a similar flavour to many of
the results in the thesis: reducing one problem to another, by
embedding arbitrary channels into channels with specific properties,
is a powerful method for obtaining results.  In fact, Fukuda has used
the same construction to show that the additivity of the Holevo
capacity and the minimum output entropy can be restricted to the
unital channels without loss of generality~\cite{Fukuda07}.
As in the case of the $\chi$-capacity, these non-additivity results leave open the question
of on which classes of channels does the additivity of the minimum
output entropy hold?  This is currently an important open problem in
quantum information, as a deeper understanding of the channels for
which the minimum output entropy is not additive may lead to a deeper
understanding of those channels for which $C_\chi$ is not additive.

\subsection{Relation to the maximum output \texorpdfstring{$p$}{p}-norm}

In an effort to better understand the question of the additivity of
the minimum output entropy, yet another question has been raised.
This problem is the multiplicativity of the maximum
output $p$-norm, which was first conjectured by Amosov, Holevo, and
Werner~\cite{AmosovH+00}.  
The conjecture
corresponding to this quantity was that it is multiplicative with
respect to the tensor product of two channels, i.e.\ that
\begin{equation}\label{intro-eqn-pnorm-mult}
  \opv_p(\Phi \tensor \Psi) \stackrel{?}{=} \opv_p(\Phi) \opv_p(\Psi).
\end{equation}
This conjecture is not true in general: channels can be constructed
for any fixed $p>1$ that falsify this conjecture~\cite{HaydenW08}.

Amosov, Holevo, and Werner have related the multiplicativity of
$\opv_p$ to the additivity of $S_{\min}$~\cite{AmosovH+00}.
This relationship can be used to show that the minimum output entropy
is additive for a pair of channels if and only if the maximum output
$p$-norm is additive on the same channels, for values of $p$ close to 1.
\begin{theorem}[Amosov, Holevo, Werner~\cite{AmosovH+00}]
  Let $\Phi_1, \Phi_2 \in \transform{H,K}$.  If for some 
  sequence $p_i \to 1$, with $p_i \geq 1$ for all $i$, it
  holds that
  \[ \opv_{p_i} (\Phi_1 \tprod \Phi_2) 
      = \opv_{p_i} (\Phi_1) \opv_{p_i}(\Phi_2),  \]
  then the minimum output entropy satisfies
  \[ S_{\min}(\Phi_1 \tprod \Phi_2) 
      = S_{\min}(\Phi_1) +  S_{\min}(\Phi_2) \]
  
  \begin{proof}
    Let the sequence $\{p_i\}$ be as in the statement of the theorem.
    The result is easy to verify by introducing the quantum R{\'e}nyi
    entropy of order $p$ of a density matrix $\sigma$
    \begin{equation}\label{meas-eqn-renyi}
      R_p(\sigma) = \frac{1}{1-p} \log \tr \sigma^p.
    \end{equation}
    The important property is that taking the limit as $i \rightarrow
    \infty$ (so that $p \rightarrow 1$ from above) results in the
    usual entropy (up to a factor due to the base of the logarithm)
    \[ \lim_{i \rightarrow \infty} R_{p_i}(\sigma) = (\log e) S(\sigma). \]
    This can be verified by invoking l'Hopital's rule on
    Equation~\eqref{meas-eqn-renyi}.  By observing that the logarithm
    of the $p$-norm can be used to recover $R_p$, we may conclude
    from the previous equation that
    \begin{align*}
       \lim_{i \rightarrow \infty} \frac{p_i}{1-p_i} \log \norm{\sigma}_{p_i}
       = \lim_{i \rightarrow \infty} R_{p_i}(\sigma)      
       = (\log e) S(\sigma).
     \end{align*}
     From which the multiplicativity of the maximum output $p$-norm
     immediately implies the additivity of the minimum output entropy.
     This follows from the fact that $\log \opv_{p_i} (\Psi)\leq 0$ as
     $\opv_{P_i}(\Psi) \leq 1$ for any channel $\Psi$.  More concretely, we have
    \begin{align*}
       S_{\min}(\Phi_1 \tprod \Phi_2)
       & = \frac{1}{\log e} \lim_{i \rightarrow \infty} \frac{p_i}{1-p_i} \log \opv_{p_i} (\Phi_1 \tprod \Phi_2) \\
       & = \frac{1}{\log e} \lim_{i \rightarrow \infty} \frac{p_i}{1-p_i} \log \opv_{p_i} (\Phi_1)
           + \frac{1}{\log e} \lim_{i \rightarrow \infty} \frac{p_i}{1-p_i} \log \opv_{p_i}(\Phi_2) \\
       & = S_{\min}(\Phi_1) + S_{\min}(\Phi_2),
     \end{align*}
     as desired.
   \end{proof}
\end{theorem}

The minimum output entropy and the maximum output $p$-norm will be
encountered again in Chapter~\ref{chap-mixed-unitary}, where it is
shown that the additivity (respectively multiplicativity) of a channel can be
rephrased in terms of the additivity (multiplicativity) of a related
set of mixed-unitary channels.
 
\subsection{Additivity and multiplicativity on classes of channels}

The questions of the additivity of the Holevo capacity (also called
the $\chi$-capacity) and the
multiplicativity of the maximum output $p$-norm have been resolved for
many of the restricted classes of channels that are studied in this thesis.

It is shown in~\cite{CubittR+08} that the additivity of degradable
channels is equivalent to the additivity of general channels, using a
result from~\cite{FukudaW07}.  Combining this result with the result
that the additivity problems are equivalent on a
class of channels and the class of complementary
channels~\cite{Holevo07, KingM+07} shows that the additivity of the
antidegradable channels is also equivalent to the general case.  The
recent result of Hastings~\cite{Hastings09} can then be used to
show that there exist degradable and antidegradable channels that are
not additive.  It is perhaps not a surprise that these channels are
well-behaved: the degradable and antidegradable channels cannot
be used to transmit quantum information to the environment and
receiver, respectively, because to do so would violate the principle
of no-cloning.

It is also perhaps not a surprise that on the entanglement breaking
channels the minimum output entropy is
additive~\cite{Shor02} and the maximum
output $p$-norm is multiplicative~\cite{King03}.
The problem of distinguishing channels of this class,
however, is quite interesting and remains open.

The unital channels cannot decrease the entropy~\cite{KingR01}.  
This property makes them interesting from the
perspective of additivity, as channels that do not reduce entropy 
would seem to be a natural noise model.  Fukuda has shown how to construct a
unital channel from a general channel, without changing the minimum
output entropy or the maximum output $p$-norm~\cite{Fukuda07}, using a
the same construction used by Shor to show that the additivity of
$S_{\min}$ implies the additivity of $C_\chi$~\cite{Shor04}.  This
implies that for a set of channels the question of
additivity can rephrased in terms of the additivity of a related set of unital channels.

The mixed-unitary channels are a subclass of the unital channels.
In the case of qubit mixed-unitary
channels, both additivity and multiplicativity are known to
hold~\cite{King02}.  For general mixed-unitary channels,
both additivity~\cite{Hastings09} and 
multiplicativity~\cite{HaydenW08} are known to fail.  In fact, all of
the recent counterexamples to additivity and multiplicativity are
obtained by choosing a random mixed-unitary channel from some
distribution.  This makes these channels very interesting.  It is
shown in Chapter~\ref{chap-mixed-unitary} that the additivity or
multiplicativity for a general channel can be reduced
to the approximate additivity or multiplicativity of a mixed-unitary channel.

\section{The trace norm}\label{meas-scn-tnorm}

The trace norm is perhaps the most important measure of size and
distance in quantum information.  The trace norm of the difference of
two states measures how distinguishable the two states are, which
makes this an essential quantity for the problems considered in this
thesis.  The remainder of this section surveys some properties of this
norm, further background on the trace norm can be obtained
in the books~\cite{NielsenC00} and~\cite{Bhatia97}.

We have already encountered the trace norm:
it is simply the $p=1$ case of the Schatten $p$-norm discussed in Section~\ref{meas-scn-pnorms}.
This implies that $\tnorm{X}$ is given by the sum of the singular
values of $X$, though it is often useful to define this norm by
an explicit formula.  Such a formula can be obtained by noticing that
the singular values of $X$ are exactly the square roots of the
eigenvalues of the positive operator $X^* X$.  Using this observation,
the trace norm can equivalently be defined as
\begin{equation}
  \tnorm{X} = \norm{X}_1 = \tr \sqrt{X^* X}.
\end{equation}
\index{trace norm}%
\nomenclature[anormt]{$\tnorm{\cdot}$}{trace norm}%
The trace norm inherits many properties from the $p$-norm.  The
triangle inequality is the $k=n$ case of Fan's Theorem
(\ref{meas-thm-kyfan}), and unitary invariance is given by
Equation~\eqref{meas-eqn-pnorm-unitary-invar}.
One other convenient property is that $\tnorm{\rho} = 1$ for any
density matrix $\rho$, which is implied by the fact that density
operators are positive semidefinite operators with unit trace, so that
the eigenvalues are all positive and sum to one.
Several other
properties of the trace norm can be easily derived from the following
characterization.
\begin{lemma}\label{meas-lem-tnorm-max}
  For any $X \in \linear{H}$,
  \begin{equation*}
    \tnorm{X} = \max_{U \in \unitary{H}} \abs{ \tr X U }
  \end{equation*}
  
  \begin{proof}
    Let $X$ have a singular value decomposition given by $X = \sum_i
    s_i \ket{\phi_i}\bra{\psi_i}$, where $\{\ket{\phi_i}\}$ and
    $\{\ket{\psi_i}\}$ are orthonormal bases for $\mathcal{H}$.  Then
    for any unitary $U \in \unitary{H}$
    \begin{align}
      \abs{ \tr XU }
      &= \abs{ \sum_i s_i \bra{\psi_i} U \ket{\phi_i} }
      \leq \sum_i s_i \abs{ \bra{\psi_i} U \ket{\phi_i} }
      \leq \sum_i s_i
      = \tnorm{X}.\label{meas-eqn-tnorm-max-unitary}
    \end{align}
    If the unitary $U$ is chosen such that $U \ket{\phi_i} =
    \ket{\psi_i}$, then 
    \begin{equation*}
      \bra{\psi_i} U \ket{\phi_i} = \braket{\psi_i}{\psi_i} = 1,
    \end{equation*}
    for all $i$.  In this case equality is achieved in
    Equation~\eqref{meas-eqn-tnorm-max-unitary}.
  \end{proof}
\end{lemma}

From this characterization it is easy to see that the trace norm does
not increase under the partial trace, and in fact, does not increase
under the application of any channel.  This is intuitively obvious:
applying any potentially noisy operation to two states cannot help to
distinguish them.
\index{trace norm!monotonicity}%
\begin{theorem}\label{meas-thm-tnorm-monotonicity}
  Let $\Phi \in \transform{H,K}$, then for any $\rho, \sigma \in \density{H}$
  \[ \tnorm{\Phi(\rho) - \Phi(\sigma)} \leq \tnorm{\rho - \sigma}. \]

  \begin{proof}
    We first prove this for the case that $\Phi \in \transform{K
      \tprod B, K}$ is the partial trace over the space $\mathcal{B}$.
    For any $\rho, \sigma \in \density{K \tprod B}$, this follows
    directly from Lemma~\ref{meas-lem-tnorm-max}, since
    \begin{equation*}
      \tnorm{ \ptr{B} \rho - \ptr{B} \sigma }
     = \max_{U \in \unitary{K}} \abs{ \tr \left[(\sigma - \rho)(U \tprod \identity{B} )\right] }
      \leq \max_{U \in \unitary{K \tprod B}} \abs{ \tr (\sigma - \rho)U }
      = \tnorm{\rho - \sigma}.
    \end{equation*}

    To see the general case, let $\Phi \in \transform{H,K}$ have Strinespring
    representation given by $\Phi(\rho) = \ptr{B} U (\rho \tprod \ket 0 \bra
    0) U^*$.  Then the previous equation and the unitary invariance
    of the trace norm imply that
    \begin{align*}
      \tnorm{\Phi(\rho) - \Phi(\sigma)}
      &= \tnorm{ \ptr{B} U \left[ (\rho -\sigma)\tprod \ket 0 \bra 0 \right] U^* }\\
      &\leq \tnorm{ U \left[ (\rho -\sigma)\tprod \ket 0 \bra 0 \right] U^* }\\
      &=\tnorm{ (\rho -\sigma)\tprod \ket 0 \bra 0 }\\
      &=\tnorm{\rho -\sigma},
    \end{align*}
    as required.
  \end{proof}
\end{theorem}

The following theorem due to Helstrom~\cite{Helstrom67} formalizes the
notion that the trace norm of the difference of two density matrices
represents how well they can be distinguished by a measurement.  This
result underlies the definition of the quantum circuit
distinguishability problem in Chapter~\ref{chap-distinguishability},
as the problem of distinguishing channels is exactly the problem of
distinguishing the outputs of the channels.  This result is easy to
generalize to the case that the two density matrices are not chosen
with equal probabilities, but doing so unnecessarily complicates the
argument.  
\index{trace norm!Helstrom measurement}%
\index{minimum error state distinguishability|see{Helstrom Measurement}}%
\index{Helstrom Measurement}%
\begin{theorem}[Helstrom~\cite{Helstrom67}]\label{meas-thm-Helstrom}
  The optimal probability that an unknown state $\xi \in \density{H}$ that is chosen
  uniformly at random from the set $\{\rho, \sigma\}$ can be correctly
  identified is given by
  \[ \frac{1}{2} + \frac{\tnorm{\rho - \sigma}}{4}. \]

  \begin{proof}
    The optimal strategy consists of some two-outcome POVM
    measurement.  By Naimark's theorem it may be assumed that the
    optimal measurement is a projective measurement $\Pi_\rho,
    \Pi_\sigma$ with $\Pi_\rho + \Pi_\sigma = \identity{H}$, since the
    operation that embeds $\rho$ and $\sigma$ into a larger space is
    an isometry, which will not affect the trace norm, by unitary
    invariance.  The probability that this measurement succeeds is
    \begin{equation*}
      p_{\mathrm{succ}} = \frac{1}{2} \tr( \Pi_\rho \rho ) + \frac{1}{2} \tr( \Pi_\sigma \sigma).
    \end{equation*}
    Similarly, the probability of failure is
    \begin{equation*}
      p_{\mathrm{fail}} = \frac{1}{2} \tr( \Pi_\sigma \rho ) + \frac{1}{2} \tr( \Pi_\rho \sigma).
    \end{equation*}
    Subtracting the probability of failure from the probability of
    success gives the bound
    \begin{equation}\label{meas-eqn-Helstrom-bound}
      p_{\mathrm{succ}} - p_{\mathrm{fail}}
      = \frac{1}{2} \tr \left( (\Pi_\rho - \Pi_\sigma)(\rho - \sigma)\right)
      \leq \frac{1}{2} \tnorm{\rho - \sigma}, 
    \end{equation}
    where the inequality follows from Lemma~\ref{meas-lem-tnorm-max}
    by the fact that $\Pi_\rho - \Pi_\sigma$ is a unitary operator.
    Adding this equation to the equation $p_{\mathrm{succ}} + p_{\mathrm{fail}} = 1$
    results in the bound
    \begin{equation}\label{meas-eqn-Helstrom-bound-prob}
      2 p_{\mathrm{succ}} \leq 1 + \frac{1}{2} \tnorm{\rho - \sigma}.
    \end{equation}
    This is the probability given in the statement of the theorem, and
    so it remains only to show that it can be achieved.

    To see this, consider the projectors
    $\Pi_+$ and $\Pi_-$ that project onto the positive and non-positive
    eigenspaces of $\rho - \sigma$.  Re-examining
    Equation~\eqref{meas-eqn-Helstrom-bound} with this measurement
    results in
   \begin{equation*}
     p_{\mathrm{succ}} - p_{\mathrm{fail}}
     = \frac{1}{2} \tr \left( (\Pi_+ - \Pi_-)(\rho - \sigma)\right)
     = \frac{1}{2} \tr \abs{\rho - \sigma}
     = \frac{1}{2} \tnorm{\rho - \sigma},
     \end{equation*}
      since the eigenvalues of the Hermitian operator $(\Pi_+ -
    \Pi_-)(\rho - \sigma)$ are the absolute values of the eigenvalues
    of $\rho - \sigma$.  This demonstrates that
    this measurement achieves the bound in
    Equation~\eqref{meas-eqn-Helstrom-bound-prob}.
  \end{proof}
\end{theorem}

There is one final property of the trace norm that is needed in
Section~\ref{meas-scn-dnorm-diff} and
Chapter~\ref{chap-distinguishability}.  This property relates the
trace norm of an operator to a Hermitian block matrix that is closely
related to it.  This construction is often useful: it is one way to
take a general linear operator and construct a Hermitian operator on a
space including one additional qubit.
The proof of this relationship is not
difficult or particularly illuminating, but it is included as this
result is critical to the some of the proofs that follow.
\begin{lemma}\label{meas-lem-block-tnorm}
  Let $A \in \linear{H}$ be a linear operator, then
  \[ \tnorm{ \ket 0 \bra 1 \tprod A + \ket 1 \bra 0 \tprod A^* }
     = 2 \tnorm{A}. \]  
  \begin{proof}
    Let $r$ be the rank of $A$ and let $n = \dm{H}$.
    If $r = 0$ the result is trivial, so
    we may assume that $r>0$.
    Let $\hat{A} =  \ket 0 \bra 1 \tprod A + \ket 1 \bra 0 \tprod
    A^*$.  Written as a block matrix, this operator is
    \[ \hat{A} =
       \begin{pmatrix}
         0   & A \\
         A^* & 0
       \end{pmatrix}. \]
    For the evaluation of the trace norm of $\hat{A}$ it suffices to consider
    the eigenvalues, as this operator is Hermitian by construction.
    To compute these eigenvalues, let $A$ have singular value
    decomposition $A = \sum_{i=1}^r s_i \ket{\psi_i} \bra{\phi_i}$, where
    $\{\ket{\psi_i}\}$ and $\{\ket{\phi_i}\}$ are orthonormal sets of
    vectors.  Let the notation $[ \phi, \psi ]^\top$ denote the vector
    of length $2n$ whose first $n$ entries are the entries of $\phi$
    and whose second $n$ entries are the entries of $\psi$.    
    As observed in~\cite[Section II.1]{Bhatia97}, the $2r$ 
    nonzero eigenvalues of $\hat{A}$ are given by
    \begin{align*}
       \begin{pmatrix}
         0   & A \\
         A^* & 0
       \end{pmatrix}
       \begin{pmatrix}
         \phi_i \\
         \psi_i
       \end{pmatrix}
       &= s_i
       \begin{pmatrix}
         \phi_i \\
         \psi_i
       \end{pmatrix}
       &
       \begin{pmatrix}
         0   & A \\
         A^* & 0
       \end{pmatrix}
       \begin{pmatrix}
         -\phi_i \\
         \psi_i
       \end{pmatrix}
       &= -s_i
       \begin{pmatrix}
         -\phi_i \\
         \psi_i
       \end{pmatrix},
    \end{align*}
    for $i \in \{1, \ldots, r\}$.
    This implies that
    \[ \tnorm{\hat{A}}
       = 2 \sum_{i=1}^r \abs{s_i}
       = 2 \tnorm{A}, \]
    as desired.
  \end{proof}
\end{lemma}

The trace norm can be extended from states to channels, though some
care must be taken in doing so to ensure that the resulting norm
retains the desirable  properties of the trace norm, such as the
relationship to the optimal distinguishing probability.  This
extension is the focus the next section.

\section{The diamond norm}\label{meas-scn-dnorm}

In this section the diamond norm is introduced and studied.  This norm
defines the distance measure that is central to the results that
follow on the distinguishability of channels.  The norm is introduced
and some of the basic properties of the norm that can be found in the
literature are discussed.  Further background on the diamond norm can
be found in~\cite{Kitaev97, AharonovK+98} and~\cite{KitaevS+02}.

The diamond norm is a norm on channels with similar properties to the
trace norm on quantum states.  It will give a numerical value to the
distance between quantum channels, and, more importantly, as in the
case of the trace norm, it will be closely related to how well two
channels can be distinguished.

The straightforward way to extend the trace norm to quantum channels
is to do as we have done for the entropy: optimize
the output over all input states.  Doing this for the
trace norm results in the norm given defined by
\index{trace norm!of channels}%
\begin{equation}\label{meas-eqn-tnorm-channel-defn}
  \tnorm{\Phi} = \max_{X \in \linear{H}, X \neq 0}
  \frac{\tnorm{\Phi(X)}}{\tnorm{X}}.
\end{equation}
This results in a norm, as it inherits many of the properties of the
trace norm directly.  As an example, it is easy to see that this norm
obeys the triangle inequality.  One of the other properties inherited
by this norm is \emph{submultiplicativity}.  This property will be
useful, and so a short proof of this fact is given below.
\begin{lemma}\label{meas-lem-dnorm-submult}
  For any $\Phi \in \transform{K,F}$ and any $\Psi \in
  \transform{H,K}$,
  \[ \tnorm{\Phi \circ \Psi} \leq \tnorm{\Phi} \tnorm{\Psi}. \]

  \begin{proof}
    Let $\Phi, \Psi$ be as in the statement of the theorem.  Using the
    definition of the trace norm on channels in
    Equation~\eqref{meas-eqn-tnorm-channel-defn} we have
    \begin{align*}
      \tnorm{\Phi} \tnorm{\Psi}
      &= \max_{X \in \linear{K}} \max_{Y \in \linear{H}}
        \frac{\tnorm{ \Phi(X) } \tnorm{ \Psi(Y) }} {\tnorm{X} \tnorm{Y}} \\
      &\geq \max_{Y \in \linear{H}}
        \frac{\tnorm{ \Phi(\Psi(Y)) } \tnorm{ \Psi(Y) }} {\tnorm{\Psi(Y)} \tnorm{Y}} \\
      &= \max_{Y \in \linear{H}}
        \frac{\tnorm{ \Phi(\Psi(Y)) }} {\tnorm{Y}} \\
      &=\tnorm{\Phi \circ \Psi},
   \end{align*}
   as in the statement of the lemma.
  \end{proof}
\end{lemma}

One other useful fact related to this norm is that it is always
achieved on an input operator of the form $\ket\phi\bra\psi$.  This
follows from a direct convexity argument.
\begin{proposition}\label{meas-prop-tnorm-conv}
  For any $\Phi\colon \linear{H} \to \linear{K}$, there are pure
  states $\ket\phi$ and $\ket\psi$ such that
  \[ \tnorm{\Phi} = \tnorm{\Phi(\ket\phi\bra\psi)}. \]

  \begin{proof}
    Let $X \in \linear{H}$ achieve the maximum in the definition of
    the norm, and let $X = \sum_i s_i \ket{\phi_i}\bra{\psi_i}$ be a
    singular value decomposition of $X$.  Applying the
    triangle inequality to the definition of the operator trace norm
    (Equation~\eqref{meas-eqn-tnorm-channel-defn}) results in
    \begin{align*}
      \tnorm{\Phi}
      &= \frac{\tnorm{\Phi(X)}}{\tnorm{X}}
      = \frac{\tnorm{\sum_i s_i
          \Phi(\ket{\phi_i}\bra{\psi_i})}}{\tnorm{X}}
      \leq \frac{\sum_i s_i \tnorm{\Phi(\ket{\phi_i}\bra{\psi_i})}}{\tnorm{X}}      
    \end{align*}
    Then, since the definition of the
    trace norm implies that $\tnorm{X} = \sum_i s_i$, we may view this
    as a weighted average over terms
    $\tnorm{\Phi(\ket{\phi_i}\bra{\psi_i})}$.  Since at least one of
    these terms must be at least as large as the average, there exists
    an $i$ for which
    \[ \tnorm{\Phi} \leq \tnorm{\Phi(\ket{\phi_i}\bra{\psi_i})}, \]
    which implies that the trace norm is achieved on the state 
    $\ket{\phi_i}\bra{\psi_i}$.
  \end{proof}
\end{proposition}

Unfortunately, this extension of the trace norm to channels does not
have all of the properties that we might like it to have.  The most
important of these is \emph{stability}, which is, the norm of an
operator should not depend on the existence of a reference system, or
more concretely, the norm of the map $\Phi$ should not be smaller
than the norm of the map $\Phi \tprod \tidentity{}$.  An example
due to Watrous~\cite{Watrous08distinguishing} provides two channels
$\Phi, \Psi$ on $d$-dimensional states such that $\tnorm{\Phi \tprod
  \tidentity{} - \Psi \tprod \tidentity{}} = 2$ for an appropriate
reference system, but $\tnorm{\Phi - \Psi} \in O(1/d)$.  Phrased in
terms of distinguishability, these are two channels that are perfectly
distinguishable with a reference system but almost identical without it.

For this reason, we make use of the \emph{diamond norm}, introduced by
Kitaev~\cite{Kitaev97}.  This norm defines in the reference system,
which stabilizes the super-operator trace norm.
\begin{defn}\label{meas-defn-dnorm}
  For a linear map $\Phi$ taking $\linear{H}$ to $\linear{K}$, the
  diamond norm of $\Phi$ is
  \begin{equation*}
    \dnorm{\Phi} 
    = \tnorm{\Phi \tprod \tidentity{H}}
    = \max_{X \in \linear{H \tprod H}, X \neq 0} \frac{\tnorm{ (\Phi \tprod \tidentity{H})(X) }}{\tnorm{X}}.
  \end{equation*}
\end{defn}
\index{diamond norm}%
\nomenclature[anormd]{$\dnorm{\cdot}$}{diamond norm}
This norm is closely related to the completely bounded norm studied in
operator algebra.  If $\Phi$ maps $\linear{H}$ to $\linear{K}$ and
$\Phi^*$ is the adjoint map defined by $\tr( A^* \Phi(B)) = \tr(
  (\Phi^*(A))^* B )$, then $\dnorm{\Phi} =
\norm{\Phi^*}_{\mathrm{cb}}$.  More information on the completely
bounded norm can be found in~\cite{Paulsen02}.

As in the case of the super-operator trace norm, this norm inherits
many properties from the trace norm, such as the triangle inequality
and invariance under unitary operations.  From a computational
perspective, the optimization in the definition can be cast as a
semidefinite program~\cite{Watrous09semidefinite} or as a more general convex
optimization problem~\cite{Ben-AroyaT-S09}.  The
paper~\cite{JohnstonK+09} also gives a heuristic for evaluating this norm.

It is not too difficult to see that this norm is stable.  This was
first noted by Kitaev~\cite{Kitaev97} for the diamond norm, and by
Smith~\cite{Smith83} for the equivalent case of the completely bounded
norm.  The simpler proof used here can be found in~\cite{Watrous05},
though a similar argument appears in~\cite{GilchristL+05}, for
the case that the maximization in the
definition of the diamond norm is restricted to the density operators.
\index{diamond norm!stabilization}%
\begin{theorem}[Kitaev~\cite{Kitaev97}, Smith~\cite{Smith83}]
\label{meas-thm-dnorm-stabilize}
  Let $\Phi$ be a linear map from $\linear{H}$ to $\linear{K}$.  For
  any space $\mathcal{F}$
  \[ \dnorm{\Phi} = \tnorm{\Phi \tprod \tidentity{H}} \geq \tnorm{\Phi
    \tprod \tidentity{F}}. \]

  \begin{proof}
    In the case that $\dm{F} < \dm{H}$ the statement of the theorem is
    clear: the maximization in the definition of the super-operator
    trace norm is being taken over a smaller space.

    In the case that $\dm{F} \geq \dm{H}$, Proposition~\ref{meas-prop-tnorm-conv}
    implies that there exist vectors $\ket\phi$ and
    $\ket\psi$ such that
    \begin{equation*}
      \tnorm{\Phi \tprod \tidentity{F}} 
      =\tnorm{(\Phi \tprod \tidentity{F}) (\ket \phi \bra \psi)}.
    \end{equation*} 
    If we take Schmidt decompositions of these vectors, they can have
    at most $d = \min \{ \dm{F}, \dm{H} \} = \dm{H}$ terms.  Doing so, we
    have
    \begin{align}
      \ket\phi &= \sum_{i=1}^d \lambda_i \ket{a_i} \ket{x_i}, &
      \ket\psi &= \sum_{i=1}^d \gamma_i \ket{b_i} \ket{y_i},
      \label{meas-eqn-stab-schmidt}
    \end{align}
    where $\{\ket{a_i}\}, \{\ket{b_i}\}$ are orthonormal bases for
    $\mathcal{H}$, and $\{\ket{x_i}\}, \{\ket{y_i}\}$ are bases for
    $d$-dimensional subspaces of $\mathcal{F}$.  The remainder of the
    proof involves the straightforward but technical argument based on
    the Schmidt decomposition that we
    can embed these subspaces into a space of dimension $d$ with no
    loss in the value of the norm.  This is simply a formalization of
    the observation that since the states $\ket\phi$ and $\ket\psi$
    live in a $d$-dimensional subspace of $\mathcal{F}$, we do not
    need the auxiliary space in the diamond norm to have more than
    this dimension.

    To formalize this, let $U, V \in \unitary{H,F}$ be the isometries that take
    the standard basis $\{\ket{i}\}$ of $\mathcal{H}$ to the bases
    $\{x_i\}$ and $\{y_i\}$, respectively.  The maps $UU^*$ and $VV^*$
    are then the projections onto the spaces spanned by $\{x_i\}$ and
    $\{y_i\}$, which do not affect $\ket\phi$ and $\ket\psi$ by
    Equation~\eqref{meas-eqn-stab-schmidt}.  Using this notation
    \begin{align*}
      \tnorm{\Phi \tprod \tidentity{H}}
      &\geq \tnorm{(\Phi \tprod \tidentity{H}) 
        [(\identity{H} \tprod U^*) \ket \phi \bra \psi (\identity{H} \tprod V)]} \\
      &= \tnorm{(\identity{K} \tprod U)(\Phi \tprod \tidentity{H}) 
        [(\identity{H} \tprod U^*) \ket \phi \bra \psi (\identity{H}
        \tprod V)](\identity{K} \tprod V^*)} \\
      &= \tnorm{(\Phi \tprod \tidentity{F}) 
        [(\identity{H} \tprod UU^*) \ket \phi \bra \psi (\identity{H}
        \tprod VV^*)]} \\
      &= \tnorm{(\Phi \tprod \tidentity{F}) 
        (\ket \phi \bra \psi)}\\
      &= \tnorm{\Phi \tprod \tidentity{F}},
    \end{align*}
    as desired.    
  \end{proof}
\end{theorem}

In addition to stability, this norm has several other convenient
properties.  One of these is multiplicativity with respect to the
tensor product.  This is not true of the maximum output $p$-norm for
any $p>1$~\cite{HaydenW08}, but in the case of the diamond norm, it is
a direct consequence of the submultiplicativity of the trace norm.
\index{diamond norm!multiplicativity}%
\begin{theorem}\label{meas-thm-dnorm-mult}
  Let $\Phi \in \transform{H,K}$ and $\Psi \in \transform{F,G}$.  Then
  \[ \dnorm{\Phi \tprod \Psi} = \dnorm{\Phi} \dnorm{\Psi}. \]

  \begin{proof}
    One direction follows immediately from the multiplicativity of the
    trace norm with respect to the tensor product
    (Equation~\eqref{meas-eqn-pnorm-mult}), since
    \begin{align*}
      \dnorm{\Phi \tprod \Psi}
      &= \max_{X \in \linear{H \tprod F}} \frac{\tnorm{(\Phi \tprod \Psi)(X)}}{\tnorm{X}}\\
      &\geq \max_{X \in \linear{H}, Y \in \linear{F}} 
        \frac{\tnorm{\Phi(X) \tprod \Psi(Y)}}{\tnorm{X \tprod Y}}
%      &= \max_{X \in \linear{H}, Y \in \linear{F}} 
%       \frac{\tnorm{\Phi(X)} \tnorm{\Psi(Y)}}{\tnorm{X} \tnorm{Y}}\\
      = \dnorm{\Phi} \dnorm{\Psi}.
    \end{align*}

    The other direction is a consequence of
    Theorem~\ref{meas-thm-dnorm-stabilize} and
    Lemma~\ref{meas-lem-dnorm-submult}:
    \begin{align*}
      \dnorm{\Phi \tprod \Psi}
      &= \tnorm{\Phi \tprod \Psi \tprod \tidentity{H \tprod F}}\\
      &= \tnorm{\Phi \tprod \tidentity{F \tprod H \tprod F} \circ
                      \Psi \tprod \tidentity{H \tprod H \tprod F}}\\
      &\leq \tnorm{\Phi \tprod \tidentity{F \tprod H \tprod F}}
               \tnorm{\Psi \tprod \tidentity{H \tprod H \tprod F}}\\
      &= \tnorm{\Phi \tprod \tidentity{H}}
               \tnorm{\Psi \tprod \tidentity{F}}\\
      &= \dnorm{\Phi} \dnorm{\Psi},   
    \end{align*}
    which completes the proof.
  \end{proof}
\end{theorem}

As promised, the diamond norm of $\Phi - \Psi$ gives the probability
that an unknown channel in $\{\Phi, \Psi\}$ can be correctly
identified with only a single use of the channel.  This gives an
important operational characterization of the diamond norm that has
many useful applications to quantum error correction and other fields.
The proof follows directly from the definition of the diamond norm and
Helstrom's result on the minimum error distinguishability for two
states (Theorem~\ref{meas-thm-Helstrom}).

\index{diamond norm!relationship to distinguishability}%
\begin{corollary}\label{meas-cor-dnorm-Helstrom}
  The optimal probability that an unknown channel $\Phi \in
  \transform{H,K}$
  chosen uniformly at random from $\{\Phi_1, \Phi_2\}$ can be correctly
  identified given a single use is given by
  \[ \frac{1}{2} + \frac{\dnorm{\Phi_1 - \Phi_2}}{4}. \]

  \begin{proof}
    By Theorem~\ref{meas-thm-dnorm-difference} that follows (and is
    proven independently of this result)
    the maximization in the
    definition of the diamond norm may be taken over a pure state, so
    that for some space $\mathcal{F}$, there exists a pure state $\ket
    \psi \in \mathcal{H \tprod F}$ such that, the value in the
    statement of the theorem is equal to
    \begin{align}\label{meas-eqn-dnorm-Helstrom}
      \frac{1}{4} \left(2 + \tnorm{
            (\Phi_1 \tprod \tidentity{F})(\ket \psi \bra \psi)
          - (\Phi_2 \tprod \tidentity{F})(\ket \psi \bra \psi)
         }\right).
    \end{align}
    This expression is simply the optimal
    probability of identifying the state 
    \[ (\Phi \tprod \tidentity{F})(\ket \psi \bra \psi) \] 
    from the set
    \[ \{ (\Phi_1 \tprod \tidentity{F})(\ket \psi \bra \psi), 
          (\Phi_2 \tprod \tidentity{F})(\ket \psi \bra \psi) \}, \] 
    by Theorem~\ref{meas-thm-Helstrom}.

    Given only one use of $\Phi$, there is no other strategy than
    applying it to (a portion of) some optimal distinguishing state and then
    measuring the result.  Theorem~\ref{meas-thm-dnorm-difference}
    implies that there is such a state, and so the optimal probability
    is given by Equation~\eqref{meas-eqn-dnorm-Helstrom}, as required.
  \end{proof}
\end{corollary}

With this corollary in mind, we will define the computational problem
of distinguishing two channels in terms of the diamond norm of their
difference.  This is the main problem studied in this thesis, and so
the properties of the diamond norm that we have so far defined will be
useful throughout.

\subsection{Maximization on a pure state for the difference of channels}\label{meas-scn-dnorm-diff}

In this section it is shown that when applied to the difference of two
channels the diamond norm is achieved on a pure state.  The result
is technical, but it has many applications in the remainder of the
thesis.  This is the product of joint work with John Watrous~\cite{RosgenW05}.

The theorem applies to the maps $\Phi$ that are the difference of two
completely positive maps.  This property implies another simple
property: there exist completely positive $\Psi, \Gamma$ such that $\Phi = \Psi -
\Gamma$ if and only if $\Phi(X^*) = \Phi(X)^*$ for all $X$.
We will only need one direction of this equivalence.  If $\Phi = \Psi
- \Gamma$, taking Kraus operator decompositions of the
completely positive maps $\Psi$ and $\Gamma$ implies that
\begin{equation*}
  \Phi(X^*) 
  = \Psi(X^*) - \Gamma(X^*)
  = \sum_i A_i X^* A_i^* - B_i X^* B_i^*
  = \left( \sum_i A_i X A_i^* - B_i X B_i^* \right)^*
  =\Phi(X)^*.
\end{equation*}
This implication will be used in the proof of the main theorem of the
section.  In the construction used for the proof of this theorem, the
space $\mathcal{F}$ has dimension $2 \dm{H}$.  As explained following
the theorem, this can be achieved with a space of dimension $\dm{H}$,
though the argument is not included in the proof of the theorem for simplicity.
\index{diamond norm!condition for maximization on pure states}
\begin{theorem}\label{meas-thm-dnorm-difference}
  Let $\Phi \colon \linear{H} \to \linear{K}$ be the difference of two
  completely positive maps.  There exists a Hilbert space
  $\mathcal{F}$ and a unit vector
  $\ket{\psi} \in \mathcal{H \tprod F}$ such that
  \begin{equation*}
    \dnorm{\Phi} 
    = \tnorm{(\Phi \tprod \tidentity{F})(\ket{\psi}\bra{\psi})}.
  \end{equation*}
  \begin{proof}
    By the definition of the diamond norm
    \begin{equation*}
      \dnorm{\Phi} 
      = \tnorm{\Phi \tprod \tidentity{H}}
      = \max\{\tnorm{ (\Phi \tensor \tidentity{H})(X) } : \tnorm{X} = 1\}.
    \end{equation*}
    Let $X \in \linear{H \tprod H}$ be a state that achieves this
    maximum and let $\mathcal{C}$ be a Hilbert space of dimension two
    (i.e.\ a single qubit).  Consider the Hermitian operator $Y \in \linear{H
      \tprod H \tprod C}$ given by
    \begin{equation*}
      Y = \frac{1}{2} X \tprod \ket{0}\bra{1}+\frac{1}{2} X^* \tprod \ket{1}\bra{0}.
    \end{equation*}
    Notice also that $\tnorm{Y} = \tnorm{X} = 1$ by
    Lemma~\ref{meas-lem-block-tnorm}.
    
    As observed above, the condition that $\Phi$ is the difference of
    two complete positive transformations implies that $\Phi(X^*) =
    \Phi(X)^*$ for all $X$.  Using this, as well as Lemma~\ref{meas-lem-block-tnorm}
    \begin{align*}
      \tnorm{(\Phi \tprod \tidentity{H \tprod C})(Y)}
      & = \frac{1}{2} \tnorm{ (\Phi \tprod \tidentity{H})(X) \tprod \ket{0}\bra{1} 
        + (\Phi\tprod \tidentity{H})(X^*) \tprod \ket{1}\bra{0}} \\
      & = \frac{1}{2} \tnorm{ (\Phi \tprod \tidentity{H})(X) \tprod \ket{0}\bra{1} 
        + (\Phi\tprod \tidentity{H})(X)^* \tprod \ket{1}\bra{0}} \\
     & = \tnorm{ (\Phi \tprod \tidentity{H})(X) } \\
     & = \dnorm{\Phi}.
    \end{align*}
    This implies that the maximum is achieved on a Hermitian matrix
    $Y$.

    It is not hard to see that this implies that the maximum is
    achieved on a pure state.  To do so, note that since
    $Y$ is Hermitian, it has a spectral decomposition.  Let such a
    decomposition be given by
    \begin{equation*}
      Y = \sum_i \lambda_i \ket{\psi_i}\bra{\psi_i},
    \end{equation*}    
    where $\{ \ket{\psi_i}\}$ is an orthonormal basis of eigenvectors
    with real eigenvalues $\{\lambda_i\}$.
    In addition, because $\tnorm{Y} = 1$, it is the case that
    $\sum_i |\lambda_i| = 1$.
    By the linearity of $\Phi$, as well as the triangle inequality and
    the homogeneity of the trace norm,
    \begin{equation*}
      \tnorm{(\Phi \tprod \tidentity{H \tprod C})(Y)}
      \leq \sum_i \abs{ \lambda_i } 
        \tnorm{(\Phi \tprod \tidentity{H \tprod C})(\ket{\psi_i}\bra{\psi_i})}.
    \end{equation*}
    Because $\sum_i \abs{\lambda_i} = 1$, it follows that at
    least one term in the average achieves the bound, i.e.\ that
    \begin{equation*}
      \tnorm{(\Phi \tprod \tidentity{H \tprod C})(\ket{\psi_i}\bra{\psi_i})}
      \geq \dnorm{\Phi}
    \end{equation*}
    for some value of $i$.  For this value of $i$ we have, by
    Theorem~\ref{meas-thm-dnorm-stabilize},   
    \begin{equation*}
      \tnorm{(\Phi \tprod \tidentity{H \tprod C})(\ket{\psi_i}\bra{\psi_i})} 
      \leq \dnorm{\Phi},
    \end{equation*}
    which implies that $\tnorm{(\Phi \tprod \tidentity{H \tprod
        C})(\ket{\psi_i}\bra{\psi_i})} = \dnorm{\Phi}$ as required.
  \end{proof}
\end{theorem}

This theorem does not hold for the trace norm on super-operators, by
an example due to Watrous~\cite{Watrous05}.  It may seem odd that in
the proof of this theorem the space $\mathcal{F} = \mathcal{H \tprod
  C}$ has larger dimension than is required to achieve the maximum,
since $\dm{H \tprod C} = 2 \dm{H}$.  By examining the proof of
Theorem~\ref{meas-thm-dnorm-stabilize}, however, it can be seen that
this need not be the case.  Applying this theorem in the case that the
maximum is Hermitian produces a Hermitian state in $\density{H \tprod
  H}$ that achieves the maximum.  Applying the convexity argument made
at the end of the proof of Theorem~\ref{meas-thm-dnorm-difference}
yields a pure state in $\mathcal{H \tprod H}$ on which the maximum is
achieved.

One convenient consequence of this theorem is that the diamond norm of
any quantum channel is equal to one.  This is implied by the
definition, since for any $\Phi \in \transform{H,K}$ the theorem
implies that there is a state $\ket \psi$ such that
\begin{equation}\label{meas-eqn-dnorm-unity}
  \dnorm{\Phi}
  = \tnorm{(\Phi \tprod \tidentity{F})(\ket{\psi}\bra{\psi})}
  = \tnorm{\rho}
  = 1,
\end{equation}
where $\rho$ is the density matrix that results from applying $\Phi
\tprod \tidentity{F}$, which has trace norm one because it is
normalized.  An alternate proof of this fact, not using the above
result, can be found in~\cite{AharonovK+98}.

There is one further result on the diamond norm demonstrated in
Section~\ref{meas-scn-polarization}.  This result is a procedure for
polarizing this norm, in the sense that if the diamond norm of the
difference of the input channels is small, the result is two channels with extremely small
norm.  Similarly, if the input channels have large norm, then the
resulting channels are almost perfectly distinguishable.  Results of
this type can have powerful applications for error reduction.
The discussion of this result is
postponed until Section~\ref{meas-scn-polarization} so that the fidelity
can be introduced, as it is used in the proof of the polarization result.
The fidelity is the topic of the next section.

\section{Fidelity}\label{meas-scn-fidelity}

One of the most important tools in quantum information is the
fidelity, which provides a way to determine how close two states are
together.  For pure states $\ket \psi$ and $\ket \phi$ the fidelity
has a simple expression
\[ \F( \ket\psi, \ket \phi) = \abs{ \braket{\psi}{\phi} }. \]
This can be generalized to
the case of mixed quantum states $\rho, \sigma \in \density{H}$ by
\begin{equation}\label{meas-eqn-fidelity}
  \F(\rho,\sigma) = \tr \sqrt{\sqrt{\rho} \sigma \sqrt{\rho}}.
\end{equation}
\index{fidelity}%
\nomenclature[aF]{$\F(\rho,\sigma)$}{fidelity of states $\rho$ and $\sigma$}%
This quantity and its generalization to mixed states are due to Uhlmann~\cite{Uhlmann76}.
The fidelity ranges between $\F(\rho,\sigma) = 1$ when $\rho = \sigma$
and $\F(\rho,\sigma) = 0$ when $\rho$ and $\sigma$ have orthogonal
support.  The remainder of this section is a survey of some of the
most important properties of the fidelity.  A more complete
introduction to this quantity can be found in~\cite{NielsenC00}.

One property that is convenient to show from
Equation~\eqref{meas-eqn-fidelity} is multiplicativity with respect to
the tensor product.  Following~\cite{Jozsa94}, this is an easy
consequence of the fact that $\sqrt{\rho \tprod \sigma} = \sqrt{\rho}
\tprod \sqrt{\sigma}$.  This implies that
\index{fidelity!multiplicativity}%
\begin{align}
  \F(\rho_1 \tprod \rho_2,\sigma_1 \tprod \sigma_2)
  &= \tr \sqrt{\sqrt{\rho_1 \tprod \rho_2} (\sigma_1 \tprod \sigma_2)
    \sqrt{\rho_1 \tprod \rho_2}} \nonumber \\
  &= \left(\tr  \sqrt{\sqrt{\rho_1} \sigma_1 \sqrt{\rho_1}}\right)
  \left(\tr \sqrt{\sqrt{\rho_2} \sigma_2 \sqrt{\rho_2}}\right) \nonumber \\
  &=  \F(\rho_1, \sigma_1) \F(\rho_2,\sigma_2).\label{meas-eqn-fidelity-mult}
\end{align}
This is one of the few properties that is easy to prove from
Equation~\eqref{meas-eqn-fidelity}.  For example,
it is not clear from this equation that $\F(\rho,\sigma)
= \F(\sigma, \rho)$.  This property follows directly from a characterization of
the fidelity known as Uhlmann's theorem.  This characterization is
extremely useful, often being used as the definition of the fidelity,
and is presented as the following theorem.
\begin{theorem}[Uhlmann's Theorem \cite{Uhlmann76}]%
  \label{meas-thm-uhlmanns}%
  \index{fidelity!Uhlmann's Theorem}%
  \index{Uhlmann's Theorem}%
  Let $\rho, \sigma \in \density{H}$, and let $\mathcal{K}$ be any
  space large enough to admit purifications of $\rho$ and $\sigma$.  Then
  \[ \F(\rho,\sigma) = \max \{ \abs{ \braket{\phi}{\psi} } : 
  \ket\phi, \ket\psi \in \mathcal{H \tprod K}, 
  \ptr{K} \ket\phi\bra\phi = \rho, 
  \ptr{K} \ket\psi\bra\psi = \sigma\}. \]
\end{theorem}

Uhlmann's theorem restricted to the finite dimensional case (once
again, the cited result is much more general than has been applied
here) allows the derivation of several nice properties of the
fidelity.  These properties are summarized in the following proposition, many of
which are observed by Jozsa~\cite{Jozsa94}.

\begin{proposition}\label{meas-prop-fid-prop}
  For $\rho, \sigma \in \density{H}$ and $U \in \unitary{H,K}$, the
  fidelity satisfies
  \begin{enumerate}[(a)]
  \item $0 \leq F(\rho, \sigma) \leq 1$
    \label{meas-enum-fid1}
  \item $\F(\rho, \sigma) = \F(\sigma,\rho)$
    \label{meas-enum-fid2}
  \item $\F(U \rho U^*, U \sigma U^*) = \F(\rho, \sigma)$
    \label{meas-enum-fid3}
  \item For $\rho, \sigma \in \density{H \tprod K}$, $\F(\ptr{K}
    \rho, \ptr{K} \sigma) \geq \F(\rho, \sigma)$
    \label{meas-enum-fid4}
  \end{enumerate}

  \begin{proof}
    All four of these properties are simple corollaries of Theorem~\ref{meas-thm-uhlmanns}.
    Properties~(\ref{meas-enum-fid1}) and~(\ref{meas-enum-fid2})
    follow immediately.
    Property~(\ref{meas-enum-fid3}) follows from the fact that if
    $\ket\psi, \ket\phi \in \mathcal{H \tprod F}$ are purifications of
    $\rho$ and $\sigma$ achieving the maximum in
    the theorem, then $(U \tprod \identity{F})
    \ket\psi$ and $(U \tprod \identity{F}) \ket \phi$ are
    purifications of $U \rho U^*$ and $U \sigma U^*$, respectively,
    and
    \[ \abs{ \bra\psi (U \tprod \identity{F})^* (U \tprod \identity{F}) \ket \phi } 
        = \abs{ \braket{\psi}{\phi} } =\F(\rho,\sigma). \]
    Property~(\ref{meas-enum-fid4}) is a consequence of the fact that
    if $\ket \psi \in \mathcal{H \tprod K \tprod F}$ is a purification
    of $\rho$, then it is also a purification of $\ptr{K} \rho$.
 \end{proof}
\end{proposition}

One further property of the fidelity will be quite important: the
monotonicity under the application of a quantum channel.  We have seen
two special cases of this, unitary operations and the partial trace,
as part of the previous proposition.  Extending these cases to the set
of all channels is a simple consequence of the Stinespring
representation for channels.

This proof is due to Josza~\cite{Jozsa94} (see
also~\cite{NielsenC00}), though it is not difficult to derive from Theorem~\ref{meas-thm-uhlmanns}.
\index{fidelity!monotonicity}%
\begin{theorem}\label{meas-thm-fidelity-monotonicity}
  Let $\Phi \in \transform{H,K}$, then for any $\rho, \sigma \in \density{H}$%
  \[ \F(\rho, \sigma) \leq \F(\Phi(\rho), \Phi(\sigma)). \]
  \begin{proof}
    By Theorem~\ref{meas-thm-uhlmanns}, let $\ket\phi, \ket\psi \in
    \mathcal{H \tprod F}$ be purifications of $\rho$ and $\sigma$
    such that
    \[ \F(\rho,\sigma) = \abs{\braket{\phi}{\psi}}. \]
    Additionally, let $\Phi$ have
    a Stinespring representation given by $\Phi(X) = \ptr{B} U
    (X \tprod \ket 0 \bra 0) U^*$.  For brevity, let $\hat{U} = U
    \tprod \identity{F}$.
    Notice that $\hat{U} \ket\phi \ket
    0$ purifies $\Phi(\rho)$ and that $\hat{U} \ket\psi \ket
    0$ purifies $\Phi(\sigma)$.  Using this notation along with
    Theorem~\ref{meas-thm-uhlmanns},
    \begin{align*}
      \F(\Phi(\rho), \Phi(\sigma)) 
      &\geq \F( \hat{U} (\ket\phi \bra \phi \tprod \ket 0 \bra 0 ) \hat{U}^*,
                \hat{U} (\ket\psi \bra \psi \tprod \ket 0 \bra 0 ) \hat{U}^*) \\
      &= \abs{ \bra 0 \bra \phi \hat{U}^* \hat{U} \ket \psi \ket 0 } 
      = \abs{ \braket{\phi}{\psi} } 
      = \F(\rho,\sigma),      
    \end{align*}
    as required.
  \end{proof}
\end{theorem}

The special case of the monotonicity of the fidelity for the partial
trace (item~(\ref{meas-enum-fid4}) in
Proposition~\ref{meas-prop-fid-prop}) will be
particularly important.  This case implies that the fidelity
can only increase when it is taken over only part of a system, where
the remainder of the system has been traced out.  This property will
be essential to the results in Chapter~\ref{chap-close-images}.

\subsection{Relation to the trace norm}
\index{fidelity!relation to trace norm}%
\index{trace norm!relation to fidelity}%

The fidelity and the trace norm are two of the most useful quantities
for determining how close two quantum states are to each other.
Despite the similarities between them, it is often much more
convenient to work with one or the other of these quantities, and so
relationships between them are very useful.  This section presents two
such relationships that we will make use of later in the thesis.

Uhlmann's Theorem (\ref{meas-thm-uhlmanns}) can also be used to
characterize the fidelity using the trace norm.
This result is not hard to prove, but will be central to a couple of
proofs that appear later.
\begin{lemma}\label{meas-lemma-fidelity-tracenorm}
  Let $\rho, \xi \in \density{H}$.
  Then for arbitrary purifications
  $\ket\psi, \ket\phi \in \mathcal{H \tprod A}$
  of $\rho$ and $\xi$, respectively, we have
  $\tnorm{ \ptr{H} \ket\psi \bra\phi } = \F(\rho,\xi)$.
  \begin{proof}
    Using Lemma~\ref{meas-lem-tnorm-max}
    together with Theorem~\ref{meas-thm-uhlmanns} and the fact that
    all the purifications of $\rho, \sigma$ are unitarily equivalent
    using a unitary on the space $\mathcal{A}$, we have
    \begin{align*}
      \tnorm{ \ptr{H} \ket\psi \bra\phi }
      & = \max_{U \in \unitary{A}} \abs{\tr\left(\ptr{H}\ket\psi\bra\phi\right) U} \\
      & = \max_{U \in \unitary{A}} \abs{ \tr \ket\psi \bra\phi
      (\identity{H} \tprod U) }\\
      & = \max_{U \in \unitary{A}} \abs{ \bra\phi (\identity{H} \tprod U)
      \ket\psi }\\
      & = F(\rho,\xi)
    \end{align*}
    as claimed.
  \end{proof}
\end{lemma}

A very useful relationship between the fidelity and the trace norm is
given by the Fuchs-van de
Graaf Inequalities that relate the trace
norm and the fidelity.  These inequalities show that, up to polynomial
factors, the fidelity and the trace norm are equivalent.  This is
helpful, since it is often much easier to work with one or the other
of these quantities.
\index{Fuchs-van de Graaf Inequalities}%
\begin{theorem}[Fuchs and van de Graaf~\cite{FuchsG99}]
  \label{meas-thm-fuchs-van-de-graaf} 
  For any $\rho, \sigma \in \density{H}$
  \begin{equation*}
    1 - \F(\rho,\sigma)
    \leq \frac{1}{2} \tnorm{\rho - \sigma}
    \leq \sqrt{1 - \F(\rho,\sigma)^2}.
  \end{equation*}
\end{theorem}
The second inequality is not hard to prove.  Following~\cite{NielsenC00},
let $\ket \phi, \ket \psi \in
\mathcal{H \tprod A}$ be purifications of $\rho$ and $\sigma$
achieving the bound $\F(\rho,\sigma) = \abs{\braket{\phi}{\psi}}$
in Uhlmann's Theorem (\ref{meas-thm-uhlmanns}).  Using the
monotonicity of the trace norm under the partial trace
(Theorem~\ref{meas-thm-tnorm-monotonicity}), we have
\begin{align*}
  \tnorm{ \rho - \sigma }
  &= \tnorm{\ptr{A}(\ket\phi \bra \phi - \ket\psi \bra\psi)}\\
  &\leq \tnorm{\ket\phi \bra \phi - \ket\psi \bra\psi} \\
  &= 2 \sqrt{1 - \abs{\braket{\phi}{\psi}}^2}\\
  &\leq 2 \sqrt{1 - \F(\rho, \sigma)^2}.
\end{align*}
The first inequality is more difficult: it requires either
characterizing the fidelity and the trace norm in terms of classical
variants on the outcomes of measurements, as is done
in~\cite{FuchsG99, NielsenC00}, or proving a technical result on the
trace norm of the difference of two positive operators, as is done
in~\cite{KitaevS+02}.  Neither of these techniques are used in the
remainder of the thesis, and so the proof of this inequality is omitted.

The Fuchs-van de Graaf Inequalities may be equivalently rephrased in
terms of upper and lower bounds on the fidelity in terms of the trace
norm.  These bounds are
\begin{equation}\label{meas-eqn-reverse-fuchs-vdg}
  1 - \frac{1}{2} \tnorm{\rho - \sigma}
  \leq \F(\rho, \sigma) 
  \leq \sqrt{ 1 - \frac{1}{4} \tnorm{\rho - \sigma}^2 },
\end{equation}
and can be derived by simple manipulations of Theorem~\ref{meas-thm-fuchs-van-de-graaf}.
  
\subsection{Maximum output fidelity for channels}\label{meas-scn-chanfidelity}

The fidelity can be extended to quantum channels in much the same way
as the previous quantities that we have considered.  The \emph{maximum
  output fidelity} of two channels $\Phi_1, \Phi_2 \in
\transform{H,K}$ can be defined as
\begin{equation*}
  \F_{\max}(\Phi_1, \Phi_2) 
  = \max_{\rho,\sigma \in \density{H}} \F(\Phi_1(\rho), \Phi_2(\sigma))
\end{equation*}
\index{fidelity!maximum output fidelity}%
\nomenclature[aFmax]{$\F_{\max}(\Phi, \Psi)$}{Maximum output fidelity of channels $\Phi$ and $\Psi$}%
This quantity will is essential to the \prob{Close Images} problem
that is the focus of Chapter~\ref{chap-close-images}.  The property of
primary importance for this application is the multiplicativity of
$\F_{\max}$ with respect to the tensor product of two channels.  This
will be essential for error reduction on instances of \prob{Close
  Images}.  This result is used implicitly by Kitaev and
Watrous~\cite{KitaevW00}, and the main thrust of it can also be  found 
in~\cite{KitaevS+02} (see Problem~11.10).
Due to its importance, a
complete proof is presented here.  The method of proof used here is
due to John Watrous\footnote{John Watrous, private communication},
though it is similar to the techniques used in~\cite{KitaevW00,
  KitaevS+02}.
This proof makes use of the diamond norm, and specifically the
multiplicativity of the diamond norm with respect to tensor products,
which was introduced in Section~\ref{meas-scn-dnorm}.

The first part of the proof is a relationship between the maximum
output fidelity of two channels and the diamond norm
of a certain completely positive super-operator.
\begin{lemma}[Kitaev and Watrous~\cite{KitaevW00}]\label{meas-lem-fmax-dnorm}
  Let $\Phi, \Psi \in \transform{H,K}$, and the linear map $\Gamma \colon
  \linear{H} \to \linear{B}$ be given by
  \begin{align*}
    \Phi(X) &= \ptr{B} UXU^* \\
    \Psi(X) &= \ptr{B} VXV^* \\
    \Gamma(X) &= \ptr{K} UXV^*,
  \end{align*}
  where $U,V \in \unitary{H, B \tprod K}$.  Using this notation,
  \begin{equation*}
    \F_{\max}(\Phi, \Psi) = \dnorm{\Gamma}.
  \end{equation*}

  \begin{proof}
    Let $\mathcal{A}$ be a space with $\dm{A} = \dm{H}$ to allow
    purifications of states in $\density{H}$, and let $\hat{U} = U
    \tprod \identity{A}$ and $\hat{V} = V \tprod \identity{A}$, then
    \begin{align*}
      \F_{\max}(\Phi,\Psi)
      &= \max_{\rho,\sigma \in \density{H}} 
            \F(\Phi(\rho), \Psi(\sigma)) \\
      &= \max_{\ket\phi, \ket\psi \in \mathcal{H \tprod A}}
            \F(\ptr{A \tprod B} \hat{U} \ket\phi \bra \phi \hat{U}^*, 
                 \ptr{A \tprod B} \hat{V} \ket\psi \bra \psi \hat{V}^*),
    \end{align*}
    where $\ket\phi$ and $\ket\psi$ are purifications of $\rho$ and
    $\sigma$.  Applying Lemma~\ref{meas-lemma-fidelity-tracenorm} to
    this, since $\hat{U} \ket\phi$ purifies $\Phi(\rho)$
    and $\hat{V} \ket\psi$ purifies $\Psi)\sigma)$, results in
    \begin{align*}
      \max_{\ket\phi, \ket\psi \in \mathcal{H \tprod A}}
        \tnorm{ \ptr{K} \hat{U} \ket \phi \bra \psi \hat{V}^* }
      &= \max_{\ket\phi, \ket\psi \in \mathcal{H \tprod A}}
        \tnorm{ (\Gamma \tprod \tidentity{A}) (\ket \phi \bra \psi)}
      = \dnorm{\Gamma},
    \end{align*}
    where the last inequality is an application of Proposition~\ref{meas-prop-tnorm-conv}.
  \end{proof}
\end{lemma}

Using this lemma the desired result on the multiplicativity of
$\F_{\max}$ follows immediately from the multiplicativity of the
diamond norm with respect to the tensor product.

\begin{theorem}[Kitaev and Watrous~\cite{KitaevW00}]
  \label{meas-thm-outfid-mult}
  For any $\Phi_1, \Psi_1 \in \transform{H,K}$ and any $\Phi_2,
  \Psi_2 \in \transform{J,L}$
  \begin{equation*}
    \F_{\max}(\Phi_1 \tprod  \Phi_2, \Psi_1 \tprod \Psi_2)
    = \prod_{i=1,2} \F_{\max}(\Phi_i, \Psi_i).
  \end{equation*}

  \begin{proof}
    Let $\Phi_i(X) = \ptr{B} U_i X U_i^*$, $\Psi_i = \ptr{B} V_i X V_i^*$ 
    be Stinespring representations of the channels $\Phi_i,
    \Psi_i$ for $i = 1,2$, where for notational convenience the
    introduction of ancillary qubits has been merged into the
    isometries $U_i$ and $V_i$.  Then, setting $\Gamma_i(X) = \ptr{K}
    U_i X V_i^*$ we are in exactly the situation of
    Lemma~\ref{meas-lem-fmax-dnorm}.  Applying this lemma, as well as
    the multiplicativity of the diamond norm, gives
    \begin{equation*}
      \F_{\max}(\Phi_1 \tprod  \Phi_2, \Psi_1 \tprod \Psi_2)
      = \dnorm{\Gamma_1 \tprod \Gamma_2}
      = \dnorm{\Gamma_1} \dnorm{\Gamma_2}
      = \F_{\max}(\Phi_1, \Psi_1) \F_{\max}(\Phi_2, \Psi_2),
    \end{equation*}
    as claimed.
    \end{proof}
\end{theorem}

\section{Polarization of the diamond norm}\label{meas-scn-polarization}

This section describes a method for ``polarizing'' the diamond norm of
two channels.  This is a technique that, starting with two channels
$\Phi_1, \Phi_2 \in \transform{H,K}$, and constants $0 < b < a < 2$ such
that $2b < a^2$, creates channels $\Psi_1$ and $\Psi_2$ satisfying
\begin{align*}
  \dnorm{\Phi_1 - \Phi_2} & \leq b  \implies \dnorm{\Psi_1 - \Psi_2} \leq 2^{-k} \\
  \dnorm{\Phi_1 - \Phi_2} & \geq a  \implies \dnorm{\Psi_1 - \Psi_2} \geq 2 - 2^{-k}.
\end{align*}
The constructed channels $\Psi_1$ and $\Psi_2$ belong to
$\transform{H^{\tprod\text{$r$}}, K^{\tprod \text{$r$}}}$, where $r
\in O(k)$, i.e.\ the size of the resulting channels depends only
linearly on the error parameter $k$.
This provides a powerful technique for reducing the error
in many settings.  It provides one way to see that any promise problem
defined with a promise on the diamond norm difference of two channels
can be reduced to the same problem with a weaker gap, since the
instance with the weaker gap can be polarized using this technique.
The method is not perfect, however, as it depends on the technical
condition that $2b < a^2$.

This construction generalizes the polarization technique of Sahai and
Vadhan for the case of the $\ell_1$ norm of efficiently samplable probability
distributions~\cite{SahaiV03}.  This construction was generalized by
Watrous to the case of quantum states that can be efficiently
prepared~\cite{Watrous02}.  The further generalization given here to quantum
channels does not require any conceptual changes: the details work out
in almost exactly the same way as in the case of states.  This result
is the product of joint work with John Watrous, and has been published in~\cite{RosgenW05}.

In order that the polarization technique is useful in the setting of
computational hardness it must satisfy one further significant
property.  The construction must be efficient.  That is, given access
to polynomial-time circuits (or black boxes) for the original channels
$\Phi_1, \Phi_2$, circuits that implement the output
channels $\Psi_1$ and $\Psi_2$ can be efficiently constructed.  That
the polarization technique has this property will be easy to observe
from the construction given in the proof.

The proof of the polarization theorem makes use of two constructions.
One of these constructions increases the diamond norm and the other
reduces it.  Applying these constructions in the correct sequence will
result in transformations with the desired properties.  These two
constructions mirror the proof of the classical result due to Sahai
and Vadhan~\cite{SahaiV03}.

The first construction is a technique for increasing the diamond norm
of two channels.  The idea is simple: it is much easier to distinguish
$k$ copies of the channels than it is to distinguish one copy.  The
channels constructed using this procedure are simply $\Phi_i^{\tprod
  k}$.  The argument must be carefully made, however, to show that
entanglement across the multiple uses of the channels does not
increase the diamond norm too much.  The following direct product lemma gives bounds
on the diamond norm for the difference of $k$ copies of the two
channels.
\index{diamond norm!direct product lemma}%
\begin{lemma}\label{meas-lemma_direct_product}
  Let $\Phi_1, \Phi_2 \in \transform{H,K}$ have
  $\dnorm{\Phi_1 - \Phi_2} = \delta > 0$.  Then for any positive integer
  $k$
  \begin{equation*}
    2 - 2e^{ \frac{-k \delta^2}{8} }
    < \dnorm{ \Phi_1^{\tprod k} - \Phi_2^{\tprod k}}
    \leq  k \delta.
  \end{equation*}
  
  \begin{proof}
    To prove the first inequality,
    let $\rho \in \density{H \tprod H}$ achieve the
    maximum in the diamond norm, i.e.\ let
    \begin{equation*}
      \tnorm{(\Phi_1\tprod \tidentity{H})(\rho) -
        (\Phi_2\tprod \tidentity{H})(\rho)}
      = \dnorm{\Phi_1-\Phi_2} = \delta.
    \end{equation*}
    Such a state exists by Theorem~\ref{meas-thm-dnorm-difference}.
    As the trace norm is multiplicative with respect to the tensor
    product (by Equation~\ref{meas-eqn-pnorm-mult}), 
    $\tnorm{\rho^{\tprod k}} = \tnorm{\rho}^k = 1$.  Evaluating the maximum
    in the definition of the diamond norm on this state, we find that
    \begin{equation}\label{meas-eqn-direct-prod-ref-tnorm}
      \dnorm{ \Phi_1^{\tprod k} - \Phi_2^{\tprod k}}
      \geq \tnorm{\left((\Phi_1 \tprod \tidentity{H})(\rho)
        \right)^{\tprod k} - \left((\Phi_2 \tprod \tidentity{H})(\rho)
        \right)^{\tprod k}}.
    \end{equation}
    We can then apply the bound from~\cite{Watrous02} to the two
    states $\rho = (\Phi_1 \tprod \tidentity{H})(\rho)$ and $\sigma = (\Phi_2 \tprod
    \tidentity{H})(\rho)$ having trace distance $\delta$ to obtain the
    desired inequality.  For completeness, the proof of this bound
    follows.

    Since the fidelity is multiplicative with respect to the tensor
    product of two states (Equation~\eqref{meas-eqn-fidelity-mult})
    we can use the Fuchs-van de Graaf inequalities
    (Theorem~\ref{meas-thm-fuchs-van-de-graaf}) to obtain
    \begin{multline*}
      \tnorm{\rho^{\tprod k} - \sigma^{\tprod k}}
      \geq 2\left( 1 - \F(\rho^{\tprod k},\sigma^{\tprod k})\right)
      = 2\left( 1 - \F(\rho,\sigma)^k \right) \\
      \geq 2 - 2 \left(\sqrt{1 - \tnorm{\rho - \sigma}^2/4} \right)^k
      = 2 - 2 \left(1 - (\delta/2)^2\right)^{k/2}.
    \end{multline*}
    We can then bound this quantity using the inequality $(1-x)^k <
    e^{-kx}$, which holds for all nonzero $-1 < x < 1$.  This can be verified by
    taking logarithms and considering a Taylor series for $\ln(1-x)$.
    In our case, $x = \delta/2 < 1$, so we have
    \begin{equation*}
      2 - 2 \left(1 - (\delta/2)^2\right)^{k/2}
      > 2 - 2 \exp\left( \frac{-\delta^2}{4} \cdot \frac{k}{2}\right)
      =2 - 2e^{\frac{-k \delta^2}{8}}.
    \end{equation*}
    Combining this with
    Equation~\eqref{meas-eqn-direct-prod-ref-tnorm} proves the first
    inequality.

    The second inequality follows by induction on $k$.  The case of $k=1$
    leaves nothing to prove.  For $k > 1$, let
    $\Psi_i = \Phi_i^{\tprod{(k-1)}}$ for simplicity.  Using this
    notation, as well as the triangle inequality, we have
    \begin{align*}
      \dnorm{ \Phi_1^{\tprod k} - \Phi_2^{\tprod k}}
      & = \dnorm{ \Psi_1 \tprod \Phi_1 - \Psi_2 \tprod \Phi_2 } \\
      & = \dnorm{ \Psi_1 \tprod \Phi_1 - \Psi_2 \tprod \Phi_1 +
        \Psi_2 \tprod \Phi_1 - \Psi_2 \tprod \Phi_2 } \\
      & \leq \dnorm{(\Psi_1 - \Psi_2) \tprod \Phi_1 } +
      \dnorm{\Psi_2 \tprod (\Phi_1 - \Phi_2) } \\
      & = \dnorm{\Psi_1-\Psi_2} \dnorm{\Phi_1} + \dnorm{\Psi_2} 
      \dnorm{\Phi_1 - \Phi_2}.
    \end{align*}
    The final equality follows from the multiplicativity of the
    diamond norm, given by Theorem~\ref{meas-thm-dnorm-mult}.  Since
    the diamond norm of any channel is one
    (Equation~\eqref{meas-eqn-dnorm-unity}), the inductive hypothesis implies that
    \begin{equation*}
      \dnorm{ \Psi_1 - \Psi_2 } \dnorm{\Phi_1} + \dnorm{\Psi_2}
      \dnorm{\Phi_1 - \Phi_2}
      \leq (k - 1) \delta + \delta = k \delta
    \end{equation*}
    as required.
  \end{proof}
\end{lemma}

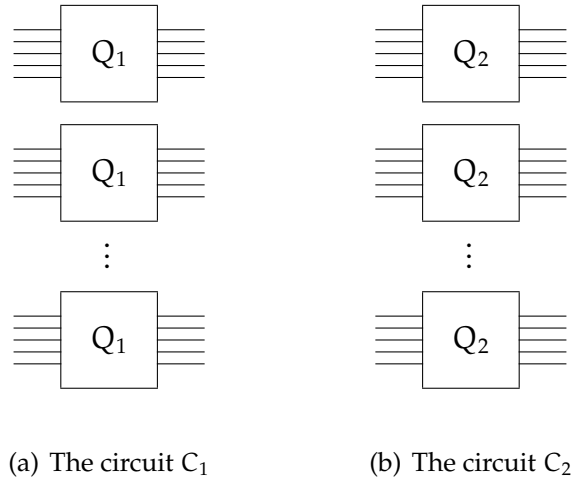
\begin{figure}
 \begin{center}
    \subfigure[The circuit $C_1$]{\setlength{\unitlength}{3947sp}%
\begingroup\makeatletter\ifx\SetFigFont\undefined%
\gdef\SetFigFont#1#2#3#4#5{%
  \reset@font\fontsize{#1}{#2pt}%
  \fontfamily{#3}\fontseries{#4}\fontshape{#5}%
  \selectfont}%
\fi\endgroup%
\begin{picture}(1824,2799)(589,-2998)
\put(1464,-2011){\makebox(0,0)[lb]{\smash{{\SetFigFont{12}{14.4}{\rmdefault}{\mddefault}{\updefault}{\color[rgb]{0,0,0}$\vdots$}%
}}}}
\thinlines
{\color[rgb]{0,0,0}\put(901,-1411){\line( 1, 0){300}}
}%
{\color[rgb]{0,0,0}\put(901,-1261){\line( 1, 0){300}}
}%
{\color[rgb]{0,0,0}\put(901,-1336){\line( 1, 0){300}}
}%
{\color[rgb]{0,0,0}\put(901,-1486){\line( 1, 0){300}}
}%
{\color[rgb]{0,0,0}\put(1801,-1561){\line( 1, 0){300}}
}%
{\color[rgb]{0,0,0}\put(2101,-1411){\line(-1, 0){300}}
}%
{\color[rgb]{0,0,0}\put(1801,-1261){\line( 1, 0){300}}
}%
{\color[rgb]{0,0,0}\put(1801,-1336){\line( 1, 0){300}}
}%
{\color[rgb]{0,0,0}\put(1801,-1486){\line( 1, 0){300}}
}%
{\color[rgb]{0,0,0}\put(901,-811){\line( 1, 0){300}}
}%
{\color[rgb]{0,0,0}\put(901,-661){\line( 1, 0){300}}
}%
{\color[rgb]{0,0,0}\put(901,-511){\line( 1, 0){300}}
}%
{\color[rgb]{0,0,0}\put(901,-586){\line( 1, 0){300}}
}%
{\color[rgb]{0,0,0}\put(901,-736){\line( 1, 0){300}}
}%
{\color[rgb]{0,0,0}\put(1801,-811){\line( 1, 0){300}}
}%
{\color[rgb]{0,0,0}\put(2101,-661){\line(-1, 0){300}}
}%
{\color[rgb]{0,0,0}\put(1801,-511){\line( 1, 0){300}}
}%
{\color[rgb]{0,0,0}\put(1801,-586){\line( 1, 0){300}}
}%
{\color[rgb]{0,0,0}\put(1801,-736){\line( 1, 0){300}}
}%
{\color[rgb]{0,0,0}\put(1201,-961){\framebox(600,600){$Q_1$}}
}%
{\color[rgb]{0,0,0}\put(901,-2611){\line( 1, 0){300}}
}%
{\color[rgb]{0,0,0}\put(901,-2461){\line( 1, 0){300}}
}%
{\color[rgb]{0,0,0}\put(901,-2311){\line( 1, 0){300}}
}%
{\color[rgb]{0,0,0}\put(901,-2386){\line( 1, 0){300}}
}%
{\color[rgb]{0,0,0}\put(901,-2536){\line( 1, 0){300}}
}%
{\color[rgb]{0,0,0}\put(1801,-2611){\line( 1, 0){300}}
}%
{\color[rgb]{0,0,0}\put(2101,-2461){\line(-1, 0){300}}
}%
{\color[rgb]{0,0,0}\put(1801,-2311){\line( 1, 0){300}}
}%
{\color[rgb]{0,0,0}\put(1801,-2386){\line( 1, 0){300}}
}%
{\color[rgb]{0,0,0}\put(1801,-2536){\line( 1, 0){300}}
}%
{\color[rgb]{0,0,0}\put(1201,-2761){\framebox(600,600){$Q_1$}}
}%
{\color[rgb]{0,0,0}\put(1201,-1711){\framebox(600,600){$Q_1$}}
}%
{\color[rgb]{0,0,0}\put(901,-1561){\line( 1, 0){300}}
}%
\end{picture}%
%
    \label{meas-fig-pol-dprod1}}
    \qquad
    \subfigure[The circuit $C_2$]{\setlength{\unitlength}{3947sp}%
\begingroup\makeatletter\ifx\SetFigFont\undefined%
\gdef\SetFigFont#1#2#3#4#5{%
  \reset@font\fontsize{#1}{#2pt}%
  \fontfamily{#3}\fontseries{#4}\fontshape{#5}%
  \selectfont}%
\fi\endgroup%
\begin{picture}(1824,2799)(589,-2998)
\put(1464,-2011){\makebox(0,0)[lb]{\smash{{\SetFigFont{12}{14.4}{\rmdefault}{\mddefault}{\updefault}{\color[rgb]{0,0,0}$\vdots$}%
}}}}
\thinlines
{\color[rgb]{0,0,0}\put(901,-1411){\line( 1, 0){300}}
}%
{\color[rgb]{0,0,0}\put(901,-1261){\line( 1, 0){300}}
}%
{\color[rgb]{0,0,0}\put(901,-1336){\line( 1, 0){300}}
}%
{\color[rgb]{0,0,0}\put(901,-1486){\line( 1, 0){300}}
}%
{\color[rgb]{0,0,0}\put(1801,-1561){\line( 1, 0){300}}
}%
{\color[rgb]{0,0,0}\put(2101,-1411){\line(-1, 0){300}}
}%
{\color[rgb]{0,0,0}\put(1801,-1261){\line( 1, 0){300}}
}%
{\color[rgb]{0,0,0}\put(1801,-1336){\line( 1, 0){300}}
}%
{\color[rgb]{0,0,0}\put(1801,-1486){\line( 1, 0){300}}
}%
{\color[rgb]{0,0,0}\put(901,-811){\line( 1, 0){300}}
}%
{\color[rgb]{0,0,0}\put(901,-661){\line( 1, 0){300}}
}%
{\color[rgb]{0,0,0}\put(901,-511){\line( 1, 0){300}}
}%
{\color[rgb]{0,0,0}\put(901,-586){\line( 1, 0){300}}
}%
{\color[rgb]{0,0,0}\put(901,-736){\line( 1, 0){300}}
}%
{\color[rgb]{0,0,0}\put(1801,-811){\line( 1, 0){300}}
}%
{\color[rgb]{0,0,0}\put(2101,-661){\line(-1, 0){300}}
}%
{\color[rgb]{0,0,0}\put(1801,-511){\line( 1, 0){300}}
}%
{\color[rgb]{0,0,0}\put(1801,-586){\line( 1, 0){300}}
}%
{\color[rgb]{0,0,0}\put(1801,-736){\line( 1, 0){300}}
}%
{\color[rgb]{0,0,0}\put(1201,-961){\framebox(600,600){$Q_2$}}
}%
{\color[rgb]{0,0,0}\put(901,-2611){\line( 1, 0){300}}
}%
{\color[rgb]{0,0,0}\put(901,-2461){\line( 1, 0){300}}
}%
{\color[rgb]{0,0,0}\put(901,-2311){\line( 1, 0){300}}
}%
{\color[rgb]{0,0,0}\put(901,-2386){\line( 1, 0){300}}
}%
{\color[rgb]{0,0,0}\put(901,-2536){\line( 1, 0){300}}
}%
{\color[rgb]{0,0,0}\put(1801,-2611){\line( 1, 0){300}}
}%
{\color[rgb]{0,0,0}\put(2101,-2461){\line(-1, 0){300}}
}%
{\color[rgb]{0,0,0}\put(1801,-2311){\line( 1, 0){300}}
}%
{\color[rgb]{0,0,0}\put(1801,-2386){\line( 1, 0){300}}
}%
{\color[rgb]{0,0,0}\put(1801,-2536){\line( 1, 0){300}}
}%
{\color[rgb]{0,0,0}\put(1201,-2761){\framebox(600,600){$Q_2$}}
}%
{\color[rgb]{0,0,0}\put(1201,-1711){\framebox(600,600){$Q_2$}}
}%
{\color[rgb]{0,0,0}\put(901,-1561){\line( 1, 0){300}}
}%
\end{picture}%
%
   \label{meas-fig-pol-dprod2}}
 \end{center}
 \caption[Circuits output by the construction in Lemma~\ref{meas-lemma_grow}]{Circuits $C_1$ and $C_2$ output by the
   construction in Lemma~\ref{meas-lemma_grow}.  Each circuit $C_i$
   contains $r$ independent copies of the circuit $Q_i$.}
 \label{meas-fig-pol-dprod}
\end{figure}

The lemma implies the existence of an efficient procedure to increase the diamond
norm of two channels, which will form half of the construction used to
polarize the diamond norm.  The key to this procedure is that while
this procedure increases the norm, it does so much faster when the
norm of the original circuits is large.  The circuits produced by this
procedure are demonstrated in Figure~\ref{meas-fig-pol-dprod}.
\begin{lemma}\label{meas-lemma_grow}
  There is a polynomial-time deterministic procedure that, on input
  $(Q_1, Q_2, 1^r)$, where $Q_1, Q_2$ are descriptions of mixed-state
  quantum circuits, produces as output descriptions of two quantum
  circuits, $(C_1, C_2)$ satisfying
  \begin{equation*}
    2 - 2 \exp\left( - \frac{r}{8} \dnorm{Q_1 - Q_2}^2 \right)
    <  \dnorm{C_1 - C_2} 
    \leq r \dnorm{Q_1 - Q_2}.
  \end{equation*}

  \begin{proof}
    For $i=1,2$, the circuit $C_i$ is constructed from $r$ parallel
    copies of the circuit $Q_i$.  This results in $C_i = Q_i^{\tprod
      r}$, so that the bounds in the statement of the lemma follow
    from Lemma~\ref{meas-lemma_direct_product}.
\end{proof}
\end{lemma}

This procedure to increase the diamond norm of the difference of two
channels is used in Chapter~\ref{chap-degr}, as it preserves the
degradability or antidegradability of the input channels.  This will
not be true of the remainder of the polarization procedure.

The second procedure that is used in the polarization construction is
used to reduce the diamond norm of the difference of two channels.
Before outlining the procedure, however, we prove the following simple
property of the norm.

\begin{proposition}\label{meas-prop_dnorm_split}
  Let $\Phi_1, \Phi_2 \in \transform{H,K}$ and 
  $\Psi_1, \Psi_2 \in \transform{F,G}$.
  Let
  \begin{align*}
    \Xi_1 & = \frac{1}{2} \Phi_1 \tprod \Psi_1 + 
    \frac{1}{2} \Phi_2 \tprod \Psi_2, \\
    \Xi_2 & = \frac{1}{2} \Phi_1 \tprod \Psi_2 + 
    \frac{1}{2} \Phi_2 \tprod \Psi_1.
  \end{align*}
  Then
  $\dnorm{\Xi_1-\Xi_2} = \frac{1}{2}\dnorm{\Phi_1-\Phi_2}
  \dnorm{\Psi_1 - \Psi_2}$.

  \begin{proof}
    The diamond norm is multiplicative with respect to tensor products
    (Theorem~\ref{meas-thm-dnorm-mult}), so that
    \begin{align*}
      \dnorm{\Xi_1 - \Xi_2}
      & =  \dnorm{ \frac{1}{2} (\Phi_1 - \Phi_2) \tprod (\Psi_1 - \Psi_2) } 
         =  \frac{1}{2} \dnorm{ \Phi_1 - \Phi_2 } \dnorm{\Psi_1 - \Psi_2}
    \end{align*}
    as required.
  \end{proof}
\end{proposition}

This property is useful in the proof that the technique for reducing
the diamond norm works correctly.  The idea behind this procedure is
that even if $Q_1$ and $Q_2$ are easy to distinguish, then
the channel $C_1$ is constructed by taking the tensor
product of $r$ channels, each chosen from $\{Q_1, Q_2\}$
uniformly at random, with the restriction that $Q_1$ appears an odd
number of times should be very hard to distinguish from the channel
$C_2$ constructed in the same way, except that $Q_1$ appears an
even number of times in $C_2$.  In effect, a procedure that
distinguishes $C_1$ and $C_2$ must succeed for all $r$ embedded
channels: this is because the goal is to determine the parity of the
number of times that $Q_1$ appears, and the parity is affected by
even a single mistake made by the distinguishing procedure.  This
construction mirrors that used on states in~\cite{Watrous02}, which
itself mirrors that used on probability distributions in~\cite{SahaiV03}.
The circuits produced by this
procedure are illustrated in Figure~\ref{meas-fig-pol-xor}.
\begin{figure}
 \begin{center}
    \subfigure[The circuit $C_1$]{\setlength{\unitlength}{3947sp}%
\begingroup\makeatletter\ifx\SetFigFont\undefined%
\gdef\SetFigFont#1#2#3#4#5{%
  \reset@font\fontsize{#1}{#2pt}%
  \fontfamily{#3}\fontseries{#4}\fontshape{#5}%
  \selectfont}%
\fi\endgroup%
\begin{picture}(1824,2799)(589,-2998)
\put(1464,-2011){\makebox(0,0)[lb]{\smash{{\SetFigFont{12}{14.4}{\rmdefault}{\mddefault}{\updefault}{\color[rgb]{0,0,0}$\vdots$}%
}}}}
\thinlines
{\color[rgb]{0,0,0}\put(901,-1411){\line( 1, 0){300}}
}%
{\color[rgb]{0,0,0}\put(901,-1261){\line( 1, 0){300}}
}%
{\color[rgb]{0,0,0}\put(901,-1336){\line( 1, 0){300}}
}%
{\color[rgb]{0,0,0}\put(901,-1486){\line( 1, 0){300}}
}%
{\color[rgb]{0,0,0}\put(1801,-1561){\line( 1, 0){300}}
}%
{\color[rgb]{0,0,0}\put(2101,-1411){\line(-1, 0){300}}
}%
{\color[rgb]{0,0,0}\put(1801,-1261){\line( 1, 0){300}}
}%
{\color[rgb]{0,0,0}\put(1801,-1336){\line( 1, 0){300}}
}%
{\color[rgb]{0,0,0}\put(1801,-1486){\line( 1, 0){300}}
}%
{\color[rgb]{0,0,0}\put(901,-811){\line( 1, 0){300}}
}%
{\color[rgb]{0,0,0}\put(901,-661){\line( 1, 0){300}}
}%
{\color[rgb]{0,0,0}\put(901,-511){\line( 1, 0){300}}
}%
{\color[rgb]{0,0,0}\put(901,-586){\line( 1, 0){300}}
}%
{\color[rgb]{0,0,0}\put(901,-736){\line( 1, 0){300}}
}%
{\color[rgb]{0,0,0}\put(1801,-811){\line( 1, 0){300}}
}%
{\color[rgb]{0,0,0}\put(2101,-661){\line(-1, 0){300}}
}%
{\color[rgb]{0,0,0}\put(1801,-511){\line( 1, 0){300}}
}%
{\color[rgb]{0,0,0}\put(1801,-586){\line( 1, 0){300}}
}%
{\color[rgb]{0,0,0}\put(1801,-736){\line( 1, 0){300}}
}%
{\color[rgb]{0,0,0}\put(1201,-961){\framebox(600,600){$Q_{x_1}$}}
}%
{\color[rgb]{0,0,0}\put(901,-2611){\line( 1, 0){300}}
}%
{\color[rgb]{0,0,0}\put(901,-2461){\line( 1, 0){300}}
}%
{\color[rgb]{0,0,0}\put(901,-2311){\line( 1, 0){300}}
}%
{\color[rgb]{0,0,0}\put(901,-2386){\line( 1, 0){300}}
}%
{\color[rgb]{0,0,0}\put(901,-2536){\line( 1, 0){300}}
}%
{\color[rgb]{0,0,0}\put(1801,-2611){\line( 1, 0){300}}
}%
{\color[rgb]{0,0,0}\put(2101,-2461){\line(-1, 0){300}}
}%
{\color[rgb]{0,0,0}\put(1801,-2311){\line( 1, 0){300}}
}%
{\color[rgb]{0,0,0}\put(1801,-2386){\line( 1, 0){300}}
}%
{\color[rgb]{0,0,0}\put(1801,-2536){\line( 1, 0){300}}
}%
{\color[rgb]{0,0,0}\put(1201,-2761){\framebox(600,600){$Q_{y_{\mathrm{odd}}}$}}
}%
{\color[rgb]{0,0,0}\put(1201,-1711){\framebox(600,600){$Q_{x_2}$}}
}%
{\color[rgb]{0,0,0}\put(901,-1561){\line( 1, 0){300}}
}%
\end{picture}%
%
    \label{meas-fig-pol-xor1}}
    \qquad
    \subfigure[The circuit $C_2$]{\setlength{\unitlength}{3947sp}%
\begingroup\makeatletter\ifx\SetFigFont\undefined%
\gdef\SetFigFont#1#2#3#4#5{%
  \reset@font\fontsize{#1}{#2pt}%
  \fontfamily{#3}\fontseries{#4}\fontshape{#5}%
  \selectfont}%
\fi\endgroup%
\begin{picture}(1824,2799)(589,-2998)
\put(1464,-2011){\makebox(0,0)[lb]{\smash{{\SetFigFont{12}{14.4}{\rmdefault}{\mddefault}{\updefault}{\color[rgb]{0,0,0}$\vdots$}%
}}}}
\thinlines
{\color[rgb]{0,0,0}\put(901,-1411){\line( 1, 0){300}}
}%
{\color[rgb]{0,0,0}\put(901,-1261){\line( 1, 0){300}}
}%
{\color[rgb]{0,0,0}\put(901,-1336){\line( 1, 0){300}}
}%
{\color[rgb]{0,0,0}\put(901,-1486){\line( 1, 0){300}}
}%
{\color[rgb]{0,0,0}\put(1801,-1561){\line( 1, 0){300}}
}%
{\color[rgb]{0,0,0}\put(2101,-1411){\line(-1, 0){300}}
}%
{\color[rgb]{0,0,0}\put(1801,-1261){\line( 1, 0){300}}
}%
{\color[rgb]{0,0,0}\put(1801,-1336){\line( 1, 0){300}}
}%
{\color[rgb]{0,0,0}\put(1801,-1486){\line( 1, 0){300}}
}%
{\color[rgb]{0,0,0}\put(901,-811){\line( 1, 0){300}}
}%
{\color[rgb]{0,0,0}\put(901,-661){\line( 1, 0){300}}
}%
{\color[rgb]{0,0,0}\put(901,-511){\line( 1, 0){300}}
}%
{\color[rgb]{0,0,0}\put(901,-586){\line( 1, 0){300}}
}%
{\color[rgb]{0,0,0}\put(901,-736){\line( 1, 0){300}}
}%
{\color[rgb]{0,0,0}\put(1801,-811){\line( 1, 0){300}}
}%
{\color[rgb]{0,0,0}\put(2101,-661){\line(-1, 0){300}}
}%
{\color[rgb]{0,0,0}\put(1801,-511){\line( 1, 0){300}}
}%
{\color[rgb]{0,0,0}\put(1801,-586){\line( 1, 0){300}}
}%
{\color[rgb]{0,0,0}\put(1801,-736){\line( 1, 0){300}}
}%
{\color[rgb]{0,0,0}\put(1201,-961){\framebox(600,600){$Q_{x_1}$}}
}%
{\color[rgb]{0,0,0}\put(901,-2611){\line( 1, 0){300}}
}%
{\color[rgb]{0,0,0}\put(901,-2461){\line( 1, 0){300}}
}%
{\color[rgb]{0,0,0}\put(901,-2311){\line( 1, 0){300}}
}%
{\color[rgb]{0,0,0}\put(901,-2386){\line( 1, 0){300}}
}%
{\color[rgb]{0,0,0}\put(901,-2536){\line( 1, 0){300}}
}%
{\color[rgb]{0,0,0}\put(1801,-2611){\line( 1, 0){300}}
}%
{\color[rgb]{0,0,0}\put(2101,-2461){\line(-1, 0){300}}
}%
{\color[rgb]{0,0,0}\put(1801,-2311){\line( 1, 0){300}}
}%
{\color[rgb]{0,0,0}\put(1801,-2386){\line( 1, 0){300}}
}%
{\color[rgb]{0,0,0}\put(1801,-2536){\line( 1, 0){300}}
}%
{\color[rgb]{0,0,0}\put(1201,-2761){\framebox(600,600){$Q_{y_{\mathrm{even}}}$}}
}%
{\color[rgb]{0,0,0}\put(1201,-1711){\framebox(600,600){$Q_{x_2}$}}
}%
{\color[rgb]{0,0,0}\put(901,-1561){\line( 1, 0){300}}
}%
\end{picture}%
%
   \label{meas-fig-pol-xor2}}
 \end{center}
 \caption[Circuits output by the construction in Lemma~\ref{meas-lemma_shrink}]{Circuits $C_1$ and $C_2$ output by the
   construction in Lemma~\ref{meas-lemma_shrink}.  The circuit $C_i$
   consists of $r-1$ independent circuits $Q_{x_i}$, each chosen
   randomly at run-time from
   $\{Q_1, Q_2\}$, and one final circuit chosen so that the parity of
   the indices of the chosen circuits is odd in the case $i=1$ and
   even in the case that $i=2$.  }
 \label{meas-fig-pol-xor}
\end{figure}
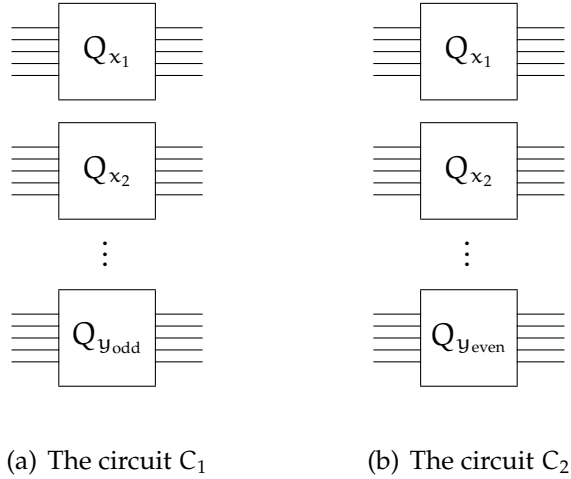
\begin{lemma}\label{meas-lemma_shrink}
  There is a deterministic polynomial-time procedure that,
  on input $(Q_1, Q_2, 1^r)$, where $Q_1, Q_2$ are descriptions of
  mixed-state quantum circuits, produces as output descriptions of two
  quantum circuits $(C_1, C_2)$ satisfying
  \[
  \dnorm{C_1 - C_2} = 2 \left( \frac{\dnorm{Q_1 - Q_2}}{2} \right)^r.
  \]
  \begin{proof}
    We use the circuits $C_1$ and $C_2$ outlined above.
    The circuit $C_1$ performs the transformation defined as
    \begin{equation*}
      C_1 =
      \frac{1}{2^{r-1}}\sum_{\stackrel{\scriptstyle x_1,\ldots,x_r\in\{1,2\}}
        {\scriptstyle x_1 + \cdots + x_r \equiv 1 \!\!\!\! \pmod 2}}
      Q_{x_1}\tprod\cdots\tprod Q_{x_r}
    \end{equation*}
    while $C_2$ performs a similar transformation defined as
    \begin{equation*}
      C_2 =
      \frac{1}{2^{r-1}}\sum_{\stackrel{\scriptstyle x_1,\ldots,x_r\in\{0,1\}}
        {\scriptstyle x_1 + \cdots + x_r \equiv 0 \!\!\!\! \pmod 2}}
      Q_{x_1}\tprod\cdots\tprod Q_{x_r}.
    \end{equation*}
    These circuits run $r$ copies of $Q_1$ and/or $Q_2$
    in parallel, where the choice of $Q_1$ or $Q_2$ determined uniformly at
    random subject to the constraint that $C_1$ applies an odd number
    of copies of $Q_1$ while $C_2$ applies an even number.
    Such circuits may be constructed in time polynomial in the sizes of
    $Q_1$ and $Q_2$ by using ancillary qubits with Hadamard and
    dephasing gates to generate the randomness.

    A proof by induction based on Proposition
    \ref{meas-prop_dnorm_split} establishes the desired equality.
    This proof is included here for completeness.  The base case,
    $r=1$, leaves nothing to prove.  Let $r > 1$, and let $D_1, D_2$
    be the channels $C_1$ and $C_2$ for the case $r -1$.  Notice that
    \begin{equation*}
      \dnorm{C_1 - C_2}
      = \frac{1}{2} \dnorm{Q_1 \tprod D_2 + Q_2 \tprod D_1 - Q_1
        \tprod D_1 - Q_2 \tprod D_2},
    \end{equation*}
    which can be observed from the construction of $C_1$ and $C_2$
    by considering the case that the first transformation is $Q_1$ or
    $Q_2$ and applying the parity conditions.  Applying
    Proposition~\ref{meas-prop_dnorm_split} to this, we have
    \begin{equation*}
      \dnorm{C_1 - C_2}
      =  \frac{1}{2} \dnorm{Q_1 - Q_2} 
         \dnorm{D_1 - D_2} 
      = 2 \left( \frac{\dnorm{Q_1 - Q_2}}{2} \right)^r,
      \end{equation*}
      where the last equality is by the induction hypothesis on
      $\dnorm{D_1 - D_2}$.
    \end{proof}
\end{lemma}

These two constructions, taken together, suffice to prove the
polarization theorem.  The proof consists of an application of
Lemma~\ref{meas-lemma_shrink}, followed by an application of
Lemma~\ref{meas-lemma_grow}, followed by one more application of
Lemma~\ref{meas-lemma_shrink}.  The proof is intuitively simple,
though it is quite technical: the value of $r$ used in each transformation must be
chosen very carefully.

\index{diamond norm!polarization}%
\begin{theorem}\label{meas-thm-polarize}
  Let the constants $a,b \in (0,2)$ satisfy $2 b < a^2$.
  There exists a deterministic polynomial-time procedure that, given
  input $(Q_1, Q_2, 1^n)$, where $Q_1$ and $Q_2$ are mixed-state quantum
  circuits, outputs quantum circuits $(C_1, C_2)$ such that
  \begin{align*}
    \dnorm{Q_1 - Q_2} & \leq b  \implies \dnorm{C_1 - C_2} < 2^{-n} \\
    \dnorm{Q_1 - Q_2} & \geq a  \implies \dnorm{C_1 - C_2} > 2 - 2^{-n}.
  \end{align*}

  \begin{proof}
    First, we apply Lemma \ref{meas-lemma_shrink} to the input $(Q_1, Q_2, 1^r)$, with
    \begin{equation*}
      r = \ceil{\log(16n) / \log( a^2 / (2 b))}.
    \end{equation*}
    This result in circuits $(Q_1', Q_2')$ such that 
    \begin{align*}
      \dnorm{Q_1 - Q_2} \leq b
      & \implies \dnorm{Q_1' - Q_2'} \leq 2 (b/2)^r, \\
      \dnorm{Q_1 - Q_2} \geq a
      & \implies \dnorm{Q_1' - Q_2'} \geq 2 (a/2)^r.
    \end{align*}
    Next, we apply Lemma \ref{meas-lemma_grow} to
    the input $(Q_1', Q_2', 1^s)$, where 
    \[ s = \floor{ (b / 2)^{-r} / 4}.\]
    This procedure produces circuits $(Q_1'', Q_2'')$ satisfying
    \begin{align*}
      \dnorm{Q_1 - Q_2} \leq b
      & \implies \dnorm{Q_1'' - Q_2''} \leq 2 (b/2)^r (b/2)^{-r}/4 = 1/2, \\
      \dnorm{Q_1 - Q_2} \geq a
      & \implies \dnorm{Q_1'' - Q_2''} >  2 - 2
      \exp(-\frac{s}{2}(a/2)^{2r} ) \geq 2 - 2 e^{-2n + 1}.
    \end{align*}
    The last inequality is due to the fact that
    \begin{equation*}
      \frac{s}{2}(a/2)^{2r} + 1
      \geq \frac{1}{8} (b/2)^{-r} (a/2)^{2r}
      \geq \frac{1}{8} \left(\frac{a^2}{2 b} \right)^r,
    \end{equation*}
    where the $+1$ term on the left is due to the floor in the
    definition of $s$.
    Taking logarithms of both sides, this is
    \begin{equation*}
      \log \left(\frac{s}{2}(a/2)^{2r} + 1\right)
      \geq \log \frac{1}{8} \left(\frac{a^2}{2 b} \right)^r
      \geq r \log \frac{a^2}{2 b} - 3
      \geq \frac{\log 16n}{\log a^2/(2b)} \log \frac{a^2}{2b} - 3
      = \log 2n.
    \end{equation*}
    This implies that $2 - 2\exp(-\frac{s}{2}(a/2)^{2r} ) \geq 2 - 2
    e^{-2n + 1}$, as required.
    
    Finally, applying Lemma \ref{meas-lemma_shrink} once more, this time
    to $(Q_1'', Q_2'', 1^t)$, where
    \[ t = \ceil{(n+1)/2}, \] 
    we obtain circuits $(C_1, C_2)$ such that
    \begin{align*}
      \dnorm{Q_1 - Q_2} \leq b
      & \implies \dnorm{C_1 - C_2} \leq (1/2)^{(n+1)/2} (1/2)^{(n - 1)/2} = 2^{-n} \\
      \dnorm{Q_1 - Q_2} \geq a
      & \implies
      \dnorm{C_1 - C_2}
      > (2 - 2 e^{-2n + 1})^{\ceil{(n+1)/2}}(1/2)^{\ceil{(n+1)/2} - 1}
      > 2 - 2^{-n}.
    \end{align*}
    The final inequality is due to the fact that by Bernoulli's
    inequality
    \begin{equation*}
      2 (1  - e^{-2n + 1})^{\ceil{(n+1)/2}}
      \geq 2 \left( 1 - \ceil{(n+1)/2} e^{-2n + 1} \right)
      > 2 - 2^{-n}.
    \end{equation*} 
    
    The circuits $(C_1, C_2)$ have size $rst$ times the size of the
    original circuits $(Q_1, Q_2)$.  By inspecting these quantities we
    find that $r,t \in O(n)$ and $s \in O(n^c)$ for $c$ a constant
    depending on the constants $a,b$.  This implies that the
    construction can be implemented in time polynomial in $n$ and the
    size of the original circuits.
  \end{proof}
\end{theorem}

\section{Conclusion}

In this chapter several different measures on quantum states and
channels have been introduced.  Many of the important properties of
these measures have also been defined.  This forms the basis for
the remainder of the thesis: the quantities described here, and their
properties, will find use throughout the problems studied later.  For
this reason it is hoped that this chapter will stand as a useful
reference for these concepts.

There are two new results contained within this chapter.  The first of
these is the theorem in
Section~\ref{meas-scn-dnorm-diff} that demonstrates that the maximum
in the diamond norm on the difference of two channels is achieved on a
pure quantum state, as opposed to a general linear operator.  This
property will be extremely useful when the diamond norm is later used
as a way to quantify the distinguishability of two quantum channels,
which is the central problem considered in this thesis.
The second new result is the polarization technique for the diamond
norm in Theorem~\ref{meas-thm-polarize}.  Part of this result is used in Chapter~\ref{chap-degr} to
reduce the error in the reductions of the distinguishability problem to
the degradable and the antidegradable channels.  Both of these results
are joint work with John Watrous~\cite{RosgenW05}.

%%% Local Variables: 
%%% mode: latex
%%% TeX-master: "thesis"
%%% End: 

\chapter{The Close Images Problem}\label{chap-close-images}

Given two quantum channels it is natural to ask how close the outputs
of the two channels can be.  When these channels are given as
mixed-state quantum circuits this becomes the computational problem
\prob{Close Images} that is considered in this chapter.  This problem
is~\class{QIP}-complete, as it is a restatement of the
definition the complexity class.  This result is due to Kitaev and
Watrous~\cite{KitaevW00}, but it is included here because it is the
result that all of the other hardness results in the thesis depend on.

The main result of this chapter is that restricting this problem to input circuits of
logarithmic depth does not reduce the computational difficulty.
This is shown by constructing a reduction from an
instance of \prob{Close Images} to an instance on log-depth circuits.
This provides further evidence for the computational power of
log-depth circuits.  The reduction that proves this result involves a simulation of
the two input circuits by log-depth circuits.  The maximum output
fidelity of the constructed circuits is related to the maximum output
fidelity for the original two circuits, so that this reduction
preserves the structure of the close images problem.

The results in this chapter on log-depth circuits have been published
in~\cite{Rosgen08distinguishing}.

\minitoc

\section{Log-depth mixed-state quantum circuits}\label{clim-scn-logdepth-circuits}

A significant practical problem in quantum information is that quantum
systems quickly decohere when allowed to interact with the
environment.  This process severely limits the length of quantum
computations that can be experimentally realized.  Short quantum
circuits provide a model of computation that can capture the kinds of
computation that we can perform under this type of time limit.  For
this reason it is of significant interest to find short quantum
circuits for important problems.

Log-depth quantum circuits have been found for several significant problems
including the approximate quantum Fourier transform~\cite{CleveW00}
and the encoding and decoding operations for many quantum error
correcting codes~\cite{MooreN02}.
In addition to these applications, a procedure for parallelizing to log-depth
a large class of quantum circuits is known~\cite{BroadbentK07}.  
These examples demonstrate the
surprising power of short quantum circuits.
It has been conjectured by Jozsa that any quantum algorithm can
be performed with logarithmic quantum circuit depth interspersed with
polynomial time classical computation~\cite{Jozsa06}.

The standard circuit model of quantum computation is the unitary
circuit model applied to pure quantum states.  In this thesis we
consider the more general model of mixed-state quantum computation
introduced in Section~\ref{compl-scn-circuit-model}.  While much of
the previous work on short quantum circuits has been in the
unitary circuit
model~\cite{FennerG+05, GreenH+02}, there has also been work outside
of this model~\cite{TerhalD04}.  The primary advantage of considering
this more general model is that the mixed state model is able to
capture any physically realizable quantum operation, and so results on
this model may have implications for experimental quantum information.

In this chapter it is shown that the apparent power of short quantum
computations comes with a price: the close images problem on
logarithmic depth quantum circuits is \emph{exactly} as difficult as
the general problem on polynomial depth circuits.
This result will be used, in Chapter~\ref{chap-distinguishability}, to
show that the problem of distinguishing mixed state circuits is also
no easier when restricted to log-depth circuits.

The remainder of this chapter is organized as follows.  In the next
section, the close images problem is discussed, and the result due to
Kitaev and Watrous~\cite{KitaevW00} that the
close images problem is complete for \class{QIP} is detailed.
In Section~\ref{clim-scn-swap-test} a key component of the reduction to
log-depth circuits is considered: the Swap Test.  This procedure can be
used to ensure that two pure quantum states are close together, and as
such is a key component of many quantum algorithms.
In Section~\ref{clim-scn-construction} a Karp reduction from the
polynomial depth to logarithmic depth versions of the close images
problem is presented in detail.
The correctness of this reduction is shown formally in Section~\ref{clim-scn-soundness}.

\section{\class{QIP} completeness of close images}\label{clim-scn-intro}

In this section an overview is given of the close images problem as it relates
to the complexity class \class{QIP}.  \prob{Close Images} is essentially a
restatement of the acceptance condition for the verifier in a quantum
interactive proof system, and so it will be important to review this
connection, as this connects all of the other computational
problems studied in the thesis to the class~\class{QIP}.

In order to model the hardness of the class of problems having quantum
interactive proof systems, Kitaev and Watrous introduced the close
images problem~\cite{KitaevW00}.  This problem can be given the
following formal definition.
\begin{problem}[Close Images]\label{clim-prob-ci}
  For constants $0 < b < a \leq 1$, the input consists of quantum
  circuits $Q_1$ and $Q_2$ that implement transformations
  in $\transform{H,K}$.
  The promise problem is to distinguish the two cases:
  \begin{description}
    \item[Yes:] $\F(Q_1(\rho), Q_2(\xi)) \geq a$ for some $\rho, \xi
      \in \density{H}$,
    \item[No:] $\F(Q_1(\rho), Q_2(\xi)) \leq b$ for all $\rho, \xi
    \in \density{H}$.
  \end{description}
\end{problem}
This is simply the problem of determining if there are inputs to $Q_1$
and $Q_2$ that cause them to output states that are nearly the same.
It will be helpful to abbreviate this problem as $\prob{CI}_{a,b}$
when the constants $a$ and $b$ will be significant.
\index{Close Images!problem definition}%
\nomenclature[PCI]{$\prob{CI}$}{Close Images Problem}%
It is the aim of the present chapter to prove that this problem remains
complete for \class{QIP} when restricted to circuits $Q_1$ and $Q_2$
that are of depth logarithmic in the number of input qubits.  This
will be achieved in the case of perfect soundness error, i.e.\ $a = 1$
in the above problem definition.  As discussed below, this problem
remains complete for \class{QIP} in this case.  This restriction
serves only to simplify the problem, as distinguishing the two
cases for a weaker promise can only be more difficult, so a hardness
result on this case
will also imply the hardness of the more general problem.
For the sake of brevity, the log-depth version of this problem will be
referred to as $\prob{Log-depth CI}_{a,b}$ and since this problem is a
restriction of a problem in $\class{QIP}$, as argued below, it is clear that it is also
in \class{QIP}.
\index{Close Images!log-depth}%
Similarly, the abbreviation $\prob{Const-depth CI}_{a,b}$ will be used
to denote the version of this problem on constant-depth circuits.
\index{Close Images!constant-depth}%

Although this problem was introduced by Kitaev and Watrous to show
that $\class{QIP} \subseteq \class{EXP}$, it was not
explicitly defined in~\cite{KitaevW00}.  For this reason the reduction
from the value of a quantum interactive proof system to the close
images problem is repeated here.  The hardness of this problem is
significant for the thesis: it
is from this reduction that all of the other \class{QIP}-hardness results
in the thesis follow.

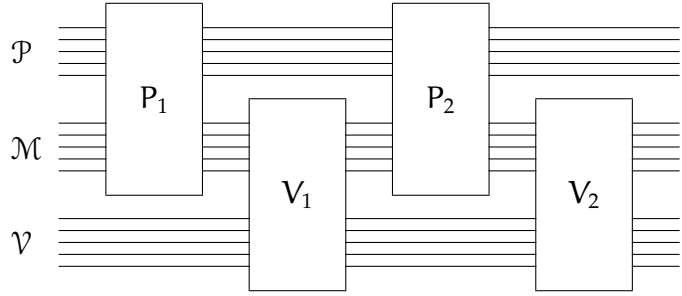
\begin{figure}
  \begin{center}
    \setlength{\unitlength}{3947sp}%
\begingroup\makeatletter\ifx\SetFigFont\undefined%
\gdef\SetFigFont#1#2#3#4#5{%
  \reset@font\fontsize{#1}{#2pt}%
  \fontfamily{#3}\fontseries{#4}\fontshape{#5}%
  \selectfont}%
\fi\endgroup%
\begin{picture}(4227,1824)(586,-2173)
\put(601,-736){\makebox(0,0)[lb]{\smash{{\SetFigFont{12}{14.4}{\rmdefault}{\mddefault}{\updefault}{\color[rgb]{0,0,0}$\mathcal{P}$}%
}}}}
\thinlines
{\color[rgb]{0,0,0}\put(1801,-661){\line( 1, 0){1200}}
}%
{\color[rgb]{0,0,0}\put(1801,-511){\line( 1, 0){1200}}
}%
{\color[rgb]{0,0,0}\put(1801,-586){\line( 1, 0){1200}}
}%
{\color[rgb]{0,0,0}\put(1801,-736){\line( 1, 0){1200}}
}%
{\color[rgb]{0,0,0}\put(1801,-1411){\line( 1, 0){300}}
}%
{\color[rgb]{0,0,0}\put(2101,-1261){\line(-1, 0){300}}
}%
{\color[rgb]{0,0,0}\put(1801,-1111){\line( 1, 0){300}}
}%
{\color[rgb]{0,0,0}\put(1801,-1186){\line( 1, 0){300}}
}%
{\color[rgb]{0,0,0}\put(1801,-1336){\line( 1, 0){300}}
}%
{\color[rgb]{0,0,0}\put(901,-811){\line( 1, 0){300}}
}%
{\color[rgb]{0,0,0}\put(901,-661){\line( 1, 0){300}}
}%
{\color[rgb]{0,0,0}\put(901,-511){\line( 1, 0){300}}
}%
{\color[rgb]{0,0,0}\put(901,-586){\line( 1, 0){300}}
}%
{\color[rgb]{0,0,0}\put(901,-736){\line( 1, 0){300}}
}%
{\color[rgb]{0,0,0}\put(901,-1411){\line( 1, 0){300}}
}%
{\color[rgb]{0,0,0}\put(901,-1261){\line( 1, 0){300}}
}%
{\color[rgb]{0,0,0}\put(901,-1111){\line( 1, 0){300}}
}%
{\color[rgb]{0,0,0}\put(901,-1186){\line( 1, 0){300}}
}%
{\color[rgb]{0,0,0}\put(901,-1336){\line( 1, 0){300}}
}%
{\color[rgb]{0,0,0}\put(901,-2011){\line( 1, 0){1200}}
}%
{\color[rgb]{0,0,0}\put(901,-1861){\line( 1, 0){1200}}
}%
{\color[rgb]{0,0,0}\put(901,-1711){\line( 1, 0){1200}}
}%
{\color[rgb]{0,0,0}\put(901,-1786){\line( 1, 0){1200}}
}%
{\color[rgb]{0,0,0}\put(901,-1936){\line( 1, 0){1200}}
}%
{\color[rgb]{0,0,0}\put(2701,-1411){\line( 1, 0){300}}
}%
{\color[rgb]{0,0,0}\put(3001,-1261){\line(-1, 0){300}}
}%
{\color[rgb]{0,0,0}\put(2701,-1111){\line( 1, 0){300}}
}%
{\color[rgb]{0,0,0}\put(2701,-1186){\line( 1, 0){300}}
}%
{\color[rgb]{0,0,0}\put(2701,-1336){\line( 1, 0){300}}
}%
{\color[rgb]{0,0,0}\put(3601,-1411){\line( 1, 0){300}}
}%
{\color[rgb]{0,0,0}\put(3901,-1261){\line(-1, 0){300}}
}%
{\color[rgb]{0,0,0}\put(3601,-1111){\line( 1, 0){300}}
}%
{\color[rgb]{0,0,0}\put(3601,-1186){\line( 1, 0){300}}
}%
{\color[rgb]{0,0,0}\put(3601,-1336){\line( 1, 0){300}}
}%
{\color[rgb]{0,0,0}\put(4501,-1411){\line( 1, 0){300}}
}%
{\color[rgb]{0,0,0}\put(4801,-1261){\line(-1, 0){300}}
}%
{\color[rgb]{0,0,0}\put(4501,-1111){\line( 1, 0){300}}
}%
{\color[rgb]{0,0,0}\put(4501,-1186){\line( 1, 0){300}}
}%
{\color[rgb]{0,0,0}\put(4501,-1336){\line( 1, 0){300}}
}%
{\color[rgb]{0,0,0}\put(4501,-2011){\line( 1, 0){300}}
}%
{\color[rgb]{0,0,0}\put(4801,-1861){\line(-1, 0){300}}
}%
{\color[rgb]{0,0,0}\put(4501,-1711){\line( 1, 0){300}}
}%
{\color[rgb]{0,0,0}\put(4501,-1786){\line( 1, 0){300}}
}%
{\color[rgb]{0,0,0}\put(4501,-1936){\line( 1, 0){300}}
}%
{\color[rgb]{0,0,0}\put(2701,-2011){\line( 1, 0){1200}}
}%
{\color[rgb]{0,0,0}\put(2701,-1861){\line( 1, 0){1200}}
}%
{\color[rgb]{0,0,0}\put(2701,-1711){\line( 1, 0){1200}}
}%
{\color[rgb]{0,0,0}\put(2701,-1786){\line( 1, 0){1200}}
}%
{\color[rgb]{0,0,0}\put(2701,-1936){\line( 1, 0){1200}}
}%
{\color[rgb]{0,0,0}\put(3601,-811){\line( 1, 0){1200}}
}%
{\color[rgb]{0,0,0}\put(3601,-661){\line( 1, 0){1200}}
}%
{\color[rgb]{0,0,0}\put(3601,-511){\line( 1, 0){1200}}
}%
{\color[rgb]{0,0,0}\put(3601,-586){\line( 1, 0){1200}}
}%
{\color[rgb]{0,0,0}\put(3601,-736){\line( 1, 0){1200}}
}%
{\color[rgb]{0,0,0}\put(1201,-1561){\framebox(600,1200){$P_1$}}
}%
{\color[rgb]{0,0,0}\put(2101,-2161){\framebox(600,1200){$V_1$}}
}%
{\color[rgb]{0,0,0}\put(3001,-1561){\framebox(600,1200){$P_2$}}
}%
{\color[rgb]{0,0,0}\put(3901,-2161){\framebox(600,1200){$V_2$}}
}%
\put(601,-1936){\makebox(0,0)[lb]{\smash{{\SetFigFont{12}{14.4}{\rmdefault}{\mddefault}{\updefault}{\color[rgb]{0,0,0}$\mathcal{V}$}%
}}}}
\put(601,-1336){\makebox(0,0)[lb]{\smash{{\SetFigFont{12}{14.4}{\rmdefault}{\mddefault}{\updefault}{\color[rgb]{0,0,0}$\mathcal{M}$}%
}}}}
{\color[rgb]{0,0,0}\put(1801,-811){\line( 1, 0){1200}}
}%
\end{picture}%
  \end{center}
  \caption[Transformations in a quantum interactive proof system]{The
    operations and Hilbert spaces corresponding to a three message
    quantum interactive proof system.}
  \label{clim-fig-qip}
\end{figure}

Recall that, due to the results of Kitaev and
Watrous~\cite{KitaevW00}, any quantum interactive proof system can be
parallelized to three messages.  For any input $x$, this results in
unitary circuits $P_1, V_1, P_2,$ and $V_2$ acting on systems
$\mathcal{V, M, P}$ for the verifier's private space, the message
space, and the prover's private space respectively.  These spaces and
transformations are illustrated in Figure~\ref{clim-fig-qip}.  Recall
from Section~\ref{compl-scn-complexity-classes}
that these circuits depend on the input, but the verifier's circuits
$V_1$ and $V_2$ must be generated in polynomial time in the length of
$x$.  For this reason the input string does not appear in the
description of the protocol: it is ``hard-coded'' into the circuits of
the two parties.
 As the verifier
accepts depending on the result of a measurement on one of the message
qubits at the end of the protocol, the value of the quantum
interactive proof system for fixed transformations $P_i$ is given by
\[ \tr \left[ \Pi \left( V_2 \circ P_2 \circ V_1 \circ P_1( \ket 0
    \bra 0 ) \right) \right], \]
where $\Pi$ is the projector onto the verifier's accepting subspace.
The prover can make the verifier accept if there exist circuits
$P_1$ and $P_2$ such that this probability is large.

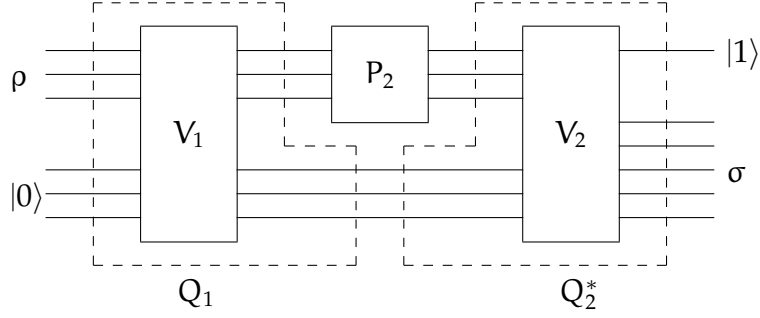
\begin{figure}
  \begin{center}
    \setlength{\unitlength}{3947sp}%
\begingroup\makeatletter\ifx\SetFigFont\undefined%
\gdef\SetFigFont#1#2#3#4#5{%
  \reset@font\fontsize{#1}{#2pt}%
  \fontfamily{#3}\fontseries{#4}\fontshape{#5}%
  \selectfont}%
\fi\endgroup%
\begin{picture}(4530,1965)(1561,-2614)
\put(6076,-1036){\makebox(0,0)[lb]{\smash{{\SetFigFont{12}{14.4}{\rmdefault}{\mddefault}{\updefault}{\color[rgb]{0,0,0}$\ket 1$}%
}}}}
\thinlines
{\color[rgb]{0,0,0}\put(1801,-1861){\line( 1, 0){600}}
}%
{\color[rgb]{0,0,0}\put(1801,-1711){\line( 1, 0){600}}
}%
{\color[rgb]{0,0,0}\put(2401,-1261){\line(-1, 0){600}}
}%
{\color[rgb]{0,0,0}\put(1801,-1111){\line( 1, 0){600}}
}%
{\color[rgb]{0,0,0}\put(3001,-2011){\line( 1, 0){1800}}
}%
{\color[rgb]{0,0,0}\put(3001,-1861){\line( 1, 0){1800}}
}%
{\color[rgb]{0,0,0}\put(3001,-1711){\line( 1, 0){1800}}
}%
{\color[rgb]{0,0,0}\put(3001,-1261){\line( 1, 0){600}}
}%
{\color[rgb]{0,0,0}\put(3001,-1111){\line( 1, 0){600}}
}%
{\color[rgb]{0,0,0}\put(4201,-1261){\line( 1, 0){600}}
}%
{\color[rgb]{0,0,0}\put(4201,-1111){\line( 1, 0){600}}
}%
{\color[rgb]{0,0,0}\put(6001,-2011){\line(-1, 0){600}}
}%
{\color[rgb]{0,0,0}\put(5401,-1861){\line( 1, 0){600}}
}%
{\color[rgb]{0,0,0}\put(5401,-1711){\line( 1, 0){600}}
}%
{\color[rgb]{0,0,0}\put(5401,-1411){\line( 1, 0){600}}
}%
{\color[rgb]{0,0,0}\put(5401,-1561){\line( 1, 0){600}}
}%
{\color[rgb]{0,0,0}\put(3601,-1411){\framebox(600,600){$P_2$}}
}%
{\color[rgb]{0,0,0}\put(3001,-961){\line( 1, 0){600}}
}%
{\color[rgb]{0,0,0}\put(4201,-961){\line( 1, 0){600}}
}%
{\color[rgb]{0,0,0}\put(5401,-961){\line( 1, 0){600}}
}%
{\color[rgb]{0,0,0}\put(2401,-2161){\framebox(600,1350){$V_1$}}
}%
{\color[rgb]{0,0,0}\put(4801,-2161){\framebox(600,1350){$V_2$}}
}%
{\color[rgb]{0,0,0}\multiput(2101,-661)(0.00000,-122.22222){14}{\line( 0,-1){ 61.111}}
\multiput(2101,-2311)(122.22222,0.00000){14}{\line( 1, 0){ 61.111}}
\multiput(3751,-2311)(0.00000,115.38462){7}{\line( 0, 1){ 57.692}}
\multiput(3751,-1561)(-128.57143,0.00000){4}{\line(-1, 0){ 64.286}}
\multiput(3301,-1561)(0.00000,120.00000){8}{\line( 0, 1){ 60.000}}
\multiput(3301,-661)(-114.28571,0.00000){11}{\line(-1, 0){ 57.143}}
}%
{\color[rgb]{0,0,0}\multiput(4501,-661)(114.28571,0.00000){11}{\line( 1, 0){ 57.143}}
\multiput(5701,-661)(0.00000,-122.22222){14}{\line( 0,-1){ 61.111}}
\multiput(5701,-2311)(-122.22222,0.00000){14}{\line(-1, 0){ 61.111}}
\multiput(4051,-2311)(0.00000,115.38462){7}{\line( 0, 1){ 57.692}}
\multiput(4051,-1561)(128.57143,0.00000){4}{\line( 1, 0){ 64.286}}
\multiput(4501,-1561)(0.00000,120.00000){8}{\line( 0, 1){ 60.000}}
}%
{\color[rgb]{0,0,0}\put(1801,-961){\line( 1, 0){600}}
}%
\put(1576,-1936){\makebox(0,0)[lb]{\smash{{\SetFigFont{12}{14.4}{\rmdefault}{\mddefault}{\updefault}{\color[rgb]{0,0,0}$\ket 0$}%
}}}}
\put(6076,-1786){\makebox(0,0)[lb]{\smash{{\SetFigFont{12}{14.4}{\rmdefault}{\mddefault}{\updefault}{\color[rgb]{0,0,0}$\sigma$}%
}}}}
\put(2626,-2536){\makebox(0,0)[lb]{\smash{{\SetFigFont{12}{14.4}{\familydefault}{\mddefault}{\updefault}{\color[rgb]{0,0,0}$Q_1$}%
}}}}
\put(5026,-2536){\makebox(0,0)[lb]{\smash{{\SetFigFont{12}{14.4}{\familydefault}{\mddefault}{\updefault}{\color[rgb]{0,0,0}$Q_2^*$}%
}}}}
\put(1576,-1186){\makebox(0,0)[lb]{\smash{{\SetFigFont{12}{14.4}{\rmdefault}{\mddefault}{\updefault}{\color[rgb]{0,0,0}$\rho$}%
}}}}
{\color[rgb]{0,0,0}\put(1801,-2011){\line( 1, 0){600}}
}%
\end{picture}%
  \end{center}
  \caption[Reduction from \class{QIP} to \prob{Close
      Images}]{Construction of the circuits $Q_1$ and $Q_2$ in the
    reduction from three message quantum interactive proof system to
    an instance of \prob{Close Images}.  The space $\mathcal{P}$ of
    the prover's private system is not shown and $\rho$ represents the
    prover's first message.}
  \label{clim-fig-qip-to-closeim}
\end{figure}

To reduce this problem to an instance of \prob{CI} we must find
transformations $Q_1$ and $Q_2$ that have close images if and only if
the verifier can be made to accept.  The construction of these two
transformations is outlined in Figure~\ref{clim-fig-qip-to-closeim}.
The transformation $Q_1$ is the first half of the protocol, consisting
of the unitary circuit $V_1$ applied to the prover's first message.
The transformation $Q_2$ represents the second half of the protocol
run in
reverse: starting from an accepting state $\ket 1 \bra 1 \tensor
\sigma$ and performing $V_2^*$, which is the inverse of the unitary
circuit $V_2$.  These transformations are given in~\cite{KitaevW00} 
more formally as
\begin{equation}\label{clim-eqn-q-defn}
  \begin{split}
    Q_1(\rho) &= \ptr{M} V_1 \left( \ket 0 \bra 0 \tensor \rho \right) V_1^*, \\
    Q_2(\sigma) &= \ptr{M} V_2^* \left( \ket 1 \bra 1 \tensor \sigma \right) V_2.
  \end{split}
\end{equation}
These transformations do not take inputs of the same dimension, but
this can easily be fixed by padding the input space of $Q_1$ with
qubits that will later be traced out.
The idea is that the verifier will accept in this
protocol if and only if there are states $\rho$ and $\sigma$
that are consistent with a transcript of the \class{QIP} protocol
where the verifier accepts.  In this case, these states exist if and
only if the verifier's private qubits do not change between $V_1$ and
$V_2$, which happens exactly when there are $\rho$ and $\sigma$ such
that $\F(Q_1(\rho),Q_2(\sigma))$ is large.  To see this more formally,
if the verifier accepts with certainty, then the output of $Q_1$ on
the input state the proof system is exactly the output of $Q_2$ on
some accepting configuration of the proof system.  This is because
there is a strategy for the prover that both causes the verifier to
accept with probability one and does not change the state of the
verifier's private space (because this is not allowed in the
\class{QIP} model).  Therefore, if the verifier accepts with
certainty, there exist $\rho, \sigma$ such that
\[ Q_1(\rho) = Q_2(\sigma), \]
so that we have a valid instance of $\prob{CI}_{1,b}$ in this case.

To see this more formally, the following lemma of Kitaev and
Watrous~\cite{KitaevW00} characterizes the probability that the
verifier can be made to accept in a three-message quantum interactive
proof system.
\begin{lemma}[Kitaev and Watrous~\cite{KitaevW00}]\label{clim-lem-accept-prob}
  Let $V_1$ and $V_2$ describe a verifier
  in a quantum interactive proof system, as shown in
  Figure~\ref{clim-fig-qip}, and let $Q_1$ and $Q_2$ be as given in
  Equation~\eqref{clim-eqn-q-defn}.  The maximum probability that the
  verifier can be made to accept is
  \begin{equation*}
    \F_{\max}(Q_1, Q_2)^2.
  \end{equation*}

  \begin{proof}
    By the definition of the model, the acceptance probability is
    given by a projector $\Pi_{\mathrm{acc}}$ onto the subspace of the
    verifier's private qubits with the first qubit in the state $\ket
    1$.  By re-adding the prover's private space $\mathcal{P}$, we may assume that all states during the protocol are
    pure..  The maximum acceptance probability is defined as a
    maximum over the prover's strategy $P$ and the initial state $\ket \phi$
    corresponding to the prover's first message, which is in the $\ket
    0$ state for all but the message qubits and the prover's private
    qubits.  Doing so, this quantity is
    \begin{align}
      \max_{P, \ket \phi} \tr( \Pi_{\mathrm{acc}} V_2 P V_1 \ket \phi \bra \phi V_1^*P^* V_2^*)
      &= \max_{P, \ket \phi, \ket \psi} \abs{
        (\bra \psi \Pi_{\mathrm{acc}}) V_2^* P V_1 \ket \phi}^2
      \nonumber \\
      &= \max_{P, \ket \phi, \ket \nu} \abs{ \tr P
        (V_1 \ket{\phi} \bra{\nu} V_2) }^2
      \label{clim-eqn-accept-prob-1}
    \end{align}
    where $\ket\nu$ is restricted to be an accepting state, i.e.\ the first of the
    verifier's private qubits is in the $\ket 1$ state, and $\ket\phi$
    is restricted to be an initial state, i.e.\ the verifier's private qubits are in
    the $\ket 0$ state.  If we first trace out the space $\mathcal{V}$ in
    this equation, then by Lemma~\ref{meas-lem-tnorm-max} the
    resulting quantity is equal to 
    \begin{equation*} 
      \max_{P, \ket \phi, \ket \nu} \abs{ \tr \ptr{V} P
        (V_1 \ket{\phi} \bra{\nu} V_2) }^2
      = \max_{\ket \phi, \ket \nu} \tnorm{ \ptr{V} V_1 \ket\phi
       \bra\nu V_2}^2
      = \max_{\ket \phi, \ket \nu} \tnorm{ \ptr{M \tprod P} V_1 \ket\phi
       \bra\nu V_2}^2,
    \end{equation*}
    where the final equality follows from the fact that the
    complementary reduced states of a pure state have the same
    singular values.
    Combining this with Equation~\eqref{clim-eqn-accept-prob-1}
    implies that the verifier can be made to accept with probability
    \begin{equation*}
      \max_{\ket \phi, \ket \nu} \tnorm{ \ptr{M \tprod P} V_1 \ket\phi
        \bra\nu V_2}^2
      = \max_{\rho, \sigma} \F(Q_1(\rho), Q_2(\sigma))^2
      = \F_{\max}(Q_1,Q_2)^2
    \end{equation*}
    where the first equality is by
    Lemma~\ref{meas-lemma-fidelity-tracenorm}.  Recall that the state
    $\ket \phi = \ket 0 \tprod \ket{\phi'}$ is a valid initial state
    and that $\ket \nu = \ket 1 \tprod \ket{\nu'}$ is a valid
    accepting state for the verifier: these conditions on the two pure
    states conform exactly to the states in the definition of $Q_1$
    and $Q_2$ in Equation~\eqref{clim-eqn-q-defn}.
  \end{proof}
\end{lemma}

This lemma implies directly that $\prob{CI}_{a,b}$ is \class{QIP}-hard
for any probabilities $a,b$ that suffice for the definition of
\class{QIP} in Section~\ref{compl-scn-complexity-classes}.  As it is
known that these parameters may be any values such that $0 < b < a
\leq 1$ with at least an inverse polynomial gap between $a$ and $b$,
this implies that $\prob{CI}_{a,b}$ is hard for these same values.

To see that this problem is in~\class{QIP}, consider the following
protocol for the verifier, due to Kitaev and Watrous\footnote{John
  Watrous, private communication}.  The verifier starts with two
circuits implementing transformations $Q_1$ and $Q_2$ with
\[ Q_i(\rho) = \ptr{B} U_i (\rho \tprod \ket 0 \bra 0) U_i^*. \]
As a first step, the prover sends a state $\rho$, promised to be a
such $Q_1(\rho)$ is close to a state in the image of $Q_2$. 
The verifier computes 
\[ U_1 (\rho \tprod \ket 0 \bra 0) U_1^* \]
and sends the part of the state in $\mathcal{B}$ to the prover.  If
there is a state $\sigma$ such that $Q_2(\sigma) = Q_1(\rho)$, then
the prover and verifier together hold a purification of this state.
In this case, the prover can apply a unitary to his portion of the
system to obtain a state corresponding to the purification that would
have been obtained had the verifier instead evaluated 
\[ U_2 (\sigma \tprod \ket 0 \bra 0) U_2^*. \]
The prover performs such a computation, and sends the state in
$\mathcal{B}$ back to the verifier, who applies $U_2^*$ and checks to
see that the result is a valid initial state (i.e.\ his private qubits
are in the $\ket 0$ state).

The above argument implies that when $Q_1(\rho) = Q_2(\sigma)$ the
prover can succeed with certainty.  In the general case, the
maximum probability that the verifier can be made to accept is given
by Lemma~\ref{clim-lem-accept-prob}, which in the case of this proof
system is exactly $\F_{\max}(Q_1, Q_2)^2$.  This argument shows that
$\prob{CI}_{a,b}$ is in \class{QIP} for all $0 < b < a \leq 1$ with at
least an inverse polynomial gap between $a$ and $b$.

The preceding arguments imply that problem is complete for
\class{QIP}.  This argument appears implicitly in~\cite{KitaevW00},
where it is used to show that $\class{QIP} \subseteq \class{EXP}$.
\begin{theorem}[Kitaev and Watrous~\cite{KitaevW00}]\label{clim-thm-ci-complete}
  For any $0 < b < a \leq 1$, the problem $\prob{CI}_{a,b}$ is
  \class{QIP}-complete.
\end{theorem}

\section{The swap test}\label{clim-scn-swap-test}

The swap test provides a simple way to detect if two pure quantum
states are the same.  It was introduced by Buhrman et al. in the
context of quantum communication complexity~\cite{BuhrmanC+01}, but it has
also found applications in error correction~\cite{BarencoB+97} and in
the estimation of various properties of quantum
states~\cite{EkertA+02}.  Generalizations of the swap test to more
than two inputs have also been considered~\cite{KadaN+08}, though they
will not be needed here.

An essential component of the swap test is the operator
$W \in \unitary{H \tensor H}$ that
swaps the states in the two spaces, i.e.\  $W \ket{a} \ket{b} = \ket{b}
\ket{a}$ for all $\ket a, \ket b \in \mathcal{H}$.
\nomenclature[FW]{$W$}{Swap operation: $W \ket{a} \ket{b} = \ket{b} \ket{a}$}%
\index{swap operation}%
\index{W operation@$W$ operation|see{swap operation}}%
Expressing $W$ in the computational basis gives
\[ W = \sum_{i,j} \ket{j}\bra{i} \tprod \ket{i} \bra{j}, \]
from which it is clear that $W$ is both Hermitian and unitary.

\begin{figure}
  \begin{center}
    \input{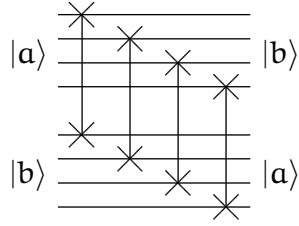}
  \end{center}
  \caption[Constant-depth implementation of a swap gate]{A
    Constant-depth implementation of the $W$ gate.}
  \label{clim-fig-Wcircuit}
\end{figure}

As circuit depth is one of the primary considerations of this chapter,
notice that as a circuit on $2n$ qubits, the operation $W$ can be 
implemented in constant
depth.  Such an implementation can be given by $n$ independent two-qubit
swaps, as shown in Figure~\ref{clim-fig-Wcircuit}.  These two qubit
swaps can each be implemented in the usual basis of quantum gates 
using three controlled-not gates, as shown in
Figure~\ref{compl-fig-swap-cnot} of Section~\ref{compl-scn-circuit-model}.

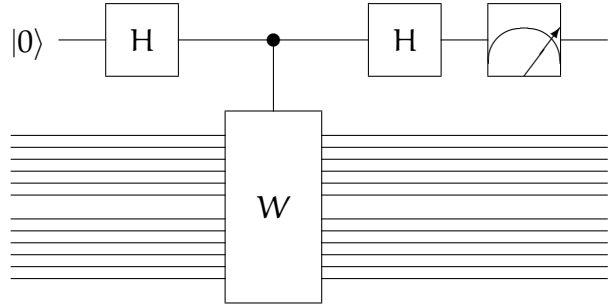
\begin{figure}
  \begin{center}
    \setlength{\unitlength}{3947sp}%
\begingroup\makeatletter\ifx\SetFigFont\undefined%
\gdef\SetFigFont#1#2#3#4#5{%
  \reset@font\fontsize{#1}{#2pt}%
  \fontfamily{#3}\fontseries{#4}\fontshape{#5}%
  \selectfont}%
\fi\endgroup%
\begin{picture}(3777,1899)(1336,-1423)
\put(1351,164){\makebox(0,0)[lb]{\smash{{\SetFigFont{12}{14.4}{\familydefault}{\mddefault}{\updefault}{\color[rgb]{0,0,0}$\ket 0$}%
}}}}
\thinlines
{\color[rgb]{0,0,0}\put(1351,-361){\line( 1, 0){1350}}
}%
{\color[rgb]{0,0,0}\put(1351,-511){\line( 1, 0){1350}}
}%
{\color[rgb]{0,0,0}\put(1351,-586){\line( 1, 0){1350}}
}%
{\color[rgb]{0,0,0}\put(1351,-661){\line( 1, 0){1350}}
}%
{\color[rgb]{0,0,0}\put(1351,-736){\line( 1, 0){1350}}
}%
{\color[rgb]{0,0,0}\put(1351,-961){\line( 1, 0){1350}}
}%
{\color[rgb]{0,0,0}\put(1351,-886){\line( 1, 0){1350}}
}%
{\color[rgb]{0,0,0}\put(1351,-1036){\line( 1, 0){1350}}
}%
{\color[rgb]{0,0,0}\put(1351,-1111){\line( 1, 0){1350}}
}%
{\color[rgb]{0,0,0}\put(1351,-1186){\line( 1, 0){1350}}
}%
{\color[rgb]{0,0,0}\put(1351,-1261){\line( 1, 0){1350}}
}%
{\color[rgb]{0,0,0}\put(3301,-961){\line( 1, 0){1800}}
}%
{\color[rgb]{0,0,0}\put(3301,-886){\line( 1, 0){1800}}
}%
{\color[rgb]{0,0,0}\put(3301,-1036){\line( 1, 0){1800}}
}%
{\color[rgb]{0,0,0}\put(3301,-1111){\line( 1, 0){1800}}
}%
{\color[rgb]{0,0,0}\put(3301,-1186){\line( 1, 0){1800}}
}%
{\color[rgb]{0,0,0}\put(3301,-1261){\line( 1, 0){1800}}
}%
{\color[rgb]{0,0,0}\put(3301,-436){\line( 1, 0){1800}}
}%
{\color[rgb]{0,0,0}\put(3301,-361){\line( 1, 0){1800}}
}%
{\color[rgb]{0,0,0}\put(3301,-511){\line( 1, 0){1800}}
}%
{\color[rgb]{0,0,0}\put(3301,-586){\line( 1, 0){1800}}
}%
{\color[rgb]{0,0,0}\put(3301,-661){\line( 1, 0){1800}}
}%
{\color[rgb]{0,0,0}\put(3301,-736){\line( 1, 0){1800}}
}%
{\color[rgb]{0,0,0}\put(4576, 89){\oval(450,450)[tr]}
\put(4576, 89){\oval(450,450)[tl]}
}%
{\color[rgb]{0,0,0}\put(3001,239){\circle*{76}}
}%
{\color[rgb]{0,0,0}\put(2701,-1411){\framebox(600,1200){$W$}}
}%
{\color[rgb]{0,0,0}\put(3601,239){\line(-1, 0){1200}}
}%
{\color[rgb]{0,0,0}\put(3001,239){\line( 0,-1){450}}
}%
{\color[rgb]{0,0,0}\put(1951, 14){\framebox(450,450){$H$}}
}%
{\color[rgb]{0,0,0}\put(3601, 14){\framebox(450,450){$H$}}
}%
{\color[rgb]{0,0,0}\put(4051,239){\line( 1, 0){300}}
}%
{\color[rgb]{0,0,0}\put(1651,239){\line( 1, 0){300}}
}%
{\color[rgb]{0,0,0}\put(4801,239){\line( 1, 0){300}}
}%
{\color[rgb]{0,0,0}\put(4576, 14){\vector( 3, 4){225}}
}%
{\color[rgb]{0,0,0}\put(4351, 14){\framebox(450,450){}}
}%
{\color[rgb]{0,0,0}\put(1351,-436){\line( 1, 0){1350}}
}%
\end{picture}%
  \end{center}
  \caption[Circuit implementation of the swap test]{A circuit implementing the swap test.}
  \label{clim-fig-swaptest}
\end{figure}

The swap test is built using a controlled-$W$ operator to
determine how close two states are to each other.
A circuit performing the swap test is given in
Figure~\ref{clim-fig-swaptest}.

\index{swap test}%

An alternate characterization of the swap test is simply as a
projective measurement onto the symmetric and antisymmetric subspaces
of the systems it is applied to.
\index{symmetric subspace}\index{antisymmetric subspace}%
Let $\{\ket i : 1 \leq i \leq d\}$ be a basis for $\mathcal{H}$.  On $\mathcal{H \tprod H}$, the symmetric and antisymmetric subspaces
are defined as
\begin{align*}
  (\mathcal{H \tprod H})_{\mathrm{sym}}
    &= \{ \ket\psi \in \mathcal{H \tprod H}: W \ket \psi = \ket \psi \} \\
    &= \lspan \{ \ket i \ket j + \ket j \ket i : i \leq j\} \\
  (\mathcal{H \tprod H})_{\mathrm{asym}} &= 
    \{ \ket\psi \in \mathcal{H \tprod H}: W \ket \psi = - \ket \psi \}  \\
    &= \lspan \{ \ket i \ket j - \ket j \ket i : i < j\},
\end{align*}
where these two subspaces represent the
$\binom{d+1}{2}$ dimensional subspace corresponding to the $+1$
eigenvalues of $W$ and the $\binom{d}{2}$ dimensional
subspace corresponding to the $-1$ eigenvalues of $W$.
From this representation it is easy to see that these two subspaces
make up the whole space, i.e.\ that 
\[ \mathcal{H \tensor H} = 
   (\mathcal{H \tprod H})_{\mathrm{sym}} 
   \oplus   (\mathcal{H \tprod H})_{\mathrm{asym}}, \]
and that the projections onto these subspaces are given by
$(\id + W)/2$ and $(\id-W)/2$.

To see that this formulation of the swap test is equivalent to the circuit
presented in Figure~\ref{clim-fig-swaptest}, consider the result of
the measurement on the control qubit and work through the circuit in
reverse.
If this measurement result is $\ket 0$, then the state of the control
qubit after the controlled-$W$ operation is $\ket 0 + \ket 1$, up to
normalization.  As this is also the state of this qubit before the
controlled-$W$ operation, then applying $W$ did not change the phase
of the system, which implies that the input has been projected onto the
symmetric subspace.  On the other hand, if the measurement result is
$\ket 1$, then the state of the control qubit after the controlled-$W$
operation is $\ket 0 - \ket 1$.  In this case, the system has been
projected onto the subspace where applying $W$ results in a phase of
$-1$, which is exactly the antisymmetric subspace of $\mathcal{H
  \tprod H}$. Thus the circuit in
Figure~\ref{clim-fig-swaptest} applies the projective measurement
given by $(\id + W)/2$ and $(\id - W)/2$, exactly as required.

This characterization immediately gives the
probability that the swap test returns the antisymmetric outcome when
applied to two pure states.
To see this, observe that on pure states $\ket \psi$ and $\ket \phi$
this occurs with probability
\begin{align}
  \frac{1}{2} \tr ((\id - W) \ket \psi \bra\psi \tprod \ket\phi \bra\phi)
  &= \frac{1}{2} \left( \tr \ket \psi \bra \psi \tprod \ket \phi \bra \phi
     - \tr \ket \phi \bra \psi \tprod \ket \psi \bra \phi \right)
     \nonumber \\
  &= \frac{1}{2}\left( 1 - \abs{\braket{\phi}{\psi}}^2 \right). \label{clim-eqn-asym-prob-pure}
\end{align}
This result can be found in~\cite{BuhrmanC+01}.

The results that follow will make use of a generalization of this
equation to the case of two (potentially entangled) mixed states.
In the process of making this generalization the square in
Equation~\eqref{clim-eqn-asym-prob-pure} is lost, and so we will only
show a lower bound.  Notice also that the requirement in the lemma
that the two input states be reduced states of each other can be made
in full generality, by applying the theorem to $\rho = \sigma \tensor
\xi$ in the case that the input states are not entangled.
\begin{lemma}\label{clim-lem-swap-fidelity}
  If $\rho \in \density{A \tensor B}$ then a swap test
  on $\mathcal{A \tensor B}$ returns the antisymmetric outcome
  with probability at least
  \[ \frac{1}{2} - \frac{1}{2}\F(\ptr{A} \rho, \ptr{B} \rho). \]

  \begin{proof}
    Let $\ket \psi \in \mathcal{A \tensor B \tensor C}$ be a
    purification of $\rho$, where $\mathcal{C}$ is an arbitrary space
    with $\dm{C} \geq \dm{A} \dm{B}$ to allow such a purification.
    The swap test measures the state on $\mathcal{A \tensor B}$ with
    the projectors $\frac{1}{2}(\id - W)$ and $\frac{1}{2}(\id + W)$.
    As $W$ is Hermitian and $W^2 = \id$, 
    the antisymmetric outcome occurs with probability given by
    \begin{align}
      \frac{1}{2} \tr \left[ (\identity{A \tprod B \tprod C} - W
        \tensor \identity{C}) \ket\psi \bra\psi \right]
     =  \frac{1}{2} - \frac{1}{2}
        \bra\psi W \tensor \identity{C} \ket\psi. \label{meas-eqn-asym-prob}
    \end{align}
    The operator $W$ is also unitary and so the states $\ket\psi$
    and $(W \tprod \identity{C})\ket\psi$ each purify both $\ptr{A \tensor C}
    \ket\psi\bra\psi$ and $\ptr{B \tensor C} \ket\psi\bra\psi$,
    and so by Uhlmann's Theorem (Theorem~\ref{meas-thm-uhlmanns})
    Equation~\eqref{meas-eqn-asym-prob} implies
    \begin{align}\label{meas-eqn-asym-result}
      \frac{1}{2} - \frac{1}{2} \bra\psi W \tensor \identity{C} \ket\psi
      & \geq \frac{1}{2} - \frac{1}{2} \F(\ptr{A \tensor C}
        \ket\psi\bra\psi, \ptr{B \tensor C} \ket\psi\bra\psi).
    \end{align}
    Finally, by observing that
    \begin{align*}
      \ptr{A \tprod C} \ket\psi\bra\psi
      &= \ptr{A} (\ptr{C} \ket\psi\bra\psi)
      = \ptr{A} \rho
    \end{align*}
    and
    \begin{align*}
      \ptr{B \tprod C} \ket\psi\bra\psi
      &= \ptr{B} (\ptr{C} \ket\psi\bra\psi)
      = \ptr{B} \rho,
    \end{align*}
    Equation~\eqref{meas-eqn-asym-result} is the lower bound in the
    statement of the lemma.
 \end{proof}
\end{lemma}
To see that this lemma generalizes
Equation~\eqref{clim-eqn-asym-prob-pure} up to a square, consider the
input state $\rho = \ket \phi \bra \phi \tensor \ket \psi \bra \psi$
and apply the definition of the Fidelity.

The main result of this chapter concerns log-depth circuits, and so it
is important to note that the swap test can be performed in log-depth.
As discussed in Proposition~\ref{compl-prop-log-control}, controlled
operations on $n$ qubits can be implemented by adding only log-depth
overhead.  This implies that the swap test can be implemented with a
log-depth circuit, as with the exception of the overhead for the
controlled operation, Figure~\ref{clim-fig-swaptest} provides a
constant depth circuit.
Additionally, this implies that if the unbounded fan-out gate is allowed
into the model of computation, this overhead can be reduced to
constant-depth, and so in this case the swap test can be performed
with a constant depth circuit.
As it is not at all clear that including this gate produces a
reasonable circuit model, any results that depend on the addition of
this gate to the circuit model are clearly marked with this
requirement.

%=============================================================================%

\section{Reduction to logarithmic depth}\label{clim-scn-construction}

In this section the reduction from the general close images problem
to the log-depth restriction of the problem is described.  This is
done in the case of one-sided error, i.e.\ the problem
$\prob{CI}_{1,b}$, where the two circuits either have intersecting
images or the largest fidelity between any two outputs is at most
$b$.  The fact that this restricted version of the problem is hard
implies that we obtain the desired hardness result even when we assume that
the input instance has $a = 0$.

The general idea behind the construction is to simply slice
the circuits of an instance of $\prob{CI}_{1,b}$ into
constant-depth pieces and run them in parallel.  These circuits
will have much larger input spaces than the original circuit, but
they are able to simulate the original circuit.
This is due to the fact that if for each of the constant depth pieces, 
the input to one piece of the
circuit is identical to the output of the previous piece, then the
output of the final piece of the circuit will be equal to the output
of the original circuit.
This need not be
the case if the intermediate inputs are not the outputs of the
previous pieces, and so additional tests that ensure
these inputs are at least close to the desired states are required.
The swap test will be used extensively to perform these tests, though
care must be taken to ensure that the resulting circuits have
logarithmic depth.

This construction is similar to an idea of Gottesman and
Chuang~\cite{GottesmanC99} in which a circuit is sliced into
constant depth pieces with teleportation used to transfer the
information between the pieces.  Conditioned on all of the
teleportations not requiring a Pauli correction, this process produces
a constant depth simulation (as a mixed state circuit) of the original 
circuit.  This process, however, does not perform any verification
that the teleportation operations were successful, and so the
resulting simulation is only accurate with exponentially small
probability.  This technique was used by Terhal and
DiVincenzo~\cite{TerhalD04} to show that exactly simulating these
circuits in polynomial time leads to unexpected complexity-theoretic
results (\class{P} = \class{PP}).
The circuit used in the paper does not conform to our circuit model,
however, as an infinite set of gates is allowed into the model.  This
difficulty was eliminated by Fenner et. al~\cite{FennerG+05} who
implemented the construction in a circuit model equivalent to the one
used here.  This paper also provides an approximate simulation of
the constant depth circuits in classical polynomial time, which
suggests that extremely short quantum circuits are not interesting
from a computational perspective.

For the reduction of \prob{Close Images} to circuits of logarithmic
depth we use a similar technique to that of Gottesman and
Chuang~\cite{GottesmanC99} of slicing the circuit into pieces.  In the
place of teleportation, however, we demand that the inputs to each of
the pieces are provided before the start of the computation.  These
intermediate inputs are then verified using a verification procedure
to ensure that the constructed circuit faithfully simulates the
original circuit.

To describe the reduction, let $Q_1$ and $Q_2$ be the circuits from
an instance of $\prob{CI}_{1,b}$, and let $n$ be the size of $Q_1$ and
$Q_2$, by padding the smaller circuit, if necessary.
In order to slice the circuits into pieces it is assumed
that $Q_1$ and $Q_2$ first introduce any necessary
ancillary qubits, then apply local unitary gates, and finally trace
out any qubits that are not part of the output.  This form for a
circuit is shown in Figure~\ref{compl-fig-unitary-simulation}, and, as
discussed in Section~\ref{compl-scn-circuit-model}, this can be
assumed with no loss of generality with only polynomial overhead, by
delaying any partial trace operations until the end of the computation and introducing
any needed ancillary qubits at the start of the computation.

\begin{figure}
  \begin{center}
    \setlength{\unitlength}{3947sp}%
\begingroup\makeatletter\ifx\SetFigFont\undefined%
\gdef\SetFigFont#1#2#3#4#5{%
  \reset@font\fontsize{#1}{#2pt}%
  \fontfamily{#3}\fontseries{#4}\fontshape{#5}%
  \selectfont}%
\fi\endgroup%
\begin{picture}(4155,1524)(511,-1573)
\put(4651,-1186){\makebox(0,0)[lb]{\smash{{\SetFigFont{12}{14.4}{\familydefault}{\mddefault}{\updefault}{\color[rgb]{0,0,0}$Q_i(\rho)$}%
}}}}
\thinlines
{\color[rgb]{0,0,0}\put(3301,-436){\line( 1, 0){300}}
}%
{\color[rgb]{0,0,0}\put(3301,-511){\line( 1, 0){300}}
}%
{\color[rgb]{0,0,0}\put(3301,-586){\line( 1, 0){300}}
}%
{\color[rgb]{0,0,0}\put(3301,-661){\line( 1, 0){300}}
}%
{\color[rgb]{0,0,0}\put(3301,-736){\line( 1, 0){300}}
}%
{\color[rgb]{0,0,0}\put(4051,-361){\line( 1, 0){300}}
}%
{\color[rgb]{0,0,0}\put(4051,-436){\line( 1, 0){300}}
}%
{\color[rgb]{0,0,0}\put(4051,-511){\line( 1, 0){300}}
}%
{\color[rgb]{0,0,0}\put(4051,-586){\line( 1, 0){300}}
}%
{\color[rgb]{0,0,0}\put(4051,-661){\line( 1, 0){300}}
}%
{\color[rgb]{0,0,0}\put(4051,-736){\line( 1, 0){300}}
}%
{\color[rgb]{0,0,0}\put(976,-361){\line( 1, 0){225}}
}%
{\color[rgb]{0,0,0}\put(976,-436){\line( 1, 0){225}}
}%
{\color[rgb]{0,0,0}\put(976,-511){\line( 1, 0){225}}
}%
{\color[rgb]{0,0,0}\put(976,-586){\line( 1, 0){225}}
}%
{\color[rgb]{0,0,0}\put(976,-661){\line( 1, 0){225}}
}%
{\color[rgb]{0,0,0}\put(976,-736){\line( 1, 0){225}}
}%
{\color[rgb]{0,0,0}\put(751,-361){\line( 1, 0){225}}
}%
{\color[rgb]{0,0,0}\put(751,-511){\line( 1, 0){225}}
}%
{\color[rgb]{0,0,0}\put(751,-436){\line( 1, 0){225}}
}%
{\color[rgb]{0,0,0}\put(751,-586){\line( 1, 0){225}}
}%
{\color[rgb]{0,0,0}\put(751,-661){\line( 1, 0){225}}
}%
{\color[rgb]{0,0,0}\put(751,-736){\line( 1, 0){225}}
}%
{\color[rgb]{0,0,0}\put(4314,-1036){\line( 1, 0){262}}
}%
{\color[rgb]{0,0,0}\put(4314,-961){\line( 1, 0){262}}
}%
{\color[rgb]{0,0,0}\put(4314,-1111){\line( 1, 0){262}}
}%
{\color[rgb]{0,0,0}\put(4314,-1186){\line( 1, 0){262}}
}%
{\color[rgb]{0,0,0}\put(4314,-1261){\line( 1, 0){262}}
}%
{\color[rgb]{0,0,0}\put(4051,-961){\line( 1, 0){263}}
}%
{\color[rgb]{0,0,0}\put(4051,-1036){\line( 1, 0){263}}
}%
{\color[rgb]{0,0,0}\put(4051,-1111){\line( 1, 0){263}}
}%
{\color[rgb]{0,0,0}\put(4051,-1186){\line( 1, 0){263}}
}%
{\color[rgb]{0,0,0}\put(4051,-1261){\line( 1, 0){263}}
}%
{\color[rgb]{0,0,0}\put(1201,-1561){\framebox(450,1500){$U_1$}}
}%
{\color[rgb]{0,0,0}\put(1951,-1561){\framebox(450,1500){$U_2$}}
}%
{\color[rgb]{0,0,0}\put(3601,-1561){\framebox(450,1500){$U_n$}}
}%
{\color[rgb]{0,0,0}\put(1651,-361){\line( 1, 0){300}}
}%
{\color[rgb]{0,0,0}\put(1651,-436){\line( 1, 0){300}}
}%
{\color[rgb]{0,0,0}\put(1651,-511){\line( 1, 0){300}}
}%
{\color[rgb]{0,0,0}\put(1651,-586){\line( 1, 0){300}}
}%
{\color[rgb]{0,0,0}\put(1651,-661){\line( 1, 0){300}}
}%
{\color[rgb]{0,0,0}\put(1651,-736){\line( 1, 0){300}}
}%
{\color[rgb]{0,0,0}\put(1651,-961){\line( 1, 0){300}}
}%
{\color[rgb]{0,0,0}\put(1651,-1036){\line( 1, 0){300}}
}%
{\color[rgb]{0,0,0}\put(1651,-1111){\line( 1, 0){300}}
}%
{\color[rgb]{0,0,0}\put(1651,-1186){\line( 1, 0){300}}
}%
{\color[rgb]{0,0,0}\put(1651,-1261){\line( 1, 0){300}}
}%
{\color[rgb]{0,0,0}\put(2401,-361){\line( 1, 0){300}}
}%
{\color[rgb]{0,0,0}\put(2401,-436){\line( 1, 0){300}}
}%
{\color[rgb]{0,0,0}\put(2401,-511){\line( 1, 0){300}}
}%
{\color[rgb]{0,0,0}\put(2401,-586){\line( 1, 0){300}}
}%
{\color[rgb]{0,0,0}\put(2401,-661){\line( 1, 0){300}}
}%
{\color[rgb]{0,0,0}\put(2401,-736){\line( 1, 0){300}}
}%
{\color[rgb]{0,0,0}\put(2401,-961){\line( 1, 0){300}}
}%
{\color[rgb]{0,0,0}\put(2401,-1036){\line( 1, 0){300}}
}%
{\color[rgb]{0,0,0}\put(2401,-1111){\line( 1, 0){300}}
}%
{\color[rgb]{0,0,0}\put(2401,-1186){\line( 1, 0){300}}
}%
{\color[rgb]{0,0,0}\put(2401,-1261){\line( 1, 0){300}}
}%
{\color[rgb]{0,0,0}\put(3301,-961){\line( 1, 0){300}}
}%
{\color[rgb]{0,0,0}\put(3301,-1036){\line( 1, 0){300}}
}%
{\color[rgb]{0,0,0}\put(3301,-1111){\line( 1, 0){300}}
}%
{\color[rgb]{0,0,0}\put(3301,-1186){\line( 1, 0){300}}
}%
{\color[rgb]{0,0,0}\put(3301,-1261){\line( 1, 0){300}}
}%
{\color[rgb]{0,0,0}\put(4351,-361){\vector( 0,-1){525}}
}%
\put(2851,-886){\makebox(0,0)[lb]{\smash{{\SetFigFont{12}{14.4}{\familydefault}{\mddefault}{\updefault}{\color[rgb]{0,0,0}$\cdots$}%
}}}}
\put(526,-586){\makebox(0,0)[lb]{\smash{{\SetFigFont{12}{14.4}{\familydefault}{\mddefault}{\updefault}{\color[rgb]{0,0,0}$\rho$}%
}}}}
{\color[rgb]{0,0,0}\put(3301,-361){\line( 1, 0){300}}
}%
\end{picture}%
  \end{center}
  \caption[Decomposition of a circuit into constant depth pieces]{The
    original circuit $Q_i$ decomposed into constant depth unitary
    circuits.}\label{clim-fig-original}
\end{figure}
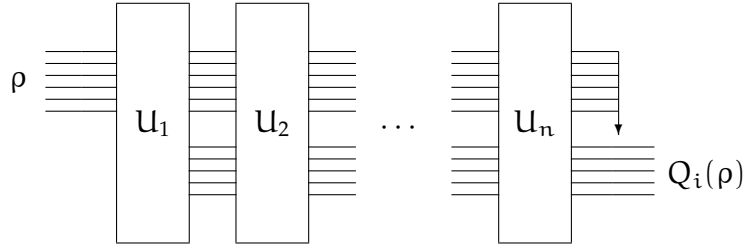

A simple way to decompose $Q_1$ into constant depth pieces is to
simply let each gate of $Q_1$ be a single piece in the decomposition.
Let $U_1, U_2, \ldots, U_n$ be these
pieces, with the additional complication that the operation $U_1$ both adds the
ancillary qubits and performs the first gate of the circuit.
In a similar way, $Q_2$ can be decomposed into constant depth pieces $V_1,
V_2, \ldots, V_n$.
Such a decomposition is shown in Figure~\ref{clim-fig-original}.
If the circuits $Q_1$ and $Q_2$ implement transformations in
$\transform{H,K}$, then as we have assumed that they are in
Stinespring form, these circuits first introduce ancillary qubits in
some space $\mathcal{A}$, apply some unitary in $\unitary{H \tprod A,
  B \tprod K}$, and finally trace out the space $\mathcal{B}$.  It can
be assumed that the spaces $\mathcal{A}$ and $\mathcal{B}$ are of the
same dimension for both $Q_1$ and $Q_2$, once again by padding the
smaller circuit with unused ancillary qubits that are later traced out.
This implies that the spaces
$\mathcal{H \tensor A}$ and $\mathcal{B \tensor K}$ are isomorphic.
Using these spaces, and implicitly this isomorphism, we have
\begin{align*}
  U_1, V_1 & \in \unitarynomc{\mc{H}_1, \mc{B}_1 \tprod \mc{K}_1} \\
  U_i, V_i & \in \unitarynomc{\mc{H}_i \tprod \mc{A}_i, 
                              \mc{B}_i \tprod \mc{K}_i}
      \quad \text{for $2 \leq i \leq n$},
\end{align*}
where the subscripted spaces are copies of the
non-subscripted spaces that hold the input or output of one of the
pieces of the original circuits.
As an example of this notation, if $\rho \in \density{H}$, then the
output of the circuit $Q_1$ on $\rho$ is given by
\begin{equation}\label{clim:eqn:orig-output}
  \tr_{\mc{B}_n} U_n U_{n-1} \cdots U_1 \rho U_1^* U_2^* \cdots U_n^*,
\end{equation}
and the output of $Q_2$ is given by the same expression with
the $V_i$ in place of the unitaries $U_i$.

This decomposition of $Q_1$ and $Q_2$ will be used to construct
circuits $C_1$ and $C_2$ that have logarithmic depth and still, in
some sense, faithfully implement $Q_1$ and $Q_2$.  
This is done by placing the
circuits corresponding to $U_1, \ldots, U_n$ in parallel, and tracing
out all the qubits that are not in the output of $U_n$.  Such a
circuit is constant depth, but does not necessarily output a state in
the image of $Q_1$, as the input to $U_{i+1}$ is not necessarily close to the
output from $U_i$.
This problem is solved by comparing the 
output of $U_i$ to the input to $U_{i}$ using the swap test.  
The swap test will fail to detect the case that the two inputs are different
with some probability, but in Section~\ref{clim-scn-soundness} it is
shown that this probability can be upper bounded by an expression
involving the trace norm of the two states.

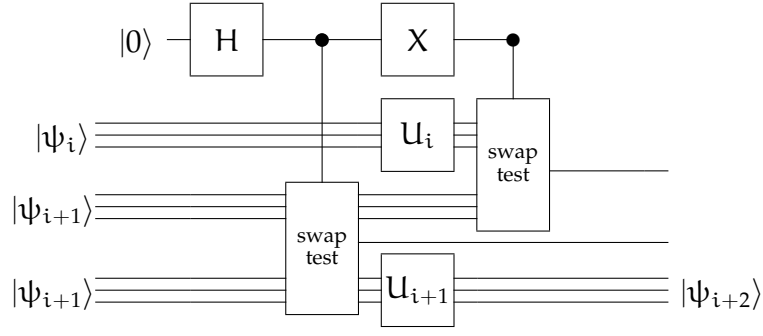
\begin{figure}
  \begin{center}
    \setlength{\unitlength}{3947sp}%
\begingroup\makeatletter\ifx\SetFigFont\undefined%
\gdef\SetFigFont#1#2#3#4#5{%
  \reset@font\fontsize{#1}{#2pt}%
  \fontfamily{#3}\fontseries{#4}\fontshape{#5}%
  \selectfont}%
\fi\endgroup%
\begin{picture}(4200,2049)(-524,-2623)
\put(-374,-1486){\makebox(0,0)[lb]{\smash{{\SetFigFont{12}{14.4}{\familydefault}{\mddefault}{\updefault}{\color[rgb]{0,0,0}$\ket{\psi_i}$}%
}}}}
\thinlines
{\color[rgb]{0,0,0}\put(601,-1861){\line( 1, 0){600}}
}%
{\color[rgb]{0,0,0}\put(601,-1936){\line( 1, 0){600}}
}%
{\color[rgb]{0,0,0}\put(601,-1786){\line(-1, 0){600}}
}%
{\color[rgb]{0,0,0}\put(601,-1861){\line(-1, 0){600}}
}%
{\color[rgb]{0,0,0}\put(601,-1936){\line(-1, 0){600}}
}%
{\color[rgb]{0,0,0}\put(601,-2311){\line( 1, 0){600}}
}%
{\color[rgb]{0,0,0}\put(601,-2386){\line( 1, 0){600}}
}%
{\color[rgb]{0,0,0}\put(601,-2461){\line( 1, 0){600}}
}%
{\color[rgb]{0,0,0}\put(601,-2311){\line(-1, 0){600}}
}%
{\color[rgb]{0,0,0}\put(601,-2386){\line(-1, 0){600}}
}%
{\color[rgb]{0,0,0}\put(601,-2461){\line(-1, 0){600}}
}%
{\color[rgb]{0,0,0}\put(2626,-811){\circle*{76}}
}%
{\color[rgb]{0,0,0}\put(1426,-811){\circle*{76}}
}%
{\color[rgb]{0,0,0}\put(1651,-1336){\line( 1, 0){150}}
}%
{\color[rgb]{0,0,0}\put(1651,-1411){\line( 1, 0){150}}
}%
{\color[rgb]{0,0,0}\put(1651,-1486){\line( 1, 0){150}}
}%
{\color[rgb]{0,0,0}\put(1801,-1636){\framebox(450,450){$U_i$}}
}%
{\color[rgb]{0,0,0}\put(2401,-2011){\framebox(450,825){\parbox{0.6in}{\scriptsize\centering{swap\\[-1mm] test}}}}
}%
{\color[rgb]{0,0,0}\put(2851,-1636){\line( 1, 0){600}}
}%
{\color[rgb]{0,0,0}\put(3451,-1636){\line( 1, 0){150}}
}%
{\color[rgb]{0,0,0}\put(1651,-1786){\line( 1, 0){750}}
}%
{\color[rgb]{0,0,0}\put(1651,-1861){\line( 1, 0){750}}
}%
{\color[rgb]{0,0,0}\put(1651,-1936){\line( 1, 0){750}}
}%
{\color[rgb]{0,0,0}\put(2251,-1336){\line( 1, 0){150}}
}%
{\color[rgb]{0,0,0}\put(2251,-1411){\line( 1, 0){150}}
}%
{\color[rgb]{0,0,0}\put(2251,-1486){\line( 1, 0){150}}
}%
{\color[rgb]{0,0,0}\put(1651,-2086){\line( 1, 0){1950}}
}%
{\color[rgb]{0,0,0}\put(1801,-2611){\framebox(450,450){$U_{i+1}$}}
}%
{\color[rgb]{0,0,0}\put(1651,-2311){\line( 1, 0){150}}
}%
{\color[rgb]{0,0,0}\put(1651,-2386){\line( 1, 0){150}}
}%
{\color[rgb]{0,0,0}\put(1651,-2461){\line( 1, 0){150}}
}%
{\color[rgb]{0,0,0}\put(2251,-2311){\line( 1, 0){150}}
}%
{\color[rgb]{0,0,0}\put(2251,-2386){\line( 1, 0){150}}
}%
{\color[rgb]{0,0,0}\put(2251,-2461){\line( 1, 0){150}}
}%
{\color[rgb]{0,0,0}\put(1201,-2536){\framebox(450,825){\parbox{0.6in}{\scriptsize\centering{swap\\[-1mm] test}}}}
}%
{\color[rgb]{0,0,0}\put(1651,-1336){\line(-1, 0){1650}}
}%
{\color[rgb]{0,0,0}\put(1651,-1411){\line(-1, 0){1650}}
}%
{\color[rgb]{0,0,0}\put(  1,-1486){\line( 1, 0){1650}}
}%
{\color[rgb]{0,0,0}\put(2401,-2311){\line( 1, 0){1200}}
}%
{\color[rgb]{0,0,0}\put(2401,-2386){\line( 1, 0){1200}}
}%
{\color[rgb]{0,0,0}\put(2401,-2461){\line( 1, 0){1200}}
}%
%{\color[rgb]{0,0,0}\put(601,-811){\line( 1, 0){2025}}
%}%
{\color[rgb]{0,0,0}\put(2626,-1186){\line( 0, 1){375}}
}%
{\color[rgb]{0,0,0}\put(1426,-1711){\line( 0, 1){900}}
}%
{\color[rgb]{0,0,0}\put(601,-1786){\line( 1, 0){600}}
}%
{\color[rgb]{0,0,0}\put(1801,-1036){\framebox(450,450){$X$}}
}%
{\color[rgb]{0,0,0}\put(601,-1036){\framebox(450,450){$H$}}
}%
{\color[rgb]{0,0,0}\put(451,-811){\line( 1, 0){150}}
}%
{\color[rgb]{0,0,0}\put(1051,-811){\line( 1, 0){750}}
}%
{\color[rgb]{0,0,0}\put(2251,-811){\line( 1, 0){375}}
}%
\put(151,-886){\makebox(0,0)[lb]{\smash{{\SetFigFont{12}{14.4}{\familydefault}{\mddefault}{\updefault}{\color[rgb]{0,0,0}$\ket 0$}%
}}}}
\put(3676,-2461){\makebox(0,0)[lb]{\smash{{\SetFigFont{12}{14.4}{\familydefault}{\mddefault}{\updefault}{\color[rgb]{0,0,0}$\ket{\psi_{i+2}}$}%
}}}}
\put(-524,-1936){\makebox(0,0)[lb]{\smash{{\SetFigFont{12}{14.4}{\familydefault}{\mddefault}{\updefault}{\color[rgb]{0,0,0}$\ket{\psi_{i+1}}$}%
}}}}
\put(-524,-2461){\makebox(0,0)[lb]{\smash{{\SetFigFont{12}{14.4}{\familydefault}{\mddefault}{\updefault}{\color[rgb]{0,0,0}$\ket{\psi_{i+1}}$}%
}}}}
{\color[rgb]{0,0,0}\put(1426,-811){\circle*{76}}
}%
\end{picture}%
  \end{center}
  \caption[Testing procedure used in reduction to log-depth
    circuits]{Testing that the output of $U_i$ is close to the input
    of $U_{i+1}$.  The inputs $\ket{\psi_j}$ are the ideal inputs to
    $U_j$, and are labelled for clarity only -- no assumptions are
    made about these states.  Qubits that do not reach the right edge
    are traced out, but for clarity these operations are not shown in
    the figure.}\label{clim-fig-test}
\end{figure}
In order for this comparison procedure to be done in log-depth an
auxiliary input is first compared against the input to $U_{i+1}$ and then
held in reserve to compare to the output of $U_i$.  This
strategy avoids the comparing the input to $U_{i+1}$ directly to the
output of $U_i$, which leads to a circuit of linear depth.  
This depth reduction comes at a cost, however, as the two states
are always compared through an intermediary state, which can at worst 
halve the probability of detecting when these two states differ, since 
one test is replaced with two.  This constant
loss will not affect the main result in a significant way.  
An example of the construction
used to ensure that the output of $U_i$ agrees with the input to
$U_{i+1}$ is given in Figure~\ref{clim-fig-test}.

To simplify the analysis of the constructed circuits these two tests
are controlled so that exactly one of the two tests is performed.
This will increase the failure probability by another factor of two, but
allows the analysis of each swap test to ignore the effect of the
other.
To implement this scheme a control
qubit is used so that either the first or the second test is performed
between every two pieces $U_i, U_{i+1}$ of the circuit.  
If a test is not performed, then the value of the output qubit of the
swap test is left unchanged, and so the result of the test is a qubit
in the $\ket 0$ state.  In the case that a test is performed, the
output is either $\ket 0$ for the symmetric subspace (i.e.\ the two
states are the same) or $\ket 1$ for the antisymmetric subspace
(i.e.\ the two states differ).  These outputs are classical values, but
they are treated as the two orthogonal quantum states $\ket 0$ and
$\ket 1$ for convenience.
Controlled application of these swap tests can be implemented in
log-depth using the techniques described in 
Proposition~\ref{compl-prop-log-control}.

\begin{figure}
  \begin{center}
    \setlength{\unitlength}{3947sp}%
\begingroup\makeatletter\ifx\SetFigFont\undefined%
\gdef\SetFigFont#1#2#3#4#5{%
  \reset@font\fontsize{#1}{#2pt}%
  \fontfamily{#3}\fontseries{#4}\fontshape{#5}%
  \selectfont}%
\fi\endgroup%
\begin{picture}(5561,1824)(1114,-2473)
\put(2551,-2386){\makebox(0,0)[lb]{\smash{{\SetFigFont{12}{14.4}{\familydefault}{\mddefault}{\updefault}{\color[rgb]{0,0,0}Output of $U_n$}%
}}}}
\thinlines
{\color[rgb]{0,0,0}\put(4051,-811){\line( 1, 0){375}}
}%
{\color[rgb]{0,0,0}\put(4051,-886){\line( 1, 0){375}}
}%
{\color[rgb]{0,0,0}\put(4051,-1036){\line( 1, 0){375}}
}%
{\color[rgb]{0,0,0}\put(4051,-1111){\line( 1, 0){375}}
}%
{\color[rgb]{0,0,0}\put(4051,-1186){\line( 1, 0){375}}
}%
{\color[rgb]{0,0,0}\put(4051,-1336){\line( 1, 0){375}}
}%
{\color[rgb]{0,0,0}\put(4051,-1411){\line( 1, 0){375}}
}%
{\color[rgb]{0,0,0}\put(4051,-1486){\line( 1, 0){375}}
}%
{\color[rgb]{0,0,0}\put(4051,-2236){\line( 1, 0){375}}
}%
{\color[rgb]{0,0,0}\put(4051,-2311){\line( 1, 0){375}}
}%
{\color[rgb]{0,0,0}\put(4051,-2386){\line( 1, 0){375}}
}%
{\color[rgb]{0,0,0}\put(5026,-736){\line( 1, 0){375}}
}%
{\color[rgb]{0,0,0}\put(5026,-811){\line( 1, 0){375}}
}%
{\color[rgb]{0,0,0}\put(5026,-886){\line( 1, 0){375}}
}%
{\color[rgb]{0,0,0}\put(5026,-1111){\line( 1, 0){375}}
}%
{\color[rgb]{0,0,0}\put(5026,-1186){\line( 1, 0){375}}
}%
{\color[rgb]{0,0,0}\put(5026,-1261){\line( 1, 0){375}}
}%
{\color[rgb]{0,0,0}\put(5026,-1486){\line( 1, 0){375}}
}%
{\color[rgb]{0,0,0}\put(5026,-1561){\line( 1, 0){375}}
}%
{\color[rgb]{0,0,0}\put(5026,-1636){\line( 1, 0){375}}
}%
{\color[rgb]{0,0,0}\put(5026,-1861){\line( 1, 0){375}}
}%
{\color[rgb]{0,0,0}\put(5026,-1936){\line( 1, 0){375}}
}%
{\color[rgb]{0,0,0}\put(5026,-2011){\line( 1, 0){375}}
}%
{\color[rgb]{0,0,0}\put(5026,-2236){\line( 1, 0){375}}
}%
{\color[rgb]{0,0,0}\put(5026,-2311){\line( 1, 0){375}}
}%
{\color[rgb]{0,0,0}\put(5026,-2386){\line( 1, 0){375}}
}%
{\color[rgb]{0,0,0}\put(1126,-736){\line( 1, 0){375}}
}%
{\color[rgb]{0,0,0}\put(1126,-811){\line( 1, 0){375}}
}%
{\color[rgb]{0,0,0}\put(1126,-886){\line( 1, 0){375}}
}%
{\color[rgb]{0,0,0}\put(1126,-1036){\line( 1, 0){375}}
}%
{\color[rgb]{0,0,0}\put(1126,-1111){\line( 1, 0){375}}
}%
{\color[rgb]{0,0,0}\put(1126,-1186){\line( 1, 0){375}}
}%
{\color[rgb]{0,0,0}\put(1126,-1336){\line( 1, 0){375}}
}%
{\color[rgb]{0,0,0}\put(1126,-1411){\line( 1, 0){375}}
}%
{\color[rgb]{0,0,0}\put(1126,-1486){\line( 1, 0){375}}
}%
{\color[rgb]{0,0,0}\put(1126,-2236){\line( 1, 0){375}}
}%
{\color[rgb]{0,0,0}\put(1126,-2311){\line( 1, 0){375}}
}%
{\color[rgb]{0,0,0}\put(1126,-2386){\line( 1, 0){375}}
}%
{\color[rgb]{0,0,0}\put(2101,-736){\line( 1, 0){375}}
}%
{\color[rgb]{0,0,0}\put(2101,-811){\line( 1, 0){375}}
}%
{\color[rgb]{0,0,0}\put(2101,-886){\line( 1, 0){375}}
}%
{\color[rgb]{0,0,0}\put(2101,-1111){\line( 1, 0){375}}
}%
{\color[rgb]{0,0,0}\put(2101,-1186){\line( 1, 0){375}}
}%
{\color[rgb]{0,0,0}\put(2101,-1261){\line( 1, 0){375}}
}%
{\color[rgb]{0,0,0}\put(2101,-1486){\line( 1, 0){375}}
}%
{\color[rgb]{0,0,0}\put(2101,-1561){\line( 1, 0){375}}
}%
{\color[rgb]{0,0,0}\put(2101,-1636){\line( 1, 0){375}}
}%
{\color[rgb]{0,0,0}\put(2101,-1861){\line( 1, 0){375}}
}%
{\color[rgb]{0,0,0}\put(2101,-1936){\line( 1, 0){375}}
}%
{\color[rgb]{0,0,0}\put(2101,-2011){\line( 1, 0){375}}
}%
{\color[rgb]{0,0,0}\put(2101,-2236){\line( 1, 0){375}}
}%
{\color[rgb]{0,0,0}\put(2101,-2311){\line( 1, 0){375}}
}%
{\color[rgb]{0,0,0}\put(2101,-2386){\line( 1, 0){375}}
}%
{\color[rgb]{0,0,0}\put(4426,-2461){\framebox(600,1800){$C_2$}}
}%
{\color[rgb]{0,0,0}\put(1501,-2461){\framebox(600,1800){$C_1$}}
}%
\put(4201,-1936){\makebox(0,0)[lb]{\smash{{\SetFigFont{12}{14.4}{\familydefault}{\mddefault}{\updefault}{\color[rgb]{0,0,0}$\vdots$}%
}}}}
\put(5176,-1074){\makebox(0,0)[lb]{\smash{{\SetFigFont{12}{14.4}{\familydefault}{\mddefault}{\updefault}{\color[rgb]{0,0,0}$\vdots$}%
}}}}
\put(5176,-1824){\makebox(0,0)[lb]{\smash{{\SetFigFont{12}{14.4}{\familydefault}{\mddefault}{\updefault}{\color[rgb]{0,0,0}$\vdots$}%
}}}}
\put(1276,-1936){\makebox(0,0)[lb]{\smash{{\SetFigFont{12}{14.4}{\familydefault}{\mddefault}{\updefault}{\color[rgb]{0,0,0}$\vdots$}%
}}}}
\put(2251,-1074){\makebox(0,0)[lb]{\smash{{\SetFigFont{12}{14.4}{\familydefault}{\mddefault}{\updefault}{\color[rgb]{0,0,0}$\vdots$}%
}}}}
\put(2251,-1824){\makebox(0,0)[lb]{\smash{{\SetFigFont{12}{14.4}{\familydefault}{\mddefault}{\updefault}{\color[rgb]{0,0,0}$\vdots$}%
}}}}
\put(2551,-1036){\makebox(0,0)[lb]{\smash{{\SetFigFont{12}{14.4}{\familydefault}{\mddefault}{\updefault}{\color[rgb]{0,0,0}$\Bigg\} {\ket 0}^{\tensor n}$}%
}}}}
\put(2551,-1786){\makebox(0,0)[lb]{\smash{{\SetFigFont{12}{14.4}{\familydefault}{\mddefault}{\updefault}{\color[rgb]{0,0,0}$\Bigg\}$ Swap tests}%
}}}}
\put(5476,-1036){\makebox(0,0)[lb]{\smash{{\SetFigFont{12}{14.4}{\familydefault}{\mddefault}{\updefault}{\color[rgb]{0,0,0}$\Bigg\}$ Swap tests}%
}}}}
\put(5476,-1786){\makebox(0,0)[lb]{\smash{{\SetFigFont{12}{14.4}{\familydefault}{\mddefault}{\updefault}{\color[rgb]{0,0,0}$\Bigg\} {\ket 0}^{\tensor n}$}%
}}}}
\put(5476,-2386){\makebox(0,0)[lb]{\smash{{\SetFigFont{12}{14.4}{\familydefault}{\mddefault}{\updefault}{\color[rgb]{0,0,0}Output of $V_n$}%
}}}}
{\color[rgb]{0,0,0}\put(4051,-736){\line( 1, 0){375}}
}%
\end{picture}%
  \end{center}
  \caption[Overview of the output spaces of the constructed log-depth
  circuits]{The outputs of $C_1$ and $C_2$.  The dummy $\ket 0$ qubits
    of one circuit line up with the outputs of the swap tests of the
    other.}\label{clim-fig-dummy}
\end{figure}
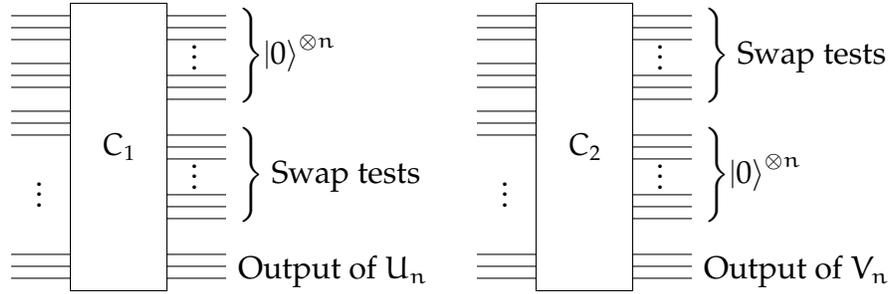
After adding these two tests between each piece of the circuit there is
one final modification to obtain the circuits $C_1$ and $C_2$.
If any of the swap tests fail, i.e.\ detect
states in the antisymmetric subspace, then they
will output qubits in the $\ket 1$ state.  As yes instances of
$\prob{CI}_{1,b}$ have outputs that are close together,
we can ensure that if any of the swap tests fail then the outputs of
the constructed circuits are far apart 
by adding dummy qubits in the $\ket 0$ state to be compared
to the outputs of the swap tests in the other circuit.  The
arrangement of these dummy
qubits is shown in Figure~\ref{clim-fig-dummy}.

\begin{figure}
  \begin{center}
    \setlength{\unitlength}{3947sp}%
\begingroup\makeatletter\ifx\SetFigFont\undefined%
\gdef\SetFigFont#1#2#3#4#5{%
  \reset@font\fontsize{#1}{#2pt}%
  \fontfamily{#3}\fontseries{#4}\fontshape{#5}%
  \selectfont}%
\fi\endgroup%
\begin{picture}(3624,5349)(-11,-4348)
\put(301, 14){\makebox(0,0)[lb]{\smash{{\SetFigFont{12}{14.4}{\familydefault}{\mddefault}{\updefault}{\color[rgb]{0,0,0}$\ket 0$}%
}}}}
\thinlines
{\color[rgb]{0,0,0}\put(601,-1411){\line( 1, 0){600}}
}%
{\color[rgb]{0,0,0}\put(601,-1486){\line( 1, 0){600}}
}%
{\color[rgb]{0,0,0}\put(601,-1336){\line(-1, 0){600}}
}%
{\color[rgb]{0,0,0}\put(601,-1411){\line(-1, 0){600}}
}%
{\color[rgb]{0,0,0}\put(601,-1486){\line(-1, 0){600}}
}%
{\color[rgb]{0,0,0}\put(601,-1786){\line( 1, 0){600}}
}%
{\color[rgb]{0,0,0}\put(601,-1861){\line( 1, 0){600}}
}%
{\color[rgb]{0,0,0}\put(601,-1936){\line( 1, 0){600}}
}%
{\color[rgb]{0,0,0}\put(601,-1786){\line(-1, 0){600}}
}%
{\color[rgb]{0,0,0}\put(601,-1861){\line(-1, 0){600}}
}%
{\color[rgb]{0,0,0}\put(601,-1936){\line(-1, 0){600}}
}%
{\color[rgb]{0,0,0}\put(601,-2311){\line( 1, 0){600}}
}%
{\color[rgb]{0,0,0}\put(601,-2386){\line( 1, 0){600}}
}%
{\color[rgb]{0,0,0}\put(601,-2461){\line( 1, 0){600}}
}%
{\color[rgb]{0,0,0}\put(601,-2311){\line(-1, 0){600}}
}%
{\color[rgb]{0,0,0}\put(601,-2386){\line(-1, 0){600}}
}%
{\color[rgb]{0,0,0}\put(601,-2461){\line(-1, 0){600}}
}%
{\color[rgb]{0,0,0}\put(1426, 89){\circle*{76}}
}%
{\color[rgb]{0,0,0}\put(2626, 89){\circle*{76}}
}%
{\color[rgb]{0,0,0}\put(901,-361){\line( 1, 0){300}}
}%
{\color[rgb]{0,0,0}\put(901,-436){\line( 1, 0){300}}
}%
{\color[rgb]{0,0,0}\put(601,-361){\line( 1, 0){300}}
}%
{\color[rgb]{0,0,0}\put(601,-436){\line( 1, 0){300}}
}%
{\color[rgb]{0,0,0}\put(1801,-661){\framebox(450,450){$U_1$}}
}%
{\color[rgb]{0,0,0}\put(1201,-361){\line( 1, 0){600}}
}%
{\color[rgb]{0,0,0}\put(1201,-436){\line( 1, 0){600}}
}%
{\color[rgb]{0,0,0}\put(2251,-361){\line( 1, 0){150}}
}%
{\color[rgb]{0,0,0}\put(2251,-436){\line( 1, 0){150}}
}%
{\color[rgb]{0,0,0}\put(2251,-511){\line( 1, 0){150}}
}%
{\color[rgb]{0,0,0}\put(2851,-661){\line( 1, 0){600}}
}%
{\color[rgb]{0,0,0}\put(3451,-661){\line( 1, 0){150}}
}%
{\color[rgb]{0,0,0}\put(1426,-736){\line( 0, 1){825}}
}%
{\color[rgb]{0,0,0}\put(1201, 89){\line( 1, 0){600}}
}%
{\color[rgb]{0,0,0}\put(2251, 89){\line( 1, 0){375}}
}%
{\color[rgb]{0,0,0}\put(751,-136){\framebox(450,450){$H$}}
}%
{\color[rgb]{0,0,0}\put(1801,-136){\framebox(450,450){$X$}}
}%
{\color[rgb]{0,0,0}\put(751, 89){\line(-1, 0){150}}
}%
{\color[rgb]{0,0,0}\put(601,-361){\line(-1, 0){600}}
}%
{\color[rgb]{0,0,0}\put(601,-436){\line(-1, 0){600}}
}%
{\color[rgb]{0,0,0}\put(601,539){\line( 1, 0){3000}}
}%
{\color[rgb]{0,0,0}\put(601,839){\line( 1, 0){3000}}
}%
{\color[rgb]{0,0,0}\put(601,989){\line( 1, 0){3000}}
}%
{\color[rgb]{0,0,0}\put(601,914){\line( 1, 0){3000}}
}%
{\color[rgb]{0,0,0}\put(2401,-1036){\framebox(450,825){\parbox{0.6in}{\scriptsize\centering{swap\\[-1mm] test}}}}
}%
{\color[rgb]{0,0,0}\put(601,-3136){\line(-1, 0){600}}
}%
{\color[rgb]{0,0,0}\put(601,-3211){\line(-1, 0){600}}
}%
{\color[rgb]{0,0,0}\put(601,-3286){\line(-1, 0){600}}
}%
{\color[rgb]{0,0,0}\put(601,-3136){\line( 1, 0){600}}
}%
{\color[rgb]{0,0,0}\put(601,-3286){\line( 1, 0){600}}
}%
{\color[rgb]{0,0,0}\put(601,-3211){\line( 1, 0){600}}
}%
{\color[rgb]{0,0,0}\put(1651,-3661){\line( 1, 0){150}}
}%
{\color[rgb]{0,0,0}\put(1651,-3736){\line( 1, 0){150}}
}%
{\color[rgb]{0,0,0}\put(1651,-3811){\line( 1, 0){150}}
}%
{\color[rgb]{0,0,0}\put(1801,-3961){\framebox(450,450){$U_n$}}
}%
{\color[rgb]{0,0,0}\put(2401,-4336){\framebox(450,825){\parbox{0.6in}{\scriptsize\centering{swap\\[-1mm] test}}}}
}%
{\color[rgb]{0,0,0}\put(2851,-3961){\line( 1, 0){600}}
}%
{\color[rgb]{0,0,0}\put(3451,-3961){\line( 1, 0){150}}
}%
{\color[rgb]{0,0,0}\put(1651,-4111){\line( 1, 0){750}}
}%
{\color[rgb]{0,0,0}\put(1651,-4186){\line( 1, 0){750}}
}%
{\color[rgb]{0,0,0}\put(1651,-4261){\line( 1, 0){750}}
}%
{\color[rgb]{0,0,0}\put(2251,-3661){\line( 1, 0){150}}
}%
{\color[rgb]{0,0,0}\put(2251,-3736){\line( 1, 0){150}}
}%
{\color[rgb]{0,0,0}\put(2251,-3811){\line( 1, 0){150}}
}%
{\color[rgb]{0,0,0}\put(601,-3661){\line( 1, 0){600}}
}%
{\color[rgb]{0,0,0}\put(601,-3736){\line( 1, 0){600}}
}%
{\color[rgb]{0,0,0}\put(601,-3811){\line( 1, 0){600}}
}%
{\color[rgb]{0,0,0}\put(601,-3661){\line(-1, 0){600}}
}%
{\color[rgb]{0,0,0}\put(601,-3736){\line(-1, 0){600}}
}%
{\color[rgb]{0,0,0}\put(601,-3811){\line(-1, 0){600}}
}%
{\color[rgb]{0,0,0}\put(601,-4111){\line( 1, 0){600}}
}%
{\color[rgb]{0,0,0}\put(601,-4186){\line( 1, 0){600}}
}%
{\color[rgb]{0,0,0}\put(601,-4261){\line( 1, 0){600}}
}%
{\color[rgb]{0,0,0}\put(601,-4111){\line(-1, 0){600}}
}%
{\color[rgb]{0,0,0}\put(601,-4186){\line(-1, 0){600}}
}%
{\color[rgb]{0,0,0}\put(601,-4261){\line(-1, 0){600}}
}%
{\color[rgb]{0,0,0}\put(2851,-4111){\line( 1, 0){750}}
}%
{\color[rgb]{0,0,0}\put(2851,-4186){\line( 1, 0){750}}
}%
{\color[rgb]{0,0,0}\put(2851,-4261){\line( 1, 0){750}}
}%
{\color[rgb]{0,0,0}\put(1201,-4111){\line( 1, 0){450}}
}%
{\color[rgb]{0,0,0}\put(1201,-4186){\line( 1, 0){450}}
}%
{\color[rgb]{0,0,0}\put(1201,-4261){\line( 1, 0){450}}
}%
{\color[rgb]{0,0,0}\put(1426,-2911){\line( 0,-1){150}}
}%
{\color[rgb]{0,0,0}\put(2626,-1036){\line( 0,-1){150}}
}%
{\color[rgb]{0,0,0}\put(2626,-2011){\line( 0,-1){150}}
}%
{\color[rgb]{0,0,0}\put(1651,-961){\line( 1, 0){750}}
}%
{\color[rgb]{0,0,0}\put(1651,-811){\line( 1, 0){750}}
}%
{\color[rgb]{0,0,0}\put(1651,-886){\line( 1, 0){750}}
}%
{\color[rgb]{0,0,0}\put(601,-811){\line(-1, 0){600}}
}%
{\color[rgb]{0,0,0}\put(601,-886){\line(-1, 0){600}}
}%
{\color[rgb]{0,0,0}\put(601,-961){\line(-1, 0){600}}
}%
{\color[rgb]{0,0,0}\put(601,-811){\line( 1, 0){600}}
}%
{\color[rgb]{0,0,0}\put(601,-961){\line( 1, 0){600}}
}%
{\color[rgb]{0,0,0}\put(601,-886){\line( 1, 0){600}}
}%
{\color[rgb]{0,0,0}\put(1651,-1336){\line( 1, 0){150}}
}%
{\color[rgb]{0,0,0}\put(1651,-1411){\line( 1, 0){150}}
}%
{\color[rgb]{0,0,0}\put(1651,-1486){\line( 1, 0){150}}
}%
{\color[rgb]{0,0,0}\put(1651,-1111){\line( 1, 0){1800}}
}%
{\color[rgb]{0,0,0}\put(1801,-1636){\framebox(450,450){$U_2$}}
}%
{\color[rgb]{0,0,0}\put(2401,-2011){\framebox(450,825){\parbox{0.6in}{\scriptsize\centering{swap\\[-1mm] test}}}}
}%
{\color[rgb]{0,0,0}\put(2851,-1636){\line( 1, 0){600}}
}%
{\color[rgb]{0,0,0}\put(3451,-1636){\line( 1, 0){150}}
}%
{\color[rgb]{0,0,0}\put(3451,-1111){\line( 1, 0){150}}
}%
{\color[rgb]{0,0,0}\put(1651,-1786){\line( 1, 0){750}}
}%
{\color[rgb]{0,0,0}\put(1651,-1861){\line( 1, 0){750}}
}%
{\color[rgb]{0,0,0}\put(1651,-1936){\line( 1, 0){750}}
}%
{\color[rgb]{0,0,0}\put(2251,-1336){\line( 1, 0){150}}
}%
{\color[rgb]{0,0,0}\put(2251,-1411){\line( 1, 0){150}}
}%
{\color[rgb]{0,0,0}\put(2251,-1486){\line( 1, 0){150}}
}%
{\color[rgb]{0,0,0}\put(1651,-2086){\line( 1, 0){1950}}
}%
{\color[rgb]{0,0,0}\put(1801,-2611){\framebox(450,450){$U_3$}}
}%
{\color[rgb]{0,0,0}\put(2401,-2986){\framebox(450,825){\parbox{0.6in}{\scriptsize\centering{swap\\[-1mm] test}}}}
}%
{\color[rgb]{0,0,0}\put(1651,-2311){\line( 1, 0){150}}
}%
{\color[rgb]{0,0,0}\put(1651,-2386){\line( 1, 0){150}}
}%
{\color[rgb]{0,0,0}\put(1651,-2461){\line( 1, 0){150}}
}%
{\color[rgb]{0,0,0}\put(2251,-2311){\line( 1, 0){150}}
}%
{\color[rgb]{0,0,0}\put(2251,-2386){\line( 1, 0){150}}
}%
{\color[rgb]{0,0,0}\put(2251,-2461){\line( 1, 0){150}}
}%
{\color[rgb]{0,0,0}\put(2851,-2611){\line( 1, 0){750}}
}%
{\color[rgb]{0,0,0}\put(1201,-1561){\framebox(450,825){\parbox{0.6in}{\scriptsize\centering{swap\\[-1mm] test}}}}
}%
{\color[rgb]{0,0,0}\put(1201,-3886){\framebox(450,825){\parbox{0.6in}{\scriptsize\centering{swap\\[-1mm] test}}}}
}%
{\color[rgb]{0,0,0}\put(1426,-2536){\line( 0,-1){150}}
}%
{\color[rgb]{0,0,0}\put(1201,-2536){\framebox(450,825){\parbox{0.6in}{\scriptsize\centering{swap\\[-1mm] test}}}}
}%
{\color[rgb]{0,0,0}\put(1426,-1561){\line( 0,-1){150}}
}%
{\color[rgb]{0,0,0}\put(2626, 89){\line( 0,-1){300}}
}%
{\color[rgb]{0,0,0}\put(2626,-2986){\line( 0,-1){150}}
}%
{\color[rgb]{0,0,0}\put(2626,-3361){\line( 0,-1){150}}
}%
\put(2026,614){\makebox(0,0)[lb]{\smash{{\SetFigFont{12}{14.4}{\familydefault}{\mddefault}{\updefault}{\color[rgb]{0,0,0}$\vdots$}%
}}}}
\put(1951,-3061){\makebox(0,0)[lb]{\smash{{\SetFigFont{12}{14.4}{\familydefault}{\mddefault}{\updefault}{\color[rgb]{0,0,0}$\vdots$}%
}}}}
\put(301,764){\makebox(0,0)[lb]{\smash{{\SetFigFont{12}{14.4}{\familydefault}{\mddefault}{\updefault}{\color[rgb]{0,0,0}$\ket 0$}%
}}}}
{\color[rgb]{0,0,0}\put(601,-1336){\line( 1, 0){600}}
}%
\end{picture}%
  \end{center}
  \caption[The output of the reduction to log-depth circuits]{The
    constructed circuit $C_1$.  In the circuit $C_2$ the dummy zero output
    qubits are swapped with the qubits containing the results of the
    swap tests.  All qubits that do not reach the right edge of the
    figure are traced out, but this is for notational convenience
    only: the constructed circuits are in Stinespring form.}\label{clim-fig-logdepth}
\end{figure}
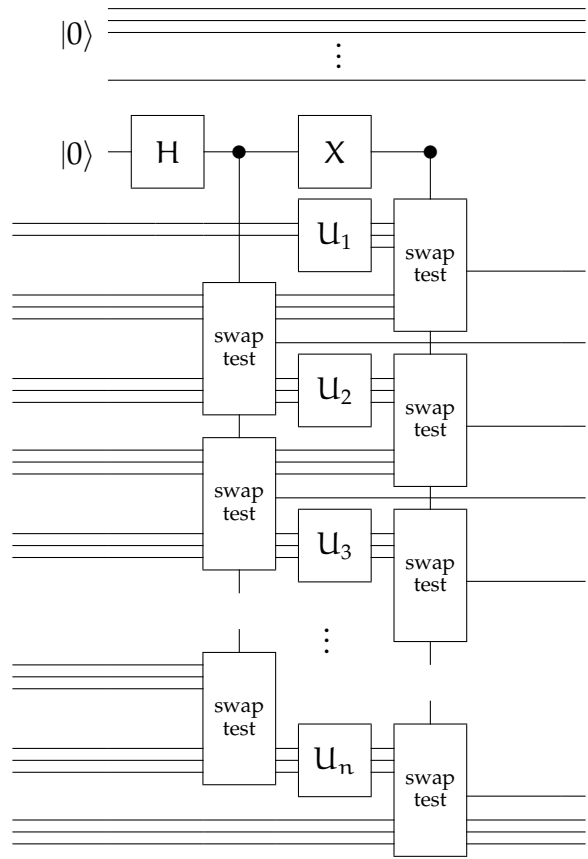
The constructed circuits $C_1$ and $C_2$ are obtained by decomposing
$Q_1$ and $Q_2$ into constant depth pieces, inserting the swap tests shown in
Figure~\ref{clim-fig-test}, and adding dummy qubits to ensure that the
swap tests in the other circuit do not fail.
The final circuit $C_1$ constructed from $Q_1$, including these dummy qubits, 
is shown in Figure~\ref{clim-fig-logdepth}, the circuit $C_2$ is
similar, with the exception that the qubits
corresponding to the swap test outputs and dummy qubits have been
swapped, as shown in~\ref{clim-fig-dummy}.
At the end of these circuits, all qubits are traced out except the
output (in the space $\mathcal{K}_n$) of $U_n$ or $V_n$, the output of
the swap tests, and the dummy zero qubits.  Notice that the circuit $C_i$
can be computed from $Q_i$ in polynomial time, as it is
simply a rearrangement of the gates of the original circuit with the
addition of a linear number of extra gates.

If the outputs of the circuits $C_1$
and $C_2$ are close together, then at an intuitive level the output of the swap
tests in each circuit must be close to zero and the output of $U_n$
and $V_n$ must also be close.  If the swap tests do not fail with high
probability (i.e.\ the outputs are close to zero), then these circuits
will more or less faithfully reproduce the output of $Q_1$ and $Q_2$.
Thus, in the case that the outputs of $C_1$ and $C_2$ can be made close, we
will be able to argue that the output of $Q_1$ and $Q_2$ can also be
made close.  Proving that this picture is accurate forms the
content of Section~\ref{clim-scn-soundness}.

In the other direction, it is much simpler to argue that if
there are states $\rho, \xi \in
\density{H}$ such that $Q_1(\rho) = Q_2(\xi)$, then there are similar
states for the constructed circuits $C_1$ and $C_2$.  This is the
content of the following proposition.

\begin{proposition}\label{clim-prop-simulate}
  If there exist states $\rho, \xi$ such that $Q_1(\rho) = Q_2(\xi)$, then
  there exist states $\rho', \xi'$ such that $C_1(\rho') = C_2(\xi')$.

  \begin{proof}
    To prove the proposition, states $\rho'$ and $\xi'$ are constructed so that
    \begin{equation}\label{clim-eqn-simulation}
      C_1(\rho') = \ket 0 \bra 0 \tprod Q_1(\rho)
       = \ket 0 \bra 0 \tprod Q_2(\xi) = C_2(\xi').
    \end{equation}
    To find these states, notice that both the output fidelity and
    construction of the circuits $C_i$ do not change
    if additional ancillary qubits are added to
    the circuits $Q_i$ to allow purification of the input states,
    so long as these extra qubits are traced out at the end of the
    circuit.
    These purifications are pure states and all operations performed
    during the circuit $Q_i$ are unitary, which implies that
    the intermediate states of the circuits are also pure.
    
    If a purification of the state $\rho$ is $\ket\psi$, then by
    providing the pure state
    \begin{equation}\label{cim-eqn-optimal-input}
      \ket\gamma = 
      \ket\psi 
      \tensor (U_1 \ket\psi)^{\tprod 2} 
      \tensor (U_2 U_1 \ket \psi)^{\tprod 2} \tprod \cdots 
      \tensor (U_{n-1} U_{n-2} \cdots U_1 \ket\psi)^{\tprod 2}
    \end{equation}
    as input to $C_1$, the output of each block of the circuit will
    be identical to the input to the next block, by construction.
    All but the first piece of this state is repeated twice: this is
    to provide the correct intermediate inputs that are used by the
    swap tests to compare the output of one block to the input of the
    next, as shown in Figure~\ref{clim-fig-test}.
    This ensures that all
    the swap tests will succeed with probability one, which can be
    seen from Equation~\eqref{clim-eqn-asym-prob-pure}.  Let this
    constructed state $\rho'$ be given by $\rho' = \ket \gamma \bra \gamma$.

    It remains to check that on $\rho'$ that $C_1$
    simulates $Q_1$ on $\rho$.
    By the construction of $C_1$ and $\rho'$, the output is exactly
    \[ C_1(\rho') = \ket 0 \bra 0 \tprod 
       \tr_{\mathcal{B}_n} U_n U_{n-1} \cdots U_1 
       \rho U_1^* U_2^* \cdots U_n^*, \]
    which is equal to the output of $Q_1$ on $\rho$, up to
    a number of qubits in the $\ket 0$ state, which correspond to the
    dummy qubits and the outputs of the swap tests.
    By the symmetry of the construction, a state $\xi'$ for the
    circuit $C_2$ can be constructed from $\xi$ in the same way, and for these
    constructed $\rho'$ and $\xi'$ Equation~\ref{clim-eqn-simulation}
    is satisfied, which completes the proof.
  \end{proof}
\end{proposition}
This proposition implies that the reduction presented in this section 
maps yes instances of
$\prob{CI}_{1,b}$ to yes instances of $\prob{Log-depth CI}_{1,b}$.
The remaining direction is considerably more intricate, and forms the
content of the next section.

%=============================================================================%

\section{Correctness of the reduction}\label{clim-scn-soundness}

In this section it is argued formally that the reduction presented in
the previous section maps negative instances $(Q_1, Q_2)$ of the close images
problem, i.e.\ those for which $Q_1(\rho)$ and $Q_2(\sigma)$ are far
apart for all states $\rho$ and $\sigma$,  to negative instances of log-depth
close images.  
Less formally, it is shown that if the images of the original circuits
$Q_1$ and $Q_2$ are far apart then so must be the images of the
constructed circuits $C_1$ and $C_2$.
This argument is technical but fairly
straightforward: the basic idea is to transform the fidelity to the
trace norm using the Fuchs-van de Graaf Inequalities
(Theorem~\ref{meas-thm-fuchs-van-de-graaf}), apply the triangle
inequality to reduce the problem to individual blocks of the circuits
$C_1$ and $C_2$, and finally return to the fidelity with another
application of the Fuchs-van de Graaf Inequalities.  As might be
expected, this technique results in poor error bounds: the value $b$,
which is the maximum output fidelity allowed in a no instance of the
close images problem, ends up polynomially close to 1.  This is dealt
with using a parallelization technique due to Kitaev and
Watrous~\cite{KitaevW00} that improves the value of $b$ to any
constant $b>0$.
 
As the constructed circuits $C_1$ and $C_2$ can be used 
to simulate $Q_1$ and $Q_2$, by Proposition~\ref{clim-prop-simulate},
the result is obtained by arguing that either the outputs of $C_1$ and $C_2$ 
correspond to the outputs of $Q_1$ and $Q_2$, respectively, or the
outputs of $C_1$ and $C_2$ are far apart.
In the case that this simulation is not faithful it is shown that
some swap test fails with non-negligible probability.  This implies
that outputs of the constructed circuits are far apart, as the
failing swap test produces a state of the form $(1-p) \ket 0 \bra 0 +
p \ket 1 \bra 1$ that has low fidelity with the
corresponding dummy zero qubit of the other circuit.

Lemma~\ref{clim-lem-swap-fidelity} which describes the behaviour of
the swap test on mixed states does not immediately apply
to the circuits $C_1$ and $C_2$.  This is because
in these circuits the output of one block of the circuit is not directly
compared to the input to the next block, but instead each of these
states are with probability $1/2$ compared to some intermediate value.  
In order to deal with this difficulty, we use the Fuchs-van de Graaf
Inequalities (Theorem~\ref{meas-thm-fuchs-van-de-graaf}) to
translate the fidelity to a relation involving the trace norm, which
we can then apply the triangle inequality to.
The triangle inequality shows that at least one of the
two swap tests fails with probability bounded below by an expression
involving the fidelity, which is lower bounded by 
Lemma~\ref{clim-lem-swap-fidelity}.  In the proof of the following
corollary the
reduced states of various parts of the input to either of the circuits
$C_1$ or $C_2$ are used, but no assumption is made on the form of the
input state, i.e.\ it is not assumed that the input is separable across
the block boundaries of the circuit.
For instance, the density matrices
$\rho_i, \sigma_i,$ and $\xi_i$ that appear in the lemma may be part
of some larger entangled pure state, so that the failure probabilities
of the two swap tests need not be independent.  To clarify the
notation used below, the state $\xi_i$ is the output of the $(i-1)$st block
(i.e.\ the output of $U_{i-1}$ in
Figure~\ref{clim-fig-qip-to-closeim}), the state $\rho_i$ is the input
to the $i$th block, and the state $\sigma_i$ is the intermediate state
used to indirectly compare $\rho_i$ and $\xi_i$.

\begin{corollary}\label{clim-cor-failprob}
  If $\ket\psi$ is input to the circuit $C_a$ for $a \in \{1,2\}$, 
  with $\rho_i$ the reduced state of $\ket\psi\bra\psi$ on
  $\mathcal{H}_i \tensor \mathcal{A}_i$, then
  at least one of the swap tests on the $i$th block
  of $C_a$ fails with probability at least
  \[ \frac{1}{128} \tnorm{U_i \rho_{i-1} U_i^* - \rho_i}^2. \]

  \begin{proof}
    In the $i$th block of $C_a$ there are two inputs to the first swap
    test: let the reduced density operators of these inputs be
    $\rho_i$ and $\sigma_i$, as discussed above.
    The inputs to the second swap test are then given by $\sigma_i$
    and $U_i \rho_{i-1} U_i^* = \xi_i$.
    As exactly one of these tests is performed we do not need to
    consider the effect of the first test on the state when
    considering the second test, and so the same input state
    $\sigma_i$ is used for both swap tests.

    By Lemma~\ref{clim-lem-swap-fidelity}, the failure probability of
    the first
    and second tests, when performed, are at least
    $\frac{1}{2}(1 - \F(\rho_i,\sigma_i))$ and
    $\frac{1}{2}(1 - \F(\sigma_i,\xi_i))$, respectively.  
    Thus, the probability $p$ that at least one of these tests
    fails, given that each of them is performed with probability
    $1/2$, is at least
    \[p \geq \frac{1}{2} \max \left\{ \frac{1}{2}(1 -\F(\sigma_i,\xi_i)),
           \frac{1}{2}(1 -\F(\rho_i, \sigma_i)) \right\}
      = \frac{1}{4} \left(1 - \min\{\F(\sigma_i, \xi_i)),
           \F(\rho_i, \sigma_i)\}\right). \]
    By the Fuchs-van de Graaf inequalities
    (Theorem~\ref{meas-thm-fuchs-van-de-graaf}), 
    this fidelity may be
    replaced by the trace norm.  Doing so, we obtain
    \[ p \geq \frac{1}{32} \max( \tnorm{\sigma_i - \xi_i}^2,
           \tnorm{\rho_i - \sigma_i}^2 ), \]
    where we have made use of Bernoulli's inequality to show that
    $\sqrt{1 - x} \leq 1 - x/2$ when simplifying this equation.
    Finally, as this maximum must be at least the average of the two
    values,
    \[ p \geq \frac{1}{32} \left(\frac{\tnorm{\sigma_i - \xi_i}}{2}
           + \frac{\tnorm{\rho_i - \sigma_i}}{2} \right)^2
       \geq \frac{1}{128} \tnorm{\rho_i - \xi_i}^2, \]
    where the last inequality is the
    triangle inequality.
  \end{proof}
\end{corollary}

By repeatedly applying some of the properties of the trace norm
discussed in Chapter~\ref{chap-meas} it is somewhat tedious but not
difficult to use Corollary~\ref{clim-cor-failprob} to bound the
distance between the images of the constructed circuits in terms of
the distance between the images of the original circuits.
In the following theorem $n$ is the size of the circuits $Q_1$ and
$Q_2$, as in Section~\ref{clim-scn-construction}.  Informally, this
theorem states that ``no'' instances of the problem $\prob{CI}_{1,b}$ are
mapped to ``no'' instances of the problem 
$\prob{Log-depth CI}_{1,b'}$ with the resulting value $b'$ only
polynomially closer to 1 than the value $b$, which shows
the \class{QIP}-completeness of log-depth close images for these large
values of $b$.

\begin{theorem}\label{clim-thm-soundness}
  If $\F(Q_1(\rho_0), Q_2(\xi_0)) < 1 - c$ for all $\rho_0, \xi_0 \in \density{H}$ then 
  \[ \F(C_1(\rho), C_2(\xi)) <  1 - \frac{c^2}{576 n^2} \]
  for all $\rho, \xi \in \density{H_0 \tprod
    \bigotimes_{\mathrm{i=2}}^{\mathrm{2n}} H_{\mathrm{i}} \tensor A_{\mathrm{i}}}$

  \begin{proof}
    Let $\rho$ and $\xi$ be inputs to $C_1$ and $C_2$, and let
    $\rho_i, \xi_i$ be the reduced states of these inputs on 
    $\mathcal{H}_i \tensor \mathcal{A}_i$ for $1
    \leq i \leq 2n$, where the states for $i > n$ are the
    inputs that are only used by the swap tests, which will not
    be referred to explicitly.
    That is, $\rho_i$ and $\xi_i$ for $1 \le i \le n$ 
    are the portions of the state that are
    input to the unitaries $U_i$ and $V_i$ that make up the circuits
    $Q_1$ and $Q_2$, as shown in Figure~\ref{clim-fig-qip-to-closeim}.
    The output of the circuits $C_1$ and $C_2$ is
    given by the output qubits corresponding to the swap tests
    as well as the states $\tr_{\mathcal{B}_n}
    \rho_n$ and $\tr_{\mathcal{B}_n} \xi_n$, where $\mathcal{B}_n$ is
    simply the space that is traced out to obtain the output from the
    unitary representations of the original circuits.
    In this notation, $\rho_1$ and $\xi_1$ are the inputs to the first
    pieces $U_1$ and $V_1$ of the constructed circuits $C_1$ and
    $C_2$.  These two states are density matrices in
    $\density{H_\mathrm{1}} \cong \density{H}$, and we can also
    consider them as potential inputs to the original circuits $Q_1$
    and $Q_2$.

   By the condition on the fidelity of $Q_1$ and $Q_2$ in the
    statement of the theorem, as well as the
    Fuchs-van de Graaf inequalities
    (Theorem~\ref{meas-thm-fuchs-van-de-graaf}), we have
    \[ 2c < \tnorm{ Q_1(\rho_1) - Q_2(\xi_1) }. \]
    Using the triangle inequality we can relate this to the
    distance between the constructed circuits.  By adding terms and
    simplifying, we obtain
    \begin{align*}
      2c &< \tnorm{ Q_1(\rho_1) 
	           -\tr_{\mathcal{B}_n}\rho_n 
		   +\tr_{\mathcal{B}_n}\xi_n 
		   - Q_2(\xi_1)
	           +\tr_{\mathcal{B}_n}\rho_n 
		   -\tr_{\mathcal{B}_n}\xi_n } \\
        &\le \tnorm{ Q_1(\rho_1)
	           -\tr_{\mathcal{B}_n}\rho_n }
	  +\tnorm{ \tr_{\mathcal{B}_n}\xi_n 
		   - Q_2(\xi_1)}
	  +\tnorm{ \tr_{\mathcal{B}_n}\rho_n 
		   -\tr_{\mathcal{B}_n}\xi_n } .
    \end{align*}
    We now observe that $\tnorm{ \tr_{\mathcal{B}_n}\rho_n
    -\tr_{\mathcal{B}_n}\xi_n } \leq \tnorm{ C_1(\rho) - C_2(\xi) }$
    by the monotonicity of the trace norm under quantum operations 
    (Theorem~\ref{meas-thm-tnorm-monotonicity}),
    since the former can be obtained from the later by applying the
    parital trace on the appropriate space.  Using this we have
    \begin{align}
    2c < \tnorm{ Q_1(\rho_1)
	           -\tr_{\mathcal{B}_n}\rho_n }
	  +\tnorm{ \tr_{\mathcal{B}_n}\xi_n 
		   - Q_2(\xi_1)}
	  +\tnorm{ C_1(\rho) - C_2(\xi) } \label{clim-eqn-three-terms-2c}
    \end{align}
    As the three terms on the right are nonnegative, at least one of them
    must be larger than the average $2c/3$.  
    If $\tnorm{ C_1(\rho) - C_2(\xi) } > 2c/3$
    then $\F(C_1(\rho), C_2(\xi)) < 1 - c^2/18$, 
    by Theorem~\ref{meas-thm-fuchs-van-de-graaf}, 
    and there is nothing left to prove.

    The cases where one of the first two terms of
    \eqref{clim-eqn-three-terms-2c} exceeds $2c/3$
    are symmetric, and so we can consider only the first term.
    Expanding $Q_1(\rho_1)$ in terms of the $U_i$, we obtain
    \begin{align*}
      \frac{2c}{3} 
      &< \tnorm{ Q_1(\rho_1) - \tr_{\mathcal{B}_n}\rho_n }\\
      &= \tnorm{ 
	    \tr_{\mathcal{B}_n} 
          U_n U_{n-1} \cdots U_1 \rho_1 U_1^* U_2^* \cdots U_n^*
 	   -\tr_{\mathcal{B}_n} \rho_n } \\
      &\le \tnorm{ U_n U_{n-1} \cdots U_1 \rho_1 U_1^* U_2^* \cdots U_n^*
	    -\rho_n },
    \end{align*}
    where once again the monotonicity of the trace norm
    under the partial trace (Theorem~\ref{meas-thm-tnorm-monotonicity})
    has been used.
    By adding and subtracting the term $U_n \cdots U_2 \rho_1 U_2^*
    \cdots U_n^*$ inside the norm, and then applying both the
    triangle inequality and the unitary invariance of the trace norm, 
    we have
    \begin{align*}
      \frac{2c}{3}
      &< \tnorm{U_1 \rho_1 U_1^* - \rho_2}
       + \tnorm{U_{n} U_{n-1} \cdots U_2 \rho_2 U_2^* U_3^* \cdots U_{n}^*
	    -\rho_n}.
    \end{align*}
    Here the unitary invariance of the trace norm
    has been used
    to discard the operators $U_2, \ldots U_n$ from the first term.
    Repeating this strategy, by adding and subtracting the term
    $U_n \cdots U_3 \rho_3 U_3^* \cdots U_n^*$ and once again applying
    the triangle inequality results in
    \begin{align*}
      \frac{2c}{3}
      &< \tnorm{U_1 \rho_1 U_1^* - \rho_2}
       + \tnorm{U_2 \rho_2 U_2^* - \rho_3}
       + \tnorm{U_{n} U_{n-1} \cdots U_3 \rho_3 U_3^* U_4^* \cdots U_{n}^*
	    -\rho_n}.
    \end{align*}
    Continuing in this fashion we have
    \begin{align*}
      \frac{2c}{3}
      &< \sum_{i=1}^{n-1} \tnorm{U_i \rho_{i} U_i^* - \rho_{i+1}}.
    \end{align*}
    As all terms in this sum are nonnegative,
    there must be at least one term in the sum that exceeds
    $2c / (3n)$, as this is a lower bound on the average of all terms.
    Thus, for some value of $i$, we have
    $ \tnorm{U_i \rho_{i} U_i^* - \rho_{i+1}} > 2c / (3n), $
    and so by Corollary~\ref{clim-cor-failprob} one of the
    corresponding swap tests fails with probability
    $p > c^2 / (288 n^2)$.  The qubit representing the output value of
    this swap test is then of the form $(1-p) \ket 0 \bra 0 + p \ket 1
    \bra 1$, and so, by the monotonicity of the fidelity under
    the partial trace (Theorem~\ref{meas-thm-fidelity-monotonicity}),
    \[ \F(C_1(\rho), C_2(\xi)) 
       \leq \F((1-p) \ket 0 \bra 0 + p \ket 1 \bra 1, \ket 0 \bra 0)
       = \sqrt{1-p}
       < 1 - \frac{c^2}{576 n^2}, \]
    as in the statement of the theorem.
  \end{proof}
\end{theorem}

By combining Theorem~\ref{clim-thm-soundness} with
Proposition~\ref{clim-prop-simulate}, and the multiplicativity of the
maximum output fidelity of two transformations, given as
Theorem~\ref{meas-thm-outfid-mult}, we obtain the main result of this
chapter.

\begin{corollary}\label{clim-cor-log-result}
  $\prob{Log-depth CI}_{a,b}$ is \class{QIP}-complete
  for any constants $0 < b < a \leq 1$.

  \begin{proof}
    Theorem~\ref{clim-thm-soundness} together with
    Proposition~\ref{clim-prop-simulate} establish the result that 
    $\prob{CI}_{1,b}$ reduces to
    $\prob{Log-depth CI}_{1,b'}$ for $b' \geq 1 - (1-b)^2/(576 n^2)$, 
    where $n$ is an upper bound on the size of the circuits.

    The value of $b'$ can be improved using
    Theorem~\ref{meas-thm-outfid-mult} of Kitaev, Shen, and
    Vyalyi~\cite{KitaevS+02}, which shows that if the circuits $C_1$
    and $C_2$ are repeated $r$ times in parallel, then the maximum
    output fidelity is
    \[ \max_{\rho, \xi} \F(C_1^{\tprod r}(\rho), C_2^{\tprod r}(\xi))
       = \max_{\rho, \xi} \F(C_1(\rho), C_2(\xi))^r. \]
    This implies that $\prob{CI}_{1,b}$ reduces to $\prob{Log-depth CI}_{1,b'}$
    for all
    \[ b' \geq \left( 1 - \frac{(1-b)^2}{576 n^2} \right)^r, \]
    and so, by taking $r$ polynomially large in $n$ and $b$, we may
    take $b' \leq b$, which implies that
    $\prob{CI}_{1,b}$ reduces to $\prob{Log-depth CI}_{1,b}$.

    This shows that $\prob{Log-depth CI}_{1,b}$ is then
    \class{QIP}-complete for any constant $0 < b < 1$, as by
    Theorem~\ref{clim-thm-ci-complete} due to Kitaev and
    Watrous~\cite{KitaevW00}, $\prob{CI}_{a,b}$ is
    \class{QIP}-complete for all $0 < b < a \leq 1$.
    Generalizing the log depth close images problem for all values of
    $a$ gives the
    problem $\prob{Log-depth CI}_{a,b}$ for $0 < b < a \leq 1$, which
    is also complete for \class{QIP} as a it can be obtained by
    weakening the promise.  This more general problem is in
    \class{QIP} as it is a restriction of $\prob{CI}_{a,b}$ to
    log-depth circuits.
  \end{proof}
\end{corollary}

As the circuits constructed by the reduction only make use of
logarithmic depth when performing swap tests, and the controlled swap
operations performed by these tests can be accomplished in constant
depth using unbounded fan-out gates, as described in
Proposition~\ref{compl-prop-const-control}, the following Corollary follows
immediately from the previous one.

\begin{corollary}\label{clim-cor-const-result}
  The problem $\prob{Const-depth CI}_{a,b}$ on circuits with the
  unbounded fan-out gate is \class{QIP}-complete for
  any constants $0 < b < a \leq 1$.
\end{corollary}

\section{Conclusion}

In this chapter, the problem \prob{Close Images} has been introduced.
This problem asks: given two quantum channels, as mixed state quantum
circuits, how close are the images of the two channels?  More
concretely, how large is the minimum distance of any two outputs of
the channels, where the fidelity is used as the notion of distance.
This problem is complete for the class \class{QIP}~\cite{KitaevW00}.

The main result of the chapter is a reduction of this problem to the
case of logarithmic depth circuits.  This reduction works only for the
case that the two circuits are promised to either have intersecting
images or images that are far apart, but a hardness result on this
special case also implies the hardness of the general problem.  This
restriction is necessary to the proof that the reduction is correct as
it enables the use of a parallel repetition technique to strengthen the
promise of the class of instances that is shown to be hard.

This hardness result is the base for the main result of the next
chapter, which is that the computational problem of distinguishing
quantum circuits is also \class{QIP}-hard.  The result of this chapter
enables the hardness of this distinguishability problem to be extended
even to the case of channels implemented by log-depth circuits.

% LocalWords:  Jozsa DiVincenzo 

%%% Local Variables: 
%%% mode: latex
%%% TeX-master: "thesis"
%%% End: 

\chapter{Distinguishability of Quantum Computations}
\label{chap-distinguishability}

A natural problem in quantum information is to discriminate between
two quantum channels.  In the model where channels are represented as
quantum circuits, this is the computational distinguishability problem
on channels, though the
difficulty of the problem does not change if the circuits are replaced
by black boxes that can be performed, but not inspected.  The main
result of this chapter is that this problem is computationally very
difficult, as it is complete for the class \class{QIP} of problems
having quantum interactive proof systems.  This also
implies~\cite{JainJ+09} that this problem is complete for
\class{PSPACE}, which gives a new quantum characterization for a
classical complexity class.

The majority of the results in this chapter are in collaboration with
John Watrous, and have been published in~\cite{RosgenW05}.  The
results in Section~\ref{dist-scn-logdepth} have appeared
in~\cite{Rosgen08distinguishing}.

\minitoc

\section{Overview of distinguishability problems}\label{dist-scn-intro}

The problem of distinguishing two computations is central to computer
science.  This is the problem that asks, given two computations
represented in some way, do the two computations always produce
results that are close together, or are there inputs on which they act
differently?  This is an important problem both theoretically and
practically: determining if some process has been correctly
implemented is one of the most important tasks in experimental 
quantum computing.

The problem of estimating an unknown quantum channel is known as \emph{process
  tomography}~\cite{ChuangN97, PoyatosC+97}.  All known approaches for
approximate process tomography unfortunately require exponential
time.  This is not a surprise, however, as the complete
characterization of a quantum channel on $n$ qubits requires an
exponential number of parameters.  The main result of this chapter is
that even the simpler task of \emph{distinguishing} two quantum
channels, given as mixed-state circuits, is computationally
intractable.  That the distinguishability problem reduces to process
tomography is clear: one way to solve the problem is characterize the
two channels with enough accuracy to detect the case that they are far
apart.

Returning to classical complexity theory, the most basic
distinguishability problem asks: given two classical deterministic
circuits, is there an
input on which they produce different outputs?
This problem is in \class{NP}, as given such an input a verifier can
both simulate the two circuits and check to see if they agree, all in
polynomial time.  This problem is also complete for \class{NP}.  To
see this, notice that a circuit is satisfiable if and only if it is
distinguishable from the circuit that always outputs false.  The
satisfiability problem is the original \class{NP}-complete
problem~\cite{Cook71}, and so the problem of distinguishing classical
deterministic computations must also be \class{NP}-complete.

Adding randomness to the circuit model, in the form of gates that
produce unbiased coin flips, appears to increase the difficulty of the problem.
Averaged over the values of the coin flips, the two circuits in the
distinguishability problem produce probability distributions --
distinguishing these distributions is also a nontrivial problem.
To avoid the problem of distinguishing randomized circuits being
artificially difficult, the additional promise is given that the two probabilistic
circuits to be distinguished either produce output
distributions that are very close for any input, or
that there exists some input on which the distributions produced are
far apart.  The usual measure of distance on probability distributions
is the difference in the $\ell_1$ norm, 
that is defined, for distributions $p, q$, as
$ \norm{p - q}_{\ell_1} = \sum_x \abs{ p(x) - q(x) }. $
This norm is simply the classical analogue of the trace norm of the
difference of two density operators.
Even when given an input on which they produce maximally distant
output distributions, distinguishing two randomized circuits
is complete for the class \class{SZK} of problems with
statistical zero-knowledge proof systems~\cite{SahaiV03}.  This is in
sharp contrast to the deterministic case where a verifier can check, given
an input, whether or not two deterministic
circuits produce the same output.  The best complexity theoretic upper
bound known for the randomized circuit 
distinguishability problem is \class{AM}, by
simply having the prover first produce an optimal distinguishing input
for the two circuits and then performing the \class{SZK} protocol due
to Sahai and Vadhan~\cite{SahaiV03} for the statistical difference
problem that remains.  This problem is not known to be complete for
\class{AM}.

Extending this problem in the direction of quantum information leads
first to the natural problem of distinguishing unitary circuits.  
As in the case of randomized classical circuits, the outputs of
unitary quantum circuits are nontrivial to compare.  Once again, a
promise is used to keep the problem from being artificially
difficult.  In this case, the distance measure can be either the
diamond norm or the trace norm, as it is known that
they agree on unitary transformations~\cite{AharonovK+98, 
ChildsP+00}, and the promise is that the distance between the
two given transformations, which is given by
\[ \max_{\ket \psi \in \mathcal{H}} \tnorm{U \ket\psi \bra\psi U^* - V
  \ket\psi\bra\psi V^*}, \] is
either close to zero or close to two.  The input in this equation may
be restricted to pure states by
Theorem~\ref{meas-thm-dnorm-difference}.  This problem has been shown
to be complete for \class{QMA} by Janzing, Wocjan, and
Beth~\cite{JanzingW+05}.  This result implies that, given an optimal
input state, a verifier with access to a quantum computer can solve
the distinguishability problem on unitary circuits.  This is not
unexpected, as unitary circuits are similar to deterministic quantum
computation.

\begin{figure}
  \begin{center}
    \setlength{\unitlength}{3947sp}%
\begingroup\makeatletter\ifx\SetFigFont\undefined%
\gdef\SetFigFont#1#2#3#4#5{%
  \reset@font\fontsize{#1}{#2pt}%
  \fontfamily{#3}\fontseries{#4}\fontshape{#5}%
  \selectfont}%
\fi\endgroup%
\begin{picture}(6302,2099)(-1350,-1409)
\put(1876, 14){\makebox(0,0)[lb]{\smash{{\SetFigFont{12}{14.4}{\familydefault}{\mddefault}{\updefault}{\color[rgb]{0,0,0}\class{EXP}}%
}}}}
{\color[rgb]{0,0,0}\thinlines
\put(2251,-811){\circle*{76}}
}%
{\color[rgb]{0,0,0}\put(1351,-811){\circle*{76}}
}%
{\color[rgb]{0,0,0}\put(1801, 89){\circle*{76}}
}%
{\color[rgb]{0,0,0}\put(1801,-361){\circle*{76}}
}%
{\color[rgb]{0,0,0}\put(1801,-1261){\line( 1, 1){450}}
}%
{\color[rgb]{0,0,0}\put(1801,-1261){\line(-1, 1){450}}
}%
{\color[rgb]{0,0,0}\put(1351,-811){\line( 1, 1){450}}
}%
{\color[rgb]{0,0,0}\put(2251,-811){\line(-1, 1){450}}
}%
{\color[rgb]{0,0,0}\put(1801,-361){\line( 0, 1){450}}
}%
\put(2326,-886){\makebox(0,0)[lb]{\smash{{\SetFigFont{12}{14.4}{\familydefault}{\mddefault}{\updefault}{\color[rgb]{0,0,0}\class{QMA}, (Unitary Quantum)}%
}}}}
\put(1876,-1336){\makebox(0,0)[lb]{\smash{{\SetFigFont{12}{14.4}{\familydefault}{\mddefault}{\updefault}{\color[rgb]{0,0,0}\class{NP}, (Deterministic Classical)}%
}}}}
\put(-1274,-886){\makebox(0,0)[lb]{\smash{{\SetFigFont{12}{14.4}{\familydefault}{\mddefault}{\updefault}{\color[rgb]{0,0,0}\class{AM}, (Randomized Classical)}%
}}}}
\put(1876,-436){\makebox(0,0)[lb]{\smash{{\SetFigFont{12}{14.4}{\familydefault}{\mddefault}{\updefault}{\color[rgb]{0,0,0}\class{QIP}, (Mixed-state Quantum)}%
}}}}
{\color[rgb]{0,0,0}\put(1801,-1261){\circle*{76}}
}%
\end{picture}%
  \end{center}
  \caption[Complexity classes and distinguishability problems]{Complexity classes and distinguishability problems.}
  \label{dist-fig-complexity-classes}
\end{figure}
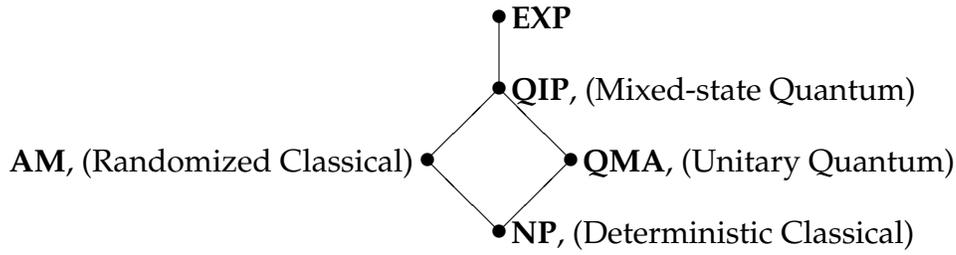
Adding both the elements of randomness and quantumness to the
computations being distinguished results in a significantly more
difficult problem than adding just one of the two elements.
This is the distinguishability problem for mixed-state quantum circuits,
as defined in Section~\ref{compl-scn-circuit-model}.
Once again we require a promise to avoid an artificially difficult
problem: the circuits to be distinguished can be assumed to have a
diamond norm difference that is either close to zero or close to two.
It is surprising that this problem is
\class{QIP}-complete.  The class \class{QIP} is
believed to be much larger than the classes \class{QMA} and
\class{AM}.  As evidence of this, the polynomial hierarchy collapses
if \class{QIP} is contained in either of these classes.  These
complexity classes, known inclusions among them, and the
distinguishability problems related to them, 
are summarized in Figure~\ref{dist-fig-complexity-classes}.
Removing either the quantum computing, leaving randomized circuits, or
randomness, leaving unitary quantum circuits, seems to change the
essential character of the problem: the hardness appears to lie in the
combination of both of these ingredients.

The main result of this section is showing the
\class{QIP}-completeness for the mixed-state circuit
 distinguishability problem
using a Karp reduction from the problem \prob{Close Images} of
Chapter~\ref{chap-close-images}.
The main technique is a scheme for using two
transformations that are close together for some inputs and producing
from them two transformations that act very differently on a
specific input state, whereas when the scheme is applied to
transformations that are far apart, the resulting transformations are
very close together.  In some sense this reduction inverts the
distance between two circuits: circuits that are close together are
mapped to circuits that are far apart, and vice versa, though the
definitions of distance used in the close images and
distinguishability problems are not the same.

\section{Quantum circuit distinguishability}\label{dist-scn-problem}

The problem of distinguishing mixed state quantum circuits can be
stated more intuitively in the following way: given a black box that
is promised to implement one of two known quantum channels, with what
probability can the channel be identified with only a single use of
the black box?  As was discussed in Section~\ref{meas-scn-dnorm}, the
maximum probability that the correct channel can be identified is
given by
\[ \frac{1}{2} + \frac{\dnorm{\Phi_1 - \Phi_2}}{4}, \]
where the channels $\Phi_i$ represent the two known channels.  This
implies that the problem of estimating the diamond norm of the
difference of two channels is equivalent to estimating the probability
that the black box can be correctly identified with a single use,
given descriptions of the two channels $\Phi_1, \Phi_2$.

To obtain a computational problem, let these channels be given by
mixed-state quantum circuits: this results in the quantum circuit
distinguishability problem that is the focus of this chapter.  The
main result of the chapter is to show that this problem is
complete for \class{QIP}.  As a by-product, the definition of this
problem implies that simply deciding if the two channels are close
together for all inputs or far apart on some input state is equivalent
to determining with what probability the correct channel is identified
in the black box problem.  The formal definition of the
distinguishability (promise) problem is given below.
\begin{problem}[Quantum Circuit Distinguishability]\label{dist-prob-qcd}
  For constants $0 \leq b < a \leq 2$, the input consists of quantum
  circuits $Q_1$ and $Q_2$ that implement transformations in
  $\transform{H,K}$.
  The promise problem is to distinguish the two cases:
  \begin{description}
    \item[Yes:] $\dnorm{Q_1 - Q_2} \geq a$,
    \item[No:] $\dnorm{Q_1 - Q_2} \leq b$.
  \end{description}
\end{problem}
\index{Quantum Circuit Distinguishability!problem definition}%
Less formally, this problem asks: is there an input density matrix
$\rho$ on which the circuits $Q_1$ and $Q_2$ can be made to act
differently?
Theorem~\ref{meas-thm-dnorm-difference} implies that this problem can
be stated in terms of pure state inputs.
\nomenclature[PQCD]{$\prob{QCD}$}{Quantum Circuit Distinguishability problem}%
For notational convenience, this problem will be referred to as
$\prob{QCD}_{a,b}$, with the logarithmic and constant-depth variants
referred to as $\prob{Log-depth QCD}_{a,b}$ and 
$\prob{Const-depth QCD}_{a,b}$, though they will not be encountered until
Section~\ref{dist-scn-logdepth}.

The Quantum Circuit Distinguishability problem appears on the surface
to be very similar to the Close Images problem considered in
Chapter~\ref{chap-close-images}, but a closer inspection reveals that
this is not the case.  Given two circuits $Q_1$ and $Q_2$, the close
images problem asks if there are two inputs $\rho$ and $\sigma$ on
which the two circuits act the same, i.e.\ $Q_1(\rho) \approx
Q_2(\sigma)$.  On the other hand, the circuit distinguishability
problem asks if there is one input $\rho$ for which the states
$Q_1(\rho)$ and $Q_2(\rho)$ are nearly orthogonal.  The two problems
ask for the two channels to have significantly different properties,
though the problems do not appear to be dual to each other in any real
sense.

In addition to this, the circuit distinguishability problem has an
operational meaning in terms of how well an unknown quantum process chosen
from a set of two known channels can be identified.  An alternate
characterization of the problem is, given two quantum channels, are
they almost the same, or are there inputs on which they differ significantly.
This is a simplification of quantum process
tomography~\cite{ChuangN97, PoyatosC+97}, which has many applications
in quantum information, but is unfortunately intractable in the
computational sense.  In contrast, the Close Images problem, as
discussed in Section~\ref{clim-scn-intro}, is quite closely related to
whether or not the verifier can be made to accept in a quantum
interactive proof system.  This makes the circuit distinguishability
problem interesting for the study of the class \class{QIP}, as it
gives a quantum information theoretic characterization of the class
that is not a restatement of the definition.

\section{\class{QIP} protocol}\label{dist-scn-protocol}

The aim of this section is to present and analyze a protocol that puts
the circuit distinguishability problem inside of \class{QIP}.  This is
an essential step in showing that this problem is complete for
this class.

The basic idea of the protocol used to achieve this is to have the
prover send a state on which the two circuits are maximally
distinguishable, apply one of the two circuits at random, and then ask
the prover to determine which circuit has been applied.  It is not
hard to see that by playing
honestly, the prover will be able to succeed with probability related
to the diamond norm of the difference of the two circuits.
It is only slightly more difficult to see that this is also the optimal
strategy for a dishonest prover.  A more complete description of the
protocol follows.

\index{Quantum Circuit Distinguishability!\class{QIP} protocol}
\begin{proto}[Quantum Circuit Distinguishability]\label{dist-tnd_proto}
  As input, both the prover $P$ and the verifier $V$ receive circuits $Q_1,
  Q_2 \in \transform{H,K}$ of size at most $n$.  The three steps of
  the protocol are:
  \begin{enumerate}
  \item\label{dist-step_send_rho}
    $V$ receives from $P$ a state $\rho \in \density{H}$.

  \item\label{dist-step_send_back}
    $V$ chooses $i \in \{1,2\}$ uniformly at random and sends 
    $Q_i(\rho)$ back to $P$.

  \item\label{dist-step_challenge}
    $V$ receives from $P$ some $j \in \{1,2\}$, accepts if $i = j$, and rejects
    otherwise.
  \end{enumerate}
\end{proto}

The idea behind this protocol is that if the two circuits are far
apart the prover can find an input state on which they are
distinguishable.  The prover is then asked to perform
this distinguishing task.  Both choosing the state $\rho$ and
performing the measurement to distinguish the output states
$Q_1(\rho)$ and $Q_2(\rho)$ may be computationally intractable and so
in the protocol the prover is required to perform them: the verifier
only needs to flip a coin and apply one of the two circuits, which can
be done in polynomial time, given circuit descriptions.

Step~\ref{dist-step_challenge} of this protocol does not strictly fit
into the model of quantum interactive proof systems.  This is because
the prover sends classical information to the verifier, and not a
quantum message.  This difficulty can be avoided by allowing the prover to
send a qubit to the verifier, who immediately measures it in the
computational basis.  Such a modification does not change the
protocol, as any state sent during this step gives the prover no way
to do better than simply sending either $\ket 1$ or $\ket 2$.

To show that this protocol puts the distinguishability problem into
\class{QIP}, it remains to show that the prover can succeed with
probability $p$ on yes instances and with probability at most $q$ on
no instances, for some values of $p,q$ that are at least polynomially
far apart.  Error reduction for \class{QIP} allows this to be
amplified to any constant gap, as discussed in
Section~\ref{compl-scn-complexity-classes}.  This is the content of
the following theorem.

\begin{theorem}\label{dist-theorem-QCD_in_QIP}
For any constants $a,b$ with $0 \leq b < a \leq 2$,
$\prob{QCD}_{a,b} \in \class{QIP}$.

  \begin{proof}
    To show that the verifier of Protocol~\ref{dist-tnd_proto} forms a
    quantum interactive proof system for $\prob{QCD}_{a,b}$ bounds
    must be placed on the error probability of the protocol for both
    positive and negative instances of the problem.  This will be done
    by showing that in either case, the maximum acceptance probability
    of the verifier is given by
    \[ \frac{1}{2} + \frac{1}{4}\dnorm{Q_1 - Q_2}, \]
    which is simply the optimal probability that a black box can be
    identified as either $Q_1$ or $Q_2$ with only a single use, as
    discussed in Corollary~\ref{meas-cor-dnorm-Helstrom}.  It is
    not hard to see that this is exactly the task faced by the prover
    in the protocol.

    By Theorem~\ref{meas-thm-dnorm-difference} there exists a Hilbert space
    $\mathcal{F}$ and a pure state
    $\ket{\psi} \in \mathcal{H} \tprod \mathcal{F}$ such that
    \[ \dnorm{Q_1 - Q_2}
       = \tnorm{(Q_1 \tprod \tidentity{F})(\ket{\psi}\bra{\psi})
          - (Q_2 \tprod \tidentity{F})(\ket{\psi}\bra{\psi})}. \]
    For this state $\ket{\psi}$, let
    \begin{align*}
      \rho_1 & = (Q_1 \tprod \tidentity{F})(\ket{\psi}\bra{\psi}), \\
      \rho_2 & = (Q_2 \tprod \tidentity{F})(\ket{\psi}\bra{\psi}).
    \end{align*}
    Let $\Pi_1$ and $\Pi_2 = \id - \Pi_1$ be projection operators on
    $\mathcal{K} \tprod \mathcal{F}$ that specify an optimal
    projective measurement for distinguishing $\rho_1$ from $\rho_2$.
    These projection operators form the Helstrom measurement, which
    is discussed in Theorem~\ref{meas-thm-Helstrom}.
    Such a measurement satisfies
    \[ \tr \Pi_1 (\rho_1 - \rho_2) = \tr \Pi_2 (\rho_2 - \rho_1) 
       = \frac{1}{2}\tnorm{\rho_1 - \rho_2}, \]
    as $\Pi_1$ is the projector onto the positive eigenvalues of
    $\rho_1 - \rho_2$ and $\tr( \rho_1 - \rho_2) = 0$.
    
    A strategy for the prover that convinces the verifier to accept with
    probability
    \[ \frac{1}{2} + \frac{1}{4}\dnorm{Q_1 - Q_2} \]
    is as follows.
    The prover prepares the state $\ket{\psi}$ and sends the reduced state
    $\rho = \ptr{F} \ket\psi \bra\psi$ to the verifier, keeping the
    portion of the state on $\mathcal{F}$ in reserve.

    Upon receiving $\sigma = Q_i(\rho)$ from the verifier, the prover measures
    the state on $\mathcal{K} \tprod \mathcal{F}$
    with the measurement $\{\Pi_1,\Pi_2\}$ and
    returns the result to the verifier.
    By Theorem~\ref{meas-thm-Helstrom}, 
    this measurement correctly determines $i$ with probability
    \[ \frac{1}{2} + \frac{1}{4}\tnorm{\rho_1 - \rho_2} =
       \frac{1}{2} + \frac{1}{4}\dnorm{Q_1 - Q_2}. \]

    That this strategy is optimal can be argued as follows.
    Let $\xi \in \density{H \tensor F}$ be the state of the system immediately
    after the first message is sent, where the space $\mathcal{F}$
    represents the private space of the prover (which need not be the same
    size as the space $\mathcal{F}$ considered above).
    The verifier applies either $Q_1$ or $Q_2$ to the space $\mathcal{H}$,
    which results in the global state
    $(Q_1 \tprod \tidentity{F})(\xi)$ with probability 1/2 and
    $(Q_2 \tprod \tidentity{F})(\xi)$ with probability 1/2.
    This state is, after step~\ref{dist-step_send_back} of the protocol,
    in the possession of the prover.
    The prover's final message to the verifier is immediately 
    measured by the verifier, resulting in a single bit.
    This process may be viewed as a two-valued measurement on 
    $\mathcal{K \tprod F}$.
    The probability that the optimal measurement of this type 
    is correct is given by Theorem~\ref{meas-thm-Helstrom}, and so
    \[ \frac{1}{2} + \frac{1}{4}\tnorm{
          (Q_1 \tprod \tidentity{F})(\xi) -
          (Q_2 \tprod \tidentity{F})(\xi)}
       \leq \frac{1}{2} + \frac{1}{4}\dnorm{Q_1 - Q_2} \]
    is an upper bound on the success probability of the prover, as required.
    
    This gives a quantum interactive proof system for $\prob{QCD}_{a,b}$
    that accepts yes instances with probability $1/2 + a/4$ and accepts no
    instances with probability at most $1/2 + b/4$.  This proves that
    $\prob{QCD}_{a,b} \in \class{QIP}$ as $b < a$ with at least a
    polynomial gap between them, 
    by the definition of the distinguishability problem.
  \end{proof}
\end{theorem}

As discussed in Section~\ref{dist-scn-intro}, the version of this
problem defined on classical randomized circuits is contained in the
complexity class~\class{AM}.  One way to see this is to consider
Protocol~\ref{dist-tnd_proto} with all of the quantum information
removed.  The prover can still send an optimal distinguishing input in
Step~\ref{dist-step_send_rho} and decide which distribution the sample
is from in Step~\ref{dist-step_challenge}.  The analysis of this
classical protocol is virtually identical to the quantum one: this
generalizes a result of Sahai and Vadhan~\cite{SahaiV03} on the
statistical difference problem to the case of circuits that take input
states.

To complete the proof that $\prob{QCD}_{a,b}$ is \class{QIP}-complete
the Close Images problem is reduced to it.  The next section contains
a description of this reduction.

\section{Reduction from Close Images}\label{dist-scn-reduction}

This section presents a reduction from the close images problem to the
circuit distinguishability problem that will be used to show that
$\prob{QCD}_{a,b}$ is \class{QIP}-hard for any constants $a$ and $b$
such that $0 < b < a \leq 2$.  This is done using a standard
polynomial-time Karp reduction: a polynomial time procedure that
transforms instances of one problem to another that outputs a yes
instance of \prob{QCD} if and only if the input instance of \prob{CI}
was also a yes instance.  The analysis of the reduction presented in
this section appears in Section~\ref{dist-scn-correctness}.

The reduction takes as input an instance of the \prob{CI} problem,
which is given by a pair $(Q_1, Q_2)$ of mixed-state quantum
circuits implementing channels in $\transform{H,K}$.  The reduction
produces as output a pair of circuits $(C_1, C_2)$ that form an
instance of the \prob{QCD} problem.

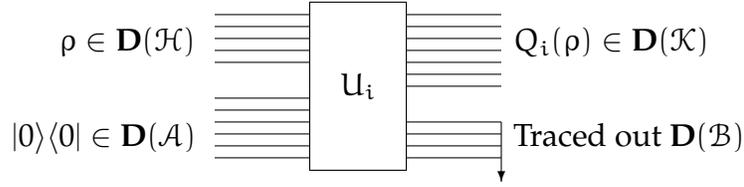
\begin{figure}
  \begin{center}
    \setlength{\unitlength}{3947sp}%
\begingroup\makeatletter\ifx\SetFigFont\undefined%
\gdef\SetFigFont#1#2#3#4#5{%
  \reset@font\fontsize{#1}{#2pt}%
  \fontfamily{#3}\fontseries{#4}\fontshape{#5}%
  \selectfont}%
\fi\endgroup%
\begin{picture}(5424,1374)(-1211,-1273)
\put(-674,-961){\makebox(0,0)[lb]{\smash{{\SetFigFont{12}{14.4}{\familydefault}{\mddefault}{\updefault}{\color[rgb]{0,0,0}$\ket 0 \bra 0 \in \density{A}$}%
}}}}
\thinlines
{\color[rgb]{0,0,0}\put(601,-211){\line( 1, 0){600}}
}%
{\color[rgb]{0,0,0}\put(601,-286){\line( 1, 0){600}}
}%
{\color[rgb]{0,0,0}\put(601,-361){\line( 1, 0){600}}
}%
{\color[rgb]{0,0,0}\put(601,-436){\line( 1, 0){600}}
}%
{\color[rgb]{0,0,0}\put(1801,-211){\line( 1, 0){600}}
}%
{\color[rgb]{0,0,0}\put(1801,-286){\line( 1, 0){600}}
}%
{\color[rgb]{0,0,0}\put(1801,-361){\line( 1, 0){600}}
}%
{\color[rgb]{0,0,0}\put(1801,-436){\line( 1, 0){600}}
}%
{\color[rgb]{0,0,0}\put(1801,-511){\line( 1, 0){600}}
}%
{\color[rgb]{0,0,0}\put(601,-811){\line( 1, 0){600}}
}%
{\color[rgb]{0,0,0}\put(601,-886){\line( 1, 0){600}}
}%
{\color[rgb]{0,0,0}\put(601,-961){\line( 1, 0){600}}
}%
{\color[rgb]{0,0,0}\put(601,-1036){\line( 1, 0){600}}
}%
{\color[rgb]{0,0,0}\put(601,-136){\line( 1, 0){600}}
}%
{\color[rgb]{0,0,0}\put(1801,-136){\line( 1, 0){600}}
}%
{\color[rgb]{0,0,0}\put(1801,-1036){\line( 1, 0){600}}
}%
{\color[rgb]{0,0,0}\put(1801,-961){\line( 1, 0){600}}
}%
{\color[rgb]{0,0,0}\put(1801,-886){\line( 1, 0){600}}
}%
{\color[rgb]{0,0,0}\put(1801,-811){\line( 1, 0){600}}
}%
{\color[rgb]{0,0,0}\put(601,-736){\line( 1, 0){600}}
}%
{\color[rgb]{0,0,0}\put(601,-661){\line( 1, 0){600}}
}%
{\color[rgb]{0,0,0}\put(1801,-586){\line( 1, 0){600}}
}%
{\color[rgb]{0,0,0}\put(2401,-811){\vector( 0,-1){375}}
}%
\put(2476,-361){\makebox(0,0)[lb]{\smash{{\SetFigFont{12}{14.4}{\familydefault}{\mddefault}{\updefault}{\color[rgb]{0,0,0}$Q_i(\rho) \in \density{K}$}%
}}}}
\put(2476,-961){\makebox(0,0)[lb]{\smash{{\SetFigFont{12}{14.4}{\familydefault}{\mddefault}{\updefault}{\color[rgb]{0,0,0}Traced out $\density{B}$}%
}}}}
\put(-374,-361){\makebox(0,0)[lb]{\smash{{\SetFigFont{12}{14.4}{\familydefault}{\mddefault}{\updefault}{\color[rgb]{0,0,0}$\rho \in \density{H}$}%
}}}}
{\color[rgb]{0,0,0}\put(1201,-1111){\framebox(600,1050){$U_i$}}
}%
\end{picture}%
  \end{center}
  \caption[Input circuits for the reduction]{The circuit $Q_i$ in
    Stinespring form, with the Hilbert spaces labelled.}
  \label{dist-fig-spaces}
\end{figure}

As in the case of the reduction in Chapter~\ref{chap-close-images}, we
may assume that the input circuits are given in Stinespring form.  In
this form the circuit consists of three parts.  First is the
introduction of any ancillary qubits in the $\ket 0$ state, second is
a unitary circuit applied to the input and ancillary qubits, and
finally, the third part is the tracing out of any qubits that are not
a part of the output space.  A general mixed-state quantum circuit can
be put into this form in polynomial time, and this assumption can be
made without
loss of generality that the input circuits are of this form.  This is
discussed in more detail in Section~\ref{compl-scn-circuit-model}.
As the reduction will modify the circuits $Q_1$ and $Q_2$, it is
helpful to identify the names of the various Hilbert spaces associated
with them.  As mentioned above, the circuit $Q_i$ implements an operation in
$\transform{H,K}$.  The spaces $\mathcal{H}$ and $\mathcal{K}$ will be
referred to as the ``input'' and ``output'' spaces of the circuit,
respectively.  Given in Stinespring form the circuits $Q_i$ makes use
of ancillary qubits.  Let the space $\mathcal{A}$ represent the space
these ancillary qubits are added in, and let $\mathcal{B}$ represent
the space that is traced out after the unitary is applied.  The space
$\mathcal{A}$ will be called the ``ancillary'' space of $Q_i$
and $\mathcal{B}$ will be called the ``environment'' space.
Furthermore, let $U_i$ be the unitary operation in $\unitary{H \tensor
  A, K \tensor B}$ that is applied as part of the circuit $Q_i$.  
As we may assume without loss of generality that each circuit uses the
same number of ancillary qubits by padding one of the circuit with
ancillary qubits that are left unused and later traced out, we take the four
Hilbert spaces $\mathcal{H,K,A,B}$ to be the same 
for each of the input circuits.
Notice also that since $U_i$ is unitary the spaces $\mathcal{H \tensor A}$
and $\mathcal{K \tensor B}$ have the same dimension, and so are isomorphic.
The various Hilbert spaces associated with the circuit $Q_i$ are
summarized in Figure~\ref{dist-fig-spaces}.

\begin{figure}
  \begin{center}
    \setlength{\unitlength}{3947sp}%
\begingroup\makeatletter\ifx\SetFigFont\undefined%
\gdef\SetFigFont#1#2#3#4#5{%
  \reset@font\fontsize{#1}{#2pt}%
  \fontfamily{#3}\fontseries{#4}\fontshape{#5}%
  \selectfont}%
\fi\endgroup%
\begin{picture}(4152,1674)(61,-1198)
\put( 76,-886){\makebox(0,0)[lb]{\smash{{\SetFigFont{12}{14.4}{\rmdefault}{\mddefault}{\updefault}{\color[rgb]{0,0,0}$\ket 0$}%
}}}}
\thinlines
{\color[rgb]{0,0,0}\put(1801,-286){\line( 1, 0){600}}
}%
{\color[rgb]{0,0,0}\put(1801,-361){\line( 1, 0){600}}
}%
{\color[rgb]{0,0,0}\put(1801,-436){\line( 1, 0){600}}
}%
{\color[rgb]{0,0,0}\put(1801,-811){\line( 1, 0){600}}
}%
{\color[rgb]{0,0,0}\put(1801,-886){\line( 1, 0){600}}
}%
{\color[rgb]{0,0,0}\put(1801,-961){\line( 1, 0){600}}
}%
{\color[rgb]{0,0,0}\put(1801,-1036){\line( 1, 0){600}}
}%
{\color[rgb]{0,0,0}\put(1801,-136){\line( 1, 0){600}}
}%
{\color[rgb]{0,0,0}\put(1801,-736){\line( 1, 0){600}}
}%
{\color[rgb]{0,0,0}\put(1801,-661){\line( 1, 0){600}}
}%
{\color[rgb]{0,0,0}\put(2401,-211){\line( 1, 0){600}}
}%
{\color[rgb]{0,0,0}\put(2401,-286){\line( 1, 0){600}}
}%
{\color[rgb]{0,0,0}\put(2401,-361){\line( 1, 0){600}}
}%
{\color[rgb]{0,0,0}\put(2401,-436){\line( 1, 0){600}}
}%
{\color[rgb]{0,0,0}\put(2401,-811){\line( 1, 0){600}}
}%
{\color[rgb]{0,0,0}\put(2401,-886){\line( 1, 0){600}}
}%
{\color[rgb]{0,0,0}\put(2401,-961){\line( 1, 0){600}}
}%
{\color[rgb]{0,0,0}\put(2401,-1036){\line( 1, 0){600}}
}%
{\color[rgb]{0,0,0}\put(2401,-136){\line( 1, 0){600}}
}%
{\color[rgb]{0,0,0}\put(2401,-736){\line( 1, 0){600}}
}%
{\color[rgb]{0,0,0}\put(2401,-661){\line( 1, 0){600}}
}%
{\color[rgb]{0,0,0}\put(1501,239){\circle*{76}}
}%
{\color[rgb]{0,0,0}\put(3301,239){\circle*{76}}
}%
{\color[rgb]{0,0,0}\put(376,-811){\line( 1, 0){825}}
}%
{\color[rgb]{0,0,0}\put(376,-886){\line( 1, 0){825}}
}%
{\color[rgb]{0,0,0}\put(376,-961){\line( 1, 0){825}}
}%
{\color[rgb]{0,0,0}\put(376,-1036){\line( 1, 0){825}}
}%
{\color[rgb]{0,0,0}\put(376,-736){\line( 1, 0){825}}
}%
{\color[rgb]{0,0,0}\put(376,-661){\line( 1, 0){825}}
}%
{\color[rgb]{0,0,0}\put( 76,-211){\line( 1, 0){1125}}
}%
{\color[rgb]{0,0,0}\put( 76,-286){\line( 1, 0){1125}}
}%
{\color[rgb]{0,0,0}\put( 76,-361){\line( 1, 0){1125}}
}%
{\color[rgb]{0,0,0}\put( 76,-436){\line( 1, 0){1125}}
}%
{\color[rgb]{0,0,0}\put( 76,-136){\line( 1, 0){1125}}
}%
{\color[rgb]{0,0,0}\put(1201,-1111){\framebox(600,1050){$U_1$}}
}%
{\color[rgb]{0,0,0}\put(3001,-1111){\framebox(600,1050){$U_2$}}
}%
{\color[rgb]{0,0,0}\put(2176, 14){\framebox(450,450){$X$}}
}%
{\color[rgb]{0,0,0}\put(826,239){\line( 1, 0){1350}}
}%
{\color[rgb]{0,0,0}\put(2626,239){\line( 1, 0){1575}}
}%
{\color[rgb]{0,0,0}\put(1501,239){\line( 0,-1){300}}
}%
{\color[rgb]{0,0,0}\put(3301,239){\line( 0,-1){300}}
}%
{\color[rgb]{0,0,0}\put(3601,-211){\line( 1, 0){600}}
}%
{\color[rgb]{0,0,0}\put(3601,-286){\line( 1, 0){600}}
}%
{\color[rgb]{0,0,0}\put(3601,-361){\line( 1, 0){600}}
}%
{\color[rgb]{0,0,0}\put(3601,-436){\line( 1, 0){600}}
}%
{\color[rgb]{0,0,0}\put(3601,-511){\line( 1, 0){600}}
}%
{\color[rgb]{0,0,0}\put(3601,-136){\line( 1, 0){600}}
}%
{\color[rgb]{0,0,0}\put(3601,-1036){\line( 1, 0){300}}
}%
{\color[rgb]{0,0,0}\put(3601,-961){\line( 1, 0){300}}
}%
{\color[rgb]{0,0,0}\put(3601,-886){\line( 1, 0){300}}
}%
{\color[rgb]{0,0,0}\put(3601,-811){\line( 1, 0){300}}
}%
{\color[rgb]{0,0,0}\put(3601,-586){\line( 1, 0){600}}
}%
{\color[rgb]{0,0,0}\put(3901,-811){\vector( 0,-1){375}}
}%
{\color[rgb]{0,0,0}\put(376, 14){\framebox(450,450){$X$}}
}%
{\color[rgb]{0,0,0}\put( 76,239){\line( 1, 0){300}}
}%
{\color[rgb]{0,0,0}\put(1801,-211){\line( 1, 0){600}}
}%
\end{picture}%
  \end{center}
  \caption[Circuit to apply either input circuit based on a control
  qubit]{A Circuit to apply $Q_1$ or $Q_2$ based on the value of a
    control qubit.}
  \label{dist-fig-controlled}
\end{figure}
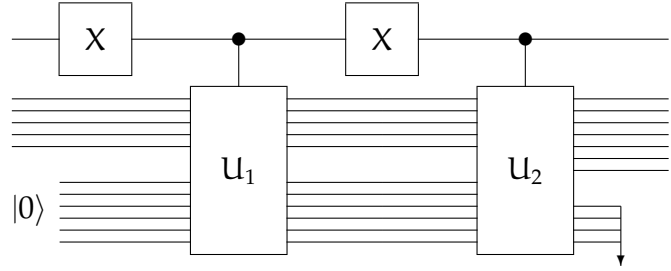

An important piece of the reduction will be a circuit that, based on
the value of a control qubit, applies either $Q_1$ or $Q_2$ to the
input state.  Such a circuit is easily constructed in polynomial time
by simply replacing
each gate of $Q_1$ and $Q_2$ with gates that are controlled by the
value of the control qubit.  These controlled gates need not be in the
family of gates in the circuit model, but since the number of gates in
the model is finite, decompositions of these controlled gates in terms
of gates in the model can be constructed efficiently.
Let, for concreteness, the circuit that implements one of the two
input circuits implement
$Q_1$ when the control qubit is $\ket 0$ and $Q_2$ when the control
qubit is $\ket 1$.  One construction for such a circuit is given in
Figure~\ref{dist-fig-controlled}.  Let $\mathcal{Q}$ be the Hilbert
space containing the control qubit, so that the constructed
transformation is a channel in $\transform{Q \tprod H, Q \tprod K}$.
There is some ambiguity in the constructed transformation: the
controlled $U_1$ operation takes an input in $\mathcal{Q \tprod H
  \tprod K}$ and produces output in $\mathcal{Q \tprod K \tprod B}$
and this is followed by a controlled $U_2$ operation.  This operation
also expects an input in the space $\mathcal{Q \tprod H \tprod K}$,
not the space $\mathcal{Q \tprod K \tprod B}$.  Fortunately both of
these spaces have the same dimension, and so by implicitly making use
of the isomorphism between them, this potential difficulty is avoided.

\begin{figure}
  \begin{center}
    \subfigure[The circuit $C_1$]{\setlength{\unitlength}{3947sp}%
\begingroup\makeatletter\ifx\SetFigFont\undefined%
\gdef\SetFigFont#1#2#3#4#5{%
  \reset@font\fontsize{#1}{#2pt}%
  \fontfamily{#3}\fontseries{#4}\fontshape{#5}%
  \selectfont}%
\fi\endgroup%
\begin{picture}(4677,1599)(61,-1123)
\put( 76,-886){\makebox(0,0)[lb]{\smash{{\SetFigFont{12}{14.4}{\rmdefault}{\mddefault}{\updefault}{\color[rgb]{0,0,0}$\ket 0$}%
}}}}
\thinlines
{\color[rgb]{0,0,0}\put(1801,-286){\line( 1, 0){600}}
}%
{\color[rgb]{0,0,0}\put(1801,-361){\line( 1, 0){600}}
}%
{\color[rgb]{0,0,0}\put(1801,-436){\line( 1, 0){600}}
}%
{\color[rgb]{0,0,0}\put(1801,-811){\line( 1, 0){600}}
}%
{\color[rgb]{0,0,0}\put(1801,-886){\line( 1, 0){600}}
}%
{\color[rgb]{0,0,0}\put(1801,-961){\line( 1, 0){600}}
}%
{\color[rgb]{0,0,0}\put(1801,-1036){\line( 1, 0){600}}
}%
{\color[rgb]{0,0,0}\put(1801,-136){\line( 1, 0){600}}
}%
{\color[rgb]{0,0,0}\put(1801,-736){\line( 1, 0){600}}
}%
{\color[rgb]{0,0,0}\put(1801,-661){\line( 1, 0){600}}
}%
{\color[rgb]{0,0,0}\put(2401,-211){\line( 1, 0){600}}
}%
{\color[rgb]{0,0,0}\put(2401,-286){\line( 1, 0){600}}
}%
{\color[rgb]{0,0,0}\put(2401,-361){\line( 1, 0){600}}
}%
{\color[rgb]{0,0,0}\put(2401,-436){\line( 1, 0){600}}
}%
{\color[rgb]{0,0,0}\put(2401,-811){\line( 1, 0){600}}
}%
{\color[rgb]{0,0,0}\put(2401,-886){\line( 1, 0){600}}
}%
{\color[rgb]{0,0,0}\put(2401,-961){\line( 1, 0){600}}
}%
{\color[rgb]{0,0,0}\put(2401,-1036){\line( 1, 0){600}}
}%
{\color[rgb]{0,0,0}\put(2401,-136){\line( 1, 0){600}}
}%
{\color[rgb]{0,0,0}\put(2401,-736){\line( 1, 0){600}}
}%
{\color[rgb]{0,0,0}\put(2401,-661){\line( 1, 0){600}}
}%
{\color[rgb]{0,0,0}\put(1501,239){\circle*{76}}
}%
{\color[rgb]{0,0,0}\put(3301,239){\circle*{76}}
}%
{\color[rgb]{0,0,0}\put(376,-811){\line( 1, 0){825}}
}%
{\color[rgb]{0,0,0}\put(376,-886){\line( 1, 0){825}}
}%
{\color[rgb]{0,0,0}\put(376,-961){\line( 1, 0){825}}
}%
{\color[rgb]{0,0,0}\put(376,-1036){\line( 1, 0){825}}
}%
{\color[rgb]{0,0,0}\put(376,-736){\line( 1, 0){825}}
}%
{\color[rgb]{0,0,0}\put(376,-661){\line( 1, 0){825}}
}%
{\color[rgb]{0,0,0}\put(3601,-1036){\line( 1, 0){1125}}
}%
{\color[rgb]{0,0,0}\put(3601,-961){\line( 1, 0){1125}}
}%
{\color[rgb]{0,0,0}\put(3601,-886){\line( 1, 0){1125}}
}%
{\color[rgb]{0,0,0}\put(3601,-811){\line( 1, 0){1125}}
}%
{\color[rgb]{0,0,0}\put(3601,-211){\line( 1, 0){600}}
}%
{\color[rgb]{0,0,0}\put(3601,-286){\line( 1, 0){600}}
}%
{\color[rgb]{0,0,0}\put(3601,-361){\line( 1, 0){600}}
}%
{\color[rgb]{0,0,0}\put(3601,-436){\line( 1, 0){600}}
}%
{\color[rgb]{0,0,0}\put(3601,-511){\line( 1, 0){600}}
}%
{\color[rgb]{0,0,0}\put(3601,-136){\line( 1, 0){600}}
}%
{\color[rgb]{0,0,0}\put(3601,-586){\line( 1, 0){600}}
}%
{\color[rgb]{0,0,0}\put( 76,-211){\line( 1, 0){1125}}
}%
{\color[rgb]{0,0,0}\put( 76,-286){\line( 1, 0){1125}}
}%
{\color[rgb]{0,0,0}\put( 76,-361){\line( 1, 0){1125}}
}%
{\color[rgb]{0,0,0}\put( 76,-436){\line( 1, 0){1125}}
}%
{\color[rgb]{0,0,0}\put( 76,-136){\line( 1, 0){1125}}
}%
{\color[rgb]{0,0,0}\put(1201,-1111){\framebox(600,1050){$U_1$}}
}%
{\color[rgb]{0,0,0}\put(3001,-1111){\framebox(600,1050){$U_2$}}
}%
{\color[rgb]{0,0,0}\put(2176, 14){\framebox(450,450){$X$}}
}%
{\color[rgb]{0,0,0}\put(826,239){\line( 1, 0){1350}}
}%
{\color[rgb]{0,0,0}\put(2626,239){\line( 1, 0){2100}}
}%
{\color[rgb]{0,0,0}\put(1501,239){\line( 0,-1){300}}
}%
{\color[rgb]{0,0,0}\put(3301,239){\line( 0,-1){300}}
}%
{\color[rgb]{0,0,0}\put(4201,-136){\vector( 0,-1){525}}
}%
{\color[rgb]{0,0,0}\put(376, 14){\framebox(450,450){$X$}}
}%
{\color[rgb]{0,0,0}\put( 76,239){\line( 1, 0){300}}
}%
{\color[rgb]{0,0,0}\put(1801,-211){\line( 1, 0){600}}
}%
\end{picture}%
%
    \label{dist-fig-output-one}}\\
    \subfigure[The circuit $C_2$]{\setlength{\unitlength}{3947sp}%
\begingroup\makeatletter\ifx\SetFigFont\undefined%
\gdef\SetFigFont#1#2#3#4#5{%
  \reset@font\fontsize{#1}{#2pt}%
  \fontfamily{#3}\fontseries{#4}\fontshape{#5}%
  \selectfont}%
\fi\endgroup%
\begin{picture}(4677,1599)(61,-1123)
\put( 76,-886){\makebox(0,0)[lb]{\smash{{\SetFigFont{12}{14.4}{\rmdefault}{\mddefault}{\updefault}{\color[rgb]{0,0,0}$\ket 0$}%
}}}}
\thinlines
{\color[rgb]{0,0,0}\put(1801,-286){\line( 1, 0){600}}
}%
{\color[rgb]{0,0,0}\put(1801,-361){\line( 1, 0){600}}
}%
{\color[rgb]{0,0,0}\put(1801,-436){\line( 1, 0){600}}
}%
{\color[rgb]{0,0,0}\put(1801,-811){\line( 1, 0){600}}
}%
{\color[rgb]{0,0,0}\put(1801,-886){\line( 1, 0){600}}
}%
{\color[rgb]{0,0,0}\put(1801,-961){\line( 1, 0){600}}
}%
{\color[rgb]{0,0,0}\put(1801,-1036){\line( 1, 0){600}}
}%
{\color[rgb]{0,0,0}\put(1801,-136){\line( 1, 0){600}}
}%
{\color[rgb]{0,0,0}\put(1801,-736){\line( 1, 0){600}}
}%
{\color[rgb]{0,0,0}\put(1801,-661){\line( 1, 0){600}}
}%
{\color[rgb]{0,0,0}\put(2401,-211){\line( 1, 0){600}}
}%
{\color[rgb]{0,0,0}\put(2401,-286){\line( 1, 0){600}}
}%
{\color[rgb]{0,0,0}\put(2401,-361){\line( 1, 0){600}}
}%
{\color[rgb]{0,0,0}\put(2401,-436){\line( 1, 0){600}}
}%
{\color[rgb]{0,0,0}\put(2401,-811){\line( 1, 0){600}}
}%
{\color[rgb]{0,0,0}\put(2401,-886){\line( 1, 0){600}}
}%
{\color[rgb]{0,0,0}\put(2401,-961){\line( 1, 0){600}}
}%
{\color[rgb]{0,0,0}\put(2401,-1036){\line( 1, 0){600}}
}%
{\color[rgb]{0,0,0}\put(2401,-136){\line( 1, 0){600}}
}%
{\color[rgb]{0,0,0}\put(2401,-736){\line( 1, 0){600}}
}%
{\color[rgb]{0,0,0}\put(2401,-661){\line( 1, 0){600}}
}%
{\color[rgb]{0,0,0}\put(1501,239){\circle*{76}}
}%
{\color[rgb]{0,0,0}\put(3301,239){\circle*{76}}
}%
{\color[rgb]{0,0,0}\put(376,-811){\line( 1, 0){825}}
}%
{\color[rgb]{0,0,0}\put(376,-886){\line( 1, 0){825}}
}%
{\color[rgb]{0,0,0}\put(376,-961){\line( 1, 0){825}}
}%
{\color[rgb]{0,0,0}\put(376,-1036){\line( 1, 0){825}}
}%
{\color[rgb]{0,0,0}\put(376,-736){\line( 1, 0){825}}
}%
{\color[rgb]{0,0,0}\put(376,-661){\line( 1, 0){825}}
}%
{\color[rgb]{0,0,0}\put(3601,-1036){\line( 1, 0){1125}}
}%
{\color[rgb]{0,0,0}\put(3601,-961){\line( 1, 0){1125}}
}%
{\color[rgb]{0,0,0}\put(3601,-886){\line( 1, 0){1125}}
}%
{\color[rgb]{0,0,0}\put(3601,-811){\line( 1, 0){1125}}
}%
{\color[rgb]{0,0,0}\put(3601,-211){\line( 1, 0){600}}
}%
{\color[rgb]{0,0,0}\put(3601,-286){\line( 1, 0){600}}
}%
{\color[rgb]{0,0,0}\put(3601,-361){\line( 1, 0){600}}
}%
{\color[rgb]{0,0,0}\put(3601,-436){\line( 1, 0){600}}
}%
{\color[rgb]{0,0,0}\put(3601,-511){\line( 1, 0){600}}
}%
{\color[rgb]{0,0,0}\put(3601,-136){\line( 1, 0){600}}
}%
{\color[rgb]{0,0,0}\put(3601,-586){\line( 1, 0){600}}
}%
{\color[rgb]{0,0,0}\put( 76,-211){\line( 1, 0){1125}}
}%
{\color[rgb]{0,0,0}\put( 76,-286){\line( 1, 0){1125}}
}%
{\color[rgb]{0,0,0}\put( 76,-361){\line( 1, 0){1125}}
}%
{\color[rgb]{0,0,0}\put( 76,-436){\line( 1, 0){1125}}
}%
{\color[rgb]{0,0,0}\put( 76,-136){\line( 1, 0){1125}}
}%
{\color[rgb]{0,0,0}\put(1201,-1111){\framebox(600,1050){$U_1$}}
}%
{\color[rgb]{0,0,0}\put(3001,-1111){\framebox(600,1050){$U_2$}}
}%
{\color[rgb]{0,0,0}\put(2176, 14){\framebox(450,450){$X$}}
}%
{\color[rgb]{0,0,0}\put(826,239){\line( 1, 0){1350}}
}%
{\color[rgb]{0,0,0}\put(2626,239){\line( 1, 0){1350}}
}%
{\color[rgb]{0,0,0}\put(1501,239){\line( 0,-1){300}}
}%
{\color[rgb]{0,0,0}\put(3301,239){\line( 0,-1){300}}
}%
{\color[rgb]{0,0,0}\put(3976, 14){\framebox(450,450){$Z$}}
}%
{\color[rgb]{0,0,0}\put(4426,239){\line( 1, 0){300}}
}%
{\color[rgb]{0,0,0}\put(4201,-136){\vector( 0,-1){525}}
}%
{\color[rgb]{0,0,0}\put(376, 14){\framebox(450,450){$X$}}
}%
{\color[rgb]{0,0,0}\put( 76,239){\line( 1, 0){300}}
}%
{\color[rgb]{0,0,0}\put(1801,-211){\line( 1, 0){600}}
}%
\end{picture}%
%
    \label{dist-fig-output-two}}
    \caption[Circuits output by the reduction]{Circuits output by the reduction.} \label{dist-fig-outputs} 
  \end{center}
\end{figure}
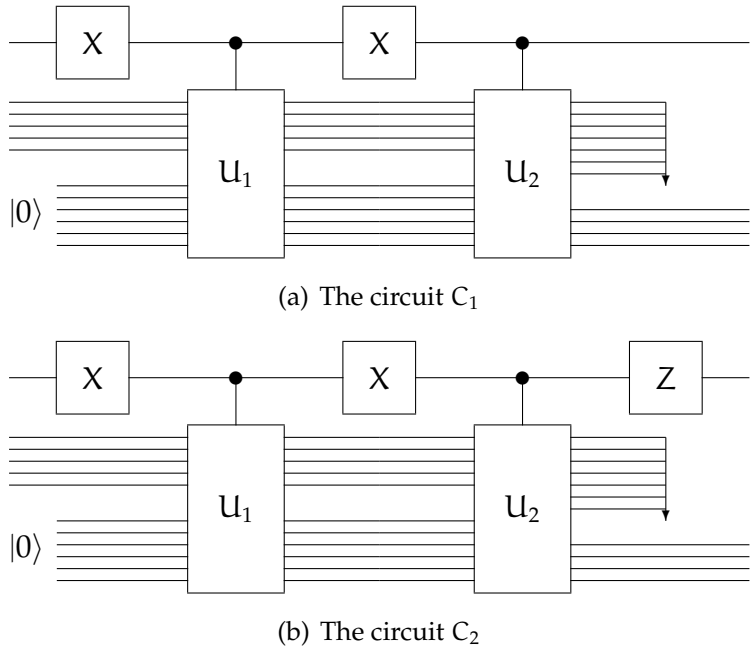

The circuit shown in Figure~\ref{dist-fig-controlled} is very close to
the circuits that will be the output of the reduction.  To obtain
these circuits one critical modification is made: instead of tracing
out the environment space $\mathcal{B}$, the ``output'' space
$\mathcal{K}$ is traced out instead.  This reversal of the purposes of
the output and environment spaces is essential to the reduction.
Taking a Stinespring
representation of a channel and tracing out the output instead of the
environment leads to what has been called a \emph{conjugate} or
\emph{complementary} channel~\cite{DevetakS05, Holevo07, KingM+07}.
Viewed in this light, this circuit simply applies a fixed conjugate of
either $Q_1$ or $Q_2$, depending on the value of a control qubit.
This is how the circuit $C_1$, one of the two circuits output by the
reduction is constructed.  This circuit is demonstrated in
Figure~\ref{dist-fig-output-one}.

The circuit $C_2$ is constructed in the same way as the circuit $C_1$,
with one difference.  This is a Pauli $Z$ operation is applied to the
control qubit after the controlled operations.  The circuit $C_2$ is
shown in Figure~\ref{dist-fig-output-two}.  This $Z$ gate will make a
substantial difference in the output of the two circuits when the
control qubit in the space $\mathcal{Q}$ has not been decohered by the
other operations of the circuit $C_2$.
The output of the
reduction is an instance of \prob{QCD} given by the pair of circuits
$(C_1, C_2)$.

The key to this reduction is that when an input is given to
either $C_1$ or $C_2$ with the control qubit in a superposition of
$\ket 0$ and $\ket 1$, then both of the circuits $Q_1$ and $Q_2$ are
run.  By tracing out the ``output'' space $\mathcal{K}$ the idea is
that if the outputs of $C_1$ and $C_2$ are sufficiently far apart, then
tracing out the space $\mathcal{K}$ is akin to measuring the control
qubit but forgetting the result.  Intuitively, if there is enough
information in $\mathcal{K}$ to identify which of the two circuits
$Q_1$ or $Q_2$ has been performed, then the control qubit will be
subject to decoherence.  In this case the Pauli $Z$ gate in $C_2$ has
no effect: the control qubit has decohered to a mixture of the form $p
\ket 0 \bra 0 + (1-p) \ket 1 \bra 1$, and applying $Z$ to such a state
has no effect, since
\[ Z \left( p \ket 0 \bra 0 + (1-p) \ket 1 \bra 1 \right) Z
   =  p Z \ket 0 \bra 0 Z + (1-p) Z \ket 1 \bra 1 Z
   = p \ket 0 \bra 0 + (1-p) \ket 1 \bra 1 \]
On the other hand, if the outputs of $Q_1$ and
$Q_2$ are sufficiently close, then there should be no information
about the control qubit in the space $\mathcal{K}$, so tracing it out
will not have any effect.  In this case the control qubit remains in
a pure state such as $(\ket 0 + \ket 1)/\sqrt 2$, so that applying the
Pauli $Z$ operation in the circuit $C_2$ results in the state
\[ Z \ket + = Z \frac{ \ket 0 + \ket 1 }{\sqrt{2}} 
   = \frac{\ket 0 - \ket  1}{\sqrt{2}} = \ket -, \]
which is orthogonal to the control qubit output by $C_1$.  In this way
the reduction effectively inverts the closeness of the circuits: if
the circuits $Q_1$ and $Q_2$ can be made to output states that are
close together, then the output of the circuits $C_1$ and $C_2$
can be made far apart.  In the other direction, if the original
circuits always output states that are distinguishable, then the
constructed circuits will always be close together, as the control
qubit is left in an incoherent mixture after tracing out the space 
$\mathcal{K}$, so that the $Z$ operation in $C_2$ has little effect.

This argument is not complete: the circuits $C_1$ and $C_2$ also
output the environment space $\mathcal{B}$ of the original
circuits, which has been ignored.  The notions of closeness used in
the problems \prob{CI} and \prob{QCD} are also not the same.
Significant care must be taken to formalize this intuitive picture.
This is the content of the next section, which contains a formal proof
that this reduction implies the \class{QIP}-hardness of the \prob{QCD}
problem.

\section{Correctness of the reduction}\label{dist-scn-correctness}

This section contains the formal proof that the reduction presented in
Section~\ref{dist-scn-reduction} implies that the \prob{QCD} problem
is \class{QIP}-hard.  Proving that this problem is \class{QIP}-hard
implies that it is also \class{QIP}-complete, as it is argued in
Section~\ref{dist-scn-protocol} that these problems belong to
\class{QIP}.

This section is quite technical.  The reader uninterested in the
details of the proof of the main result is invited to skip the proofs
of the lemmas found here: the proofs are not overly difficult but
much of the intuition has already been presented in the previous
section, so it is unlikely that a detailed study of these proofs 
will provide a clearer picture of the results.

As in the previous section, let $(Q_1, Q_2)$ be the instance of
$\prob{CI}$ provided as input to the reduction, where these
circuits implement transformations in $\transform{H,K}$ given by
\begin{align*}
  Q_1(\rho) &= \ptr{B} U_1 ( \rho \tprod \ket 0 \bra 0) U_1^*, \\
  Q_2(\rho) &= \ptr{B} U_2 ( \rho \tprod \ket 0 \bra 0) U_2^*,
\end{align*}
where $U_1, U_2 \in \unitary{H \tensor A, K \tensor B}$.  This is
summarized in Figure~\ref{dist-fig-spaces}.

From these circuits the reduction described in the previous section
produces as output $(C_1, C_2)$, a pair of circuits that form an instance of
$\prob{QCD}$.  Let, for notational convenience, the operator
$V$ be the operator that applies the operator $U_{i+1}$ when a control
qubit is in the $\ket i$ state, i.e.\ let the operation $V$ by defined
by
\begin{align*}
  V(\ket 0 \tprod \ket \psi) &= \ket 0 \tprod U_1 \ket \psi \\
  V(\ket 1 \tprod \ket \psi) &= \ket 1 \tprod U_2 \ket \psi,
\end{align*}
for all states $\ket\psi$.
One implementation for the operation $V$ is given by
Figure~\ref{dist-fig-controlled}, with the exception that the
operation $V$ does not trace out the qubits in the space
$\mathcal{B}$.

Using this notation, the circuits in
Figure~\ref{dist-fig-outputs} implement the operations in the space
$\transform{Q \tprod H, Q \tprod B}$ given by
\begin{equation}\label{dist-eqns-constructed-circuits}
  \begin{split}
    C_1(\rho) &= \ptr{K} V
                         (\rho \tprod \ket 0 \bra 0)
                         V^* \\
    C_2(\rho) &= \ptr{K} Z_{\mathcal{Q}} V
                         (\rho \tprod \ket 0 \bra 0)
                         V^* Z_{\mathcal{Q}}.
  \end{split}
\end{equation}
The operation $Z_{\mathcal{Q}}$ in this equation is simply shorthand
for the application of the Pauli $Z$ gate to the qubit represented by
$\mathcal{Q}$, i.e.\ $Z_{\mathcal{Q}} = Z \tprod \identity{K} \tprod
\identity{B}$.  This characterization of the circuits produced by the
reduction will be used to show that $C_1$ and $C_2$ are
distinguishable if and only if $Q_1$ and $Q_2$ have close images.

The main result of this section is that the maximum output fidelity of
$Q_1$ and $Q_2$ is equal to the diamond norm of the difference of
$C_1$ and $C_2$.  The proof of this is presented in two steps.  The
first, and simplest, of these steps is to show that the diamond norm
provides a lower bound on the maximum output fidelity.  This is argued
directly from the properties of the diamond norm and the constructed circuits.

\begin{lemma}\label{dist-lem-leq}
  Given circuits $Q_1$ and $Q_2$, and the circuits $C_1$ and $C_2$
  constructed from them given by~\eqref{dist-eqns-constructed-circuits}
  \[ \frac{1}{2} \dnorm{C_1 - C_2}
     \leq  \max_{\rho,\sigma \in \density{H}} \F(Q_1(\rho), Q_2(\sigma)). \]

  \begin{proof}
    By Theorem~\ref{meas-thm-dnorm-difference} the diamond norm of the
    difference of two channels is achieved on a pure state in some
    larger system.  Let this state be $\ket \psi \in \mathcal{Q \tprod
      H \tprod F}$, where $\mathcal{F}$ is the reference system
    implied by the theorem.  For this state, we have
    \begin{equation}\label{dist-eqn-leq-dnorm-tnorm}
      \dnorm{C_1 - C_2}
      = \tnorm{(C_1 \tprod \tidentity{F})(\ket \psi \bra \psi) - 
               (C_2 \tprod \tidentity{F})(\ket \psi \bra \psi)}
    \end{equation}   
    As $\ket \psi$ is a unit vector, it may be written in terms of
    components on the two subspaces where the qubit in the space
    $\mathcal{Q}$ is either $\ket 0$ or $\ket 1$.  More formally,
    there is some $p \in [0,1]$ and states $\ket{\psi_0}, \ket{\psi_1} \in
    \mathcal{H \tprod F}$ such that
    \[ \ket \psi = \sqrt{p} \ket 0 \ket{\psi_0} 
                 + \sqrt{1-p} \ket 1 \ket{\psi_1}. \]
    Using this decomposition we can evaluate the circuits $C_1$ and
    $C_2$ on the state $\ket\psi$.  The input state can be decomposed as
   \begin{equation}\label{dist-eqn-state-decomp}
      \begin{split}
        p \ket 0 \bra 0 \tprod \ket{\psi_0}\bra{\psi_0}
        &+ (1-p) \ket 1 \bra 1 \tprod \ket{\psi_1}\bra{\psi_1}\\
        &+ \sqrt{p(1-p)} \ket 0 \bra 1 \tprod \ket{\psi_0}\bra{\psi_1}\\
        &+ \sqrt{p(1-p)} \ket 1 \bra 0 \tprod
        \ket{\psi_1}\bra{\psi_0}.
      \end{split}
    \end{equation}
    For the sake of brevity, further notation is introduced.  Let
    $\ket{\phi_i} = (U_{i+1} \tprod \identity{F})\ket{\psi_i}$.  
    This notation suffices as the unitary $U_1$ is only applied to the
    state $\ket{\psi_0}$, and likewise with $U_2$ with the state $\ket{\psi_1}$.
    Making
    use of this notation, we can consider the behaviour of the
    circuits $C_1$ and $C_2$ on the terms in
    Equation~\eqref{dist-eqn-state-decomp}.
    The output of $C_1$ is given by
    \begin{equation}\label{dist-eqn-leq-c1-output}
      (C_1 \tprod \tidentity{F})
      (\ket i \bra j \tprod \ket{\psi_i} \bra{\psi_j})
      = \ket i \bra j \tprod \ptr{K} \ket{\phi_i} \bra{\phi_j},
    \end{equation}
    for all $i,j \in \{0,1\}$.
    The output of $C_2$ differs only slightly, being given by
    \begin{equation}\label{dist-eqn-leq-c2-output}
      (C_2 \tprod \tidentity{F})
      (\ket i \bra j \tprod \ket{\psi_i} \bra{\psi_j})
      = (-1)^{i + j} \ket i \bra j \tprod \ptr{K} \ket{\phi_i} \bra{\phi_j},
    \end{equation}
    where the $(-1)^{i+j}$ factor is due to the Pauli $Z$ gate in the
    circuit $C_2$.
    Notice that when $i = j$, as in the first two terms of
    Equation~\eqref{dist-eqn-state-decomp}, the two circuits produce
    identical output.  The difference between the two circuits lies in
    the behaviour on the final two terms of this equation.  On these
    two terms the circuits agree, up to a multiplicative factor of
    $-1$, as can be seen from Equations~\eqref{dist-eqn-leq-c1-output}
    and ~\eqref{dist-eqn-leq-c2-output}.  Using this observation, the
    difference between the outputs of the two circuits is given by
    \begin{equation*}\label{eqn-dist-leq-difference}
      (C_1 \tprod \tidentity{F} - C_2 \tprod \tidentity{F})
      (\ket \psi \bra \psi) 
      = 2 \sqrt{p (1-p)}
        \left( \ket 0 \bra 1 \tprod \ptr{K} \ket{\phi_0} \bra{\phi_1}
        +      \ket 1 \bra 0 \tprod \ptr{K} \ket{\phi_1} \bra{\phi_0} \right).
    \end{equation*}
    Combining this with Equation~\eqref{dist-eqn-leq-dnorm-tnorm} yields
    \begin{align*}
      \dnorm{C_1 - C_2}
      &= \tnorm{(C_1 \tprod \tidentity{F} - 
                C_2 \tprod \tidentity{F})(\ket \psi \bra \psi)} \\
      &= 2 \sqrt{p (1-p)} 
        \tnorm{ \ket 0 \bra 1 \tprod \ptr{K} \ket{\phi_0} \bra{\phi_1}
               +\ket 1 \bra 0 \tprod \ptr{K} \ket{\phi_1} \bra{\phi_0}}.      
    \end{align*}
    From this equation, as well as Lemmas~\ref{meas-lem-block-tnorm}
    and~\ref{meas-lemma-fidelity-tracenorm}, we find that
    \begin{align*}
      \dnorm{C_1 - C_2}
      &= 4 \sqrt{p (1-p)} \tnorm{\ptr{K} \ket{\phi_0} \bra{\phi_1}}\\
      &= 4 \sqrt{p (1-p)} \F(\ptr{B \tprod F} \ket{\phi_0}\bra{\phi_0},
                                         \ptr{B \tprod F} \ket{\phi_1}\bra{\phi_1}) \\
      &= 4 \sqrt{p (1-p)} \F(Q_1(\ptr{F} \ket{\psi_0}\bra{\psi_0},
                                         Q_2(\ptr{F} \ket{\psi_1}\bra{\psi_1}) \\
     &\leq 2 \max_{\rho,\sigma} \F(Q_1(\rho), Q_2(\sigma)),
    \end{align*}
    which completes the proof of the lemma.
  \end{proof}
\end{lemma}

The following lemma formalizes the intuitive picture presented in
Section~\ref{dist-scn-reduction} that the constructed circuits $C_1$
and $C_2$ are distinguishable if the original circuits
$Q_1$ and $Q_2$ have output states with high fidelity.  This is the
second direction of the proof of the main result of this chapter.

\begin{lemma}\label{dist-lem-geq}
  Given circuits $Q_1$ and $Q_2$, and the circuits $C_1$ and $C_2$
  constructed from them given by~\eqref{dist-eqns-constructed-circuits},
  \[ \frac{1}{2} \dnorm{C_1 - C_2}
     \geq \max_{\rho,\sigma \in \density{H}} \F(Q_1(\rho),
     Q_2(\sigma)). \]

  \begin{proof}
    Let $\rho_1, \rho_2 \in \density{H}$ be two arbitrary states.  For
    these states, we will show that
    \[
    \dnorm{C_1 - C_2} \geq 2 \F(Q_1(\rho_1),Q_2(\rho_2)).
    \]
    Let the states $\ket{\psi_0}, \ket{\psi_1}\in\mathcal{H \tprod F}$ be
    purifications of $\rho_1$ and $\rho_2$, where $\mathcal{F}$ is any
    Hilbert space large enough to admit these purifications.  These
    states will play a similar role to the states of the same name
    used in the proof of Lemma~\ref{dist-lem-leq}.  Following the
    notation in this lemma further, let
    $\ket{\phi_i} = (U_{i+1} \tprod \identity{F})\ket{\psi_i}$ be the
    states produced by applying the ``appropriate'' unitary to these
    states.
    Using this notation, consider the input state $\ket\psi \in
    \mathcal{Q \tprod H}$ to $C_1$ and $C_2$
    given by
    \[\ket{\psi} = \frac{1}{\sqrt{2}}\ket{0}\ket{\psi_0} + 
                  \frac{1}{\sqrt{2}}\ket{1}\ket{\psi_1}. \]
    On this state the output of the two channels is exactly as
    discussed in Lemma~\ref{dist-lem-leq} with $p = 1/2$.  In particular, the channel
    $C_1$ produces the output
    \begin{equation}\label{dist-eqn-geq-c1out}
      (C_1 \tprod \tidentity{F})(\ket\psi \bra\psi)
      =
      \frac{1}{2} \sum_{i,j \in \{0,1\}}
        \ket i \bra j \tprod \ptr{K} \ket{\phi_i}\bra{\phi_j}
    \end{equation}
    while the circuit $C_2$ produces the output
    \begin{equation}\label{dist-eqn-geq-c2out}
      (C_2 \tprod \tidentity{F})(\ket\psi \bra\psi)
      =
      \frac{1}{2} \sum_{i,j \in \{0,1\}} (-1)^{i+j}
        \ket i \bra j \tprod \ptr{K} \ket{\phi_i}\bra{\phi_j}.
    \end{equation}
    These equations follow from identical reasoning to the derivation
    of equations~\eqref{dist-eqn-leq-c1-output}
    and~\eqref{dist-eqn-leq-c2-output} of the previous lemma.

    Notice that the pure states $\ket{\phi_0},\ket{\phi_1} \in
    \mathcal{K \tprod B \tprod F}$ are purifications of $Q_1(\rho_1)$
    and $Q_2(\rho_2)$, respectively.  This allows us to use
    Lemma~\ref{meas-lemma-fidelity-tracenorm} to transform the trace
    norm of $\ptr{K} \ket{\phi_0}\bra{\phi_1}$ into the fidelity of
    the input states $\rho_1$ and $\rho_2$ to the original circuits.
    This trace norm will be essential, as the difference between the
    outputs of the circuits $C_i$ consists only of terms of this form,
    which can be seen from Equations~\eqref{dist-eqn-geq-c1out}
    and~\eqref{dist-eqn-geq-c2out}.  This, along with
    Theorem~\ref{meas-thm-dnorm-stabilize} and
    Lemma~\ref{meas-lem-block-tnorm} show that
    \begin{align*}
      \dnorm{C_1 - C_2}
      &\geq \tnorm{(C_1 \tprod \tidentity{F} - C_2 \tprod \tidentity{F})
                   (\ket \psi \bra \psi)} \\
      &= \tnorm{
           \ket 0 \bra 1 \tprod \ptr{K} \ket{\phi_0} \bra{\phi_1}
          +\ket 1 \bra 0 \tprod \ptr{K} \ket{\phi_1} \bra{\phi_0}} \\
      &= 2 \tnorm{\ptr{K} \ket{\phi_0} \bra{\phi_1}} \\
      &= 2 \F(Q_1(\rho_1), Q_2(\rho_2)).
    \end{align*}
    This completes the proof of the lemma.
  \end{proof}
\end{lemma}

With these two lemmas, most of the work in proving the main result has
been completed.  We have so far shown that
the diamond norm difference of the constructed instance $(C_1,
C_2)$ of \prob{QCD} and the maximum output fidelity of the input
instance $(Q_1, Q_2)$ of \prob{CI} are equal.  This fact is stated as
the following theorem for easy reference.  This also proves that the
reduction correctly produces ``yes'' instances of \prob{QCD} if and
only if it is given ``yes'' instances of \prob{CI}.

\begin{theorem}\label{dist-thm-reduction}
  Given circuits $Q_1$ and $Q_2$, and the circuits $C_1$ and $C_2$
  constructed from them given by~\eqref{dist-eqns-constructed-circuits},
  \[ \frac{1}{2} \dnorm{C_1 - C_2}
     = \max_{\rho,\sigma \in \density{H}} \F(Q_1(\rho),
     Q_2(\sigma)). \]
  \begin{proof}
    Lemma~\ref{dist-lem-leq} provides the upper bound and
    Lemma~\ref{dist-lem-geq} provides the lower bound.  Taken together 
    they prove the desired equation.
  \end{proof}
\end{theorem}

This theorem immediately implies the main result of the chapter:
the \class{QIP} 
hardness of the distinguishability problem for mixed-state quantum
circuits.

\begin{corollary}
  For any $0 < b < a \leq 2$ the problem
  $\prob{QCD}_{a,b}$ is \class{QIP}-complete.

  \begin{proof}
    Theorem~\ref{dist-thm-reduction} and the construction in
    Section~\ref{dist-scn-reduction} imply the reduction
    \[ \prob{CI}_{a/2,b/2} \leq_m^p \prob{QCD}_{a,b}. \]
    As $\prob{CI}_{a/2,b/2}$ is \class{QIP}-hard for any $0 < b < a
    \leq 2$ by Theorem~\ref{clim-thm-ci-complete}, which is due to
    Kitaev and Watrous~\cite{KitaevW00}, this reduction implies that
    the distinguishability problem is \class{QIP}-hard for the values
    of $a,b$ specified in the theorem.

    The distinguishability problem is also complete for \class{QIP}
    for these values of $a$ and $b$,
    as it is in \class{QIP} by Theorem~\ref{dist-theorem-QCD_in_QIP}.
  \end{proof}
\end{corollary}

\section{Distinguishing log-depth computations}\label{dist-scn-logdepth}

In this section it is discussed how to extend the hardness of the
\prob{QCD} problem to the case of input circuits that have logarithmic depth.
This can be done by simply noting that the reduction of
Section~\ref{dist-scn-reduction} can be modified to produce output
circuits of logarithmic depth, and the hardness will follow from the
hardness of the log-depth close images problem.

To see that the \class{QIP}-hardness of the problem $\prob{Log-depth
  CI}_{a/2,b/2}$ can be extended to the problem $\prob{Log-depth QCD}_{a,b}$,
observe that the reduction in Section~\ref{dist-scn-reduction} simply
takes the input circuits an produces circuits that apply them based on
the value of a control qubit.  These controlled operations can be
implemented in logarithmic depth using a tree structure with copies of
the control qubit made in the computational basis -- this is discussed
in Proposition~\ref{compl-prop-log-control}.  If this more careful
implementation of the controlled $U_1$ and $U_2$ operations is made,
then the output circuits of Figure~\ref{dist-fig-outputs} have
logarithmic depth if and only if the input circuits do.  This requires
that the circuits for the operations $U_i$ that implement the input
circuits $Q_i$ by the equation
\[ Q_i(\rho) = \ptr{B} U_i (\rho \tprod \ket 0 \bra 0) U_i^* \]
can be assumed to have logarithmic depth when the mixed-state circuits
$Q_i$ do, but these circuits can be constructed by simply delaying any
partial trace operations that are performed during the circuit.  These circuits
have the same depth as the original mixed state circuits and they can
be constructed in polynomial time.
This implies that 
\[ \prob{Log-depth CI}_{a/2,b/2} \leq_m^p \prob{Log-depth QCD}_{a,b}, \]
which, by Corollary~\ref{clim-cor-log-result} immediately implies the
following corollary, since \prob{Log-depth QCD} is in \class{QIP} by
Theorem~\ref{dist-theorem-QCD_in_QIP} as it is a restriction of the
general problem.
\begin{corollary}
  For any $0 < b < a \leq 2$ the problem
  $\prob{Log-depth QCD}_{a,b}$ is \class{QIP}-complete.
\end{corollary}

As in Chapter~\ref{chap-close-images}, the only place that this
construction requires logarithmic depth circuits are the controlled
operations.  If the unbounded fan-out gate is allowed into the basis
of computational gates, then the circuits can be reduced to constant
depth, as discussed in Proposition~\ref{compl-prop-const-control}.
This implies the following result.

\begin{corollary}\label{dist-cor-hardness}
  For any $0 < b < a \leq 2$, the problem
  $\prob{Const-depth QCD}_{a,b}$ on circuits with the
  unbounded fan-out gate is \class{QIP}-complete.
\end{corollary}

\section{Conclusion}

In this chapter, the problem \prob{Quantum Circuit Distinguishability}
has been introduced and studied. 
This is the problem of determining if two quantum channels, given as
mixed-state quantum circuits, is there an input on which they produce
nearly orthogonal output states, or are they effectively the same on
all inputs?  

The main result of the chapter is that this problem is complete for
the class \class{QIP} of problems that have quantum interactive proof
systems, when the phrases ``nearly orthogonal'' and ``effectively the
same'' are formalized as large and small diamond norm distance, respectively.
This result requires many of the results on the diamond norm from
Chapter~\ref{chap-meas}, such as
Theorem~\ref{meas-thm-dnorm-difference}, which proves that the diamond
norm of the difference of two channels is achieved on a pure state
input.  This result can also be extended to the case of channels
specified by logarithmic depth circuits, or even constant depth
circuits if the unbounded fan-out gate is included in the circuit
model.

The main result of this chapter is essential for the result in the
next two chapters that this distinguishability problem remains hard
when restricted to circuits that implement convex mixtures of unitary
channels and when restricted to the degradable or antidegradable
channels.  These results will be shown by reducing the problem
considered here to restricted versions.

%%% Local Variables: 
%%% mode: latex
%%% TeX-master: "thesis"
%%% End: 

\chapter{Degradable and Antidegradable Channels}\label{chap-degr}

The degradable and antidegradable channels are two of the most
interesting classes of quantum channels.  The degradable channels are,
informally, the channels where the output space contains more
information about the input than the
environment, in the sense that the output state can be used to
reconstruct the state of the environment.  
The antidegradable channels can be similarly thought of
as those channels whose output contains less information about the
input than the environment does.
These channels have many nice properties when
considered for the transmission of quantum information.  This is
interesting as these channels can be otherwise awkward to work with:
as an example, the set of degradable channels is not even convex!

The main result of this chapter is that the quantum circuit
distinguishability problem considered in
Chapter~\ref{chap-distinguishability} remains \class{QIP}-complete on
both the degradable and antidegradable channels.  This lends evidence to the
notion that the difficulty of distinguishing quantum channels has
little to do with how well they preserve information.

\minitoc

\section{Degradable and antidegradable channels}

As defined in Chapter~\ref{chap-intro}, a channel $\Phi \in \transform{H,K}$ is degradable
if there exists a second channel $\Delta$ that maps the output state
of $\Phi$ to the state of the environment.  More precisely, if
\[ \Phi(\rho) = \ptr{B} U( \rho \tprod \ket 0 \bra 0) U^*, \]
\index{degradable channel}\index{channel!degradable}%
then $\Phi$ is called degradable if there exists a channel $\Delta \in
\transform{K,B}$ such that
\[ (\Delta \circ \Phi)(\rho) = \ptr{K} U( \rho \tprod \ket 0 \bra 0)
U^* = \Phi^C(\rho). \]
The channel $\Phi^C$ is called the complementary channel to $\Phi$,
and it is only defined up to an isometry, since it depends on the
Stinespring representation of $\Phi$.  This does not affect the notion
of degradability, however, as this isometry can be viewed as a part of
to the degrading map $\Delta$.
Loosely, these are the channels whose output contains more information
than the environment, because the output can be degraded to give the
state of the environment.
These channels were introduced by Shor and
Devetak~\cite{DevetakS05} to study the capacity of a channel for
transmitting quantum information.  Notice that the set of degradable channels
is not convex: any unitary channel is degradable, but the completely
depolarizing channel is not, and it can be written as a convex
combination of unitary channels (see Proposition~\ref{mu-prop-depol-mu}).

A channel is called antidegradable if the complementary channel is degradable.
Alternately, a channel is $\Phi$ antidegradable is there exists a map
\index{antidegradable channel}\index{channel!antidegradable}%
$A$ such that $A \circ \Phi^C = \Phi$, where once again the channel
$\Phi^C$ is only defined up to an isometry, but this isometry can also
be part of the map $A$, so that the antidegradable channels are also
well-defined.  This class of channels has been introduced by Wolf and
P{\'e}rez-Garcia~\cite{WolfP07}, and can be informally thought of the
class of those very noisy channels that lose more information to the
environment than they preserve in the output.
A thorough discussion of the degradable and antidegradable
channels can be found in~\cite{CubittR+08}, where it is shown that,
unlike the degradable channels, the set of antidegradable channels is convex.

The degradable and antidegradable channels are very interesting from a
quantum information perspective.  A simple no-cloning argument implies
that the antidegradable channels have zero capacity for the
transmission of quantum information.  This argument was first
presented for erasure channels in~\cite{BennettD+97}, extended to lossy
bosonic channels in~\cite{GiovannettiL+03}, and finally applied to
antidegradable channels in~\cite{GiovannettiF05}.
It is also known that the coherent information is additive on
degradable channels, which implies that the quantum capacity is given by
the coherent information of a single use of the channel, i.e.\ that the
formula for the quantum capacity does not require
regularization~\cite{DevetakS05}.

As the degradable and antidegradable channels have nice properties
with respect to the transmission of quantum information, it might be
hoped that similar properties extend to the transmission of classical
information.  In the case of the Holevo (or $\chi$-)capacity, it is shown
in~\cite{CubittR+08} that the additivity of this quantity on
degradable channels is equivalent to the general case, making use of a
result from~\cite{FukudaW07}.  As it is also known that this
additivity problem is equivalent on the complementary class of
channels~\cite{Holevo07, KingM+07}, this implies that the additivity
of the antidegradable channels is also equivalent to the general case.
Finally, using the recent result of Hastings~\cite{Hastings09}, there
are degradable and antidegradable channels that do not have additive
Holevo capacity.

Interestingly, we can adapt the same construction used by Cubitt, Ruskai,
and Smith~\cite{CubittR+08} to show that the quantum circuit
distinguishability problem restricted to either the degradable or
antidegradable channels remains \class{QIP}-complete.  These results
are the focus of the remainder of this chapter.

\section{Simulation by a degradable channel}\label{degr-scn-degr-sim}

Given a circuit $Q$ implementing a transformation in
$\transform{H,K}$, the goal is to efficiently construct a circuit $C$
implementing a degradable channel in $\transform{H,K}$ that is closely
related to the original circuit $Q$.  This reduction will make use of
the results used in the case of the minimum output
entropy~\cite{CubittR+08}: the construction presented here, as well as
the proof that the resulting channel is degradable, can both be found
in this work.

To describe the channel, we assume that $\dm{H} = \dm{K}$, i.e.\ that
the circuit $Q$ has identical input and output dimension.  This may be
assumed without loss of generality by padding the smaller space with
unused $\ket 0$ qubits, since these qubits will not affect the diamond
norm used in the definition of the distinguishability problem.  Once
this padding has been completed, we may view $Q$ as an implementation
of some channel in $\transform{H,H}$.  The channel $C$ constructed
from $Q$ will make use of an additional output qubit in the space
$\mathcal{C}$ of dimension 2, so that $Q \in \transform{H, C \tprod H}$.

\begin{figure}
  \begin{center}
    \setlength{\unitlength}{3947sp}%
\begingroup\makeatletter\ifx\SetFigFont\undefined%
\gdef\SetFigFont#1#2#3#4#5{%
  \reset@font\fontsize{#1}{#2pt}%
  \fontfamily{#3}\fontseries{#4}\fontshape{#5}%
  \selectfont}%
\fi\endgroup%
\begin{picture}(2877,2124)(-164,-1648)
\put(-149,-1036){\makebox(0,0)[lb]{\smash{{\SetFigFont{12}{14.4}{\rmdefault}{\mddefault}{\updefault}{\color[rgb]{0,0,0}$\ket 0$}%
}}}}
\thinlines
{\color[rgb]{0,0,0}\put(-149,-211){\line( 1, 0){1350}}
}%
{\color[rgb]{0,0,0}\put(-149,-286){\line( 1, 0){1350}}
}%
{\color[rgb]{0,0,0}\put(-149,-361){\line( 1, 0){1350}}
}%
{\color[rgb]{0,0,0}\put(-149,-436){\line( 1, 0){1350}}
}%
{\color[rgb]{0,0,0}\put(-149,-136){\line( 1, 0){1350}}
}%
{\color[rgb]{0,0,0}\put(2101,239){\circle*{76}}
}%
{\color[rgb]{0,0,0}\put(2701,-211){\line(-1, 0){900}}
}%
{\color[rgb]{0,0,0}\put(2701,-286){\line(-1, 0){900}}
}%
{\color[rgb]{0,0,0}\put(2701,-361){\line(-1, 0){900}}
}%
{\color[rgb]{0,0,0}\put(2701,-436){\line(-1, 0){900}}
}%
{\color[rgb]{0,0,0}\put(2701,-136){\line(-1, 0){900}}
}%
{\color[rgb]{0,0,0}\put(2401,-961){\line(-1, 0){600}}
}%
{\color[rgb]{0,0,0}\put(2401,-1036){\line(-1, 0){600}}
}%
{\color[rgb]{0,0,0}\put(2401,-1111){\line(-1, 0){600}}
}%
{\color[rgb]{0,0,0}\put(2401,-1186){\line(-1, 0){600}}
}%
{\color[rgb]{0,0,0}\put(2401,-886){\line(-1, 0){600}}
}%
{\color[rgb]{0,0,0}\put(2401,-811){\line(-1, 0){600}}
}%
{\color[rgb]{0,0,0}\put(151,-961){\line( 1, 0){1050}}
}%
{\color[rgb]{0,0,0}\put(151,-1036){\line( 1, 0){1050}}
}%
{\color[rgb]{0,0,0}\put(151,-1111){\line( 1, 0){1050}}
}%
{\color[rgb]{0,0,0}\put(151,-1186){\line( 1, 0){1050}}
}%
{\color[rgb]{0,0,0}\put(151,-886){\line( 1, 0){1050}}
}%
{\color[rgb]{0,0,0}\put(151,-811){\line( 1, 0){1050}}
}%
{\color[rgb]{0,0,0}\put(2101,-1411){\circle{76}}
}%
{\color[rgb]{0,0,0}\put(901,239){\line( 1, 0){1800}}
}%
{\color[rgb]{0,0,0}\put(1501,239){\line( 0,-1){300}}
}%
{\color[rgb]{0,0,0}\put(451, 14){\framebox(450,450){$H$}}
}%
{\color[rgb]{0,0,0}\put(151,239){\line( 1, 0){300}}
}%
{\color[rgb]{0,0,0}\put(1201,-1261){\framebox(600,1200){$U$}}
}%
{\color[rgb]{0,0,0}\put(151,-1411){\line( 1, 0){2250}}
}%
{\color[rgb]{0,0,0}\put(2401,-811){\vector( 0,-1){825}}
}%
{\color[rgb]{0,0,0}\put(2101,239){\line( 0,-1){1688}}
}%
{\color[rgb]{0,0,0}\put(-149,-511){\line( 1, 0){1350}}
}%
{\color[rgb]{0,0,0}\put(1801,-511){\line( 1, 0){900}}
}%
\put(-149,164){\makebox(0,0)[lb]{\smash{{\SetFigFont{12}{14.4}{\rmdefault}{\mddefault}{\updefault}{\color[rgb]{0,0,0}$\ket 0$}%
}}}}
\put(-149,-1486){\makebox(0,0)[lb]{\smash{{\SetFigFont{12}{14.4}{\rmdefault}{\mddefault}{\updefault}{\color[rgb]{0,0,0}$\ket 0$}%
}}}}
{\color[rgb]{0,0,0}\put(1501,239){\circle*{76}}
}%
\end{picture}%
  \end{center}
  \caption[Reduction to a degradable channel]{The degradable channel
    $C$ constructed from $Q$.}
  \label{degr-fig-degrad-reduction}
\end{figure}
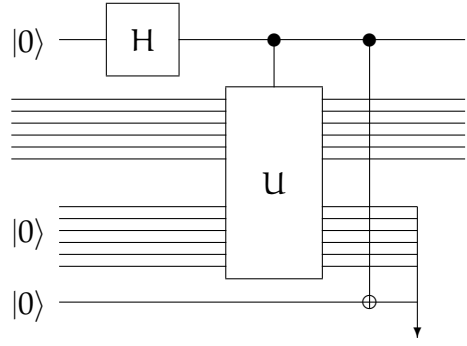

The basic idea is to implement the channel
\begin{equation}\label{degr-eqn-degr-const}
  C(\rho) = \frac{1}{2} \ket 0 \bra 0 \tprod \rho + \frac{1}{2}\ket 1 \bra 1 \tprod Q(\rho).
\end{equation}
This is just the channel that applies the circuit $Q$ with probability
$1/2$, and does nothing to the input with probability $1/2$.  If $Q$
is given in Stinespring form with unitary $U$, so that 
\[ Q(\rho) = \ptr{B} U (\rho \tprod \ket 0 \bra 0) U^*,\] 
then the channel $C$ can
be implemented as shown in Figure~\ref{degr-fig-degrad-reduction}.
The idea in this implementation is that the top ancillary qubit (which
is the qubit in the space $\mathcal{C}$) is placed in the $\ket +$
state, which results in the circuit for $Q$ being applied with
probability one-half, as the value of this qubit is `copied' onto one
of the environment qubits by the controlled-not gate.  This results in
the mixture in Equation~\eqref{degr-eqn-degr-const}.

To see that the circuit $C$ implements a degradable operation, the
degrading map that takes the output state to the environment state can
be explicitly constructed.  As the complementary channel is defined
only up to an isometry, we may construct this map for \emph{any} of
the complementary channels $C^C$ defined by $C$, as this isometry can
be added to the degrading map as required.  For this reason, we
consider the complementary channel defined by the implementation in
Figure~\ref{degr-fig-degrad-reduction}, i.e.\ the channel from the
input to all those qubits that are traced out.  This results in the
complementary channel
\begin{equation}\label{degr-eqn-degrad-compl}
  C^C(\rho) = \frac{1}{2} \ket 0 \bra 0 \tprod \ket 0 \bra 0 + \frac{1}{2} \ket 1 \bra 1 \tprod Q^C(\rho),
\end{equation}
where $Q^C$ has implementation 
\[ Q^C(\rho) = \ptr{H} U ( \rho \tprod \ket 0 \bra 0) U^*, \] 
which is obtained by tracing out the `output' space of the
original circuit.

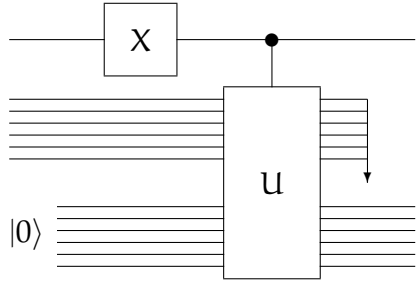
\begin{figure}
  \begin{center}
    \setlength{\unitlength}{3947sp}%
\begingroup\makeatletter\ifx\SetFigFont\undefined%
\gdef\SetFigFont#1#2#3#4#5{%
  \reset@font\fontsize{#1}{#2pt}%
  \fontfamily{#3}\fontseries{#4}\fontshape{#5}%
  \selectfont}%
\fi\endgroup%
\begin{picture}(2577,1749)(1186,-1273)
\put(1201,-1036){\makebox(0,0)[lb]{\smash{{\SetFigFont{12}{14.4}{\rmdefault}{\mddefault}{\updefault}{\color[rgb]{0,0,0}$\ket 0$}%
}}}}
\thinlines
{\color[rgb]{0,0,0}\put(1501,-1036){\line( 1, 0){1050}}
}%
{\color[rgb]{0,0,0}\put(1501,-1111){\line( 1, 0){1050}}
}%
{\color[rgb]{0,0,0}\put(1501,-1186){\line( 1, 0){1050}}
}%
{\color[rgb]{0,0,0}\put(1501,-886){\line( 1, 0){1050}}
}%
{\color[rgb]{0,0,0}\put(1501,-811){\line( 1, 0){1050}}
}%
{\color[rgb]{0,0,0}\put(2851,239){\circle*{76}}
}%
{\color[rgb]{0,0,0}\put(3151,-961){\line( 1, 0){600}}
}%
{\color[rgb]{0,0,0}\put(3151,-1036){\line( 1, 0){600}}
}%
{\color[rgb]{0,0,0}\put(3151,-1111){\line( 1, 0){600}}
}%
{\color[rgb]{0,0,0}\put(3151,-1186){\line( 1, 0){600}}
}%
{\color[rgb]{0,0,0}\put(3151,-886){\line( 1, 0){600}}
}%
{\color[rgb]{0,0,0}\put(3151,-811){\line( 1, 0){600}}
}%
{\color[rgb]{0,0,0}\put(1201,-511){\line( 1, 0){1350}}
}%
{\color[rgb]{0,0,0}\put(1201,-211){\line( 1, 0){1350}}
}%
{\color[rgb]{0,0,0}\put(1201,-286){\line( 1, 0){1350}}
}%
{\color[rgb]{0,0,0}\put(1201,-361){\line( 1, 0){1350}}
}%
{\color[rgb]{0,0,0}\put(1201,-436){\line( 1, 0){1350}}
}%
{\color[rgb]{0,0,0}\put(1201,-136){\line( 1, 0){1350}}
}%
{\color[rgb]{0,0,0}\put(3151,-286){\line( 1, 0){300}}
}%
{\color[rgb]{0,0,0}\put(3151,-361){\line( 1, 0){300}}
}%
{\color[rgb]{0,0,0}\put(3151,-436){\line( 1, 0){300}}
}%
{\color[rgb]{0,0,0}\put(3151,-511){\line( 1, 0){300}}
}%
{\color[rgb]{0,0,0}\put(3151,-211){\line( 1, 0){300}}
}%
{\color[rgb]{0,0,0}\put(3151,-136){\line( 1, 0){300}}
}%
{\color[rgb]{0,0,0}\put(2551,-1261){\framebox(600,1200){$U$}}
}%
{\color[rgb]{0,0,0}\put(1801, 14){\framebox(450,450){$X$}}
}%
{\color[rgb]{0,0,0}\put(1201,239){\line( 1, 0){600}}
}%
{\color[rgb]{0,0,0}\put(2851,-61){\line( 0, 1){300}}
}%
{\color[rgb]{0,0,0}\put(2251,239){\line( 1, 0){1500}}
}%
{\color[rgb]{0,0,0}\put(3451,-136){\vector( 0,-1){525}}
}%
{\color[rgb]{0,0,0}\put(1501,-961){\line( 1, 0){1050}}
}%
\end{picture}%
  \end{center}
  \caption[Degrading channel for the channel in
  Figure~\ref{degr-fig-degrad-reduction}]{The degrading channel $\Delta_C$
    corresponding to the channel in $C$ in
    Figure~\ref{degr-fig-degrad-reduction}.}
  \label{degr-fig-degrad-degrading}
\end{figure}

It is not hard to see how to implement the degrading map $\Delta_C$ for this channel.  
Starting with the output state of $C$
\[ C(\rho) = \frac{1}{2} \ket 0 \bra 0 \tprod \rho + \frac{1}{2} \ket 1 \bra 1 \tprod Q(\rho), \] 
as given by Equation~\eqref{degr-eqn-degr-const},
this channel can, based on the flag state in the space $\mathcal{C}$
either output $\ket 0 \bra 0$ or $Q^C(\rho)$.  More formally, when the
flag state is $\ket 0$, the state in $\mathcal{H}$ is the original
input $\rho$, so that the channel $Q^C$ can be applied to it by
performing the unitary $U$ from the circuit $Q$ and tracing out the
appropriate space.  On the other hand, when this flag state is $\ket
1$, the degrading map needs to output $\ket 0 \bra 0$, which can be
done by producing the correct number of untouched ancillary qubits as
output.  All that remains in to invert the flag qubit to get exactly
the output of $C^C$.  A circuit implementation of the channel $\Delta_C$ is
presented in Figure~\ref{degr-fig-degrad-degrading}.
We can formally verify that this map performs the required operation by observing that
\begin{align*}
  \Delta_C( C(\rho) )
  &= \frac{1}{2} \Delta_C\left(\ket 0 \bra 0 \tprod \rho + \ket 1 \bra 1 \tprod Q(\rho)\right) \\
  &= \frac{1}{2} \ket 1 \bra 1 \tprod Q^C(\rho) + \frac{1}{2} \ket 0 \bra 0 \tprod \ket 0 \bra 0 \\
  &= C^C(\rho),
\end{align*}
where the final equality is Equation~\eqref{degr-eqn-degrad-compl}.
This argument, due to Cubitt, Ruskai, and Smith~\cite{CubittR+08}
proves that the channel $C$ is degradable.  In the next section we
consider the implications of this construction for the computational
hardness of the problem of distinguishing quantum circuits.

\section{Distinguishing degradable channels}\label{degr-scn-degr-dist}

The construction in the previous section essentially embeds any
channel into a degradable channel.  This construction can be used to
show that distinguishing degradable channels is no easier than
distinguishing general channels.

As a first step towards this, a formal definition of the circuit
distinguishability problem of Chapter~\ref{chap-distinguishability} is
presented.  This is simply the general problem with the extra
restriction that the input circuits implement channels that are degradable.
\begin{problem}[Degradable Quantum Circuit Distinguishability]
  \label{degr-prob-dqcd}
  For constants $0 \leq b < a \leq 2$, the input consists of
  quantum
  circuits $C_1$ and $C_2$ that implement degradable transformations in
  $\transform{H,K}$.
  The promise problem is to distinguish the two cases:
  \begin{description}
    \item[Yes:] $\dnorm{C_1 - C_2} \geq a$,
    \item[No:] $\dnorm{C_1 - C_2} \leq b$.
  \end{description}
\end{problem}
\index{Quantum Circuit Distinguishability!degradable}%

The primary ingredient in the proof that this problem is
\class{QIP}-complete, is the result that the construction in the
previous section does not significantly affect the diamond norm of the
difference of two channels.  This is not difficult to see from the
output of the construction, given by
Equation~\eqref{degr-eqn-degr-const}, but for completeness it is argued formally in the
following lemma.

\begin{lemma}\label{degr-lem-degr-dnorm}
  Let $Q_1, Q_2$ be quantum circuits implementing transformations in
  $\transform{H,K}$.  If $C_1, C_2 \in \transform{H, C \tprod K}$
  are given by
  \[ C_i(\rho) 
      = \frac{1}{2} \ket 0 \bra 0 \tprod \rho 
      + \frac{1}{2} \ket 1 \bra 1 \tprod Q_i(\rho), \]
  for $i \in \{1,2\}$, then
  \[ \dnorm{C_1 - C_2} = \frac{1}{2} \dnorm{Q_1 - Q_2}. \]
  \begin{proof}
    Let $\rho \in \density{H \tprod F}$ be an arbitrary state.  Then
    \begin{align*}
      \tnorm{(C_1 \tprod \tidentity{F} - C_2 \tprod \tidentity{F})(\rho)}
      &= \frac{1}{2} \tnorm{ 
           \ket 0 \bra 0 \tprod (\rho - \rho) 
        + \ket 1 \bra 1 \tprod ( 
            \left[ Q_1 \tprod \tidentity{F} - Q_2 \tprod \tidentity{F}
            \right] (\rho)) } \\
      &= \frac{1}{2} \tnorm{ 
       \ket 1 \bra 1 \tprod ( 
            \left[ Q_1 \tprod \tidentity{F} - Q_2 \tprod \tidentity{F}
            \right] (\rho)) } \\
      &= \frac{1}{2} \tnorm{ 
       \left( Q_1 \tprod \tidentity{F} - Q_2 \tprod \tidentity{F} \right) (\rho) }.
    \end{align*}
    Since the diamond norm is defined as the maximization over all
    states $\rho$, this implies the statement of the lemma.
  \end{proof}
\end{lemma}

Let $(Q_1, Q_2)$ be an instance of $\prob{QCD}_{a,b}$.
As it is demonstrated in the previous section how to efficiently
construct the channels $C_i$ from the channels $Q_i$, this lemma
implies the following reduction
\[ \prob{QCD}_{a,b} \leq_m \prob{Degradable QCD}_{a/2,b/2}. \]
This implies that distinguishing degradable circuits is hard for all
$0 < b < a \leq 1$, using the hardness result for general circuits
(Corollary~\ref{dist-cor-hardness}).

This result can be strengthened using a result from
Section~\ref{meas-scn-polarization} on the polarization of the diamond
norm.  The general construction does not preserve degradability, but
for the special case of interest using only a portion of the
polarization construction will suffice.  The strategy is to take an
instance $(C_1, C_2)$ of $\prob{Degradable QCD}_{1,\epsilon}$ and
construct the instance $(C_1^{\tensor k}, C_2^{\tensor k})$.  This
second instance will have outputs that are more distinguishable, for
the simple reason that there are more copies of the states to be
distinguished available.  This will send the norm for `yes' instances
of the problem from 1 to a value close to 2, but it will also have the
property that the norm of `no' instances is not made too large.  This
is a straightforward consequence of
Lemma~\ref{meas-lemma_direct_product}, which appears as part of the
procedure for polarizing the diamond norm.

\begin{corollary}\label{degr-cor-degr-hard}
  For any constants $0 < b < a < 2$, the problem
  $\prob{Degradable QCD}_{a,b}$ is \class{QIP}-complete.

  \begin{proof}
    This problem is in \class{QIP} as it is a restriction of the
    general problem, which is in \class{QIP} by
    Theorem~\ref{dist-theorem-QCD_in_QIP}.  To see that it is
    \class{QIP}-hard, take an instance $(Q_1, Q_2)$ of the
    \class{QIP}-complete problem $\prob{QCD}_{2,\epsilon}$, for
    $\epsilon>0$ a constant.

    Applying the construction of Section~\ref{degr-scn-degr-sim}
    to $(Q_1, Q_2)$ results in the instance $(C_1, C_2)$ of
    $\prob{Degradable QCD}_{1,\epsilon/2}$, by
    Lemma~\ref{degr-lem-degr-dnorm}.  
    As the degradable channels are closed under tensor products, 
    $(C_1^{\tprod k}, C_2^{\tprod k})$ is a pair of circuits
    implementing degradable channels.  By
    Lemma~\ref{meas-lemma_direct_product}, we have the following
    implications
    \begin{align*}
      \dnorm{C_1 - C_2} \geq 1 & \implies
      \dnorm{C_1^{\tprod k} - C_2^{\tprod k}} \geq 2 - 2^{-k/8}, \\
      \dnorm{C_1 - C_2} \leq \frac{\epsilon}{2} & \implies
      \dnorm{C_1^{\tprod k} - C_2^{\tprod k}} \leq \frac{k \epsilon}{2}. 
    \end{align*}
    These equations imply that for any constants $0 < b < a < 2$,
    there are choices of the constants $k, \epsilon$ so that
    \begin{align*}
      \dnorm{Q_1 - Q_2} = 2 & \implies
      \dnorm{C_1^{\tprod k} - C_2^{\tprod k}} \geq a, \\
      \dnorm{Q_1 - Q_2} \leq \epsilon & \implies
      \dnorm{C_1^{\tprod k} - C_2^{\tprod k}} \leq b,
    \end{align*}
    which implies the \class{QIP} hardness of $\prob{Degradable QCD}_{a,b}$
  \end{proof}
\end{corollary}

\section{Simulation by an antidegradable channel}\label{degr-scn-adegr-sim}

In this section a construction very similar to that used in
Section~\ref{degr-scn-degr-sim} is presented that takes any circuit
$Q$ to a circuit $C$ implementing an antidegradable channel.  The idea
is to (with probability one-half) send the input state to the
environment, so that the channel that maps the environment state to
the output state will have a copy of the input state.  This
construction (and the proof that it produces an antidegradable
channel) is very similar to a construction used in~\cite{CubittR+08}
for degradable channels.

Once again we may assume that $Q$ implements a channel in
$\transform{H,H}$, i.e.\ that $Q$ has the same input and output
dimension, by embedding the smaller space into the larger, if
necessary.  As in Section~\ref{degr-scn-degr-sim}, the constructed
circuit $C$ will use one additional output qubit, implementing an
antidegradable transformation in $\transform{H, C \tprod H}$.

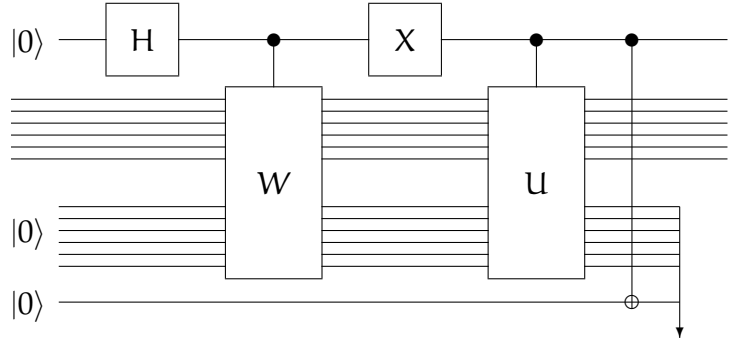
\begin{figure}
  \begin{center}
    \setlength{\unitlength}{3947sp}%
\begingroup\makeatletter\ifx\SetFigFont\undefined%
\gdef\SetFigFont#1#2#3#4#5{%
  \reset@font\fontsize{#1}{#2pt}%
  \fontfamily{#3}\fontseries{#4}\fontshape{#5}%
  \selectfont}%
\fi\endgroup%
\begin{picture}(4527,2124)(-464,-1648)
\put(-449,-1036){\makebox(0,0)[lb]{\smash{{\SetFigFont{12}{14.4}{\rmdefault}{\mddefault}{\updefault}{\color[rgb]{0,0,0}$\ket 0$}%
}}}}
\thinlines
{\color[rgb]{0,0,0}\put(-149,-1036){\line( 1, 0){1050}}
}%
{\color[rgb]{0,0,0}\put(-149,-1111){\line( 1, 0){1050}}
}%
{\color[rgb]{0,0,0}\put(-149,-1186){\line( 1, 0){1050}}
}%
{\color[rgb]{0,0,0}\put(-149,-886){\line( 1, 0){1050}}
}%
{\color[rgb]{0,0,0}\put(-149,-811){\line( 1, 0){1050}}
}%
{\color[rgb]{0,0,0}\put(1501,-961){\line( 1, 0){1050}}
}%
{\color[rgb]{0,0,0}\put(1501,-1036){\line( 1, 0){1050}}
}%
{\color[rgb]{0,0,0}\put(1501,-1111){\line( 1, 0){1050}}
}%
{\color[rgb]{0,0,0}\put(1501,-1186){\line( 1, 0){1050}}
}%
{\color[rgb]{0,0,0}\put(1501,-886){\line( 1, 0){1050}}
}%
{\color[rgb]{0,0,0}\put(1501,-811){\line( 1, 0){1050}}
}%
{\color[rgb]{0,0,0}\put(1201,239){\circle*{76}}
}%
{\color[rgb]{0,0,0}\put(2851,239){\circle*{76}}
}%
{\color[rgb]{0,0,0}\put(3151,-961){\line( 1, 0){600}}
}%
{\color[rgb]{0,0,0}\put(3151,-1036){\line( 1, 0){600}}
}%
{\color[rgb]{0,0,0}\put(3151,-1111){\line( 1, 0){600}}
}%
{\color[rgb]{0,0,0}\put(3151,-1186){\line( 1, 0){600}}
}%
{\color[rgb]{0,0,0}\put(3151,-886){\line( 1, 0){600}}
}%
{\color[rgb]{0,0,0}\put(3151,-811){\line( 1, 0){600}}
}%
{\color[rgb]{0,0,0}\put(-449,-511){\line( 1, 0){1350}}
}%
{\color[rgb]{0,0,0}\put(-449,-211){\line( 1, 0){1350}}
}%
{\color[rgb]{0,0,0}\put(-449,-286){\line( 1, 0){1350}}
}%
{\color[rgb]{0,0,0}\put(-449,-361){\line( 1, 0){1350}}
}%
{\color[rgb]{0,0,0}\put(-449,-436){\line( 1, 0){1350}}
}%
{\color[rgb]{0,0,0}\put(-449,-136){\line( 1, 0){1350}}
}%
{\color[rgb]{0,0,0}\put(1501,-511){\line( 1, 0){1050}}
}%
{\color[rgb]{0,0,0}\put(1501,-211){\line( 1, 0){1050}}
}%
{\color[rgb]{0,0,0}\put(1501,-286){\line( 1, 0){1050}}
}%
{\color[rgb]{0,0,0}\put(1501,-361){\line( 1, 0){1050}}
}%
{\color[rgb]{0,0,0}\put(1501,-436){\line( 1, 0){1050}}
}%
{\color[rgb]{0,0,0}\put(1501,-136){\line( 1, 0){1050}}
}%
{\color[rgb]{0,0,0}\put(3151,-286){\line( 1, 0){900}}
}%
{\color[rgb]{0,0,0}\put(3151,-361){\line( 1, 0){900}}
}%
{\color[rgb]{0,0,0}\put(3151,-436){\line( 1, 0){900}}
}%
{\color[rgb]{0,0,0}\put(3151,-511){\line( 1, 0){900}}
}%
{\color[rgb]{0,0,0}\put(3151,-211){\line( 1, 0){900}}
}%
{\color[rgb]{0,0,0}\put(3151,-136){\line( 1, 0){900}}
}%
{\color[rgb]{0,0,0}\put(3451,239){\circle*{76}}
}%
{\color[rgb]{0,0,0}\put(3451,-1411){\circle{76}}
}%
{\color[rgb]{0,0,0}\put(901,-1261){\framebox(600,1200){$W$}}
}%
{\color[rgb]{0,0,0}\put(1201,239){\line( 0,-1){300}}
}%
{\color[rgb]{0,0,0}\put(2551,-1261){\framebox(600,1200){$U$}}
}%
{\color[rgb]{0,0,0}\put(1801, 14){\framebox(450,450){$X$}}
}%
{\color[rgb]{0,0,0}\put(601,239){\line( 1, 0){1200}}
}%
{\color[rgb]{0,0,0}\put(2851,-61){\line( 0, 1){300}}
}%
{\color[rgb]{0,0,0}\put(2251,239){\line( 1, 0){1800}}
}%
{\color[rgb]{0,0,0}\put(151, 14){\framebox(450,450){$H$}}
}%
{\color[rgb]{0,0,0}\put(-149,239){\line( 1, 0){300}}
}%
{\color[rgb]{0,0,0}\put(-149,-1411){\line( 1, 0){3900}}
}%
{\color[rgb]{0,0,0}\put(3751,-811){\vector( 0,-1){825}}
}%
{\color[rgb]{0,0,0}\put(3451,239){\line( 0,-1){1688}}
}%
\put(-449,-1486){\makebox(0,0)[lb]{\smash{{\SetFigFont{12}{14.4}{\rmdefault}{\mddefault}{\updefault}{\color[rgb]{0,0,0}$\ket 0$}%
}}}}
\put(-449,164){\makebox(0,0)[lb]{\smash{{\SetFigFont{12}{14.4}{\rmdefault}{\mddefault}{\updefault}{\color[rgb]{0,0,0}$\ket 0$}%
}}}}
{\color[rgb]{0,0,0}\put(-149,-961){\line( 1, 0){1050}}
}%
\end{picture}%
  \end{center}
  \caption[Reduction to an antidegradable channel]{The antidegradable channel
    $C$ constructed from $Q$.}
  \label{degr-fig-adeg-reduction}
\end{figure}

Let $Q$ implement the transformation given by
\[ Q(\rho) = \ptr{B} U (\rho \tprod \ket 0 \bra 0) U^*, \] 
where, as usual, since $Q$ is assumed (without loss of generality) to
be in Stinespring form, the input specifies a circuit for computing
the unitary $U$.  The channel $C$ will be constructed as
\begin{equation}\label{degr-eqn-adeg-const}
  C(\rho) = \frac{1}{2} \ket 0 \bra 0 \tprod \ket 0 \bra 0 +
  \frac{1}{2} \ket 1 \bra 1 \tprod Q(\rho).
\end{equation}
This is just the channel that applies $Q$ with probability one-half,
outputs $\ket 0$ with probability one-half, and outputs a flag qubit
in the space $\mathcal{C}$ to indicate which case has occurred.  In a
way very similar to the construction in
Section~\ref{degr-scn-degr-sim}, this channel can be implemented using
a controlled-$U$ operation.  In this case, however, we will also need
the operation $W$ that swaps the states in two spaces (i.e.\ $W \ket a
\ket b = \ket b \ket a$).  An implementation of the channel $C$ is
given in Figure~\ref{degr-fig-adeg-reduction}.
This circuit will, depending on the value of the control qubit in the
space $\mathcal{C}$ either apply $Q$ or output the pure state
$\ket 0$, as required.

\begin{figure}
  \begin{center}
    \setlength{\unitlength}{3947sp}%
\begingroup\makeatletter\ifx\SetFigFont\undefined%
\gdef\SetFigFont#1#2#3#4#5{%
  \reset@font\fontsize{#1}{#2pt}%
  \fontfamily{#3}\fontseries{#4}\fontshape{#5}%
  \selectfont}%
\fi\endgroup%
\begin{picture}(3477,1899)(286,-1423)
\put(301,-1036){\makebox(0,0)[lb]{\smash{{\SetFigFont{12}{14.4}{\rmdefault}{\mddefault}{\updefault}{\color[rgb]{0,0,0}$\ket 0$}%
}}}}
\thinlines
{\color[rgb]{0,0,0}\put(601,-1036){\line( 1, 0){300}}
}%
{\color[rgb]{0,0,0}\put(601,-1111){\line( 1, 0){300}}
}%
{\color[rgb]{0,0,0}\put(601,-1186){\line( 1, 0){300}}
}%
{\color[rgb]{0,0,0}\put(601,-886){\line( 1, 0){300}}
}%
{\color[rgb]{0,0,0}\put(601,-811){\line( 1, 0){300}}
}%
{\color[rgb]{0,0,0}\put(1501,-961){\line( 1, 0){1050}}
}%
{\color[rgb]{0,0,0}\put(1501,-1036){\line( 1, 0){1050}}
}%
{\color[rgb]{0,0,0}\put(1501,-1111){\line( 1, 0){1050}}
}%
{\color[rgb]{0,0,0}\put(1501,-1186){\line( 1, 0){1050}}
}%
{\color[rgb]{0,0,0}\put(1501,-886){\line( 1, 0){1050}}
}%
{\color[rgb]{0,0,0}\put(1501,-811){\line( 1, 0){1050}}
}%
{\color[rgb]{0,0,0}\put(1201,239){\circle*{76}}
}%
{\color[rgb]{0,0,0}\put(2851,239){\circle*{76}}
}%
{\color[rgb]{0,0,0}\put(3151,-961){\line( 1, 0){300}}
}%
{\color[rgb]{0,0,0}\put(3151,-1036){\line( 1, 0){300}}
}%
{\color[rgb]{0,0,0}\put(3151,-1111){\line( 1, 0){300}}
}%
{\color[rgb]{0,0,0}\put(3151,-1186){\line( 1, 0){300}}
}%
{\color[rgb]{0,0,0}\put(3151,-886){\line( 1, 0){300}}
}%
{\color[rgb]{0,0,0}\put(3151,-811){\line( 1, 0){300}}
}%
{\color[rgb]{0,0,0}\put(301,-511){\line( 1, 0){600}}
}%
{\color[rgb]{0,0,0}\put(301,-211){\line( 1, 0){600}}
}%
{\color[rgb]{0,0,0}\put(301,-286){\line( 1, 0){600}}
}%
{\color[rgb]{0,0,0}\put(301,-361){\line( 1, 0){600}}
}%
{\color[rgb]{0,0,0}\put(301,-436){\line( 1, 0){600}}
}%
{\color[rgb]{0,0,0}\put(301,-136){\line( 1, 0){600}}
}%
{\color[rgb]{0,0,0}\put(1501,-511){\line( 1, 0){1050}}
}%
{\color[rgb]{0,0,0}\put(1501,-211){\line( 1, 0){1050}}
}%
{\color[rgb]{0,0,0}\put(1501,-286){\line( 1, 0){1050}}
}%
{\color[rgb]{0,0,0}\put(1501,-361){\line( 1, 0){1050}}
}%
{\color[rgb]{0,0,0}\put(1501,-436){\line( 1, 0){1050}}
}%
{\color[rgb]{0,0,0}\put(1501,-136){\line( 1, 0){1050}}
}%
{\color[rgb]{0,0,0}\put(3151,-286){\line( 1, 0){600}}
}%
{\color[rgb]{0,0,0}\put(3151,-361){\line( 1, 0){600}}
}%
{\color[rgb]{0,0,0}\put(3151,-436){\line( 1, 0){600}}
}%
{\color[rgb]{0,0,0}\put(3151,-511){\line( 1, 0){600}}
}%
{\color[rgb]{0,0,0}\put(3151,-211){\line( 1, 0){600}}
}%
{\color[rgb]{0,0,0}\put(3151,-136){\line( 1, 0){600}}
}%
{\color[rgb]{0,0,0}\put(901,-1261){\framebox(600,1200){$W$}}
}%
{\color[rgb]{0,0,0}\put(1201,239){\line( 0,-1){300}}
}%
{\color[rgb]{0,0,0}\put(2551,-1261){\framebox(600,1200){$U$}}
}%
{\color[rgb]{0,0,0}\put(1801, 14){\framebox(450,450){$X$}}
}%
{\color[rgb]{0,0,0}\put(301,239){\line( 1, 0){1500}}
}%
{\color[rgb]{0,0,0}\put(2851,-61){\line( 0, 1){300}}
}%
{\color[rgb]{0,0,0}\put(2251,239){\line( 1, 0){1500}}
}%
{\color[rgb]{0,0,0}\put(3451,-811){\vector( 0,-1){600}}
}%
{\color[rgb]{0,0,0}\put(601,-961){\line( 1, 0){300}}
}%
\end{picture}%
  \end{center}
  \caption[Anti-degrading channel for the channel in
  Figure~\ref{degr-fig-adeg-reduction}]{The anti-degrading channel
    corresponding to the channel in $C$ in
    Figure~\ref{degr-fig-adeg-reduction}.}
  \label{degr-fig-adeg-adegrading}
\end{figure}
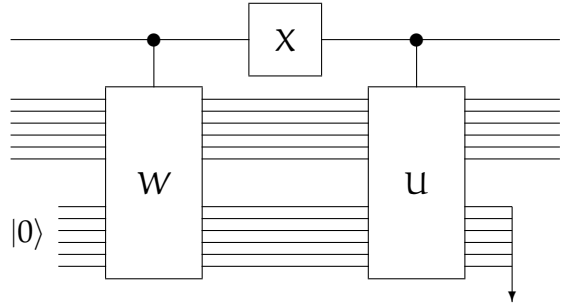

To show that the circuit $C$ implements an antidegradable channel, we
explicitly construct the map $A_C$ that maps the environment state of
$C$ to the output state.  The environment state of $C$ is once again
simply the state produced by $C^C$, the complementary channel to $C$.
As before, this channel is only defined up to an isometry, but this
is not significant as this isometry can be absorbed into the
definition of $A_C$.  One implementation of the channel $C^C$ is
obtained by considering the channel mapping the input of $C$ to the
space traced out by the circuit in
Figure~\ref{degr-fig-adeg-reduction}.  This channel is given by
\begin{equation}\label{degr-eqn-adeg-compl}
  C^C(\rho) = \frac{1}{2} \ket 0 \bra 0 \tprod \rho + \frac{1}{2} \ket 1 \bra 1 \tprod Q^C(\rho),
\end{equation}
where once again the channel $Q^C$ is given by
\[ Q^C(\rho) = \ptr{H} U ( \rho \tprod \ket 0 \bra 0) U^*. \] 
Given the state in Equation~\eqref{degr-eqn-adeg-compl} it is not hard
to see how to map it to the state in Equation~\eqref{degr-eqn-adeg-const}.
This can be done by implementing one of two operations, depending on
the value of the flag qubit in the space $\mathcal{C'}$, which is the
`copy' of the control qubit traced out in Figure~\ref{degr-fig-adeg-reduction}.
If this qubit is in the state $\ket 0$, then the other portion of the
input state is $\rho$, the original input to $C$, so that applying the
circuit for $Q$ produces the state $Q(\rho)$.  If the control qubit is
in the $\ket 1$ state, however, the remainder of the input state is
$Q^C(\rho)$.  This state can be discarded (i.e.\ traced out) and
ancillary qubits in the state $\ket 0$ can be used as the output, using the swap
operation $W$.  As before, the value of the qubit in $\mathcal{C'}$
needs to be flipped with a Pauli $X$ gate so that the state is exactly
correct.  A circuit implementing this is shown in
Figure~\ref{degr-fig-adeg-adegrading}.

To see that $A_C$ correctly implements the anti-degrading map for $C$, we may compute
\begin{align*}
  A_C( C^C(\rho) )
  &= \frac{1}{2} A_C\left(\ket 0 \bra 0 \tprod \rho + \ket 1 \bra 1 \tprod Q^C(\rho)\right) \\
  &= \frac{1}{2} \ket 1 \bra 1 \tprod Q(\rho) + \frac{1}{2} \ket 0 \bra 0 \tprod \ket 0 \bra 0 \\
  &= C(\rho),
\end{align*}
where the final equality is Equation~\ref{degr-eqn-adeg-const}.  This
demonstrates that the channel $C$ constructed from $Q$ is
antidegradable.  In the following section the implications of this
construction for the hardness of computationally distinguishing
antidegradable channels is considered.

\section{Distinguishing antidegradable channels}

In a very similar way to the degradable case, 
the construction in the previous section embeds any
channel into an antidegradable one.  In exactly the same manner as
Section~\ref{degr-scn-degr-dist}, this can be used to show the
hardness of distinguishing circuits that implement antidegradable
transformations.

As in the degradable case, the distinguishability problem in the
antidegradable case is simply the restriction of the problem to a
smaller class of channels.
\begin{problem}[Antidegradable Quantum Circuit Distinguishability]
  \label{degr-prob-adqcd}
  For constants $0 \leq b < a \leq 2$, the input consists of
  quantum
  circuits $C_1$ and $C_2$ that implement antidegradable transformations in
  $\transform{H,K}$.
  The promise problem is to distinguish the two cases:
  \begin{description}
    \item[Yes:] $\dnorm{C_1 - C_2} \geq a$,
    \item[No:] $\dnorm{C_1 - C_2} \leq b$.
  \end{description}
\end{problem}
\index{Quantum Circuit Distinguishability!antidegradable}%

Once again the key technique to proving that the problem is
\class{QIP}-complete is to place bounds on the diamond norm of the
difference of two channels that have had the construction of the
previous section applied to them.  The proof of this lemma is
identical to the proof of Lemma~\ref{degr-lem-degr-dnorm}.

\begin{lemma}\label{degr-lem-adeg-dnorm}
  Let $Q_1, Q_2$ be quantum circuits implementing transformations in
  $\transform{H,K}$.  If $C_1, C_2 \in \transform{H, C \tprod K}$
  be given by
  \[ C_i(\rho) 
      = \frac{1}{2} \ket 0 \bra 0 \tprod \ket 0 \bra 0
      + \frac{1}{2} \ket 1 \bra 1 \tprod Q_i(\rho), \]
  for $i \in \{1,2\}$, then
  \[ \dnorm{C_1 - C_2} = \frac{1}{2} \dnorm{Q_1 - Q_2}. \]
  \begin{proof}
    Let $\rho \in \density{H \tprod F}$ be arbitrary.  Then
    \begin{align*}
      \tnorm{(C_1 \tprod \tidentity{F} - C_2 \tprod \tidentity{F})(\rho)}
     &= \frac{1}{2} \tnorm{ 
       \left( Q_1 \tprod \tidentity{F} - Q_2 \tprod \tidentity{F} \right) (\rho) },
    \end{align*}
    as in the proof of Lemma~\ref{degr-lem-degr-dnorm}.  This implies
    the statement of the lemma.
  \end{proof}
\end{lemma}

Exactly as in the degradable case, this implies the
\class{QIP}-hardness of distinguishing antidegradable channels for
constants $0
< b < a \leq 1$.  Once again, this can be strengthened to any constants $0
< b < a < 2$ using the polarization techniques of
Section~\ref{meas-scn-polarization}, using the property that the
antidegradable channels are closed under tensor products.  The next
corollary follows from Lemma~\ref{degr-lem-adeg-dnorm} in
exactly the same manner that Corollary~\ref{degr-cor-degr-hard} follows
from Lemma~\ref{degr-lem-degr-dnorm}, so the proof has been omitted.

\begin{corollary}\label{degr-cor-adeg-hard}
  For any constants $0 < b < a < 2$, the problem
  $\prob{Antidegradable QCD}_{a,b}$ is \class{QIP}-complete.
\end{corollary}

\section{Conclusion}

This chapter has presented a construction for embedding an arbitrary
channel into a degradable channel due to Cubitt, Ruskai, and
Smith~\cite{CubittR+08}, as well as a closely related construction for
antidegradable channels.  These constructions can be efficiently
implemented on quantum circuits, so that instances of the quantum
circuit distinguishability problem can be mapped to degradable or
antidegradable channels.

The main result of the chapter is that the distinguishability problem
on quantum circuits remains hard when restricted to either the class
of degradable channels or the class of antidegradable channels.  The
proof of this result makes use of the diamond norm polarization
techniques of Section~\ref{meas-scn-polarization}.

%%% Local Variables: 
%%% mode: latex
%%% TeX-master: "thesis"
%%% End: 

\chapter{Mixed-Unitary Channels}\label{chap-mixed-unitary}

The mixed-unitary channels are an interesting class of quantum
operations.  These are the channels that probabilistically apply one
of a set of unitary operations.  These channels have several
interesting properties and many of the common transformations used in
quantum information are mixed-unitary.  
For these reasons the problems of determining the additivity of the
classical capacity of a mixed-unitary channel and distinguishing
circuits that implement mixed-unitary operations are important steps
toward understanding these problems.

In the distinguishability case it is shown that distinguishing
mixed-unitary channels is exactly as computationally difficult as
general channels, using a reduction that essentially 
simulates a general channel with a mixed unitary one.  In the case of
additivity, a similar reduction is used to show that given a channel,
there is a mixed-unitary channel that is approximately additive if
and only if the original channel is additive.  By sending the
approximation error to zero this produces a sequence of mixed-unitary
channels with the property that the original channel is 
additive if and only if the tail of the sequence consists of additive
mixed-unitary channels.

The results in this chapter have been published in~\cite{Rosgen08additivity}.

\minitoc

\section{Mixed-unitary channels}\label{mu-scn-intro}

As defined in Chapter~\ref{chap-intro}, a quantum channel $\Phi$ is mixed-unitary if
there exist unitary operators $U_1, \ldots, U_n$
and a probability distribution $p_1, \ldots, p_n$ such that
\index{mixed-unitary!channel}%
\index{channel!mixed-unitary}%
\begin{equation}\label{mu-eqn-rand-unitary}
  \Phi(X) = \sum_{i=1}^n p_i U_i X U_i^*.
\end{equation}
These channels have many interesting properties.  These channels are
commonly known as the \emph{random unitary} channels, but they will be
referred to as the mixed-unitary channels here to avoid confusion with
unitary operators drawn chosen at random according to some measure.
This notational choice was suggested by Watrous in~\cite{Watrous09mixing}.

It has been shown by Gregoratti
and Werner~\cite{GregorattiW03} that the mixed-unitary channels
describe exactly the noise processes that can be corrected using
classical information obtained by measuring the environment.  One way
to see that this correction is possible is to consider a
Stinespring representation for the channel in
Equation~\eqref{mu-eqn-rand-unitary}.  One such representation can be
constructed using the operations $V$ and $W$ given by
\begin{align*}
  V \ket \psi \ket i &= (U_i \ket \psi) \ket i, \\
  W \ket 0 &= \sum_i \sqrt{p_i} \ket i.
\end{align*}
The operation $V$ is a unitary operation in $\unitary{H \tprod A, K
  \tprod B}$ and the operator $W$ can be extended to a unitary
operation in $\unitary{A}$ in an arbitrary way.  A Stinespring
representation for $\Phi$ is then given by
\[ \Phi(\rho) = \ptr{B} V (\tidentity{H} \tprod W) \left( \rho \tprod \ket 0
  \bra 0 \right) (\tidentity{H} \tprod W^*) V^*, \]
where the operator $W$ prepares a weighted superposition of the
ancillary space, the operator $V$ applies the corresponding unitary
operator from Equation~\eqref{mu-eqn-rand-unitary}, and finally the
partial trace over $\mathcal{B}$ produces the desired mixture.  To see
that this can be perfectly reversed with a measurement of
$\mathcal{B}$, notice that when applied to the state $\rho$, before
the partial trace the system is in the state
\[ \sigma = \sum_{i,j} \sqrt{p_i p_j} (U_i \rho U_j^*) \tprod \ket i \bra j. \]
Measuring the second system in the computational basis gives an
outcome $a$ with probability $p_a$, leaving the system in the state
\[ \frac{(\identity{H} \tprod \bra a) 
         \sigma
         (\identity{H} \tprod \ket a)}
        {p_a}
   = U_a \rho U_a^*. \]
The original state can then be recovered by simply applying the $U_a^*$
selected by the outcome of the measurement.  This example describes in
principle any mixed-unitary channel, due to
the uniqueness of the Stinespring representation, up to an isometry on
the space $\mathcal{B}$ which corresponds to a different measurement.
The fact that the mixed-unitary channels are the only channels that
can be corrected using the strategy of measuring the environment and 
applying a correction is more complicated and can be found
in~\cite{GregorattiW03}.

One question that this correction scheme raises is how much classical
information must be recovered from the environment to correct a
mixed-unitary channel?  This corresponds to minimizing the number of
operators $U_i$ in Equation~\eqref{mu-eqn-rand-unitary}.  A simple
bound on this quantity is given by Buscemi~\cite{Buscemi06}, who shows
that the number $n$ of unitary operators in
Equation~\eqref{mu-eqn-rand-unitary} is at most the square 
of the minimum number of Kraus operators in a Kraus representation of $\Phi$.

Audenaert and Scheel have also
provided a characterization of the mixed-unitary
channels~\cite{AudenaertS08}, and used it to construct a measure of
the distance from a quantum channel to the set of mixed-unitary channels.

The remainder of this chapter provides an answer to the question: are
the problems of the additivity of the classical capacity and the
distinguishability of quantum channels simplified when restricted to
mixed-unitary channels?  This is answered in the negative, using a
method to approximate an arbitrary quantum channel by a mixed-unitary
one.  This approximation will only faithfully implement the channel on
low-entropy outputs, as the technique used will be able to decide when the
approximation would fail and instead produce a highly mixed state.
This suffices to produce a channel with the same minimum output
entropy or maximum output $p$-norm, however, as these quantities are
defined only by the low-entropy outputs of the channel.

These results extends the work of Fukuda~\cite{Fukuda07} on unital
channels to the mixed-unitary case.  The unital case is discussed in
the next section.
In the Section~\ref{mu-scn-channels} the mixed-unitary case is described.
Section~\ref{mu-scn-properties} proves some properties of the
construction that will be used in Sections~\ref{mu-scn-mult}
and~\ref{mu-scn-add} that consider the multiplicativity of the
$p$-norm and the additivity of the minimum output entropy
(respectively).  An efficient circuit construction for this reduction
is then presented in Section~\ref{mu-scn-circuits}, which is used in
Section~\ref{mu-scn-distinguish} to reduce the circuit
distinguishability problem from general channels to the mixed-unitary
channels.

\section{Unital channels}\label{mu-scn-unital}

Recall from Chapter~\ref{chap-intro} that a channel $\Phi \in
\transform{H,K}$ is doubly stochastic if $\Phi(\identity{H}) =
\identity{K}$, and unital if it is also the case that $\mathcal{H} =
\mathcal{K}$.  This section provides an overview of a result related
to the main results of the chapter.  This result is Fukuda's proof
that the additivity of the minimum output entropy or the
multiplicativity of the maximum output $p$-norm of an arbitrary
channel $\Phi$ is equivalent to the same problem on a related unital
channel $\Phi'$~\cite{Fukuda07}.

The unital channels can be characterized in a similar way to the
mixed-unitary channels.  Mendl and Wolf have shown that any unital
channel $\Phi$ can be represented in the form
\[ \Phi(\rho) = \sum_i \lambda_i U_i \rho U_i^*, \]
where the $U_i$ are unitary operators and $\sum_i \lambda_i = 1$, with
$\lambda_i \in \mathbb{R}$ for all $i$~\cite{MendlW08}.
It is also of note that
for channels on qubits the mixed-unitary channels are exactly the unital
channels~\cite{Tregub86,KuemmererM87}.  This is no longer true for
channels on systems of larger dimension, as shown by Landau and
Streater~\cite{LandauS93}.  An interesting fact is that unital but
not mixed-unitary channel that they use is, up to
unitary conjugation, the same channel used by Werner and Holevo to
find the first example of the super-multiplicativity of the $p$-norm of
a quantum channel~\cite{WernerH02}.  This might suggest that these
channels are the key to this property, but the results of this chapter
imply that the mixed-unitary channels do not hold a special place with
respect to this property of super-multiplicativity.  Indeed, Hayden and
Winter~\cite{HaydenW08} have shown that mixed-unitary channels
also exhibit this property.

The key ingredient in Fukuda's reduction to the unital case is the
addition of an extra input system that allows for an input determined
selection of one of the discrete Weyl operators $W_{i,j}$ introduced in
Chapter~\ref{chap-intro}  to be applied to the output of the channel.  
Letting $\Phi \in \transform{H,K}$, with $d = \dm{K}$, the channel $\Phi'$ is
constructed as
\begin{equation}\label{mu-eqn-fukuda}
  \Phi'(\rho \tprod \ket{i,j}\bra{i,j}) = W_{i,j} \Phi(\rho) W_{i,j}^*,
\end{equation}
for any $1 \leq i,j \leq d$.  This defines a channel $\Phi'$ on
$\transform{H \tprod K \tprod K, K}$ by linearity.  Such a channel can be
implemented by measuring the input space $\mathcal{K \tprod K}$ in the
computational basis to decide which of the Weyl operators to apply and
then tracing out the result.
This is the same construction used by Shor to prove that the
additivity of the minimum output entropy of a channel $\Phi$ implies
the additivity of the Holevo $\chi$-capacity of the channel
$\Phi'$~\cite{Shor04}.  This construction was discussed in Section~\ref{meas-scn-smin-hcap}.

To see that this channel is doubly stochastic, notice that on any
input of the form $\rho \tprod \nidentity{K \tprod K}$
\[ \Phi'(\rho \tprod \nidentity{K \tprod K}) 
   = \frac{1}{d^2} \sum_{i,j=1}^{d} W_{i,j} \Phi(\rho) W_{i,j}^* 
   = \nidentity{K}. \]
The fact that this mixture of discrete Weyl operators mixes states in
this way is shown in Proposition~\ref{mu-prop-depol-mu} as part of the
proof that this operation is mixed-unitary.
This equation implies that, on the particular input $\identity{H \tprod K \tprod K}$ the
output of $\Phi'$ is given by $\identity{K}$, as required.
This channel is not unital, but it can be made so by adding an
additional output space of the correct dimension in which the output
state is always completely mixed.  This extra mixed state will affect
the minimum output entropy or the maximum output $p$-norm by a
constant depending on $d$, and so it will have no effect on
additivity or multiplicativity.

\index{unital!reduction to additivity of $S_{min}$}%
\index{unital!reduction multiplicativity of $\opv_p$}%
Fukuda proves the following result about this construction.
\begin{theorem}[Fukuda~\cite{Fukuda07}]
  Let $\Phi \in \transform{H,K}, \Psi \in \transform{X,Y}$ and let
  $\Phi'$ be the doubly stochastic channel constructed from $\Phi$ as
  in Equation~\eqref{mu-eqn-fukuda}.  For these channels, and any $p
  \in [1,\infty]$,
  \begin{align*}
    S_{\min}(\Psi \tprod \Phi) &= S_{\min}(\Psi \tprod \Phi') \\
    \norm{\Psi \tprod \Phi}_p &= \norm{\Psi \tprod \Phi'}_p.
  \end{align*}
  
  \begin{proof}
    Only a proof of the minimum output entropy case is provided, as
    the proof for the maximum output $p$-norm is identical, with the
    concavity of the entropy replaced by the triangle inequality.

    To see that $S_{\min}(\Psi \tprod \Phi) \geq S_{\min}(\Psi \tprod
    \Phi')$, notice that since $W_{0,0} = \id$, if the input in the space
    that controls the Weyl operations is given as $\ket{0,0}$, then
    \[
      (\Psi \tprod \Phi')(\rho \tprod \ket{0,0} \bra{0,0})
      = \identity{K} (\Psi \tprod \Phi)(\rho) \identity{K} 
      = (\Psi \tprod \Phi)(\rho), \]
    from which the desired inequality follows immediately.

    In the other direction, notice that since the channel $\Phi'$ can
    be assumed to immediately measure the control input, the channel
    $\Psi \tprod \Phi'$ can be written as the probabilistic
    application of one of the discrete Weyl operators to the output of
    $\Psi \tprod \Phi$.
    To this end, let $\rho$ be an input minimizing $S((\Psi \tprod
    \Phi')(\rho))$ and let the result of measuring the control input
    in the computational basis  be $\ket{i,j}$ with
    probability $p_{i,j}$, and let $\rho_{i,j}$ be the state after the
    measurement has produced outcome ${i,j}$.  In this notation we have
    \begin{align*}
      S_{\min}(\Psi \tprod \Phi')
      &= S((\Psi \tprod \Phi')(\rho)) \\
      &= S\left( \sum_{i,j} p_{i,j} (\id \tprod W_{i,j})
                            (\Psi \tprod \Phi)(\rho_{i,j})
                            (\id \tprod W_{i,j}^*) \right) \\
      &\geq \sum_{i,j} p_{i,j} S\left( (\id \tprod W_{i,j})
                            (\Psi \tprod \Phi)(\rho_{i,j})
                            (\id \tprod W_{i,j}^*) \right) \\
      &= \sum_{i,j} p_{i,j} S\left((\Psi \tprod \Phi)(\rho_{i,j})\right) \\
      &\geq S_{\min}(\Psi \tprod \Phi),
    \end{align*}
    where the concavity of the entropy and the unitary
    invariance of the entropy have been used.
  \end{proof}
\end{theorem}

In Sections~\ref{mu-scn-add} and~\ref{mu-scn-mult} similar results are
shown for the mixed-unitary channels, though the techniques used to
prove them do not seem to be directly related to Fukuda's construction in the
unital case.

\section{Mixed-unitary approximation}\label{mu-scn-channels}

Given a representation of a channel $\Phi$ in Stinespring form, that
is, an implementation of the form
\begin{equation}\label{mu-eqn-stinespring}
  \Phi(X) = \ptr{B} U (\ket 0 \bra 0 \tprod X) U^*,
\end{equation}
there are only two operations that are not mixed-unitary.  These are
the partial trace over the system $\mathcal{B}$ and the introduction
of the ancillary system in the state $\ket 0$.
The goal of this section is to describe a method for approximating
these two operations with mixed-unitaries, so that when combined with
the circuit for the operation $U$ in
Equation~\eqref{mu-eqn-stinespring}, the result is a mixed-unitary
approximation of $\Phi$.

Though this approximation does have an efficient circuit
implementation, the discussion of mixed-state quantum circuits is
postponed to Section~\ref{mu-scn-circuits} as this allows the
construction to be described in simpler terms.  Additionally, some of
the applications of this simulation technique do not depend on
efficient circuit implementations, so this simplified exposition is
useful for readers not interested in computational complexity.
Despite this avoidance of the quantum circuit model, the figures in
this will use circuit diagrams, but this is done only for clarity: no
assumptions are made regarding implementations of the channels
depicted.

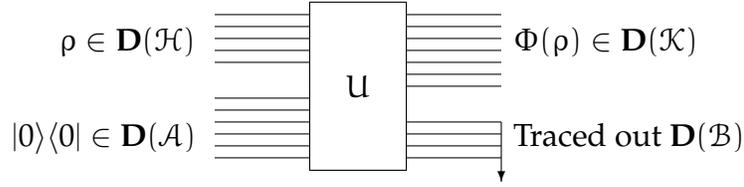
\begin{figure}
  \begin{center}
    \setlength{\unitlength}{3947sp}%
\begingroup\makeatletter\ifx\SetFigFont\undefined%
\gdef\SetFigFont#1#2#3#4#5{%
  \reset@font\fontsize{#1}{#2pt}%
  \fontfamily{#3}\fontseries{#4}\fontshape{#5}%
  \selectfont}%
\fi\endgroup%
\begin{picture}(5424,1374)(-1211,-1273)
\put(2476,-361){\makebox(0,0)[lb]{\smash{{\SetFigFont{12}{14.4}{\familydefault}{\mddefault}{\updefault}{\color[rgb]{0,0,0}$\Phi(\rho) \in \density{K}$}%
}}}}
\thinlines
{\color[rgb]{0,0,0}\put(601,-211){\line( 1, 0){600}}
}%
{\color[rgb]{0,0,0}\put(601,-286){\line( 1, 0){600}}
}%
{\color[rgb]{0,0,0}\put(601,-361){\line( 1, 0){600}}
}%
{\color[rgb]{0,0,0}\put(601,-436){\line( 1, 0){600}}
}%
{\color[rgb]{0,0,0}\put(1801,-211){\line( 1, 0){600}}
}%
{\color[rgb]{0,0,0}\put(1801,-286){\line( 1, 0){600}}
}%
{\color[rgb]{0,0,0}\put(1801,-361){\line( 1, 0){600}}
}%
{\color[rgb]{0,0,0}\put(1801,-436){\line( 1, 0){600}}
}%
{\color[rgb]{0,0,0}\put(1801,-511){\line( 1, 0){600}}
}%
{\color[rgb]{0,0,0}\put(601,-811){\line( 1, 0){600}}
}%
{\color[rgb]{0,0,0}\put(601,-886){\line( 1, 0){600}}
}%
{\color[rgb]{0,0,0}\put(601,-961){\line( 1, 0){600}}
}%
{\color[rgb]{0,0,0}\put(601,-1036){\line( 1, 0){600}}
}%
{\color[rgb]{0,0,0}\put(601,-136){\line( 1, 0){600}}
}%
{\color[rgb]{0,0,0}\put(1801,-136){\line( 1, 0){600}}
}%
{\color[rgb]{0,0,0}\put(1801,-1036){\line( 1, 0){600}}
}%
{\color[rgb]{0,0,0}\put(1801,-961){\line( 1, 0){600}}
}%
{\color[rgb]{0,0,0}\put(1801,-886){\line( 1, 0){600}}
}%
{\color[rgb]{0,0,0}\put(1801,-811){\line( 1, 0){600}}
}%
{\color[rgb]{0,0,0}\put(601,-736){\line( 1, 0){600}}
}%
{\color[rgb]{0,0,0}\put(601,-661){\line( 1, 0){600}}
}%
{\color[rgb]{0,0,0}\put(1801,-586){\line( 1, 0){600}}
}%
{\color[rgb]{0,0,0}\put(2401,-811){\vector( 0,-1){375}}
}%
\put(2476,-961){\makebox(0,0)[lb]{\smash{{\SetFigFont{12}{14.4}{\familydefault}{\mddefault}{\updefault}{\color[rgb]{0,0,0}Traced out $\density{B}$}%
}}}}
\put(-374,-361){\makebox(0,0)[lb]{\smash{{\SetFigFont{12}{14.4}{\familydefault}{\mddefault}{\updefault}{\color[rgb]{0,0,0}$\rho \in \density{H}$}%
}}}}
\put(-674,-961){\makebox(0,0)[lb]{\smash{{\SetFigFont{12}{14.4}{\familydefault}{\mddefault}{\updefault}{\color[rgb]{0,0,0}$\ket 0 \bra 0 \in \density{A}$}%
}}}}
{\color[rgb]{0,0,0}\put(1201,-1111){\framebox(600,1050){$U$}}
}%
\end{picture}%
  \end{center}
  \caption[Channel to be approximated by a mixed-unitary]{The channel
    $\Phi$ to be approximated by a mixed-unitary, in Stinespring
    form with labelled spaces.}
  \label{mu-fig-spaces}
\end{figure}

To describe this approximation we fix notation throughout the
next three sections.  To this end let $\Phi \in \transform{H,K}$ be a
quantum channel, and
let a Stinespring representation for $\Phi$ be as given in
Equation~\ref{mu-eqn-stinespring}.  In this representation let $\mathcal{A}$ be the
space containing the ancillary space starting in the $\ket 0$ state, and let
$\mathcal{B}$ be the space that is traced out.  This implies that the
operator $U$ is a unitary map from $\mathcal{A \tprod H}$ to $\mathcal{K \tprod B}$.
These spaces are summarized in Figure~\ref{mu-fig-spaces}.

\subsection{Simulating the partial trace}\label{mu-scn-ptrace}

Of the two operations in Equation~\eqref{mu-eqn-stinespring} that are
not mixed-unitary, the partial trace is the easiest to simulate with a
mixed-unitary channel.  At an intuitive level, the partial trace
represents the loss of information to the environment in a quantum
channel, but this operation is not mixed-unitary as it changes the
dimension of the system being considered.  The direct approach to
simulating this with a mixed-unitary is to model the loss of
information with a completely depolarizing channel, which avoids the
issue of the change in dimensionality.  It is not hard to prove that
this approach works, nor is it hard to see that the completely
depolarizing channel is mixed-unitary.

This depolarizing channel can be implemented as a mixture of the
discrete Weyl operators, which are also known as the generalized Pauli
operators~\cite{AmbainisM+00,BoykinR03,HaydenL+04}.  These unitary operators,
as discussed in Chapter~\ref{chap-intro}, form an orthogonal basis for
the space $\linear{A}$ of linear operators on a Hilbert space
$\mathcal{A}$ of dimension $d$.

\begin{proposition}\label{mu-prop-depol-mu}
  The completely depolarizing channel on $\mathcal{A}$ has
  implementation as a mixed-unitary channel given by
  \[ N(\rho) 
     = \frac{1}{d^2} \sum_{a,b = 0}^{d-1} W_{a,b} \rho W_{a,b}^* 
     = \nidentity{A} \]
  \begin{proof}
    Let $\rho \in \density{A}$ be a density matrix and let $d = \dim \mathcal{A}$.  By
    Equation~\ref{intro-eqn-weyl-basis} the operators $W_{a,b}$ form a
    basis of $\linear{A}$, so that $\rho$ can be decomposed as
    \begin{equation}\label{mu-rho-decomposed-weyl}
      \rho = \sum_{e,f=1}^d \lambda_{e,f} W_{e,f},
    \end{equation}
    for some coefficients $\lambda_{e,f} \in \mathbb{C}$.  Notice also that since
    $W_{a,b}$ has trace zero unless $a = b = 1$ and $W_{0,0} =
    \identity{A}$, it is the case that
    \[ \lambda_{0,0} = \frac{\tr \rho}{\tr \identity{A}} = \frac{1}{d}. \]
    Putting this
    decomposition into the proposed implementation, we obtain
    \begin{align*}
      \frac{1}{d^2} \sum_{a,b} W_{a,b} \rho W_{a,b}^*
      &= \frac{1}{d^2} \sum_{a,b,e,f} \lambda_{e,f} W_{a,b} W_{e,f} W_{a,b}^*.
    \end{align*}
    Using Equation~\ref{intro-eqn-weyl-commute} and the unitarity of
    the discrete Weyl operators to manipulate this sum gives
    \begin{align*}
      \frac{1}{d^2} \sum_{a,b,e,f} \lambda_{e,f} \omega^{be - af} W_{e,f}
      &= \sum_{e,f} \lambda_{e,f} \left( \sum_{a,b} \omega^{be - af} \right) W_{e,f}.
    \end{align*}
    Since $\omega$ is a primitive $d$th root of unity, this inner
    summation is zero unless $e = f = 0$,
    and so we have shown that
    \begin{align*}
      \frac{1}{d^2} \sum_{a,b} W_{a,b} \rho W_{a,b}^*
      &= \frac{1}{d^2} \lambda_{0,0} \sum_{a,b=0}^{d-1} W_{0,0}
        = \frac{1}{d^3} \sum_{a,b=0}^{d-1} \identity{A}
        = \nidentity{A}
        = N(\rho),
    \end{align*}
    as required.
  \end{proof}
\end{proposition}

This proves that the channel $N_{\mathcal{B}}$ that completely
depolarizes the space $\mathcal{B}$ can be implemented as a
mixed-unitary channel.  To see that this channel can be used to
replace the partial trace observe that one implementation of this
channel simply traces out the state in $\mathcal{B}$ and replaces it
with the state $\nidentity{B}$ that has been separately prepared.
From this implementation it is clear that for a state $\rho \in
\density{A \tprod B}$ it holds that
\begin{equation}\label{mu-eqn-traceout}
  N_{\mathcal{B}}(\rho) = \left( \ptr{B} \rho \right) \tensor \nidentity{B},
\end{equation}
and this property will hold no matter how $N_{\mathcal{B}}$ is implemented.
This implies that if the system to be traced out instead has
$N_{\mathcal{B}}$ applied to it, the resulting state is the same, up to
a tensor factor of a maximally mixed state in the space
$\mathcal{B}$.  By replacing the partial trace over $\mathcal{B}$ in
Equation~\ref{mu-eqn-stinespring} with this channel, the result is
\[ N_{\mathcal{B}}(U (\ket 0 \bra 0 \tprod X) U^*) = \Phi(X) \tprod
\nidentity{B}. \]
This is the best that can be hoped for, as a mixed-unitary
transformation cannot change the dimension of the system it acts on.

\subsection{Simulating the ancillary space}\label{mu-scn-ancilla}

Replacing the introduction of the ancillary space $\mathcal{A}$ with a
mixed-unitary operation is more complicated than replacing the partial
trace.  In order to do this
the input space of the transformation is to expanded to include the space
$\mathcal{A}$.  The input state of this system will not, in general,
be the desired state $\ket 0$, so additional operations are needed
to ensure that this is the case for any input state that either
maximizes distinguishability or minimizes the output entropy of the
resulting channel.

Because these quantities of interest, the minimum output
entropy and the maximum output $p$-norm, involve optimizing over input
states, the channel can be constructed so that
those inputs that achieve the optimal value have the desired property:
the input state in the space $\mathcal{A}$ is (close to) the state
$\ket 0$.  Given this property, the values of these optimizations will
stay approximately the same when taken over the mixed-unitary
simulations of the original channels.

To this end, the ideal operation $\Lambda$ to ensure this condition does not
alter any input state of the form $\ket 0 \bra 0 \tprod \sigma$, but
takes any orthogonal state to the completely mixed state $\nidentity{A
  \tprod H}$.  This operation is, unfortunately, not mixed-unitary,
as it is not unital, since
\begin{equation}\label{mu-eqn-M-nonideal}
  \Lambda( \identity{A \tprod H} ) 
    = \frac{1}{\dim \mathcal{A}} \ket 0 \bra 0 \tprod \identity{H}
    + \left( 1 - \frac{1}{\dim \mathcal{A}} \right) \identity{A \tprod H}.
\end{equation}
  Notice, however, that this channel deviates from unitality with
additive error $1 / \dim \mathcal{A}$: there is a very good unital,
and indeed mixed-unitary, approximation to this ideal channel, which
is described in the remainder of this section.

This closely related mixed-unitary channel first projects the
input state either onto the subspace $S_0 = \ket 0 \tprod \mathcal{H}$
or the orthogonal subspace $S_0^\perp = \ket{0}^\perp \tprod
\mathcal{H}$.  This projection is then be followed by a completely
depolarizing channel on the subspace $S_0^\perp$.  These operations
can be implemented using mixed-unitary channels, and the distance from
the ideal channel will go as $O(1 / \dim \mathcal{A})$, which allows
the error to be made arbitrarily small by padding $\mathcal{A}$ with
an unused ancillary space.

The mixing process on $S_0^\perp$ is be introduced first.  It is given by
the channel $M$ that does not affect
the subspace $S_0$ but completely depolarizes the space $S_0^\perp$.
More concretely, on a state $\rho = q \rho_{S_0} + (1-q)
\rho_{S_0^\perp}$ where $\rho_{S_0} = \ket 0 \bra 0 \tprod \sigma$ is
a density operator on $S_0$ and $\rho_{S_0^\perp}$ a
density operator on $S_0^\perp$, the output of $M$ is given by
\begin{align}
  M(\rho) = q M(\rho_{S_0}) + (1-q) M(\rho_{S_0^\perp})
          &= q \rho_{S_0} + (1-q) \tilde{\id}_{S_0^\perp} \nonumber \\
          &= q \ket 0 \bra 0 \tprod \sigma
            + (1-q) \frac{\identity{A} - \ket 0 \bra 0}{\dim \mathcal{A}-1}
              \tprod \nidentity{H}. \label{mu-eqn-mixing-approx}
\end{align}
Here notation has been abused somewhat: $S_0$ and $S_0^\perp$ are
Hilbert spaces, but the whole space $\mathcal{A \tprod H}$ is \emph{not} the
tensor product of these two spaces.

The channel $M$ can be implemented as a mixed-unitary channel in the
same way as the completely depolarizing channel: a uniform mixture of
the discrete Weyl operators, except here these operators are taken
over the subspace $S_0^\perp$.  These operators exist, and the whole
construction is very similar to the one given in
Proposition~\ref{mu-prop-depol-mu}.  More concretely, where $W_{a,b}$ for
$a,b \in \mathbb{Z}_d$ are the discrete Weyl operators on the space
$S_0^\perp$ and ${\id}_{S_0}$ is the identity on the space
$S_0$, the channel $M$ can be implemented as
\[ M(\rho) = \frac{1}{d^2} \sum_{a,b = 0}^{d-1} 
                    ( {\id}_{S_0} \oplus W_{a,b}) 
                    \rho
                    ( {\id}_{S_0} \oplus W_{a,b})^*, \]
where all of the operators ${\id}_{S_0} \oplus W_{a,b}$ are
mixed-unitary by construction.

As previously mentioned, this channel does not implement the ideal transformation.
If the output of $M$ on
$\rho_{S_0^\perp}$ in Equation~\eqref{mu-eqn-mixing-approx} were the
completely mixed state on $\mathcal{A \tprod H}$ and not the subspace 
$S_0^\perp$ then this process would create an essentially
error-free mixed-unitary approximation of the original channel (for
the purpose of minimizing the output entropy or maximizing distinguishability). Fortunately, the error
involved at this step can be shown, in
Lemma~\ref{mu-lem-ancilla-mixing}, to be $O(1 / \dim{\mathcal{A}})$,
which, by Equation~\ref{mu-eqn-M-nonideal}, is as close as a
mixed-unitary channel can come to the ideal case.
Fortunately this error can be made arbitrarily small by taking the
space $\mathcal{A}$ large enough, and so this construction can be used
to approximate the ideal case.

It will be helpful for the analysis of this construction to remove the
coherences between the subspaces $S_0$ and $S_0^\perp$.  The channel
$M$ does perform this operation.  This is the operation commonly known
as dephasing that, applied to a density matrix expressed in some
basis, removes the off-diagonal terms.  This aids the analysis of the
construction, because once this dephasing is applied, an equation
similar to Equation~\eqref{mu-eqn-mixing-approx} would hold for all
input states $\rho$, not just those states that have no entanglement
between the subspaces $S_0$ and $S_0^\perp$.

While we will only need to apply dephasing between the two subspaces
$S_0$ and $S_0^\perp$, a mixed-unitary construction for the general
case is provided below.
This is the channel $D$ that completely
decoheres all information not stored in the computational basis.  More
specifically, this channel implements
\begin{equation}\label{mu-eqn-dephasing}
  D(\ket i \bra j) = \delta_{ij} \ket{i} \bra{j} = 
  \begin{cases}
    \ket i \bra i & \text{if $i=j$}, \\
    0 & \text{otherwise.}
  \end{cases}
\end{equation}
That this channel is mixed-unitary is simple to prove, using a
construction based on the discrete Weyl operators, similar to that
used for the complete depolarizing channel in
Proposition~\ref{mu-prop-depol-mu}.  That this channel can be
implemented in this way has been observed in~\cite{DattaF+06}.

\begin{proposition}\label{mu-prop-dephasing-mu}
  The completely dephasing channel on $\mathcal{A}$ defined in
  Equation~\eqref{mu-eqn-dephasing} has implementation as a
  mixed-unitary channel given by
  \[ D(\rho) 
     = \frac{1}{d} \sum_{b = 0}^{d-1} W_{0,b} \rho W_{0,b}^* \]
  \begin{proof}
    Recall that $W_{0,b} \ket j = Z^b \ket j = \omega^{bj} \ket j$ as
    introduced in Chapter~\ref{chap-intro}, where $\omega$ is a $d$th
    primitive root of unity, with $d = \dim \mathcal{A}$.  To see that
    this channel has the desired effect, let $\rho = \sum_{i,j = 0}^{d-1}
    a_{ij} \ket i \bra j$, so that
    \begin{equation}\label{mu-eqn-dephasing-impl-1}
      \frac{1}{d} \sum_{b = 0}^{d-1} W_{0,b} \rho W_{0,b}^*
      = \frac{1}{d} \sum_{b = 0}^{d-1} \sum_{i,j=0}^{d-1} a_{i,j} Z^b \ket i \bra j Z^{-b}
      = \frac{1}{d} \sum_{b = 0}^{d-1} \sum_{i,j=0}^{d-1} a_{i,j} \omega^{(i-j)b} \ket i \bra j.
    \end{equation}
     Then, since $\omega$ is a $d$th root of unity
    \begin{equation*}
      \sum_{b=0}^{d-1} \omega^{(i-j)b} = d \delta_{ij} 
      = \begin{cases}
        d & \text{if $i=j$}, \\
        0 & \text{otherwise}.
      \end{cases}
    \end{equation*}
    Combining this property with
    Equation~\eqref{mu-eqn-dephasing-impl-1} gives
    \begin{equation*}
      \frac{1}{d} \sum_{b = 0}^{d-1} W_{0,b} \rho W_{0,b}^*
      = \frac{1}{d} \sum_{i=0}^{d-1} d a_{i,i} \ket i \bra i
      = \sum_{i=0}^{d-1} a_{i,i} \ket i \bra i
      = D(\rho),
    \end{equation*}       
    which is exactly the channel defined by Equation~\eqref{mu-eqn-dephasing}.
  \end{proof}
\end{proposition}

For the specific case that we will use here, a simpler construction
suffices: instead of applying this dephasing channel to the whole of
the ancillary space $\mathcal{A}$ it only needs to be applied to
remove coherences between the two orthogonal subspaces $S_0$ and
$S_0^\perp$.  If these two subspaces are viewed as a two-dimensional
Hilbert space, the construction in
Proposition~\ref{mu-prop-dephasing-mu} can be reduced to the
application of a specific unitary operator $V$ with probability one-half.
The action of this unitary $V$ on basis states is given by
\begin{align}
  V \ket i &= \begin{cases}
    \ket i  & \text{if $\ket i \in S_0$},\\
    -\ket i & \text{if $\ket i \in S_0^\perp$}.
  \end{cases}\label{mu-eqn-unitary-subspace-dephasing}
\end{align}
In other words, $V$ applies a phase of $-1$ to states in
$S_0^\perp$ and does not change states in $S_0$.  When $V$ is applied
with probability one half the result is complete dephasing between the
two subspaces.  This can be seen by restricting the construction in
Proposition~\ref{mu-prop-dephasing-mu} to the case of a
two-dimensional system with orthogonal states that
represent the subspaces $S_0$ and $S_0^\perp$.
More concretely, when this is applied to a density matrix expressed in the
computational basis, the result is, by a simple calculation, the
zeroing of the off-diagonal elements of the first row and column.
Let this simplified dephasing channel be given by
\begin{equation*} 
  D_{S_0}(\rho) = \frac{1}{2} \left[ V \rho V^* + \rho \right],
\end{equation*}
where the operator $V$ is given in Equation~\eqref{mu-eqn-unitary-subspace-dephasing}.
When this operation is applied to a density operator $\rho \in
\density{A \tprod H}$, the result is
\begin{align}\label{mu-eqn-decoherence}
  D_{S_0}(\rho) = q \rho_{S_0} + (1-q) \rho_{S_0^\perp}
          = q \ket 0 \bra 0 \tprod \sigma + (1-q) \rho_{S_0^\perp},
\end{align}
where $\rho_{S_0} = \ket 0 \bra 0 \tprod \sigma$ is a density operator on
the subspace $S_0 = \ket 0 \tprod \mathcal{H}$, $\rho_{S_0^\perp}$ is a density
operator on the orthogonal subspace $S_0^\perp$, and $0 \leq q \leq 1$
is a probability.

Combining Equations~\eqref{mu-eqn-mixing-approx}
and~\eqref{mu-eqn-decoherence}, the output of $D_{S_0}$ followed by $M$ on a
density operator $\rho$ on $\mathcal{A \tprod H}$ is given by a state
of the form
\begin{align*}
  (M \circ D_{S_0})( \rho ) &= q M(\ket 0 \bra 0 \tprod \sigma) + (1-q)
  M(\rho_{S_0^\perp}) \\
                  &= q \ket 0 \bra 0 \tprod \sigma
            + (1-q) \frac{\identity{A} - \ket 0 \bra 0}{\dim \mathcal{A}-1}
              \tprod \nidentity{H}.
\end{align*}
This operation, $M \circ D_{S_0}$, will be used as a way to force any input
that results in a low output entropy to be close to the subspace $S_0$
of inputs having the `ancilla' space $\mathcal{A}$ in the desired
$\ket 0$ state.  On these inputs the constructed mixed-unitary
channel will behave in a similar way to the original channel that is
being approximated.  On inputs that are far from this subspace, the
resulting state has high entropy, and so it will not be close to a
state minimizing the output entropy and it will not be useful for
distinguishing two channels constructed in this way.

\subsection{Mixed-unitary approximation of a general channel}

Putting these pieces together, given a channel $\Phi(\rho) =
\ptr{B} U (\rho \tprod \ket 0 \bra 0) U^*$, the mixed-unitary
approximation $\Phi'$ is constructed as
\begin{equation}\label{mu-eqn-construction}
  \Phi'(\rho) =   
  N_\mathcal{B} \left( 
    U \left[ 
      (M \circ D_{S_0})(\rho) 
    \right] U^* 
  \right),
\end{equation}
which, more plainly, is simply the application of the ancilla
simulation procedure of Section~\ref{mu-scn-ancilla}, the unitary operation from a Stinespring
dilation of $\Phi$, and finally the completely mixing channel to the
space that would have been traced out by $\Phi$, as discussed in
Section~\ref{mu-scn-ptrace}.  As the mixed-unitary channels are closed
under composition, the channel $\Phi'$ is mixed-unitary.

It will be useful to observe that the constructed channel $\Phi'$ specified in
Equation~\eqref{mu-eqn-construction} can be used to simulate the
original channel $\Phi$.  This occurs when the input $\ket 0 \bra 0 \tprod \sigma$,
i.e.\ an input in the space $S_0$, is provided to $\Phi'$.
This is argued in the following proposition.

\begin{proposition}\label{mu-prop-simulation}
  Let $\Phi \in \transform{H,K}$.  
  If $\Phi' \in \transform{A \tprod H, K \tprod B}$ is the
  mixed-unitary channel that is constructed from $\Phi$ in
  Equation~\eqref{mu-eqn-construction}, then
  \[ \Phi'(\ket 0 \bra 0 \tprod \sigma) = \Phi(\sigma) \tprod \nidentity{B}. \]  

  \begin{proof}
    Notice that both $D_{S_0}$ and $M$ do not
    affect this input: the decoherence operation $D_{S_0}$ does not affect
    the state as it is in the subspace $S_0$ and 
    $M$ does not affect the state by
    Equation~\eqref{mu-eqn-mixing-approx}.  Thus, the output of the
    channel $\Phi'$ is
    \begin{align*}
      \Phi'(\ket 0 \bra 0 \tprod \sigma) 
      &= N_\mathcal{B} \left( 
          U \left[ 
            (M \circ D_{S_0})(\ket 0 \bra 0 \tprod \sigma) 
          \right] U^* 
        \right)  \\
      &= N_\mathcal{B} \left( 
        U (\ket 0 \bra 0 \tprod \sigma) U^* 
        \right)  \\
      &= \ptr{B} \left( 
        U (\ket 0 \bra 0 \tprod \sigma) U^* 
        \right) \tprod \nidentity{B} \\
      &= \Phi(\sigma) \tprod \nidentity{B},
    \end{align*}
    where the penultimate equality is an application of
    Equation~\eqref{mu-eqn-traceout}. 
  \end{proof}
\end{proposition}

Combining this proposition with Equation~\eqref{mu-eqn-decoherence}
that demonstrates the effect of the $M \circ D_{S_0}$ on states not of this
form, and the observation that applying $M \circ D_{S_0}$
twice has no further effect than applying it once, the output of
$\Phi'$ on an arbitrary input state $\rho$ is given by
\begin{equation}\label{mu-eqn-output}
  \Phi'(\rho)
  = p \Phi'(\ket 0 \bra 0 \tprod \sigma) + (1-p) \Phi'(\rho_{S_0^\perp})
  = p \Phi(\sigma) \tprod \nidentity{B}  + (1-p) \Phi'(\rho_{S_0^\perp}),
\end{equation}
where as in Equation~\eqref{mu-eqn-decoherence} $\rho_{S_0^\perp}$ is a density
operator on the subspace $S_0^\perp$ of inputs orthogonal to those
with the state $\ket 0$ on the space $\mathcal{A}$.
The most significant portion of the technical results in the next section lies in
bounding the distance from the maximally mixed state of the second term in
this equation, from which most of the results will follow straightforwardly.

\section{Properties of the constructed channel}\label{mu-scn-properties}

This section provides the basis for the analysis of the mixed-unitary
approximation constructed in the previous section.  The main result is
a lower bound on the output entropy when the constructed channel is
applied to a state in $S_0^\perp$, the subspace of inputs where the
`ancillary' subspace $\mathcal{A}$ is not in the desired $\ket 0$
state.  This result is not difficult to show, but it will be essential
to the results that follow.

Throughout this section, and the two sections that follow, the channel
$\Phi$ will represent an arbitrary transformation, and $\Phi'$ will
represent the mixed-unitary transformation constructed from it, as in
Equation~\eqref{mu-eqn-construction}.  The names of the Hilbert spaces
that $\Phi$ acts on will be consistent with the previous section:
$\Phi$ maps mixed states on $\mathcal{H}$ to $\mathcal{K}$, using
the ancillary system $\mathcal{A}$ and tracing out the system
$\mathcal{B}$.  The constructed channel $\Phi'$ is mixed-unitary, mapping density
matrices on $\mathcal{A \tprod H}$ to $\mathcal{K \tprod B}$.

As a first step to showing that $\Phi'$ approximates $\Phi$ it is shown
that mixed-unitary channels do not increase the distance of a state
from the completely mixed state.  This lemma can be interpreted as the
statement that
the output of a mixed-unitary channel is not more pure than the
input.  The Hilbert space $\mathcal{B}$ appearing in this lemma will
correspond to a reference system needed for the results in 
Section~\ref{mu-scn-distinguish} -- this generality will not be needed
for the results on the maximum output $p$-norm or the minimum output entropy.

\begin{lemma}\label{mu-lem-ru-dist-noise}
  Let $\threenorm{\cdot}$ be a unitarily invariant norm on $\linear{A
  \tprod B}$.  If $\Psi \in \transform{A}$ is mixed-unitary,
  then for any $\rho \in \density{A \tprod B}$
  \[ \smthreenorm{ (\Psi \tprod \tidentity{B})(\rho) 
             - \nidentity{A} \tprod \ptr{A} \rho}
  \leq \smthreenorm{ \rho
              - \nidentity{A} \tprod \ptr{A} \rho} \]
  \begin{proof}
    As $\Psi$ is mixed-unitary, let $\Psi(X) = \sum_i p_i U_i X
    U_i^*$ with the $U_i$ unitary, $0 \leq p_i \leq 1$, and $\sum_i
    p_i = 1$.  For brevity, let $\hat{U}_i = U_i \tprod
    \identity{B}$ for all $i$.  Using this notation
    \begin{align}
      \smthreenorm{ (\Psi \tprod \tidentity{B})(\rho) 
        -  \nidentity{A} \tprod \ptr{A} \rho}
      &= \smthreenorm{ \sum_i p_i 
        \hat{U}_i \rho \hat{U}_i^*
        -  \nidentity{A} \tprod \ptr{A} \rho} \nonumber \\
      &\leq \sum_i p_i \smthreenorm{ 
        \hat{U}_i \rho \hat{U}_i^*
        -  \nidentity{A} \tprod \ptr{A} \rho}. \label{mu-eqn-lem-ru-dist-noise}
    \end{align}
    Notice that $U_i \nidentity{A} U_i^* = \nidentity{A}$, which
    implies that
    $ \hat{U}_i (\nidentity{A} \tprod \sigma) \hat{U}_i^* 
        =\nidentity{A} \tprod \sigma. $
    Using this fact, as well as the unitary invariance of the norm,
    Equation~\eqref{mu-eqn-lem-ru-dist-noise} becomes
   \begin{align*}
      \sum_i p_i \smthreenorm{ \hat{U}_i ( \rho -  \nidentity{A} \tprod
        \ptr{A} \rho) \hat{U}_i^*}
      &= \sum_i p_i \smthreenorm{\rho -  \nidentity{A} \tprod \ptr{A} \rho} 
        =  \smthreenorm{ \rho -  \nidentity{A} \tprod \ptr{A} \rho}.
    \end{align*}
    Combining this with Equation~\eqref{mu-eqn-lem-ru-dist-noise}
    yields the statement of the lemma.     
  \end{proof}
\end{lemma}

This lemma will be used to show not only that the ancilla simulation procedure
sends states in the subspace $S_0^\perp$ of states where
the ancillary space is not in the $\ket 0$ state
to states that are highly mixed, but that the channel $\Phi'$
also has this behaviour.  Before doing this, however, the lemma is
extended to the case of the von Neumann entropy, where the proof is essentially
identical, with the exception that the triangle inequality is replaced
by concavity.

\begin{corollary}\label{mu-cor-ru-inc-entropy}
  If $\Psi \in \transform{A}$ is mixed-unitary, and $\rho \in
  \density{A}$, then
  \[ S(\rho) \leq S(\Psi(\rho)). \]

  \begin{proof}
    Let $\Psi(\rho) = \sum_i p_i U_i \rho U_i^*$ as in the proof of
    Lemma~\ref{mu-lem-ru-dist-noise}.
    Using this
    notation, and the concavity of the von Neumann entropy
    \begin{align*}
      S(\Psi(\rho)) 
      = S \left( \sum_i p_i U_i \rho U_i^* \right)
      \geq \sum_i p_i S(U_i \rho U_i^*)
      = \sum_i p_i S(\rho)
      = S(\rho),
    \end{align*}
    where the penultimate equality is due to the unitary invariance of the entropy.
 \end{proof}
\end{corollary}

The next lemma shows that when the input is in the subspace $S_0^\perp$
the output of $\Phi'$ is
very close to completely mixed.  The distance measure used is the
trace norm, but this can be applied also to the case of the
maximum output $p$-norm due to the fact that $\tnorm{\rho} =
\norm{\rho}_1 \geq \norm{\rho}_p$ for all $p \in [1,\infty]$.
This is the key lemma in the proof of the results on the additivity and
multiplicativity conjectures, though it is not difficult to prove.

\begin{lemma}\label{mu-lem-ancilla-mixing}
  On input states $\rho \in S_0^\perp$ the output of the channel
  $\Phi'$ given in Equation~\eqref{mu-eqn-construction} satisfies
  \[\tnorm{\Phi'(\rho) - 
     \nidentity{K \tprod B}}
     \leq \frac{2}{\dim \mathcal{A}}.\]

  \begin{proof}
    On input $\rho \in S_0^\perp$
    the operation $D_{S_0}$ that introduces decoherence between the
    subspaces $S_0$ and
    $S_0^\perp$ has no effect.  This implies that the output of $M
    \circ D_{S_0}$ on $\rho$ is obtained by setting $q = 0$ in
    Equation~\eqref{mu-eqn-mixing-approx}, which is
    \begin{equation}\label{mu-eqn-state-after-mixing}
      M(D_{S_0}(\rho)) = \frac{\identity{A} - \ket 0 \bra 0}{\dim \mathcal{A} - 1} 
      \tprod \nidentity{H},
    \end{equation}
    Setting $d = \dim \mathcal{A}$, the distance from 
    from the completely mixed state on $\mathcal{A \tprod H}$ is
    \begin{align}
      \tnorm{ \frac{\identity{A} - \ket 0 \bra 0}{d - 1} 
              \tprod \nidentity{H}
            -  \frac{\identity{A}}{d} 
              \tprod \nidentity{H} }
      &= \tnorm{ \frac{\identity{A} - d \ket 0 \bra 0}{d(d-1)}} \nonumber \\
      & \leq \frac{d-1}{d(d-1)} + \frac{d-1}{d(d-1)}
      = \frac{2}{d}. \label{mu-eqn-norm-after-mixing}
    \end{align}
    Finally, by noting that the remainder of the transformation
    $\Phi'$ is mixed-unitary and (implicitly) using the isomorphism
    $\mathcal{A \tprod H} \cong \mathcal{K \tprod B}$, an application
    of Lemma~\ref{mu-lem-ru-dist-noise} yields the desired bound.
  \end{proof}
\end{lemma}

Once again we can extend this result to the case of the
entropy.  The previous lemma on the trace norm can be extended to the
case of the entropy in the standard way: using Fannes'
inequality~\cite{Fannes73} (see also~\cite{NielsenC00}), but
given the characterization in
Equation~\eqref{mu-eqn-state-after-mixing}, a better bound can be
obtained by explicitly computing the entropy.  This bound will require
that $\dm{A} \geq 2$, but this can be assumed without loss of
generality by adding an unused ancillary space.

\begin{corollary}\label{mu-cor-mixing-entropy}
  Let $\Phi'$ be given as in Equation~\eqref{mu-eqn-construction}, and let
  $\dm{A} \geq 2$ and $\rho \in S_0^\perp$, then
  \[ S(\Phi'(\rho)) \geq S(\nidentity{K \tprod B}) - \frac{1}{\dm{A}}. \]

  \begin{proof}
    In the proof of Lemma~\ref{mu-lem-ancilla-mixing}, the output
    of $M \circ D_{S_0}$ on $\rho$ is given by
    Equation~\eqref{mu-eqn-state-after-mixing}, which states that
    \begin{equation*}
      M(D_{S_0}(\rho)) = \frac{\identity{A} - \ket 0 \bra 0}{\dim \mathcal{A} - 1} 
      \tprod \nidentity{H}.
    \end{equation*}
    Letting $d = \dim{A}$, this state has $(d-1)(\dm{H})$ eigenvalues
    and each with value equal to $1 / ((d-1)(\dm{H}))$.  Using this observation, the
    entropy of this state can be computed as
    \begin{align}
      S\left(\frac{\identity{A} - \ket 0 \bra 0}{\dim \mathcal{A} - 1}
        \tprod \nidentity{H}\right)
        &= \log ((d-1) (\dm{H})) \nonumber \\
        &= \log \dm{H} + \log \dm{A} + \log\left( \frac{d-1}{d}\right) \nonumber \\
        &= S(\nidentity{A \tprod H}) - \log\left( 1 +\frac{1}{d-1}\right).
              \label{mu-eqn-cor-mixing-entropy}
    \end{align}
    For $d\geq 2$, the last term has Taylor expansion given by
    \[ \log\left( 1 + \frac{1}{d-1}\right)
        = \frac{1}{\log e} \left[ \frac{1}{d-1} - \frac{1}{2(d-1)^2} +
          \frac{1}{3(d-1)^3} - \cdots \right]
        \leq \frac{\log_e 2}{(d-1) \log e}
        \leq \frac{1}{d}. \]        
    Combining this with Equation~\eqref{mu-eqn-cor-mixing-entropy}
    gives the lower bound on the entropy provided in the statement
    of the corollary, for the state after ancilla simulation
    procedure.             
    By Corollary~\ref{mu-cor-ru-inc-entropy} applying the remainder of
    $\Phi'$ to this state cannot decrease the entropy, as
    this portion of $\Phi'$ is mixed-unitary.
 \end{proof}
\end{corollary}

This corollary and Lemma~\ref{mu-lem-ancilla-mixing} show that when
the input state to $\Phi'$ does not have large overlap with a state in
$S_0$, the output state is highly mixed.  This property will be used
to show that any state that maximizes the output norm or minimizes the
output entropy will have large overlap with the space $S_0$ of states
of the form $\ket 0 \bra 0 \tprod \sigma$, where the simulation of the
original channel is faithful.

\section{Multiplicativity of mixed-unitary transformations}\label{mu-scn-mult}

In this section the construction of Section~\ref{mu-scn-channels} is used
to show that the maximum output $p$-norm of a channel is
multiplicative if and only if the mixed-unitary approximations to it
are also multiplicative.  This will be done for all $1 \leq p <
\infty$, using the analysis of the previous section.

This property is not difficult to show once it has been established
that the mixed-unitary channel $\Phi'$ constructed from $\Phi$ in
Equation~\eqref{mu-eqn-construction} is a good approximation with
respect to the $p$-norm.  This is the content of the following theorem.

\begin{theorem}\label{mu-thm-approx}
  If $\Phi \in \transform{H,K}$, then the mixed-unitary
  $\Phi' \in \transform{A \tprod H, K \tprod B}$
  satisfies
  \[ \opv_p (\Phi)
     \leq \frac{\opv_p(\Phi')}{\smallnorm{\nidentity{B}}_p}
     \leq \opv_p(\Phi) + \frac{2 \dm{B}}{\dm{A}}. \]
  \begin{proof}
    For convenience, let $d = \dm{A}$.
    The first inequality is simple:
    Proposition~\ref{mu-prop-simulation} shows that
    \[ \Phi'(\ket 0 \bra 0 \tprod \rho) = \Phi(\rho) \tprod \nidentity{B}, \] 
    from which it follows immediately that
    \[ \opv_p (\Phi) \smallnorm{\nidentity{B}}_p \leq \opv_p(\Phi'), \]
    as $\opv_p$ is a maximization over input states, and
    $\norm{\cdot}_p$ is multiplicative with respect to the
    tensor product of two states.

    To prove the second inequality let $\rho \in \density{A \tprod H}$
    be a state such that
    \begin{equation}\label{mu-eqn-thm-approx-rho}
      \opv_p(\Phi') = \norm{\Phi'(\rho)}_p.
    \end{equation}
    Such a state exists by the compactness of $\density{A \tprod H}$.
    The output of the channel $\Phi'$ on $\rho$ is given by
    Equation~\eqref{mu-eqn-output}, applying the triangle inequality to
    this yields
      \begin{equation*}
       \smallnorm{\Phi'(\rho)}_p
       = \smallnorm{q \Phi(\sigma) \tprod \nidentity{B}  
          + (1-q) \Phi'(\rho_{S_0^\perp})}_p  
       \leq q \smallnorm{\Phi(\sigma) \tprod \nidentity{B}}_p
          + (1-q) \smallnorm{\Phi'(\rho_{S_0^\perp})}_p.  
    \end{equation*}
    Lemma~\ref{mu-lem-ancilla-mixing} provides a bound on the
    second term of this equation, which implies that
    \begin{equation*}
       \smallnorm{\Phi'(\rho)}_p
       \leq q \smallnorm{\Phi(\sigma) \tprod \nidentity{B}}_p
           + (1-q)
           \left( \smallnorm{\nidentity{K \tprod B}}_p + \frac{2}{d} \right).
    \end{equation*}
    Then, as the norm $\norm{\cdot}_p$ is multiplicative with respect
    to the tensor product of states, and $\smallnorm{\nidentity{K}}_p \leq
    \norm{\xi}_p$ for any state $\xi \in \density{K}$,
    \begin{equation*}
      \norm{\Phi'(\rho)}_p
      \leq    q \smallnorm{\Phi(\sigma)}_p \smallnorm{\nidentity{B}}_p
           + (1 - q) \left( \smallnorm{\nidentity{K}}_p
           \smallnorm{\nidentity{B}}_p + \frac{2}{d} \right)
      \leq \smallnorm{\Phi(\sigma)}_p \smallnorm{\nidentity{B}}_p
           + \frac{2}{d}.
    \end{equation*}
    Then, by the choice of the input $\rho$ in
    Equation~\eqref{mu-eqn-thm-approx-rho}, we have shown that
    \begin{equation}\label{mu-eqn-thm-approx-penultimate}
      \opv_p(\Phi') = \norm{\Phi'(\rho)}_p
      \leq \opv_p(\Phi) \smallnorm{\nidentity{B}}_p 
        + \frac{2}{d}.
    \end{equation}
    Finally, the state $\nidentity{B}$ has $\dm{B}$ eigenvalues, each
    with value $1/\dm{B}$, which implies that
    \begin{equation*}
      \smallnorm{\nidentity{B}}_p 
      = \left( \sum_{i=1}^{\dm{B}} \frac{1}{(\dm{B})^p} \right)^{1/p}
      = \dm{B}^{1/p-1} 
      \geq \frac{1}{\dm{B}}.
    \end{equation*}
    Combining this with
    Equation~\eqref{mu-eqn-thm-approx-penultimate}, and expanding $d =
    \dm{A}$, implies
    \begin{equation*}
      \frac{\opv_p(\Phi')}{\smallnorm{\nidentity{B}}_p }
      \leq \opv_p(\Phi) + \frac{2}{\smallnorm{\nidentity{B}}_p \dm{A}}
      \leq \opv_p(\Phi) + \frac{2 \dm{B}}{\dm{A}},
    \end{equation*}
    which completes the proof of the second inequality.      
  \end{proof}
\end{theorem}

With this approximation result, the main theorem on the maximum output
$p$-norm can be shown.  This extends the construction of
Section~\ref{mu-scn-unital} due to Fukuda~\cite{Fukuda07} on unital
channels to the mixed-unitary unitary case, using essentially the same
method of proof.

\begin{theorem}\label{mu-thm-pnorm-mult}
  If $\Phi, \Psi \in \transform{H,K}$ and $p \in [1,\infty)$, then 
  \begin{equation*}
    \opv_p(\Phi \tprod \Psi)  = \opv_p(\Phi) \opv_p(\Psi)
  \end{equation*}
  if
  \begin{equation*}
    \opv_p(\Phi'_d \tprod \Psi)  = \opv_p(\Phi'_d) \opv_p(\Psi),
  \end{equation*}
  for all sufficiently large $d$, where $\Phi'_d$ is the mixed-unitary
  approximation of the channel $\Phi$ obtained by applying the construction of
  Section~\ref{mu-scn-channels} to a Stinespring dilation of $\Phi$ using
  a $d$-dimensional ancillary space.

  \begin{proof}
    As adding ancillary space to $\Phi'$ increases both $\dm{A}$ and
    $\dm{B}$, by taking $d = \dm{A}$ large enough is can be assumed
    that $\dm{B} \leq 2d$.  Let $\epsilon > 0$, and choose $d$ so that
    $4 / d^{1/p} < \epsilon$.  This, along with the choice of $\dm{B}
    \leq 2d$ implies that
    \begin{equation}\label{mu-eqn-thm-pnorm-eps}
      2 \dm{B}^{1 - 1/p} / d \leq 4 / d^{1/p} < \epsilon 
    \end{equation}
   Then, as $\Phi'_d(\ket 0 \bra 0 \tprod \rho) =
    \Phi(\rho) \tprod \nidentity{B}$ by Proposition~\ref{mu-prop-simulation},
    \begin{align*}
      \opv_p(\Phi \tprod \Psi)
      &\leq \frac{\opv_p(\Phi'_d \tprod \Psi)}{\smallnorm{\nidentity{B}}_p}.
    \end{align*}
    By assumption, this second quantity is multiplicative, so that
    \begin{equation*}
      \opv_p(\Phi \tprod \Psi)
      \leq \frac{\opv_p(\Phi'_d \tprod
        \Psi)}{\smallnorm{\nidentity{B}}_p}
      = \frac{\opv_p(\Phi'_d)
        \opv_p(\Psi)}{\smallnorm{\nidentity{B}}_p}.
    \end{equation*}
    Applying Theorem~\ref{mu-thm-approx} to this quantity shows that
    \begin{equation*}
      \opv_p(\Phi \tprod \Psi)
      \leq \left[ \opv_p(\Phi) + \frac{4}{d^{1/p}} \right] \opv_p(\Psi) 
      < \opv_p(\Phi) \opv_p(\Psi) + \epsilon,
    \end{equation*}
    where the final inequality is by the choice of $d$ to satisfy
    Equation~\eqref{mu-eqn-thm-pnorm-eps}.  As epsilon was chosen
    arbitrarily, the multiplicativity of $\opv_p(\Phi'_d)$ for all
    large enough $d$ implies the multiplicativity of $\opv_p(\Phi)$.
  \end{proof}
\end{theorem}

This theorem shows that in order to show the multiplicativity of $\opv_p$ on a
class of channels it suffices to consider a related class of
mixed-unitary channels.  This problem may be more tractable for
channels of this type: many of the known counterexamples to multiplicativity
for small values of $p$ are mixed-unitary~\cite{HaydenW08}.

\section{Mixed-unitaries and minimum output entropy}\label{mu-scn-add}

The results of the previous section on the multiplicativity of the
maximum output $p$-norm can be extended directly to the additivity of
the minimum output entropy.  This is done using very similar proof
techniques as in the previous section.

The following theorem demonstrates that the mixed-unitary $\Phi'$
constructed in Equation~\eqref{mu-eqn-construction} forms a good approximation
of the original channel $\Phi$, from which the result on the
additivity will follow directly.

\begin{theorem}\label{mu-thm-entropy-approx}
  If $\Phi \in \transform{H,K}$, then the
  mixed-unitary $\Phi' \in \transform{A
    \tprod H, K \tprod B}$ satisfies
  \[ S_{\min}(\Phi) \geq S_{\min}(\Phi') - \log \dm{B} \geq S_{\min}(\Phi) -
  \frac{1}{\dm{A}}. \]

  \begin{proof}    
    Exactly as in the case of Theorem~\ref{mu-thm-approx},
    Proposition~\ref{mu-prop-simulation} implies the first inequality,
    as $\Phi'$ on a particular state can be used to simulate $\Phi$: 
    \[\Phi'(\ket 0 \bra 0 \tprod \rho) = \Phi(\rho) \tprod \nidentity{B}. \]

   Let $\rho$ be a state minimizing $S(\Phi'(\rho))$ and for
    convenience let $\delta = 1 / \dm{A}$.
    Equation~\eqref{mu-eqn-output} gives the output of $\Phi'$ on
    $\rho$.  Applying the concavity of the entropy
    (Proposition~\ref{meas-prop-entropy-concavity}) to this, we obtain
    \begin{align*}
      S_{\min}(\Phi')
      = S(\Phi'(\rho))
      \geq q S(\Phi(\sigma) \tprod \nidentity{B}) 
        + (1-q) S(\Phi'(\rho_{S_0^\perp})).
    \end{align*}      
    Applying Corollary~\ref{mu-cor-mixing-entropy} this becomes
    \begin{align*}
      S_{\min}(\Phi')    
      &\geq q S(\Phi(\sigma) \tprod \nidentity{B}) 
        +(1-q)(S(\nidentity{A \tprod H}) - \delta).
    \end{align*}
    Notice that, since $\Phi'$ is mixed-unitary,
    $\mathcal{A \tprod H}$ is isomorphic to $\mathcal{K \tprod B}$.
    This implies that $S(\nidentity{A \tprod H}) = S(\nidentity{K
      \tprod B})$.

    Two additional properties of the entropy introduced in
    Section~\ref{meas-scn-entropy} will be useful: the additivity of
    the entropy on states
    (Equation~\eqref{meas-eqn-entropy-additive}), $S(\sigma \tprod
    \xi) = S(\sigma) + S(\xi)$, for any $\sigma, \xi$; and the fact
    that the entropy is maximized on completely mixed states, $S(\xi)
    \leq \log \dm{K} = S(\nidentity{K})$ for all $\xi \in \density{K}$
    (Proposition~\ref{meas-prop-entropy-bounds}).  Using these three
    observations, in order, we find that
    \begin{align*}
      S_{\min}(\Phi')    
      &\geq q S(\Phi(\sigma) \tprod \nidentity{B}) 
        + (1 - q)(S(\nidentity{K \tprod B}) - \delta) \\
      &= q (S(\Phi(\sigma)) + S(\nidentity{B}))
        + (1 - q)(S(\nidentity{K}) + S(\nidentity{B}) - \delta) \\
      &\geq q (S(\Phi(\sigma)) + S(\nidentity{B}))
        + (1 - q)(S(\Phi(\sigma)) + S(\nidentity{B}) - \delta) \\
      &\geq S(\Phi(\sigma)) + S(\nidentity{B}) - \delta.
    \end{align*}
    Finally, since $S(\nidentity{B}) = \log \dm{B}$ and
    $S_{\min}(\Phi) \leq S(\Phi(\xi))$ for any $\xi$, we have
    \begin{align*}
      S_{\min}(\Phi')
      &\geq S_{\min}(\Phi) + \log \dm{B} - \delta,
    \end{align*}
    which completes the proof of the theorem.    
  \end{proof}
\end{theorem}

The proof that the additivity of the minimum output entropy
can be equivalently restricted to mixed-unitary channels follows from
the previous theorem in a way that is identical to the proof of
Theorem~\ref{mu-thm-pnorm-mult}, with the exception that the $p$-norm has been
replaced by the minimum output entropy.  The method of proof here
follows Fukuda's result for unital channels~\cite{Fukuda07}.

\begin{theorem}\label{mu-thm-main-result}
  If $\Phi, \Psi \in \transform{H,K}$, then 
  \[ S_{\min}(\Phi \tprod \Psi)  = S_{\min}(\Phi) + S_{\min}(\Psi) \]
  if
  \[ S_{\min}(\Phi'_d \tprod \Psi)  = S_{\min}(\Phi'_d) + S_{\min}(\Psi), \]
  for all sufficiently large $d$, where $\Phi'_d$ is the mixed-unitary
  extension of the channel obtained by applying the construction of
  Section~\ref{mu-scn-channels} to Stinespring dilation for $\Phi$ using
  an ancillary space of dimension $d$.

  \begin{proof}
    Let $\epsilon > 0$, and choose $d$ large enough so that so that $1 / d <
    \epsilon$.  Then, as $\Phi'_d(\ket 0 \bra 0 \tprod \rho) =
    \Phi(\rho) \tprod \nidentity{B}$,
    \[
      S_{\min}(\Phi \tprod \Psi)
      \geq S_{\min}(\Phi'_d \tprod \Psi) - \log \dm{B}.
    \]
    By assumption, this second quantity is additive, so that
    \begin{align*}
      S_{\min}(\Phi \tprod \Psi)
      &\geq S_{\min}(\Phi'_d \tprod \Psi) - \log \dm{B} \\
      &= S_{\min}(\Phi'_d) + S_{\min}(\Psi) - \log \dm{B} \\
      &\geq S_{\min}(\Phi) - \frac{1}{d} + S_{\min}(\Psi) \\
      &>    S_{\min}(\Phi) + S_{\min}(\Psi) - \epsilon
    \end{align*}
    where the penultimate inequality is an application of
    Theorem~\ref{mu-thm-entropy-approx}.  As $\epsilon$ was chosen
    arbitrarily, the additivity of $\Phi'_d$ for all large enough $d$
    implies the additivity of $\Phi$.
  \end{proof}
\end{theorem}

This theorem implies that in order to prove the additivity of the
minimum output entropy for a class of channels, the hopefully simpler
class of mixed-unitary approximations can instead be considered.  This
may be a fruitful approach: the only channels for which $S_{\min}$ is
known not to be additive are mixed-unitary~\cite{Hastings09} -- this
property may be simpler to check for mixed-unitaries having certain properties.

\section{Circuit constructions}\label{mu-scn-circuits}

In this section an efficient circuit construction is provided for the
mixed-unitary approximation described in Section~\ref{mu-scn-channels}.
This construction is used to extend the hardness of computationally
distinguishing quantum circuits to the case of mixed-unitary
circuits.

Before constructing these circuits, it will be important to specify
the circuit models that are being used.
The circuit model used to define the quantum circuit
distinguishability problem is the \emph{mixed state quantum circuit} model of
Aharonov, Kitaev, and Nisan~\cite{AharonovK+98}, which is described in
Section~\ref{compl-scn-circuit-model}.  As previously discussed,
we may assume that circuits in this model
first introduce any necessary ancillary qubits, then perform a unitary
operation, and finally trace out those qubits that are not part 
of the output.  
This approach is equivalent to building a circuit for the
Stinespring dilation of a channel.
As all unitary transformations can be (approximately)
implemented using one and two qubit gates there is no loss in
generality in assuming that the unitary transformations implemented in
such a circuit are composed of gates from some
finite basis of one and two qubit gates.

The second model of quantum circuits we consider is the model of
\emph{mixed-unitary quantum circuits}.  
\index{mixed-unitary!circuits}%
These circuits consist of one
and two qubit gates from the usual circuit model as well as mixed-unitary gates, which implement
a unitary gate with probability one half.  More
formally, the application of such a gate performs the operation
\[ \rho \mapsto \frac{1}{2} U \rho U^* + \frac{1}{2} \rho, \]
where $U$ is a one or two qubit unitary gate in the standard gate set.

For technical reasons, we need to assume that the Pauli $X$ and $Z$
gates, as well as controlled versions of these gates, are part of the
standard basis.  This restriction can be avoided by allowing
gates that implement 
\[ \rho \mapsto 
    \frac{1}{2} U_k \cdots U_2 U_1 \rho U_1^* U_2^* \cdots U_k^*+ \frac{1}{2} \rho, \]
where the $U_i$ are gates of the standard model.  This allows
sequences of multiple gates, such as approximations to gates not in
the basis, to be applied with probability one half.  When proving a
hardness result, the model should be as restricted as possible, and it
is not clear that this model is not more powerful than the model where each
mixed-unitary gate is applied with an independent probability of one half.

The model of mixed-unitary circuits is an extremely simple model 
that does not appear to be
universal for the class of transformations that implement mixed-unitary operations.
It is not clear that this is the correct definition of the mixed-unitary
circuit model, but since the aim of the model to prove a
hardness result, an extremely weak definition has been chosen so that
the result will apply to as large a class of circuit models as possible.

One drawback of this weak model is that the exact
construction used in Section~\ref{mu-scn-channels} cannot be
implemented.  Specifically, the operation $D$ that decoheres the
subspaces $S_0$ and $S_0^\perp$ seems to require a unitary operation
that cannot be decomposed into a series of 
one and two qubit gates, applied with probability one half.
A similar situation occurs for the implementation of the completely
depolarizing channel on the subspace $S_0^\perp$, the implementation
of which uses the discrete Weyl operators on the
subspace $S_0^\perp$.  
These operations can be implemented in a mixed-unitary way in a more
permissive circuit model, but in order to keep the circuit model as
simple as possible, a modified construction is presented here.  This
modified construction is built from pieces that perform similar tasks
to those used in Section~\ref{mu-scn-channels}, but the specific
building blocks are not exactly the same.
The construction in this section can also be applied to the additivity
and multiplicativity problems considered in Sections~\ref{mu-scn-mult}
and~\ref{mu-scn-add}, but it is somewhat more complicated than the construction
already presented.

In order to approximate a given circuit with a mixed-unitary circuit
we once again make use of three main components, which are once again
referred to as $N,D,$ and $M$.
These pieces are labelled in this way due to the fact that they play the same
roles as the components of the same names used in
Section~\ref{mu-scn-channels}, though the details differ slightly.
The first two of these
components, $N$ the completely depolarizing channel and $D$ the completely
dephasing channel, are easy to
implement as mixed-unitary operations in the chosen circuit model.
More difficult to implement is the
channel $M$, which performs a function similar to the channel
described by Equation~\eqref{mu-eqn-mixing-approx}.

The complete dephasing channel $D$ is the channel that sets to zero
all of the off-diagonal elements of a density matrix.
More formally, the action of this operator applied to the space $\mathcal{A}$, for an
input $\rho$ on $\mathcal{A \tprod H}$ is given by
\begin{equation}\label{mu-eqn-Dcircuit}
  D_{\mathcal{A}}(\rho) 
  = \sum_{i=0}^{\dm{A} - 1} p_i \ket i \bra i \tprod \rho_i,
\end{equation}
where the $p_i$ form a probability distribution.  This operation is
equivalent to measuring the space $\mathcal{A}$ 
in the computational basis and forgetting the result.
This channel is
shown to be mixed-unitary in Proposition~\ref{mu-prop-dephasing-mu}
where it is implemented as a mixture of generalized Pauli $Z$
operations.  To implement this as a mixed-unitary circuit, observe
that restricting the construction of
Proposition~\ref{mu-prop-dephasing-mu} to the case where $\mathcal{A}$
is a two dimensional space results in exactly the channel that applies
a Pauli $Z$ gate with probability one-half.  Notice also that applying
this channel to each of $n$ qubits is identical to applying the
completely dephasing channel to the whole space.
Thus, the operation $D_{\mathcal{A}}$ that applies $D$ to the space
$\mathcal{A}$ can be implemented as a mixed-unitary circuit by
applying the Pauli $Z$ operation to each qubit of $\mathcal{A}$
independently with probability 1/2.  This construction can be found
in~\cite{ChuangY97}.

The completely noisy channel $N$ is also simple to implement as a
mixed-unitary circuit.  This channel can be realized on a single qubit
by performing a uniform mixture of the Pauli operators on each qubit,
which is a consequence of Proposition~\ref{mu-prop-depol-mu} when
restricted to the case of a single qubit.  This mixture can
be implemented by, independently on each qubit, applying the Pauli $Z$
operation with probability 1/2, followed by applying the Pauli $X$
operation with probability 1/2, as shown in~\cite{BoykinR03}.  
Intuitively, the $Z$ operations will
zero the off-diagonal elements of a density matrix (viewed in the
computational basis), and the $X$ operations will scramble the
diagonal, resulting in the completely mixed state, $\id/2$, on each
qubit.  As the tensor product of two completely mixed qubits is the
completely mixed state of the larger system, applying this
construction to each qubit in a space $\mathcal{B}$ will implement the
completely depolarizing channel $N_{\mathcal{B}}$ on that space.

In Section~\ref{mu-scn-channels} the channel $M$ was implemented as a
completely depolarizing channel on the subspace $S_0^\perp$ of inputs
not in the state $\ket 0$ on the `ancillary space' $\mathcal{A}$.  While
the same channel suffices for the circuit case, it is not at all
obvious how this channel can be implemented
using only two-qubit mixed-unitary gates.
This difficulty is avoided by implementing a closely related channel
This construction is intuitively the same: it does not affect states in the
subspace $S_0$ of inputs with the $\ket 0$ state in the space
$\mathcal{A}$, and it applies depolarizing noise to states in the
space $S_0^\perp$.  The difference is exactly how this noise is
applied.  The circuit that is constructed implements the operation
$M$ defined by
\begin{equation}\label{mu-eqn-Mcircuit}
   M(\ket i \bra i \tprod \rho) = \begin{cases} 
     \frac{1}{\dim \mathcal{A}}
     (\identity{A} - \ket 0 \bra 0 + \ket{\psi_i} \bra{\psi_i})
     \tprod \nidentity{H} & \text{if $i \neq 0$},\\
     \ket 0 \bra 0 \tprod \rho& \text{if $i=0$},
   \end{cases}
\end{equation}
where $\ket{\psi_i}$ is a nonzero computational basis state that depends on $i$.  The
exact specification of this state can be extracted from the analysis
of the circuit constructed for $M$, but this is not helpful.

It is perhaps not a surprise that the transformation $M$ can
be implemented using only controlled-mixing operations.  Before
describing this implementation, notice that the controlled application
of the completely depolarizing channel $N$ to a single qubit 
can be described by a mixed-unitary circuit.  
This is because the previously discussed implementation of $N$ is given by a mixture
of the single qubit gates $X$ and $Z$.
Adding a control
qubit to each of these gates results in two qubit gates, which fit
into the model of mixed-unitary circuits used here (because we have
assumed that $X$ and $Z$, as well as controlled versions of them, are
included in the standard basis of gates -- dropping this assumption
requires the circuit model to be generalized slightly).
It is not clear that
general controlled mixed-unitary operations can be implemented as
mixed-unitary circuits in this model, but the
only controlled operation that will be needed for this construction is
the completely depolarizing channel.

At an intuitive level, the implementation of the channel $M$ consists
of the application of controlled-depolarizing operations everywhere
that this is possible.  These operations will all be controlled by the
qubits in the space $\mathcal{A}$, which ensures that in the case of a
state in $S_0$, with $\ket 0$ in the space $\mathcal{A}$, the
operation $M$ acts trivially.

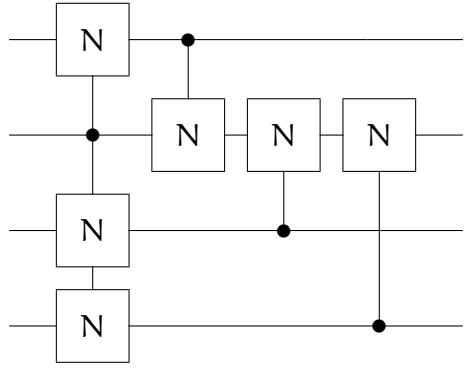
\begin{figure}
  \centering
    \setlength{\unitlength}{3947sp}%
\begingroup\makeatletter\ifx\SetFigFont\undefined%
\gdef\SetFigFont#1#2#3#4#5{%
  \reset@font\fontsize{#1}{#2pt}%
  \fontfamily{#3}\fontseries{#4}\fontshape{#5}%
  \selectfont}%
\fi\endgroup%
\begin{picture}(2874,2274)(2989,-2623)
\thinlines
{\color[rgb]{0,0,0}\put(3526,-2011){\line( 0,-1){150}}
}%
{\color[rgb]{0,0,0}\put(3751,-2386){\line( 1, 0){2100}}
}%
{\color[rgb]{0,0,0}\put(4951,-1186){\line( 1, 0){150}}
}%
{\color[rgb]{0,0,0}\put(5551,-1186){\line( 1, 0){300}}
}%
{\color[rgb]{0,0,0}\put(5251,-586){\line( 1, 0){600}}
}%
{\color[rgb]{0,0,0}\put(3301,-811){\framebox(450,450){$N$}}
}%
{\color[rgb]{0,0,0}\put(3301,-2011){\framebox(450,450){$N$}}
}%
{\color[rgb]{0,0,0}\put(3901,-1411){\framebox(450,450){$N$}}
}%
{\color[rgb]{0,0,0}\put(4501,-1411){\framebox(450,450){$N$}}
}%
{\color[rgb]{0,0,0}\put(3751,-586){\line( 1, 0){1500}}
}%
{\color[rgb]{0,0,0}\put(4351,-1186){\line( 1, 0){150}}
}%
{\color[rgb]{0,0,0}\put(3751,-1786){\line( 1, 0){2100}}
}%
{\color[rgb]{0,0,0}\put(3526,-1186){\line( 0, 1){375}}
}%
{\color[rgb]{0,0,0}\put(3526,-1186){\line( 0,-1){375}}
}%
{\color[rgb]{0,0,0}\put(4126,-586){\line( 0,-1){375}}
}%
{\color[rgb]{0,0,0}\put(4726,-1786){\line( 0, 1){375}}
}%
{\color[rgb]{0,0,0}\put(3301,-2611){\framebox(450,450){$N$}}
}%
{\color[rgb]{0,0,0}\put(5101,-1411){\framebox(450,450){$N$}}
}%
{\color[rgb]{0,0,0}\put(5326,-2386){\line( 0, 1){975}}
}%
{\color[rgb]{0,0,0}\put(3001,-2386){\line( 1, 0){300}}
}%
{\color[rgb]{0,0,0}\put(3001,-1786){\line( 1, 0){300}}
}%
{\color[rgb]{0,0,0}\put(3001,-1186){\line( 1, 0){900}}
}%
{\color[rgb]{0,0,0}\put(3001,-586){\line( 1, 0){300}}
}%
{\color[rgb]{0,0,0}\put(4126,-586){\circle*{76}}
}%
{\color[rgb]{0,0,0}\put(4726,-1786){\circle*{76}}
}%
{\color[rgb]{0,0,0}\put(5326,-2386){\circle*{76}}
}%
{\color[rgb]{0,0,0}\put(3526,-1186){\circle*{76}}
}%
\end{picture}%
    \caption[One stage of the circuit for the ancilla simulation
    procedure]{One stage of the mixing procedure on the ancillary
      qubits.  The mixing operations applied to the qubits in the
      space $\mathcal{H}$ are not shown.}
  \label{mu-fig-mixing}
\end{figure}

\begin{figure}
  \centering
    \setlength{\unitlength}{3247sp}%
\begingroup\makeatletter\ifx\SetFigFont\undefined%
\gdef\SetFigFont#1#2#3#4#5{%
  \reset@font\fontsize{#1}{#2pt}%
  \fontfamily{#3}\fontseries{#4}\fontshape{#5}%
  \selectfont}%
\fi\endgroup%
\begin{picture}(6924,3474)(289,-3673)
\thinlines
{\color[rgb]{0,0,0}\multiput(1201,-3661)(120.61856,0.00000){49}{\line( 1, 0){ 60.309}}
}%
{\color[rgb]{0,0,0}\put(5476,-1786){\line( 0,-1){675}}
}%
{\color[rgb]{0,0,0}\put(5476,-2911){\line( 0,-1){150}}
}%
{\color[rgb]{0,0,0}\put(6076,-586){\line( 0,-1){975}}
}%
{\color[rgb]{0,0,0}\put(6676,-1186){\line( 0,-1){375}}
}%
{\color[rgb]{0,0,0}\multiput(1201,-211)(0.00000,-121.05263){29}{\line( 0,-1){ 60.526}}
}%
{\color[rgb]{0,0,0}\multiput(3151,-211)(0.00000,-121.05263){29}{\line( 0,-1){ 60.526}}
}%
{\color[rgb]{0,0,0}\multiput(5101,-211)(0.00000,-121.05263){29}{\line( 0,-1){ 60.526}}
}%
{\color[rgb]{0,0,0}\multiput(7051,-211)(0.00000,-121.05263){29}{\line( 0,-1){ 60.526}}
}%
{\color[rgb]{0,0,0}\multiput(1201,-211)(120.61856,0.00000){49}{\line( 1, 0){ 60.309}}
}%
{\color[rgb]{0,0,0}\put(601,-2011){\framebox(450,450){$D$}}
}%
{\color[rgb]{0,0,0}\put(601,-1411){\framebox(450,450){$D$}}
}%
{\color[rgb]{0,0,0}\put(601,-811){\framebox(450,450){$D$}}
}%
{\color[rgb]{0,0,0}\put(1351,-1411){\framebox(450,450){$N$}}
}%
{\color[rgb]{0,0,0}\put(1351,-2011){\framebox(450,450){$N$}}
}%
{\color[rgb]{0,0,0}\put(1351,-2911){\framebox(450,450){$N$}}
}%
{\color[rgb]{0,0,0}\put(1351,-3511){\framebox(450,450){$N$}}
}%
{\color[rgb]{0,0,0}\put(1951,-811){\framebox(450,450){$N$}}
}%
{\color[rgb]{0,0,0}\put(2551,-811){\framebox(450,450){$N$}}
}%
{\color[rgb]{0,0,0}\put(3301,-811){\framebox(450,450){$N$}}
}%
{\color[rgb]{0,0,0}\put(3301,-2011){\framebox(450,450){$N$}}
}%
{\color[rgb]{0,0,0}\put(3901,-1411){\framebox(450,450){$N$}}
}%
{\color[rgb]{0,0,0}\put(4501,-1411){\framebox(450,450){$N$}}
}%
{\color[rgb]{0,0,0}\put(5251,-1411){\framebox(450,450){$N$}}
}%
{\color[rgb]{0,0,0}\put(5251,-811){\framebox(450,450){$N$}}
}%
{\color[rgb]{0,0,0}\put(5851,-2011){\framebox(450,450){$N$}}
}%
{\color[rgb]{0,0,0}\put(6451,-2011){\framebox(450,450){$N$}}
}%
{\color[rgb]{0,0,0}\put(3301,-2911){\framebox(450,450){$N$}}
}%
{\color[rgb]{0,0,0}\put(3301,-3511){\framebox(450,450){$N$}}
}%
{\color[rgb]{0,0,0}\put(5251,-2911){\framebox(450,450){$N$}}
}%
{\color[rgb]{0,0,0}\put(5251,-3511){\framebox(450,450){$N$}}
}%
{\color[rgb]{0,0,0}\put(301,-586){\line( 1, 0){300}}
}%
{\color[rgb]{0,0,0}\put(1051,-586){\line( 1, 0){900}}
}%
{\color[rgb]{0,0,0}\put(2401,-586){\line( 1, 0){150}}
}%
{\color[rgb]{0,0,0}\put(3001,-586){\line( 1, 0){300}}
}%
{\color[rgb]{0,0,0}\put(3751,-586){\line( 1, 0){1500}}
}%
{\color[rgb]{0,0,0}\put(5701,-586){\line( 1, 0){1500}}
}%
{\color[rgb]{0,0,0}\put(301,-1186){\line( 1, 0){300}}
}%
{\color[rgb]{0,0,0}\put(301,-1786){\line( 1, 0){300}}
}%
{\color[rgb]{0,0,0}\put(1051,-1786){\line( 1, 0){300}}
}%
{\color[rgb]{0,0,0}\put(1051,-1186){\line( 1, 0){300}}
}%
{\color[rgb]{0,0,0}\put(4951,-1186){\line( 1, 0){300}}
}%
{\color[rgb]{0,0,0}\put(4351,-1186){\line( 1, 0){150}}
}%
{\color[rgb]{0,0,0}\put(6301,-1786){\line( 1, 0){150}}
}%
{\color[rgb]{0,0,0}\put(1801,-1186){\line( 1, 0){2100}}
}%
{\color[rgb]{0,0,0}\put(1801,-1786){\line( 1, 0){1500}}
}%
{\color[rgb]{0,0,0}\put(3751,-1786){\line( 1, 0){2100}}
}%
{\color[rgb]{0,0,0}\put(6901,-1786){\line( 1, 0){300}}
}%
{\color[rgb]{0,0,0}\put(5701,-1186){\line( 1, 0){1500}}
}%
{\color[rgb]{0,0,0}\put(301,-2686){\line( 1, 0){1050}}
}%
{\color[rgb]{0,0,0}\put(1801,-2686){\line( 1, 0){1500}}
}%
{\color[rgb]{0,0,0}\put(3751,-2686){\line( 1, 0){1500}}
}%
{\color[rgb]{0,0,0}\put(5701,-2686){\line( 1, 0){1500}}
}%
{\color[rgb]{0,0,0}\put(301,-3286){\line( 1, 0){1050}}
}%
{\color[rgb]{0,0,0}\put(1801,-3286){\line( 1, 0){1500}}
}%
{\color[rgb]{0,0,0}\put(5701,-3286){\line( 1, 0){1500}}
}%
{\color[rgb]{0,0,0}\put(3751,-3286){\line( 1, 0){1500}}
}%
{\color[rgb]{0,0,0}\put(1576,-586){\line( 0,-1){375}}
}%
{\color[rgb]{0,0,0}\put(1576,-1411){\line( 0,-1){150}}
}%
{\color[rgb]{0,0,0}\put(1576,-2011){\line( 0,-1){450}}
}%
{\color[rgb]{0,0,0}\put(1576,-2911){\line( 0,-1){150}}
}%
{\color[rgb]{0,0,0}\put(2176,-1186){\line( 0, 1){375}}
}%
{\color[rgb]{0,0,0}\put(2776,-1786){\line( 0, 1){975}}
}%
{\color[rgb]{0,0,0}\put(3526,-1186){\line( 0, 1){375}}
}%
{\color[rgb]{0,0,0}\put(3526,-2011){\line( 0,-1){450}}
}%
{\color[rgb]{0,0,0}\put(3526,-2911){\line( 0,-1){150}}
}%
{\color[rgb]{0,0,0}\put(3526,-1186){\line( 0,-1){375}}
}%
{\color[rgb]{0,0,0}\put(4126,-586){\line( 0,-1){375}}
}%
{\color[rgb]{0,0,0}\put(4726,-1786){\line( 0, 1){375}}
}%
{\color[rgb]{0,0,0}\put(5476,-1786){\line( 0, 1){375}}
}%
{\color[rgb]{0,0,0}\put(5476,-961){\line( 0, 1){150}}
}%
{\color[rgb]{0,0,0}\put(1576,-586){\circle*{76}}
}%
{\color[rgb]{0,0,0}\put(5476,-1786){\circle*{76}}
}%
{\color[rgb]{0,0,0}\put(2176,-1186){\circle*{76}}
}%
{\color[rgb]{0,0,0}\put(2776,-1786){\circle*{76}}
}%
{\color[rgb]{0,0,0}\put(4126,-586){\circle*{76}}
}%
{\color[rgb]{0,0,0}\put(4726,-1786){\circle*{76}}
}%
{\color[rgb]{0,0,0}\put(6076,-586){\circle*{76}}
}%
{\color[rgb]{0,0,0}\put(6676,-1186){\circle*{76}}
}%
{\color[rgb]{0,0,0}\put(3526,-1186){\circle*{76}}
}%
\end{picture}%
    \caption[Circuit performing the ancilla simulation
    procedure]{Circuit performing the ancilla simulation procedure $M
      \circ D_{\mathcal{A}}$.  The top three qubits simulate the ancillary
      qubits of the original circuit in the space $\mathcal{A}$, and
      the bottom two simulate the input to the original circuit
      in the space $\mathcal{H}$.  The dashed lines separate the each
      of the three stages of the mixing procedure.}
  \label{mu-fig-ancilla}
\end{figure}
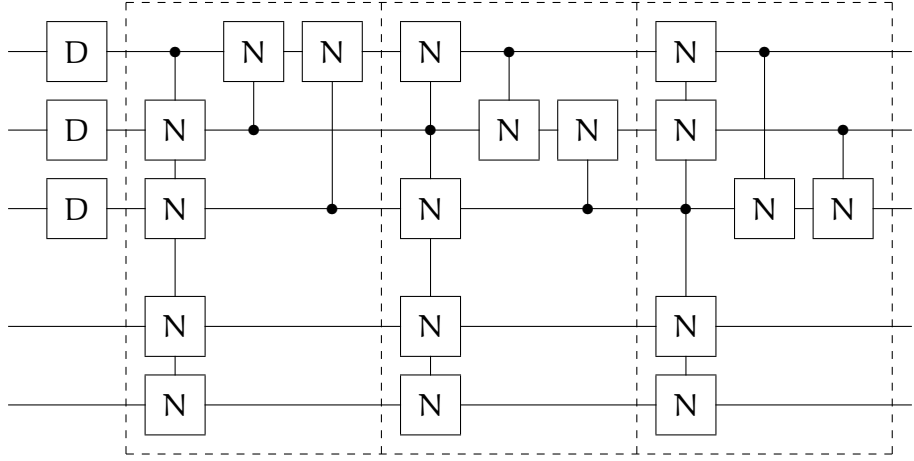

More formally, let $m$ be the number of qubits in the 
space $\mathcal{A}$ that are
given as part of the input to $M$, i.e.\ the number of ancillary qubits
used to represent the ancillary space used by the original channel.
The implementation of $M$ consists of $m$ stages, with the
$j$th stage testing that the $j$th qubit of the space $\mathcal{A}$ is
in the $\ket 0$ state, and mixing the qubits if this is not the case.
An example of one stage of the circuit
is given in Figure~\ref{mu-fig-mixing}.
The $j$th stage consists first of an
application of the controlled $N$ operation from the $j$th qubit to
each other qubit of $\mathcal{A \tprod H}$.
After these operations, stage $j$ is completed by $m-1$ further
controlled $N$ operations: each with the $j$th qubit as the target
qubit and one of the other qubits of $\mathcal{A}$ as the control qubit.  An
example of this construction with $m=3$ is presented in
Figure~\ref{mu-fig-ancilla}.

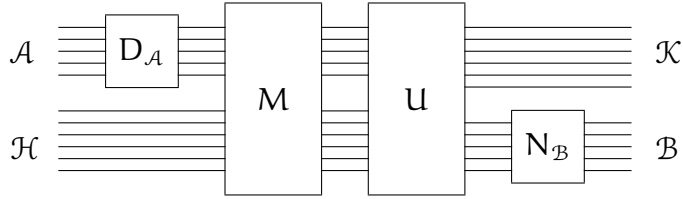
\begin{figure}
  \centering
    \setlength{\unitlength}{3947sp}%
\begingroup\makeatletter\ifx\SetFigFont\undefined%
\gdef\SetFigFont#1#2#3#4#5{%
  \reset@font\fontsize{#1}{#2pt}%
  \fontfamily{#3}\fontseries{#4}\fontshape{#5}%
  \selectfont}%
\fi\endgroup%
\begin{picture}(4080,1224)(-764,-1273)
\put(3301,-436){\makebox(0,0)[lb]{\smash{{\SetFigFont{12}{14.4}{\rmdefault}{\mddefault}{\updefault}{\color[rgb]{0,0,0}$\mathcal{K}$}%
}}}}
\thinlines
{\color[rgb]{0,0,0}\put(2101,-886){\line( 1, 0){300}}
}%
{\color[rgb]{0,0,0}\put(2101,-961){\line( 1, 0){300}}
}%
{\color[rgb]{0,0,0}\put(2101,-1111){\line( 1, 0){300}}
}%
{\color[rgb]{0,0,0}\put(2101,-1036){\line( 1, 0){300}}
}%
{\color[rgb]{0,0,0}\put(2851,-811){\line( 1, 0){300}}
}%
{\color[rgb]{0,0,0}\put(2851,-886){\line( 1, 0){300}}
}%
{\color[rgb]{0,0,0}\put(2851,-961){\line( 1, 0){300}}
}%
{\color[rgb]{0,0,0}\put(2851,-1111){\line( 1, 0){300}}
}%
{\color[rgb]{0,0,0}\put(2851,-1036){\line( 1, 0){300}}
}%
{\color[rgb]{0,0,0}\put(2101,-211){\line( 1, 0){1050}}
}%
{\color[rgb]{0,0,0}\put(2101,-286){\line( 1, 0){1050}}
}%
{\color[rgb]{0,0,0}\put(2101,-361){\line( 1, 0){1050}}
}%
{\color[rgb]{0,0,0}\put(2101,-511){\line( 1, 0){1050}}
}%
{\color[rgb]{0,0,0}\put(2101,-436){\line( 1, 0){1050}}
}%
{\color[rgb]{0,0,0}\put(2101,-586){\line( 1, 0){1050}}
}%
{\color[rgb]{0,0,0}\put(-449,-811){\line( 1, 0){1050}}
}%
{\color[rgb]{0,0,0}\put(-449,-886){\line( 1, 0){1050}}
}%
{\color[rgb]{0,0,0}\put(-449,-961){\line( 1, 0){1050}}
}%
{\color[rgb]{0,0,0}\put(-449,-1111){\line( 1, 0){1050}}
}%
{\color[rgb]{0,0,0}\put(-449,-1036){\line( 1, 0){1050}}
}%
{\color[rgb]{0,0,0}\put(-449,-736){\line( 1, 0){1050}}
}%
{\color[rgb]{0,0,0}\put(601,-1261){\framebox(600,1200){$M$}}
}%
{\color[rgb]{0,0,0}\put(1501,-1261){\framebox(600,1200){$U$}}
}%
{\color[rgb]{0,0,0}\put(2401,-1186){\framebox(450,450){$N_\mathcal{B}$}}
}%
{\color[rgb]{0,0,0}\put(1201,-811){\line( 1, 0){300}}
}%
{\color[rgb]{0,0,0}\put(1201,-886){\line( 1, 0){300}}
}%
{\color[rgb]{0,0,0}\put(1201,-961){\line( 1, 0){300}}
}%
{\color[rgb]{0,0,0}\put(1201,-1111){\line( 1, 0){300}}
}%
{\color[rgb]{0,0,0}\put(1201,-1036){\line( 1, 0){300}}
}%
{\color[rgb]{0,0,0}\put(1201,-211){\line( 1, 0){300}}
}%
{\color[rgb]{0,0,0}\put(1201,-286){\line( 1, 0){300}}
}%
{\color[rgb]{0,0,0}\put(1201,-361){\line( 1, 0){300}}
}%
{\color[rgb]{0,0,0}\put(1201,-511){\line( 1, 0){300}}
}%
{\color[rgb]{0,0,0}\put(1201,-436){\line( 1, 0){300}}
}%
{\color[rgb]{0,0,0}\put(301,-211){\line( 1, 0){300}}
}%
{\color[rgb]{0,0,0}\put(301,-286){\line( 1, 0){300}}
}%
{\color[rgb]{0,0,0}\put(301,-361){\line( 1, 0){300}}
}%
{\color[rgb]{0,0,0}\put(301,-511){\line( 1, 0){300}}
}%
{\color[rgb]{0,0,0}\put(301,-436){\line( 1, 0){300}}
}%
{\color[rgb]{0,0,0}\put(1201,-736){\line( 1, 0){300}}
}%
{\color[rgb]{0,0,0}\put(-149,-586){\framebox(450,450){$D_\mathcal{A}$}}
}%
{\color[rgb]{0,0,0}\put(-449,-211){\line( 1, 0){300}}
}%
{\color[rgb]{0,0,0}\put(-449,-286){\line( 1, 0){300}}
}%
{\color[rgb]{0,0,0}\put(-449,-361){\line( 1, 0){300}}
}%
{\color[rgb]{0,0,0}\put(-449,-511){\line( 1, 0){300}}
}%
{\color[rgb]{0,0,0}\put(-449,-436){\line( 1, 0){300}}
}%
\put(-749,-436){\makebox(0,0)[lb]{\smash{{\SetFigFont{12}{14.4}{\rmdefault}{\mddefault}{\updefault}{\color[rgb]{0,0,0}$\mathcal{A}$}%
}}}}
\put(-749,-1036){\makebox(0,0)[lb]{\smash{{\SetFigFont{12}{14.4}{\rmdefault}{\mddefault}{\updefault}{\color[rgb]{0,0,0}$\mathcal{H}$}%
}}}}
\put(3301,-1036){\makebox(0,0)[lb]{\smash{{\SetFigFont{12}{14.4}{\rmdefault}{\mddefault}{\updefault}{\color[rgb]{0,0,0}$\mathcal{B}$}%
}}}}
{\color[rgb]{0,0,0}\put(2101,-811){\line( 1, 0){300}}
}%
\end{picture}%
    \caption[The mixed-unitary circuit that simulates the original
    circuit]{The constructed mixed-unitary circuit $C$ that simulates
      the given circuit $Q$, with input and output Hilbert spaces
      marked.  The circuit $U$ is the unitary from a the
      implementation of $Q$ in Stinespring form.  The circuits
      $D_{\mathcal{A}}$, $M$, and $N_{\mathcal{B}}$ are as described
      in the text.}
  \label{mu-fig-circuit-complete}
\end{figure}

Given these circuit implementations of the three channels
$D_{\mathcal{A}}, N_{\mathcal{B}}, M$, the mixed-unitary circuit $C$
that approximates a given circuit $Q$ is constructed in exactly the
same was as in Equation~\eqref{mu-eqn-construction}.  More concretely,
let $Q$ be a circuit implementing the operation
\[ Q(\rho) = \ptr{B} U (\ket 0 \bra 0 \tprod \rho) U^*, \]
where the ancillary qubits are in the space $\mathcal{A}$.  The
circuit $C$ that approximates it is then given by
\begin{equation}\label{mu-eqn-circuit-construction}
  C(\rho) =   
  N_\mathcal{B} \left( 
    U \left[ 
      (M \circ D_\mathcal{A})(\rho) 
    \right] U^* 
  \right).
\end{equation}
This construction of the circuit $C$ is shown in Figure~\ref{mu-fig-circuit-complete}.
Notice that $C$ is constructed to be a mixed-unitary circuit, as it
the composition of smaller mixed-unitary circuits.  Since the operations
$D_{\mathcal{A}}$ and $M$ do not affect inputs of the form $\ket 0
\bra 0 \tprod \rho$ in the space $S_0$, the proof of Proposition~\ref{mu-prop-simulation}
holds also for the circuit case, so that we have
\begin{equation}\label{mu-eqn-circuit_simulation}
  C(\ket 0 \bra 0 \tprod \sigma) = Q(\sigma) \tprod \nidentity{B}.
\end{equation}
Combining this with equation~\eqref{mu-eqn-Dcircuit} and the fact that applying
$D_\mathcal{A}$ twice has no further effect, the output of
$C$ on an arbitrary input state $\rho$ is of the form
\begin{equation}\label{mu-eqn-circuit_output}
  C(\rho)
  = \sum_{i=0}^{\dm{A} - 1} p_i C(\ket i \bra i \tprod \rho_i)
  = p_0 Q(\rho_0) \tprod \nidentity{B} +\sum_{i=1}^{\dm A - 1}
        p_i C(\ket i \bra i \tprod \rho_i).
\end{equation}
In the remainder of the chapter it
is shown that this construction does not significantly alter the
distinguishability properties of quantum circuits.

As a first step towards this, it is shown that the above circuit
construction correctly implements the channel $M$ described by Equation~\ref{mu-eqn-Mcircuit}.  Much of the proof
of this lemma is similar to the proof of
Lemma~\ref{mu-lem-ancilla-mixing}, but the operation $M$ considered in
this section is slightly different and the proof must be extended to
the case where there is an additional reference system.

This system, given by the space $\mathcal{F}$, is needed in the case of
distinguishability, as a party attempting to distinguish two channels
is permitted to use a portion of a larger entangled state as input to
the channels.  This is modelled by the use of the diamond norm in the
definition of the computational problem $\prob{QCD}$, from
Section~\ref{dist-scn-problem}, the hardness of which will be extended
to the mixed-unitary case.

\begin{lemma}\label{mu-lem-ancilla-mixing-aux}
  On input states of the form $\ket k \bra k \tprod \rho \in
  \density{A \tprod H \tprod F}$ for $\ket k \bra k \in \density{A}$
  with $0 < k \leq 2^m-1$, the output of $C$ satisfies
  \[ 
     \tnorm{(C \tprod \identity{F})(\ket k \bra k \tprod \rho) - 
       \nidentity{A \tprod H}
         \tprod \ptr{H} \rho}
     \leq \frac{1}{2^{m-1}},\]
     where $m$ is the number of ancillary qubits used by the circuit $Q$.

  \begin{proof}
    On input of the form $\ket k \bra k \tprod \rho$ the decoherence
    operations that are applied to the qubits in $\mathcal{A}$ can be
    ignored, as they have no effect on qubits in a state of the
    computational basis.  As $k \neq 0$ at least one qubit is in the
    state $\ket 1$, and so the controlled-mixing operations in the
    implementation of the channel $M$ will have
    an effect.  Let the first nonzero qubit among the qubits of
    $\mathcal{A}$ be the $j$th one.
    The first controlled $N$ operation with nonzero control qubit
    that effects the $j$th qubit will be at the $j$th stage of the
    mixing process, where the $j$th qubit is the control qubit.  As
    this qubit is not modified before this stage (as any previous
    qubits are in the state $\ket 0$ by choice of $j$), the first
    $m-1$ gates
    in the $j$th stage will mix the remaining qubits, so that the
    state after these gates is, using Equation~\eqref{mu-eqn-traceout},
    \[ \ket 1 \bra 1 \tprod \nidentity{A'} \tprod \nidentity{H} \tprod
    \ptr{H} \rho, \]
    where for notational
    convenience the $j$th qubit has been written first, and
    $\mathcal{A'}$ is the space of all but the $j$th qubit of
    $\mathcal{A}$.
    The remainder of the $j$th stage of the mixing process consists of
    $m-1$ controlled $N$ gates with the $j$th qubit as the
    target, each controlled by one of the $m-1$ qubits in
    $\mathcal{A'}$.  Considering the state $\identity{A'}/2^{m-1}$ on $\mathcal{A'}$ in
    the computational basis, the only term for which qubit $j$ is not
    mixed by these operations is the all zero term.  With this
    observation, the state after the $j$th stage is
    \begin{align*}
      \frac{1}{2^{m-1}} & \left[ 
        \ket{1}\bra{1} \tprod (\ket 0 \bra 0)^{\tprod m-1}
        + \frac{\ket 0 \bra 0 + \ket 1 \bra 1}{2} \tprod
        (\identity{A'} -  (\ket 0 \bra 0)^{\tprod m-1})
        \right] 
        \tprod \nidentity{H}
        \tprod \ptr{H} \rho \nonumber \\
      &= \frac{\identity{A} 
           + \ket{1}\bra{1} \tprod (\ket 0 \bra 0)^{\tprod m-1}
           - (\ket 0 \bra 0)^{\tprod m}}{2^m}
         \tprod \nidentity{H}
         \tprod \ptr{H} \rho.
    \end{align*}
    This proves that the circuit implementing the channel $M$ does so
    correctly, as this quantity is exactly
    the state given in Equation~\eqref{mu-eqn-circuit_output}
    with the addition of $\ptr{H} \rho$ in the reference system.

    As in the proof of Lemma~\ref{mu-lem-ancilla-mixing}, let this state
    be $\sigma$.  Computing the distance from this state to the
    desired one, we have
    \begin{equation*}
      \tnorm{ \sigma - \nidentity{A} \tprod \nidentity{H}
         \tprod \ptr{H} \rho}
      = \frac{1}{2^m} 
         \tnorm{\ket{1}\bra{1} \tprod (\ket 0 \bra 0)^{\tprod m-1} 
           - (\ket 0 \bra 0)^{\tprod m}}
      = \frac{1}{2^{m-1}}.
    \end{equation*}
    Finally, by noting that the remainder of the circuit $C$ is
    mixed-unitary, Lemma~\ref{mu-lem-ru-dist-noise} implies that the
    rest of the circuit cannot increase the norm.
  \end{proof}
\end{lemma}

In the next section this lemma is used to show that the hardness of
the computational problem of distinguishing mixed-state circuits does
not change when restricted to the mixed-unitary circuits.

\section{\class{QIP}-completeness of distinguishing mixed-unitary
  circuits}\label{mu-scn-distinguish}

The construction outlined in the previous section can be used to 
find mixed-unitary approximations to general quantum circuits, with
the property that the diamond norm of the difference of two such
circuits is approximately preserved.  This property leads immediately
to a proof that
the problem of distinguishing mixed-unitary quantum circuits is
\class{QIP}-complete, which is exactly as hard as the problem of
distinguishing general quantum circuits.
This will be done by taking the instance $(Q_1, Q_2)$ of the general quantum
circuit distinguishability problem (Problem~\ref{dist-prob-qcd}),
and constructing the instance $(C_1, C_2)$ with $C_1$ and
$C_2$ mixed-unitary, by applying the construction of
Section~\ref{mu-scn-circuits} to each of these circuits.

This technique produces an instance of the mixed-unitary quantum
circuit distinguishability problem, which is hereafter referred to as
\prob{Mixed-Unitary QCD}.  This problem is identical to \prob{QCD} 
with the exception that the input circuits are required to be
mixed-unitary circuits, in the model defined in Section~\ref{mu-scn-circuits}.
This problem is more formally defined as
\begin{problem}[Mixed-unitary Quantum Circuit Distinguishability]
  \label{mu-prob-muqcd}
  For constants $0 \leq b < a \leq 2$, the input consists of
  mixed-unitary quantum
  circuits $C_1$ and $C_2$ that implement transformations in
  $\transform{H,K}$.
  The promise problem is to distinguish the two cases:
  \begin{description}
    \item[Yes:] $\dnorm{C_1 - C_2} \geq a$,
    \item[No:] $\dnorm{C_1 - C_2} \leq b$.
  \end{description}
\end{problem}
\index{Quantum Circuit Distinguishability!mixed-unitary}%
\index{mixed-unitary!distinguishability}%
As this problem is a restriction of the more general circuit
distinguishability problem, the protocol of
Section~\ref{dist-scn-protocol} shows that it too is \class{QIP}.  The
remainder of this section is devoted to showing that
\prob{Mixed-Unitary QCD} is \class{QIP}-complete for all $0 < b < a
\leq 2$

The most important step in the proof that this restricted
distinguishability problem is \class{QIP}-complete is to show that the
construction in Section~\ref{mu-scn-circuits} does not significantly
alter the diamond norm of the difference of the two circuits.  This is
the content of the following theorem.

\begin{theorem}\label{mu-thm-reduction}
  Let $Q_1$ and $Q_2$ be arbitrary circuits implementing
  transformations in $\transform{H,K}$, and let $C_i$ be the
  mixed-unitary circuit constructed from $Q_i$ as in
  Equation~\eqref{mu-eqn-circuit-construction}.
  For any $\epsilon > 0$,
  \[ \dnorm{Q_1 - Q_2}
     \leq \dnorm{C_1 - C_2} 
     \leq \dnorm{Q_1 - Q_2} + \epsilon, \]
  where the circuits $C_1$ and $C_2$ use $O(\log 1/\epsilon)$ extra
  qubits in the space $\mathcal{A}$.

  \begin{proof}
    The first inequality is not hard to show.  Once again, the idea is
    sending the input state $(\ket 0 \bra 0)^{\tprod m} \tprod \rho$
    to the circuit $C_i$ results in a simulation of $Q_i$, by
    Equation~\ref{mu-eqn-circuit_simulation}.  This will imply that
    the distinguishability of $Q_1 $ and $Q_2$ cannot be greater than
    the distinguishability of $C_1$ and $C_2$.
    To formalize this argument, note that by the definition of the
    diamond norm
   \begin{align*}
      \dnorm{Q_1 - Q_2}
      = \sup_{\rho \in \density{H \tprod F}}
          \tnorm{(Q_1 \tprod \identity{F})(\rho) 
                -(Q_2 \tprod \identity{F})(\rho)},
    \end{align*}
    and fix $\delta>0$ and $\rho$ as a state achieving a value within
    $\delta$ of this supremum.  By Equation~\ref{mu-eqn-circuit_simulation},
    if the state
    $(\ket 0 \bra 0)^{\tprod m} \tprod \rho$ is given as input to
    the circuit $C_i$, then the output is
    $(Q_i \tprod \identity{F}) (\rho)$.
    Using this property we have
    \begin{align*} 
      \dnorm{C_1 - C_2}
      & \geq \tnorm{ 
        (C_1 \tprod \identity{F})
          ((\ket 0 \bra 0)^{\tprod m} \tprod\rho)
        - (C_2 \tprod \identity{F})
          ((\ket 0 \bra 0)^{\tprod m} \tprod\rho)} \\
      &= \tnorm{ (Q_1 \tprod \identity{F}) (\rho)
        - (Q_2 \tprod \identity{F}) (\rho)} \\
      &\geq \dnorm{Q_1 - Q_2} - \delta.
    \end{align*}
    Since this is true for any $\delta > 0$, it must be the case that
    $\dnorm{Q_1 - Q_2} \leq \dnorm{C_1 - C_2}$.

    The second inequality requires somewhat more work.  The idea is to
    once again break the input space into two subspaces: the one on which
    the circuits $C_i$ simulate the circuits $Q_i$, and the orthogonal
    subspace.  We will then use Lemma~\ref{mu-lem-ancilla-mixing-aux}
    to show that on this orthogonal subspace the output states of the
    circuits $C_i$ are almost completely mixed.  This will in turn imply that
    the diamond norm on this input subspace is exponentially small, in
    the number of ancillary qubits added in the construction of the
    circuits $C_i$.  An appeal to the decoherence operation applied as
    part of the circuit construction will validate the approach of
    treating the input state as a mixture of states from these 
    two orthogonal subspaces.

    More formally, let $m$ be the
    number of ancillary qubits (the space $\mathcal{A}$) and let $n$
    be the number of input qubits (the space $\mathcal{H}$) used by 
    the circuits $Q_i$.  It can be assumed that the circuits $Q_1$ and
    $Q_2$ both use the same number of ancillary qubits by padding one
    of the circuits with unused qubits that are later traced out.
    The values of $n$ and $m$ may also be expressed as
    $m = \ceil{\log \dm{A}}$ and $n = \ceil{\log \dm{H}}$.  
    By adding at most $3 + \log (1/\epsilon)$ extra ancillary qubits
    to the space $\mathcal{A}$, we may assume that
    \begin{equation}\label{mu-eqn-m-bound-eps}
      2^{-(m-3)} < \epsilon
    \end{equation}
    
    Let $\rho \in \density{A \tprod H \tprod F}$ be a state such that
    \begin{equation}\label{mu-eqn-input-state}
      \dnorm{C_1 - C_2} - \epsilon/2
      \leq \tnorm{(C_1 \tprod \identity{F})(\rho)
        -(C_2 \tprod \identity{F})(\rho)},
    \end{equation}
    and note that the reference system $\mathcal{F}$ need not have the same
    dimension as the space of the same name considered in the proof of
    the previous inequality.  The first gates applied in the circuit
    $C_i$ are the decoherence gates applied to $\mathcal{A}$.
    These gates produce a state of the form
    $ \sum_{i=0}^{2^m - 1} p_i \ket i \bra i \tprod
       \sigma_i, $
    for $\{p_i\}$ a probability distribution.  Since applying these
    decoherence operations twice has no further effect, the output
    of the circuits $C_1$ and $C_2$ is the same on $\rho$ as it is
    on this state.  Applying the this property and triangle
    inequality to Equation~\eqref{mu-eqn-input-state}, the quantity
    of interest becomes
    \begin{equation}\label{mu-eqn-soundness1}
      \dnorm{C_1 - C_2} - \epsilon/2
      \leq \sum_{i=0}^{2^m - 1} p_i
        \tnorm{ (C_1 \tprod \identity{F})(\ket i \bra i \tprod\sigma_i)
              - (C_2 \tprod \identity{F})(\ket i \bra i
      \tprod\sigma_i)}
    \end{equation}
    Then, by applying Lemma~\ref{mu-lem-ancilla-mixing-aux} to each term
    with $i \neq 0$, the states in the norm can be replaced with completely
    mixed states on $\mathcal{A \tprod H}$ plus a small correction
    factor.  Doing this for each of these terms we have
    \begin{multline*}
      p_i \tnorm{ 
                (C_1 \tprod \identity{F})(\ket i \bra i \tprod\sigma_i)
              - (C_2 \tprod \identity{F})(\ket i \bra i \tprod\sigma_i)}\\
      \leq p_i \left[ \frac{2}{2^{m - 1}} +
          \tnorm{
              \nidentity{A \tprod H} \tprod \ptr{H} \sigma_i
            - \nidentity{A \tprod H} \tprod \ptr{H} \sigma_i}
          \right]
      = p_i / 2^{m-2} < p_i \epsilon/2.
    \end{multline*}
    Applying this to Equation~\eqref{mu-eqn-soundness1} results in
    \begin{equation*}
      \dnorm{C_1 - C_2} - \epsilon/2
      \leq 
      p_0 \tnorm{ (C_1 \tprod \identity{F})(\ket 0 \bra 0 \tprod\sigma_0)
        - (C_2 \tprod \identity{F})(\ket 0 \bra 0 \tprod\sigma_0)}
      + \sum_{i=1}^{2^m - 1} p_i \epsilon/2.
    \end{equation*}
    By Equation~\ref{mu-eqn-circuit_simulation} the output of the circuit
    $C_i$ on this input can be replaced the output of the circuit
    $Q_i$ and a maximally mixed state.
    When this is done to the previous equation,
    the desired bound is given by
    \begin{align*}
       \dnorm{C_1 - C_2}
       & \leq p_0 \tnorm{ (Q_1 \tprod \identity{F})(\sigma_0) \tprod \nidentity{B}
              - (Q_2 \tprod \identity{F})(\sigma_0) \tprod \nidentity{B}}
       + (1 - p_0) \epsilon/2  + \epsilon/2 \\
       & \leq p_0 \tnorm{ (Q_1 \tprod \identity{F})(\sigma_0)
              - (Q_2 \tprod \identity{F})(\sigma_0)}
       + \epsilon \\
       & \leq \dnorm{Q_1 - Q_2} + \epsilon.
    \end{align*}
    This completes the proof of the theorem, since $0 \leq p_0 \leq 1$.    
  \end{proof}
\end{theorem}

The quantum circuit distinguishability problem is defined in terms of
the diamond norm of the difference of two circuits.  The bounds on
this quantity provided by the previous theorem immediately imply the
following corollary.

\begin{corollary}
  For any $0 < b < a \leq 2$ the problem $\prob{Mixed-Unitary
    QCD}_{a,b}$ is \class{QIP}-complete.

  \begin{proof}
    Starting with an instance $(Q_1, Q_2)$ of $\prob{QCD}_{a,b/2}$ and
    applying the construction of Section~\ref{mu-scn-circuits} to each
    circuit results in a pair $(C_1, C_2)$.  By
    Theorem~\ref{mu-thm-reduction} on positive instances of \prob{QCD}
    we have
    \[ \dnorm{C_1 - C_2} \geq \dnorm{Q_1 - Q_2} \geq a, \]
    and on negative instances we have
    \[ \dnorm{C_1 - C_2}  
        \leq \dnorm{Q_1 - Q_2} + \epsilon 
        \leq \frac{b}{2} + \epsilon. \]
    By adding $O( \log 1 / b)$ extra qubits in the construction of
    each of the circuits $C_i$, we can make $\epsilon < b/2$, so that
    the resulting pair $(C_1, C_2)$ is an instance of the problem
    $\prob{Mixed-Unitary QCD}_{a,b}$, provided that $b > 0$.

    To see that the reduction can be implemented efficiently, notice
    that the circuit construction can be done in polynomial time,
    since we have only added $O( \log 1/b)$ qubits, as well as a few
    operations that do not depend on the actual input circuits, just
    the number of qubits they act on.
  \end{proof}
\end{corollary}

A natural question to ask is whether this hardness result can be
extended to the case of log-depth mixed-unitary circuits.  With the
construction of Section~\ref{mu-scn-circuits} this does not appear to
be possible: the controlled-mixing operation in the procedure for
mixing the input qubits if any ancillary qubits are not in the $\ket
0$ state requires mixing operations to be applied to all other qubits,
controlled by each of the qubits in the space $\mathcal{A}$.  This
results in a linear depth circuit, since each input qubit is the
target of at least $\log \dm{A}$ mixing operations.  Another approach
might be to apply a construction similar to that used in
Chapter~\ref{chap-close-images} to a mixed-unitary \prob{Close Images}
problem.  This approach runs into an immediate problem: all
mixed-unitary circuits are unital, and so they must all output the
completely mixed state when given one as input, i.e.\ the images of any
two mixed-unitary transformations intersect at the completely mixed
state, trivializing the close images problem on mixed-unitaries.

\section{Conclusion}

In this chapter a few problems are shown to be no easier when
restricted to the class of mixed-unitary channels.  This is done using
a technique by which a channel is approximated using a mixed-unitary
channel.  While this approximation does not, in general, look very
much like the original channel, for measures based on the behaviour of
the channel on low-entropy outputs this approximation can be made
arbitrarily good by padding the input channel with unused ancillary
space.  The approximation technique overcomes two main hurdles: viewed
as a circuit, the original channel may trace out qubits and it may
introduce fresh ancillary qubits in some pure state.  The partial
trace can be easily simulated by the mixed-unitary channel that maps
any state to the completely mixed state, but a more complicated
construction is required to deal with the ancillary space.

This construction is applied to the maximum output $p$-norm and the
minimum output entropy, where it is used to show that the
multiplicativity or additivity of a channel is implied by the
additivity of multiplicativity of a related set of mixed-unitary
channels.  This can be used to show that the general multiplicativity
and additivity problems are equivalent when restricted to the
mixed-unitary channels, but this is no longer an interesting result,
since mixed-unitary channels that are not additive~\cite{Hastings09}
and not multiplicative~\cite{HaydenW08} have recently been discovered.

When applied to the computational problem of distinguishing two
transformations, this approximation scheme proves that this problem
remains \class{QIP}-hard when restricted to mixed-unitary inputs.
This is perhaps surprising: mixed-unitary channels have several nice
properties~\cite{GregorattiW03}, but computational distinguishability
does not appear to be one of them.  This can be seen as evidence that,
despite the nice properties enjoyed by these channels, they may be
sufficiently general to be a useful model of noise in quantum systems.

%%% Local Variables: 
%%% mode: latex
%%% TeX-master: "thesis"
%%% End: 

\chapter{Conclusion}

This thesis has introduced the problem \prob{Quantum Circuit
  Distinguishability}, which is a computational version of channel
distinguishability.  This problem has been shown to be hard for the
class \class{QIP} of problems that have quantum interactive proof
systems, which is equal to the more familiar class
\class{PSPACE}~\cite{JainJ+09}.
This problem gives a new quantum characterization of this class
that is not closely tied to quantum interactive proof systems.

The hardness of the distinguishability problem leads naturally to a
study of the problem on restricted classes of channels.  This can be
seen as an attempt to isolate those instances of the problem that are
hard, so that more is known about those instances on which the problem
is tractable.  This thesis has presented four important classes of
channels for which the problem remains hard: the channels implemented
by log-depth circuits, the degradable channels, the antidegradable
channels, and the mixed-unitary channels.  These special cases
demonstrate that the \prob{QCD} problem is hard on a wide array of
classes of channels, i.e.\ that the hardness of the problem is very
likely not tied to just a few hard instances.

These hardness results are shown by reducing the general problem to a
version of the problem restricted to a special class of channels.
These reductions can have applications outside of complexity theory,
as they are essentially methods for the simulation of general channels
by channels in a restricted class.  These simulation techniques can
have powerful implications throughout quantum information.  One
example of such an application is the result that the additivity of
the Holevo capacity or the multiplicativity of the maximum output
$p$-norm can be (approximately) restricted to a mixed-unitary channel.
It is hoped that the other reductions presented in the thesis will
find similar applications.

Several natural questions are left open by this thesis.  Several of
the more interesting of these questions are summarized below.

\begin{itemize}
\item There are other models of computation that involve unitary
  computations applied to mixed initial states
  (see~\cite{AmbainisS+06, KnillL98}).  How hard is the
  distinguishability problem for computations in these models?
  
\item What is the complexity of distinguishing constant depth
  quantum circuits that do not use the unbounded fan-out gate?   This
  problem is hard on constant depth circuits that have access to this
  gate, but the reduction used in the thesis does not produce
  constant-depth circuits without access to this gate.
  
\item How hard is \prob{QCD} on the entanglement-breaking channels?
  It was argued in Section~\ref{intro-scn-classes} that this problem
  is hard for channels that are exponentially close together, but this
  problem is not very interesting with such a weak promise.  It is not
  known how hard the distinguishability problem is on this subset of
  the antidegradable channels with a stronger promise.

\item The random Pauli channels, also known as the Pauli diagonal
  channels, can be expressed as convex combinations of the channels
  that apply the discrete Weyl (or generalized Pauli) operators.
  These channels are an important subclass of the mixed-unitary
  channels.  Many of the pieces used in the reduction to mixed-unitary
  channels can be expressed as channels of this form, but there is one
  major problem: the unitary $U$ from a Stinespring representation of
  a general channel needs to be converted to such a channel.  Such a
  simulation result would imply that the \prob{QCD} problem is also
  hard on this class.
\end{itemize}

%%% Local Variables: 
%%% mode: latex
%%% TeX-master: "thesis"
%%% End: 

\ifthenelse{\boolean{ElectronicVersion}}{\phantomsection}{}
\addcontentsline{toc}{chapter}{Bibliography}
\bibliographystyle{halphads}
\bibliography{thesisbib}
\cleardoublepage

\ifthenelse{\boolean{ElectronicVersion}}{\phantomsection}{}
\addcontentsline{toc}{chapter}{Index}
\printindex

\end{document}